\documentclass[12pt]{article}
\usepackage{amsmath}
\usepackage{bm}
\usepackage[authoryear]{natbib}
\usepackage{lscape}

\usepackage[resetlabels]{multibib}
\newcites{supp}{Supplementary References}

\usepackage[colorlinks,citecolor=blue,linkcolor=blue,urlcolor=blue]{hyperref}
\usepackage{latexsym, epsfig, amssymb, amsmath, amsthm, graphicx, mathrsfs, booktabs}
\usepackage{rotating, tabularx, setspace,paralist}
\usepackage{multirow,multicol}
\usepackage{anysize}

\usepackage{accents}
\usepackage{bbm} 

\usepackage[linesnumbered,ruled,vlined]{algorithm2e} 
\usepackage{algorithmic}

\usepackage{subfig}

\usepackage{mathtools}

 \usepackage[a4paper,margin=0.8in]{geometry}
\usepackage[utf8]{inputenc}
\usepackage[english]{babel}

\usepackage{authblk}

\usepackage{tikz}
\newtheorem{defn}{Definition} 
\newtheorem*{defn*}{Definition} 
\newtheorem{exmp}{Example} 

\newtheorem{condi}{Condition} 
\newtheorem*{condi*}{Condition} 

\newtheorem{theorem}{Theorem}

\newtheorem{lemma}{Lemma}
\newtheorem*{lemma*}{Lemma}

\theoremstyle{plain}
\newtheorem{remark}{Remark}

\newcommand{\overbar}[1]{\mkern 1.5mu\overline{\mkern-1.5mu#1\mkern-1.5mu}\mkern 1.5mu}

\newtheorem*{claim*}{Claim}
\newtheorem*{rwr*}{Removed When Ready}

\usepackage{enumitem}
\newlist{steps}{enumerate}{1}
\setlist[steps, 1]{label = Step \arabic*:}

\SetKwInput{KwInput}{Input}                
\SetKwInput{KwOutput}{Output}              

\DeclarePairedDelimiter{\nint}\lfloor\rceil

\marginsize{0.7in}{0.7in}{0.35in}{0.35in} %

\newcommand{\blind}{1}

\begin{document}

	\def\spacingset#1{\renewcommand{\baselinestretch}%
		{#1}\small\normalsize} \spacingset{1}

	
	\if1\blind
	{
		\title{\bf Testing for Regression Heteroskedasticity with High-Dimensional Random Forests}
		\author{Chien-Ming Chi\thanks{
							Chien-Ming Chi is Assistant Research Fellow, Institute of Statistical Science, Academia Sinica, Taipei 11529, Taiwan (E-mail: \textit{xbbchi@stat.sinica.edu.tw}). %
				This work was supported by grant 111-2118-M-001-012-MY2 from the National Science and Technology Council, Taiwan.	}
			\hspace{.2cm}\\
			Academia Sinica}
		\maketitle
	} \fi
	
	\if0\blind
	{
		\bigskip
		\bigskip
		\bigskip
		\begin{center}
			{\LARGE\bf Testing for Regression Heteroskedasticity with High-Dimensional Random Forests}
		\end{center}
		\medskip
	} \fi
	
	\bigskip

	\begin{abstract}
		Statistical inference for high-dimensional regression heteroskedasticity is an important but under-explored problem. The current paper aims at filling this gap by proposing two tests, namely the variance difference test and the variance difference Breusch-Pagan test, for assessing high-dimensional regression heteroskedasticity. The former tests whether an explanatory feature of interest is associated with the conditional variance of a response variable, while the latter tests heteroskedasticity in the regression, which is known to be the Breusch-Pagan test problem.
		To formally establish the tests, we have derived rigorous P-values and test sizes, and analyzed the test power under a nonparametric heteroskedastic data generating model with high-dimensional input features. Such a model setting takes into account   high-dimensional applications with flexible structures of heteroskedasticity and features having interaction effects on the mean of the response; these are common applications in many fields such as biology. Our methods leverage machine learning mean prediction methods such as random forests and use  knockoff variables as negative controls. Particularly,  the definition of knockoffs for our test statistics is more flexible than the original definition of knockoffs, and we give a detailed comparison of these two definitions and discuss the advantages of our knockoffs. The satisfactory empirical performance of the proposed tests is illustrated with simulation results and an HIV (Human Immunodeficiency Virus) case study. 
	\end{abstract}
	
	\noindent%

	\spacingset{1.9} 
	\section{Introduction}

	Statistical inference for heteroskedasticity has been widely applied to  estimation efficiency improvement~\citep{white1980heteroskedasticity}, financial asset allocation~\citep{mcneil2015quantitative}, genetic analysis~\citep{yang2012fto, hill2010genetic}, etc. The importance of heteroskedasticity has encouraged the development of various modeling methods~\citep{smyth1989generalized, smyth1999adjusted, su2013nonparametric, lee2006double} addressing heteroskedastic disturbance. The prediction accuracy of heteroskedasticity modeling methods has continued to be improved thanks to the emerging machine learning ideas such as ensemble trees~\citep{pratola2020heteroscedastic, hahn2020bayesian}, but rigorous inferences for regression heteroskedasticity seem to be under-explored~\citep{cleasby2011neglected, ayroles2015behavioral}. The goal of the current paper is to fill this gap by formally establishing tests for high-dimensional regression heteroskedasticity.

	Let us further clarify our motivation with an example. Suppose we want to improve the Human Immunodeficiency Virus Type 1 (HIV-1) treatment by studying the distributional relations between a specific set of genetic mutations and the HIV-1 drug resistance level for certain HIV-1 drugs~\citep{rhee2006genotypic}.
	Particularly, if we want to apply the Breusch-Pagan test~\citep{breusch1979simple} to draw the inference about whether the set of genetic mutations are associated with the conditional variance of the HIV-1 drug resistance level, we are facing at least two challenges. First, it is likely that mutations have non-linear mean effects on the drug resistance level~\citep{yu2014prediction}. Second, the number of mutations can be of the same size as or larger than the sample size. The Breusch-Pagan test may not perform well or may not be applicable in this application since it considers linear regression models with heteroskedastic disturbances and fixed feature dimensionality. Despite recent advances in making inferences for heteroskedasticity, some major limitations remain. For example, the related work reviewed below does not consider high-dimensional features with non-linear mean effects on the response variable. The proposed tests in this paper are meant to improve upon these limitations, and provide more detailed inferences about heteroskedasticity based on each feature. To appreciate our contributions, we now formally introduce the test problems as follows; the details of our HIV-1 case study are deferred to Section~\ref{Sec6}.

	Let a random variable $Y$ and a $p$-dimensional random vector $\boldsymbol{X}$ denote the response variable of interest and the explanatory feature vector, respectively. To understand how these explanatory features are associated with regression heteroskedasticity, we assume the statistical heteroskedastic model
	\begin{equation}
		\label{model.1}
		Y =m(\boldsymbol{X}) + \varepsilon\times \zeta(\boldsymbol{X}_{S^*}) 
	\end{equation}
	for some $S^*\subset\{1, \dots, p\}$,
	where  $\varepsilon$ is a model error independent of all else variables such that $\mathbb{E} (\varepsilon)= 0$ and Var$(\varepsilon) = 1$,  $m:\mathbb{R}^p\longmapsto\mathbb{R}$ is the mean function, and that $\boldsymbol{X}_{S^*} = (X_{j}, j \in S^*)^{\top}$ and  $\zeta:\mathbb{R}^{|S^*|}\longmapsto(0, \infty)$  are respectively  a small subset of relevant features and the standard deviation function such that $\textnormal{Var}\big[\zeta(\boldsymbol{X}_{S^*})\big] >0$ if and only if $S^* \not= \emptyset$. In model \eqref{model.1}, the subset of active features of the mean function is not indicated since it is of minor interest in this paper. Examples of sparse non-linear $m(\cdot)$ and monotonic $\zeta(\cdot)$ are respectively in Section~\ref{Sec3.1b} and Section~\ref{Sec4.1}. With model \eqref{model.1}, we are interested in testing
	\begin{equation}\label{null.1}
		H_{0}: j \not \in S^*
	\end{equation}
	for the $j$th feature of interest. In this paper, $X_{j}$ is a null feature if and only if $j\not \in S^*$. With $\zeta(\boldsymbol{X}_{\emptyset}) \coloneqq\zeta_{0}$ for some $\zeta_{0}>0$, we are also interested in testing 
	\begin{equation}\label{BP.1}
		H_{0}: S^* = \emptyset, 
	\end{equation}%
	which is known to be the Breusch-Pagan test.
	
	These tests  are widely used in literature. For example, recent studies have found genetic effects on the variability of a fruit fly's behavior decision~\citep{ayroles2015behavioral} and a single-nucleotide polymorphism (SNP) that is associated with the variability of body mass index and height~\citep{yang2012fto}. Their works essentially test \eqref{null.1} for each feature in their applications in order to find the significant features. A challenge there is that the selected relevant features are not associated with the underlying response variable through the mean function, which means that the existing methods such as the Lasso~\citep{tibshirani1996regression} and feature importance MDI~\citep{Breiman2001}  cannot be directly used for making inferences about \eqref{null.1} or \eqref{BP.1}. Hence, the aforementioned papers and most existing literature for genetic analysis and biological works~\citep{yang2012fto, ayroles2015behavioral, brown2014genetic, hill2010genetic} have developed statistical inference procedures for their own applications. However, these inference procedures are more or less heuristic and have no theoretical foundations. To our knowledge, the test problems \eqref{null.1} and \eqref{BP.1} have not yet been formally considered for applications with high-dimensional input features that may have complicated effects such as interaction effects on the response variable. Hence, the proposed tests, 
	namely, the variance difference (VD) and variance difference Breusch-Pagan (VDBP) tests, contribute as the first rigorous tests for the respective null hypothesis test problems \eqref{null.1} and \eqref{BP.1} in high-dimensional applications with flexible mean and standard deviation functions in  model \eqref{model.1}.

	Three advantages of the proposed tests are introduced as follows. First, since we are interested in the inference for conditional variance in model \eqref{model.1}, the first step of the VD test is to center the observed response variables by subtracting the estimated conditional means from the responses. We employ random forests~\citep{Breiman2001} as the centering method for the VD test. As a result, the VD test allows non-linear effects of the explanatory features on the conditional mean of the  response. Particularly, our simulation experiments show that the VD test is applicable when  mean function  $m(\boldsymbol{X})$ in model \eqref{model.1} consists of interaction components, which are common model components in genetic analysis~\citep{phillips2008epistasis, brown2014genetic}. Second, the VD test considers model \eqref{model.1} with a wide class of monotonic standard deviation functions. Hence, our tests are robust in practice and can reduce the chance of false positive errors due to model misspecification. Third, our tests allow high-dimensional input features. These practical advantages are also enjoyed by the VDBP test for the Breusch-Pagan test problem \eqref{BP.1}.

	On the theory side, we establish the asymptotic null distributions of the VD and VDBP tests, which enables us to calculate their test sizes and the much needed P-values. We show that the consistency of mean estimation at the centering step is crucial in order to establish the asymptotic null distributions. In addition, assuming a general model setting satisfying \eqref{model.1}, we show that the consistency of the proposed tests (i.e., test has asymptotic power one) under their respective alternative hypotheses depends on the sample size, the input feature dimensionality, the accuracy of the conditional mean estimation at the centering step, and important model parameters such as the variation level of the standard deviation function. In light of the importance of high-dimensional consistency of random forests to our tests,  some results of random forests consistency are reviewed in Section~\ref{Sec3.1b}. In fact, existing theoretical results of high-dimensional consistency under model \eqref{model.1} are mostly for random forests, which is a major reason that we consider random forests instead of other machine learning prediction methods~\citep{chen2016xgboost, chollet2015keras, friedberg2020local} as the centering method in this paper.

	Related work aiming at making inferences for general models usually assumes fixed feature dimensionality.
	To relax the linear mean function assumption made by the Breusch-Pagan test, Su \& Ullah~\citep{su2013nonparametric} used the local polynomial regression as the centering method  for testing \eqref{BP.1}, while Dumitrascu et al.~\citep{dumitrascu2019statistical} proposed a Bayesian test for the test problem~\eqref{BP.1}. Meanwhile, DGLM~\citep{smyth1989generalized} and DHGLM~\citep{lee2006double} rely on the generalized linear models for outputing heuristic P-values of the estimated heteroskedasticity coefficients for testing \eqref{null.1}; their P-values lack theoretical foundations.  Despite a broad recognition of the importance of test problems \eqref{null.1}--\eqref{BP.1}, rigorous tests for applications with high-dimensional features  are quite under-explored; particularly, the aforementioned methods~\citep{su2013nonparametric, dumitrascu2019statistical, smyth1989generalized, lee2006double} as well as most references therein all assume fixed dimentionality. Li \& Yao \citep{li2019testing} derived a rigorous high-dimensional test for \eqref{BP.1} assuming a linear mean function with the ordinary least-square method as the centering method.  In addition, there are a few high-dimensional consistent model selection methods for selecting active features in the standard deviation function~\citep{daye2012high,  chiou2020variable, doss2022nonparametric}. However, they either assumed  specific structures of mean and standard deviation functions, which may restrict their applicability, or lack theoretical foundations. Moreoever, these methods cannot be applied to test problems \eqref{null.1}--\eqref{BP.1} directly.

	It is worth mentioning that the advantages of the proposed tests are partly attributed to the use of model-X knockoff (hereafter, knockoff for short) features~\citep{CandesFanJansonLv2018}  as negative controls in establishing the VD and VDBP tests. The knockoffs for our tests are defined in a coordinate-wise fashion in Section~\ref{Sec2.1}. For example, to calculate the VD test statistic for testing \eqref{null.1}, we need the knockoff feature of the $j$th explanatory feature. The population $j$th knockoff feature, denoted by $\widetilde{X}_{j}$, is such that $(\varepsilon\times\zeta(\boldsymbol{X}_{S^*}),  X_{j})$ and 	$(\varepsilon\times\zeta(\boldsymbol{X}_{S^*}), \widetilde{X}_{j})$ have the same distribution under  \eqref{null.1}. We then use the $j$th knockoff as the negative control for establishing our tests. The definition of coordinate-wise knockoffs is more flexible than the original one in \citep{CandesFanJansonLv2018}; a detailed comparison between these two definitions is in Section~\ref{Sec2.1}.

	The rest of the paper is organized as follows. The VD and VDBP tests are introduced in Section~\ref{Sec2}, with the analysis of their test sizes and  P-values in Section~\ref{Sec3.2b} and selection power in Section~\ref{Sec4.2b}. The simulation study and HIV-1 data study are in  Sections~\ref{Sec5b}--\ref{Sec6}, respectively. Technical proofs are all deferred to the Supplementary Material.

	\subsection{Notation} \label{Sec1.1}
	Let $(\Omega, \mathcal{F}, \mathbb{P})$ denote the probability space, and $\mathcal{R}$ be the Borel $\sigma$-algebra of $\mathbb{R}$. For any vector $\vec{x} = (x_{1}, \dots, x_{p})^{\top}\in\mathbb{R}^p$, let $\vec{x}_{S}$ denote the subvector with entries in the index subset $S$, and let $\vec{x}_{-j} = (x_{1}, \dots, x_{j-1}, x_{j+1}, \dots, x_{p})^{\top}$; such expressions are also used for random vectors. This paper uses the following notation. 1)
	$|S|$ denotes the number of elements in a set $S$. 2) $\nint{x}$ denotes the closest integer of $x\in\mathbb{R}$. 3) The indicator function is denoted by $\boldsymbol{1}_{\{\cdot\}}$. 4) The symbol $\stackrel{\textnormal{D}}{\longrightarrow}$ means convergence in distribution.  In addition, we define $\zeta(\boldsymbol{X}_{S^*}) = \zeta_{0}$ when $S^* = \emptyset$ for some $\zeta_{0}>0$, where $\zeta(\boldsymbol{X}_{S^*})$ is given in model \eqref{model.1}. Moreover, we define $\zeta(\vec{z}) = \zeta_{0}$ for $\vec{z}\in\mathbb{R}^{0}$ when $S^* = \emptyset$; particularly, we have $\inf_{\vec{z}\in\mathbb{R}^{0}}\zeta(\vec{z}) = \sup_{\vec{z}\in\mathbb{R}^{0}}\zeta(\vec{z} ) = \zeta_{0}$ when $S^* = \emptyset$.

	\section{Hypothesis testing procedures}\label{Sec2}
	
	The VD and VDBP tests are presented in this section, with coordinate-wise knockoff features, the key ingredients for establishing our tests, introduced in Section~\ref{Sec2.1}. The VD test statistic for testing \eqref{null.1} involves a break $a_{j}\in\mathbb{R}$ on the $j$th coordinate, while the VDBP test statistic for \eqref{BP.1} involves $p$ breaks $a_{1}, \dots, a_{p}$ on each coordinate. To see the intuition for the use of breaks and coordinate-wise knockoffs, consider an example at population level where feature vector $\boldsymbol{X} = (X_{1}, \dots, X_{p})^{\top}$ and its coordinate-wise knockoff vector $\widetilde{\boldsymbol{X}} = (\widetilde{X}_{1}, \dots, \widetilde{X}_{p})^{\top}$ (formally defined in Section~\ref{Sec2.1}) are uniformly distributed on $[0, 1]^{2p}$, $Y= \varepsilon \times \sqrt{\exp{(X_{1})}}$ with an independent standard Gaussian model error $\varepsilon$, and constant break $a_{j}=\frac{1}{2}$ for each $j$. Then, $X_{j}$ is relevant if and only if $\big| \textnormal{Var}(Y|X_{j} >a_{j}) - \textnormal{Var}(Y|\widetilde{X}_{j} > a_{j}) \big| >0$ for each $j\in \{1, \dots, p\}$. This idea for identifying relevant features at population level is applicable to cases where $\boldsymbol{X}$ may have a general distribution, and is our basic idea for establishing the VD and VDBP tests when samples are involved. 
	
	Now, the VD and VDBP tests with given breaks are introduced in Sections~\ref{Sec2.2}--\ref{Sec2.3}, respectively. These tests are suitable for our real data study in Section~\ref{Sec6}, where $a_{j} = 0.5$ since $X_{j}$'s are binary for each $j\in \{1,\dots, p\}$. Meanwhile, for applications without given breaks, we want to select a break $\widehat{a}_{j}$ on each coordinate in a data-driven fashion so as to make our tests as powerful as possible. We introduce in Section~\ref{Sec2.4} the VD and VDBP tests with data-driven break selection. Our break selection aims at finding potential breaks for test consistency (i.e., the asymptotic test power is one), which are not necessarily the best break in terms of test power on each coordinate. Let $\Phi(\cdot)$ denote the cumulative distribution function for the standard normal distribution and $\Phi^{-1}(\cdot)$ be the inverse function of $\Phi(\cdot)$.

	\subsection{Coordinate-wise model-X knockoff features}\label{Sec2.1}
	
	\subsubsection{Definition and comparison with the original model-X knockoffs}\label{Sec2.1.1}
	The model-X coordinate-wise knockoffs are defined in Definition~\ref{knockoff.1} below. For simplicity, we refer to the original model-X knockoffs and  coordinate-wise  model-X knockoffs as  knockoffs and coordinate-wise knockoffs, respectively. 
	\begin{defn}[Coordinate-wise knockoffs]\label{knockoff.1}
		$\widetilde{X}_{j}$ is a coordinate-wise knockoff of $X_{j}$ if and only if   $(\boldsymbol{X}_{-j}, X_{j})$ and $( \boldsymbol{X}_{-j}, \widetilde{X}_{j})$ have the same distribution and  $\widetilde{X}_{j}$ is independent of $Y$ conditional on $\boldsymbol{X}$.
	\end{defn}	
	It is readily seen from Definition~\ref{knockoff.1} and Definition~\ref{knockoff.2} below that if $\widetilde{\boldsymbol{X}}$ is a knockoff vector of $\boldsymbol{X}$ satisfying Definition~\ref{knockoff.2}, then each $\widetilde{X}_{j}$ is a coordinate-wise knockoff of $X_{j}$ that satisfies Definition~\ref{knockoff.1} for $j\in\{1,\dots, p\}$, but not the other way around. This implies that a knockoff vector produced by Algorithm 1 of \citep{CandesFanJansonLv2018},  which samples $\widetilde{X}_{j}$ from the conditional distribution of $X_{j}$ on $(\boldsymbol{X}_{-j}, \widetilde{X}_{1}, \dots, \widetilde{X}_{j-1})$ from $j=1$ to $j=p$, is also a coordinate-wise knockoff vector satisfying Definition~\ref{knockoff.1}. In addition, the approximate knockoffs produced by the existing knockoff generators such as \citep{Barber2015, sesia2019gene, romano2020deep, jordon2018knockoffgan, lu2018deeppink} are also approximate coordinate-wise knockoffs. Moreover, since the assumptions on knockoffs in Definition~\ref{knockoff.1} are more flexible than those in Definition~\ref{knockoff.2}, it is possible to simplify the production procedure of coordinate-wise knockoffs. For example, to generate a vector of coordinate-wise knockoffs $\widetilde{\boldsymbol{X}}$ satisfying Definition~\ref{knockoff.1}, it suffices to
	{\small\begin{equation}\label{coordinate-wise.1}
		\textnormal{ Sample } \widetilde{X}_{j} \textnormal{ from the conditional distribution of } X_{j} \textnormal{ on } \boldsymbol{X}_{-j} \textnormal{ for each  } j\in\{1,\dots, p\},
	\end{equation}}%
which is simpler than Algorithm 1 of \citep{CandesFanJansonLv2018}. 
	\begin{defn}[Original  knockoffs in \citep{CandesFanJansonLv2018}]\label{knockoff.2}
		$\widetilde{\boldsymbol{X}} $ are knockoffs of $\boldsymbol{X}$ if and only if  \textnormal{1)} $(\boldsymbol{X}, \widetilde{\boldsymbol{X}})$ and $(\boldsymbol{X}, \widetilde{\boldsymbol{X}})_{\textnormal{swap}(S)}$ have the same distribution for each $S\subset\{1, \dots, p\}$, where $(\boldsymbol{X}, \widetilde{\boldsymbol{X}})_{\textnormal{swap}(S)}$ are obtained from $(\boldsymbol{X}, \widetilde{\boldsymbol{X}})$ by swapping the entries $X_{j}$ and $\widetilde{X}_{j}$ for each $j\in S$, and \textnormal{2)} $\widetilde{\boldsymbol{X}}$ is independent of $Y$ conditional on $\boldsymbol{X}$.
	\end{defn}

	\subsubsection{Production of approximate coordinate-wise model-X knockoffs}\label{Sec2.1.2}

	In light of Definition~\ref{knockoff.1} and \eqref{coordinate-wise.1}, we may assume a known distribution of  $\boldsymbol{X}$ for coordinate-wise knockoff production. We consider two distributions of $\boldsymbol{X}$ for our inference applications in this paper:  the multivariate Gaussian distribution~\citep{CandesFanJansonLv2018} and  the hidden Markov model (HMM)~\citep{sesia2019gene}. The former distribution for knockoffs was proposed by the original knockoffs inference papers~\citep{Barber2015, CandesFanJansonLv2018}, which is widely used for statistical inference~\citep{barber2019knockoff} because of its convenience and stable empirical performance. The latter one is a natural distributional assumption for modeling SNPs in genetic analysis, as have been introduced in \citep{sesia2019gene}. Since the approximate knockoffs for Definition~\ref{knockoff.2} are also the approximate coordinate-wise knockoffs, as have discussed in Section~\ref{Sec2.1.1},  we may generate the approximate coordinate-wise Gaussian and HMM knockoffs respectively by the R packages \texttt{knockoff}~\citep{Barber2015} and \texttt{SNPknock}~\citep{sesia2019gene}; these softwares take a sample of feature vectors as input and output the approximate knockoffs.

Besides the existing methods, we have implemented a knockoff generator for producing approximate coordinate-wise Gaussian knockoffs in a coordinate-wise  fashion based on the  ideas of  \eqref{coordinate-wise.1} and those introduced in \citep{Barber2015}. To demonstrate the advantage of our coordinate-wise knockoffs, we have  performed a numerical experiment to show that the correlation between a Gaussian variable and its coordinate-wise knockoff tend to be smaller, sometimes much smaller, than the correlation between the Gaussian variable and its knockoff generated by the R package \texttt{knockoff}~\citep{Barber2015}. To save space, details of our coordinate-wise  Gaussian knockoff generator and the results of our numerical experiments are postponed to  the Supplementary Material.

	\subsection{Variance difference test}\label{Sec2.2}
	
	In this section, we introduce the VD test for hypothesis \eqref{null.1} with some $j\in \{1, \dots, p\}$ of interest and a given break $a_{j}\in\mathbb{R}$; the subscript $j$ of  $a_{j}$ is dropped when no confusion is possible.  Let an inference sample $ \{ \boldsymbol{X}_{i}, \widetilde{\boldsymbol{X}}_{i}, Y_{i}\}_{i=1}^{n}$ with ideal coordinate-wise knockoff features satisfying Definition~\ref{knockoff.1} or approximate ones be given such that $\{ \boldsymbol{X}_{i}, \widetilde{\boldsymbol{X}}_{i}, Y_{i}\}_{i=1}^{n}$ and $(\boldsymbol{X}, \widetilde{\boldsymbol{X}}, Y)$ are i.i.d. random vectors. In addition, we are given a training sample for constructing 
	$\widehat{m}(\boldsymbol{X})$, the random forests~\citep{Breiman2001} estimate of mean function $m(\boldsymbol{X})$ in model~\eqref{model.1}. We consider two set-ups for the training sample. First, the regression trees model is trained on the inference sample; in other words, the training and inference samples are the same. In our simulation experiments and data study, we find that the VD test is easy-to-use in practice under this setting and has satisfactory empirical performance. Second, the training sample is assumed to be an independent copy of the inference sample. The analysis of the proposed tests in Sections~\ref{Sec3b}--\ref{Sec4} assumes the second setting with an independent training sample to simplify technical details. Now, we introduce the VD test in \eqref{h.1} below, which is the same testing procedure regardless of whether the training sample is an independent copy of the inference sample or not.

	For each $l\in \{1, \dots ,p\}$ and some given break $a\in\mathbb{R}$,
	\begin{equation}
		\begin{split}\label{KMtest.1}
			T_{l}(a) & = n^{-\frac{1}{2}}\sum_{i=1}^{n}(\boldsymbol{1}_{X_{il}\in(-\infty, a]}- \boldsymbol{1}_{\widetilde{X}_{il}\in(-\infty, a]}) (\widehat{\varepsilon}_{i} )^{2} ,\\
			\widehat{\sigma}_{l}^2(a) &= n^{-1}\sum_{i=1}^{n} \Big[ (\boldsymbol{1}_{X_{il} \in (-\infty, a]} -\boldsymbol{1}_{\widetilde{X}_{il} \in (-\infty, a]}) (\widehat{\varepsilon}_{i})^2- \widehat{\mu}_{l}(a) \Big]^2,
		\end{split}
	\end{equation}
	where $\widehat{\varepsilon}_{i} = Y_{i} - \widehat{m}(\boldsymbol{X}_{i})$ and $\widehat{\mu}_{l}(a) = n^{-1}\sum_{i=1}^{n}(\boldsymbol{1}_{X_{il}\in(-\infty, a]}- \boldsymbol{1}_{\widetilde{X}_{il}\in(-\infty, a]}) (\widehat{\varepsilon}_{i} )^{2} $.  The VD test statistic with break $a\in\mathbb{R}$ for the $j$th feature and a test threshold $t >0$ is given as follows.
	\begin{equation}
		\begin{split}\label{h.1}
			\textnormal{Test statistic } &: \ T_{j}(a)\big[\widehat{\sigma}_{j}(a)\big]^{-1}, \\
			\textnormal{Test } & : \textnormal{ We reject the null hypothesis in \eqref{null.1} if }  \left|T_{j}(a)\big[\widehat{\sigma}_{j}(a)\big]^{-1}\right| > t.
		\end{split}
	\end{equation}	 	
	We show in Section~\ref{Sec3.2b} that when the rejection threshold $t = -\Phi^{-1}(\alpha/2)$ for some $\alpha \in(0, 1)$, the asymptotic size is at most $\alpha$. Meanwhile, we take the frequentist interpretation of P-value, which is the probability of obtaining a test statistic at least as extreme as the observed test statistic, while assuming the null hypothesis and model regularity assumptions. In Section~\ref{Sec3.2b}, we show that the P-value of test \eqref{h.1} is estimated by $2\Phi\big\{-|T_{j}(a)\big[\widehat{\sigma}_{j}(a)\big]^{-1}|\big\}$. It is noteworthy that if we establish tests for all $1\le j\le p$, then multiplicity correction is needed in practice.

	To have some intuition for how $T_{j}(a)$ works, consider its population version as follows. For $l\in \{1,\dots, p\}$, define
	\begin{equation}
		\begin{split}\label{mu.a1}
			T_{l}^{(\star)}(a) &\coloneqq n^{-\frac{1}{2}}\sum_{i=1}^{n}(\boldsymbol{1}_{X_{il}\in(-\infty, a]} - \boldsymbol{1}_{\widetilde{X}_{il}\in (-\infty, a]} )\big[\zeta(\boldsymbol{X}_{iS^*}) \varepsilon_{i}\big]^{2}  ,\\
			\mu_{l}(a) &\coloneqq \mathbb{E}\big\{(\boldsymbol{1}_{X_{l}\in(-\infty, a]} - \boldsymbol{1}_{\widetilde{X}_{l}\in(-\infty, a]} )\big[\zeta(\boldsymbol{X}_{S^*}) \varepsilon\big]^{2}\big\}.
		\end{split}
	\end{equation}
	When the residual $\widehat{\varepsilon}_{i}$ is an 
	accurate estimate of $\zeta(\boldsymbol{X}_{iS^*}) \varepsilon_{i}$, which is the case when $\widehat{m}(\boldsymbol{X}_{i})$ is an accurate estimate of $m(\boldsymbol{X}_{i})$, it is seen from \eqref{mu.a1} that the intuitions for test statistic \eqref{testnull.1} below with $\mu_{j}(a)=0$ and the test statistic in  \eqref{h.1} above are the same. Given the form of $T_{j}^{(\star)}(a)$ and the i.i.d. observations, it is possible to show that under some regularity conditions,
	\begin{equation}
		\label{testnull.1}
		\sqrt{n}\big\{n^{-\frac{1}{2}}T_{j}^{(\star)}(a) -  \mu_{j}(a) \big\} \big[\widehat{\sigma}_{j}(a) \big]^{-1}
	\end{equation} 
	is asymptotically normally distributed with zero mean and unit variance. To make use of statistic \eqref{testnull.1} for testing \eqref{null.1}, we need to know the value of $\mu_{j}(a)$ under the null hypothesis. To this end, we rely on Lemma~\ref{lemma1} below. Lemma~\ref{lemma1} is a basic result of  exchangeability~\citep{CandesFanJansonLv2018} for null features and their knockoffs.
	\begin{lemma}\label{lemma1}
		For a null feature $X_{j}$ with $j\not\in S^*$ and its coordinate-wise knockoff $\widetilde{X}_{j}$, $(\varepsilon\times \zeta(\boldsymbol{X}_{S^*}), X_{j})$ and $(\varepsilon\times \zeta(\boldsymbol{X}_{S^*}), \widetilde{X}_{j})$ have the same distribution.	
	\end{lemma}

	By Lemma~\ref{lemma1}, for  each $j\not\in S^*$ and each $a\in\mathbb{R}$,
	\begin{equation}
		\label{key.ob}
		\mathbb{E}\big\{\boldsymbol{1}_{X_{j}\in (-\infty, a]} \big[\zeta(\boldsymbol{X}_{S^{*}})\varepsilon\big]^{2} \big\} = \mathbb{E}\big\{\boldsymbol{1}_{\widetilde{X}_{j}\in (-\infty, a]} \big[\zeta(\boldsymbol{X}_{S^{*}})\varepsilon\big]^{2} \big\},
	\end{equation}
	which implies $\mu_{j}(a) = 0$, and the desired test inference under null hypothesis  \eqref{null.1} can be established for population test statistic \eqref{testnull.1} accordingly. The result of \eqref{key.ob} is therefore the key for understanding the intuition for test statistic $T_{j}(a)$. Similar ideas have been used for testing conditional independence; for example, see~\citep{konig2021relative}.

	\subsection{Variance difference Breusch-Pagan Test}\label{Sec2.3}
	
	In this section, we introduce the VDBP test for \eqref{BP.1} with given breaks, and  assume $a_{1} = \dots = a_{p} = a$ for some $a\in\mathbb{R}$ for simplicity. As in Section~\ref{Sec2.2}, we are given an inference sample $\mathcal{X}_{n}$ of size $n$ and a training sample $\mathcal{X}_{0}$ for constructing the regression trees estimate of $m(\boldsymbol{X})$, which is denoted by $\widehat{m}(\boldsymbol{X})$. The inference sample $\mathcal{X}_{n}$ is split into $\mathcal{X}_{1}$ and $\mathcal{X}_{2}$ with $n = n_{1} + n_{2}$ and $n_{1} = \nint{\frac{1}{3} n}$. Hence, we have three samples: $\mathcal{X}_{1} = \{\boldsymbol{X}_{i}, \widetilde{\boldsymbol{X}}_{i}, Y_{i}\}_{i=1}^{n_{1}}$ for constructing the test statistics, $\mathcal{X}_{2} = \{\boldsymbol{U}_{i}, \widetilde{\boldsymbol{U}}_{i}, V_{i}\}_{i=1}^{n_{2}}$ for screening out features (see details below), and a training sample $\mathcal{X}_{0}$ for constructing  $\widehat{m}(\boldsymbol{X})$. For the training sample, the two set-ups are the same as have mentioned in Section~\ref{Sec2.2}. Now, let us introduce the  VDBP testing procedure.

	The VDBP test has two steps. First, we select the most active feature based on sample $\mathcal{X}_{2}$. For each $1\le l\le p$, let $G_{l}(a) = n_{2}^{-1}\sum_{i=1}^{n_{2}}(\boldsymbol{1}_{U_{il}\in(-\infty, a]}- \boldsymbol{1}_{\widetilde{U}_{il}\in (-\infty, a]}) (\widehat{\eta}_{i} )^{2}$ and $\widehat{l} = \arg\max_{1\le l\le p}|G_{l}(a)|$, where $\widehat{\eta}_{i} = V_{i} - \widehat{m}(\boldsymbol{U}_{i})$. Next, we perform a VD test for the selected feature $\widehat{l}$ in \eqref{h.2} below. Let $T_{l}(a)$ and $\widehat{\sigma}_{l}^2(a)$ depending on $\mathcal{X}_{1}$ for each $l\in \{1, \dots ,p\}$ be given as in \eqref{KMtest.1} in  Section~\ref{Sec2.2}; notice that  $\mathcal{X}_{1}$ instead of $\mathcal{X}_{n}$ is used at this step. The VDBP test with a given break $a\in\mathbb{R}$ and a test threshold $t >0$ is given as follows.
	\begin{equation}
		\begin{split}\label{h.2}
			\textnormal{Test statistic }&: \ T_{\widehat{l}}(a)\big[\widehat{\sigma}_{\widehat{l}}(a)\big]^{-1}, \\
			\textnormal{Test } &: \textnormal{ We reject the null hypothesis in \eqref{BP.1} if }  \left|T_{\widehat{l}}(a)\big[\widehat{\sigma}_{\widehat{l}}(a)\big]^{-1}\right| > t.
	\end{split}\end{equation}	
We show in Section~\ref{Sec3.2b} that with the rejection threshold $t = -\Phi^{-1}(\alpha/2)$ for some $\alpha \in(0, 1)$, the asymptotic size is  at most $\alpha$; meanwhile, the frequentist P-value is estimated by $2\Phi\big\{-|T_{\widehat{l}}(a)\big[\widehat{\sigma}_{\widehat{l}}(a)\big]^{-1}|\big\}$.

	\subsection{VD and VDBP tests with break selection}\label{Sec2.4}
	
	In this section, we introduce  the VD and VDBP tests with break selection  and the intuition for selecting good breaks from  break candidates. Let  $\mathcal{X}_{1} = \{Y_{i}, \boldsymbol{X}_{i}, \widetilde{\boldsymbol{X}}_{i}\}_{i=1}^{n_{1}}$, $\mathcal{X}_{2} = \{V_{i}, \boldsymbol{U}_{i}, \widetilde{\boldsymbol{U}}_{i}\}_{i=1}^{n_{2}}$, and a training sample $\mathcal{X}_{0}$ be as given in Section~\ref{Sec2.3}, with $\widehat{m}(\boldsymbol{X})$ denoting the regression trees estimate of $m(\boldsymbol{X})$ trained on $\mathcal{X}_{0}$. Let $\mathcal{X}_{n}$ denote the inference sample at hand, which is split into $\mathcal{X}_{1}$ and $\mathcal{X}_{2}$ with $n = n_{1} + n_{2}$ and $n_{1} = \nint{\gamma_{0} n}$ for some constant $0< \gamma_{0}<1$.  We set $\gamma_{0} $ to $\frac{2}{3}$ and $\frac{1}{3}$ for the VD and VDBP tests, respectively.

	Let $G_{l}(\kappa_{1,l}), \dots, G_{l}(\kappa_{R,l})$ be defined as in Section~\ref{Sec2.3} for each $l\in\{1, \dots, p\}$, where $R>1$ and   break candidates $\{\kappa_{1,l}, \dots, \kappa_{R,l}\}$  are some parameters predetermined by users with $\kappa_{r,l}<\kappa_{r+1,l}$. Practically, for each $l\in\{1, \dots ,p\}$, we set  $(\kappa_{1,l}, \kappa_{R,l})$ to be respectively the first and third quartiles of $\{ U_{il} \}_{i=1}^{n_{2}}$, and that $\kappa_{r,l}$'s are evenly distributed on $[\kappa_{1,l}, \kappa_{R,l}]$ with $R=100$. Here, for simplicity, we take $\kappa_{r,l}$'s as some predetermined constants that do not depend on sample or coordinate index, and we write $\kappa_{r} = \kappa_{r, l}$ for each $l$. The break selection is given as follows.
	$$\widehat{a}_{l} = \arg\max_{\kappa \in \{\kappa_{1},\dots, \kappa_{R}\}}|G_{l}(\kappa)| \ \ \textnormal{ for } \ \ l\in \{1,\dots, p\} \qquad \textnormal{ and } \qquad \widehat{l} = \arg\max_{1\le l \le p} |G_{l}(\widehat{a}_{l})|,$$
	where $\widehat{a}_{l}$ is the sample best break over $\{\kappa_{1}, \dots, \kappa_{R}\}$ on the $l$th coordinate. 
	
	Let us gain some insight into break selection. To achieve test consistency for VD and VDBP tests, the break selection aims at selecting $\widehat{a}_{l}$ on the $l$th coordinate such that  $|\mu_{l}(\widehat{a}_{l})|$ is uniformly bounded away from zero for each $l\in S^*$, where $\mu_{l}(x)$ for $l\in \{1, \dots, p\}$ and $x \in \mathbb{R}$ has been defined in \eqref{mu.a1}. We will see in Section~\ref{Sec4.2b} that
	$|\mu_{j}(\widehat{a}_{j})|$ has a nontrivial lower bound in a probability sense if $j\in S^*$ and 
	\begin{equation}
		\label{split.1}
		\max_{r\in\{1,\dots, R\}}|\mu_{j}(\kappa_{r})| > \underline{\mu} 
	\end{equation}
	for some signal strength $\underline{\mu}>0$ whose value may decrease as the sample size increases. We further show in  Example~\ref{dist.1} in Section~\ref{Sec4.2b} that \eqref{split.1} holds if (i) the  standard deviation function $\zeta(\cdot)$ is given as in Example~\ref{sd.exmp.4} with the variation level of $\zeta(\cdot)$ properly bounded away from below, (ii) the range between $\kappa_{1}$ and $\kappa_{R}$ is reasonably wide and $\kappa_{1}<\dots < \kappa_{R}$ are evenly distributed with some constant $R>1$, and (iii) some mild model regularity assumptions are satisfied. Such requirements on $\kappa_{r}$'s justify our data-driven choice of $\kappa_{r, l}$'s described above. 
	\begin{remark}
		Although both of our break selection and covariance change point detection~\citep{wang2021optimal, avanesov2018change, aue2009break} make inferences about the standard deviation function, the problem of break selection cannot be seen as a change point detection problem. Particularly, our break selection does not assume piecewise linear  $\zeta(\cdot)$, which precludes applications of most existing change point detection techniques to break finding  here.
		\end{remark}
	
		Next, we formally give the VD and VDBP tests with break selection. For each $l\in \{1, \dots, p\}$ and $x\in\mathbb{R}$, let $T_{l}(x)$ and $\widehat{\sigma}_{l}^2(x)$ be defined as in \eqref{KMtest.1}  with sample size $n_{1}$ in place of $n$. The VD test with break selection for the $j$th feature and a  test threshold $t >0$ is given by
	\begin{equation}
		\begin{split}\label{h.3}
			\textnormal{Test statistic } &: \ T_{j}(\widehat{a}_{j})\big[\widehat{\sigma}_{j}(\widehat{a}_{j})\big]^{-1}, \\
			\textnormal{Test } & : \textnormal{ We reject the null hypothesis in \eqref{null.1} if }  \left|T_{j}(\widehat{a}_{j})\big[\widehat{\sigma}_{j}(\widehat{a}_{j})\big]^{-1}\right| > t.
		\end{split}
	\end{equation}
	The VDBP test with break selection and a test threshold $t >0$ is given by
	\begin{equation}
		\begin{split}\label{h.4}
			\textnormal{Test statistic } & : \  T_{\widehat{l}}(\widehat{a}_{\widehat{l}})\big[\widehat{\sigma}_{\widehat{l}}(\widehat{a}_{\widehat{l}})\big]^{-1}, \\
			\textnormal{Test } & : \textnormal{ We reject the null hypothesis in \eqref{BP.1} if }  \left|T_{\widehat{l}}(\widehat{a}_{\widehat{l}})\big[\widehat{\sigma}_{\widehat{l}}(\widehat{a}_{\widehat{l}})\big]^{-1} \right| > t.
		\end{split}
	\end{equation}

	We show in Section~\ref{Sec3.2b} that when the rejection threshold $t = -\Phi^{-1}(\alpha/2)$ for some $\alpha \in(0, 1)$, both asymptotic test sizes of \eqref{h.3}--\eqref{h.4} are  at most $\alpha$, while their respective P-values are estimated by $2\Phi\big\{-|T_{j}(\widehat{a}_{j})\big[\widehat{\sigma}_{j}(\widehat{a}_{j})\big]^{-1}|\big\}$ and $2\Phi\big\{-|T_{\widehat{l}}(\widehat{a}_{\widehat{l}})\big[\widehat{\sigma}_{\widehat{l}}(\widehat{a}_{\widehat{l}})\big]^{-1}|\big\}$.

	\section{Analysis of VD and VDBP tests under the nulls}\label{Sec3b}

	We analyze four tests \eqref{h.1}, \eqref{h.2}, \eqref{h.3}, and \eqref{h.4} under their respective null hypotheses in Theorems~\ref{theorem1}--\ref{theorem4} in Section~\ref{Sec3.2b}. The asymptotic null distributions of the test statistics depend on the mean estimation accuracy, which is discussed in Section~\ref{Sec3.1b}. With these results, we can obtain the P-values and test sizes for our inference applications.

	Throughout this section,  $\{Y_{i}, \boldsymbol{X}_{i}, \widetilde{\boldsymbol{X}}_{i}, \varepsilon_{i}\}_{i=1}^{n_{1}}$, $\{V_{i}, \boldsymbol{U}_{i}, \widetilde{\boldsymbol{U}}_{i}, \eta_{i}\}_{i=1}^{n_{2}}$, $(Y, \boldsymbol{X}, \widetilde{\boldsymbol{X}}, \varepsilon)$ are i.i.d. random vectors where $(Y, \boldsymbol{X}, \varepsilon)$ follows \eqref{model.1} and $\widetilde{\boldsymbol{X}}$ is an ideal coordinate-wise knockoff  vector of $\boldsymbol{X}$ as defined in Definition~\ref{knockoff.1}. We assume to be given three independent samples: $\mathcal{X}_{1} = \{Y_{i}, \boldsymbol{X}_{i}, \widetilde{\boldsymbol{X}}_{i}\}_{i=1}^{n_{1}}$ for constructing test statistics, $\mathcal{X}_{2} = \{V_{i}, \boldsymbol{U}_{i}, \widetilde{\boldsymbol{U}}_{i}\}_{i=1}^{n_{2}}$ for selecting breaks, and a training sample $\mathcal{X}_{0}$ for training the regression trees estimate  of the mean function, which is denoted by $\widehat{m}(\cdot)$. To simplify the technical analysis in this section, we assume $n_{1}=n_{2}=n$ and that the size of the  training sample is also $n$. In addition, for tests with a given break $a_{j}\in\mathbb{R}$ on each coordinate, we assume  $a_{1} = \dots = a_{p} = a$ for some  $a\in\mathbb{R}$. Moreover, for tests with break selection, we assume the same set of constant break candidates across each coordinate, and they are denoted by $\kappa_{r}$'s with $\kappa_{1}< \dots< \kappa_{R}$ for some integer $R>1$. We use the notation $x\vee y = \max\{x, y\}$ for any $x,y\in\mathbb{R}$.
	
	\subsection{High-dimensional consistency of random forests}\label{Sec3.1b}

	The proposed tests rely on an accurate estimate of  mean function $m(\boldsymbol{X})$ in model~\eqref{model.1}. In this paper, we employ random forests~\citep{Breiman2001}  to cope with the estimation of a potentially highly non-linear sparse mean function $m(\boldsymbol{X})$ such as Example~\ref{interaction.1} below.
	\begin{exmp}\label{interaction.1}
		Let $\boldsymbol{X}$ have a uniform distribution on $[0, 1]^{p}$ and some fixed interger $0<s^*\le p$ be given. Let $m(\boldsymbol{X}) = \sum_{j=1}^{s^*}(\beta_{j}X_{j} + \sum_{l\ge j}^{s^*} \beta_{lj}X_{l}X_{j})$, in which if $\beta_{lj}\not = 0$ for some $j\le l \le s^*$, then $a\times b \ge 0$ for every $\{a, b\}\subset \{\beta_{j}, \beta_{1j}, \dots, \beta_{s^*j}\}$ where $\beta_{j_{2}j_{1}} = \beta_{j_{1}j_{2}}$; the coefficients are otherwise arbitrary.
	\end{exmp} 	
	Besides the ability to model non-linear mean functions, random forests are considered because their consistency as in Condition~\ref{consistency.3} below under heteroskedastic model \eqref{model.1} has been studied. The random forests estimate is denoted by $\widehat{m}(\boldsymbol{X})$, and the dependence of consistency rate $B_{1}$ on the training sample size $n$ is not indicated for simplicity.
	\begin{condi}\label{consistency.3}
		$\mathbb{E}\big[\widehat{m}(\boldsymbol{X}) - m(\boldsymbol{X}) \big]^{2} \le B_{1}$ for some small $B_{1}>0$.
	\end{condi}

	The consistency rate in Condition~\ref{consistency.3} of a random forests variant under heteroskedastic model \eqref{model.1} has been studied in Theorem 5 and Corollary 6 in~\citep{biau2012analysis}. There,  it is shown that $B_{1} \le C\times n^{\frac{-0.75}{s^{\star}\times(\log_{e}2) + 0.75}}$ for some constant $C>0$ and all large $n$ (training sample size) where $s^{\star}$ is the number of active features in $m(\boldsymbol{X})$, while assuming independent explanatory features, a Lipschitz continuous $m(\boldsymbol{X})$ that takes Example~\ref{interaction.1} into account, $\textnormal{Var}(Y|\boldsymbol{X}) < C_{2}$ for some constant $C_{2} >0$ almost surely, and a simplified splitting rule of random forests. In addition, the feature dimensionality $p$ is allowed to be much larger than $s^{\star}$ as long as the
	splitting procedure can track and split the strong coordinates of active features. Besides, recent results~\citep{chi2020asymptotic, syrgkanis2020estimation, klusowski2021universal} have analyzed the high-dimensional consistency rate of random forests while assuming the original CART splitting rule and homoscedastic models with general mean functions such as Example~\ref{interaction.1}. It is possible to extend these results to cases with heteroskedastic models. 
	
	\subsection{P-values and test sizes under null hypotheses}\label{Sec3.2b}
	
	Let some nondecreasing  real sequence $\{\overbar{M}_{i}\}_{i\ge 1}$ with $\overbar{M}_{1}\ge1$ and some $c>0$ be given for the following regularity conditions.
	\begin{condi}\label{bound}
		 $|m(\boldsymbol{X})| \le \overbar{M}_{1}$ almost surely and $|\widehat{m}(\boldsymbol{X})|\le \overbar{M}_{n}$ almost surely for each $n$.
	\end{condi}

	\begin{condi}\label{non.trivial.2}
		$\mathbb{P}(\{X_{j} \le \kappa_{1}\}\cap\{\widetilde{X}_{j} \le \kappa_{1}\}) > c$ and $\mathbb{P}(\{X_{j} \le \kappa_{1}\}\cap\{\widetilde{X}_{j} >\kappa_{R}\}) > c$.
	\end{condi}

	\begin{condi}\label{A4}
		The training sample $\mathcal{X}_{0}$ for  $\widehat{m}(\cdot)$ is independent of $\{V_{i}, \boldsymbol{U}_{i}, \widetilde{\boldsymbol{U}}_{i}, \eta_{i}\}_{i=1}^{n}$, $\{Y_{i}, \boldsymbol{X}_{i}, \widetilde{\boldsymbol{X}}_{i}, \varepsilon_{i}\}_{i=1}^{n}$, and $(Y, \boldsymbol{X}, \widetilde{\boldsymbol{X}}, \varepsilon)$.
	\end{condi}

	Some comments on these conditions are given as follows. The assumption of an independent training sample in Condition~\ref{A4} is assumed to simplify the technical analysis for our main results. Condition~\ref{bound} with $\overbar{M}_{n} = \log{(n)}$ is a common assumption for analyzing nonparametric and machine learning predictors~\citep{kohler2021rate}. Condition~\ref{non.trivial.2} controls the  dependence of $\widetilde{X}_{j}$ on $X_{j}$ conditional on $\boldsymbol{X}_{-j}$. Condition~\ref{non.trivial.2} with a proper choice of $(c,\kappa_{1}, \kappa_{R})$ is satisfied if, for example, $(X_{j}, \widetilde{X}_{j})$ has a bivariate normal distribution with a non-singular covariance matrix.

	Theorem~\ref{theorem1} below analyzes the asymptotic properties for the VD test \eqref{h.1}. Based on the results of Theorem~\ref{theorem1}, we can obtain the much needed P-value and test size. Recall that in Section~\ref{Sec1.1}, we have defined $\zeta(\vec{z}) = \zeta_{0}$ for $\vec{z}\in\mathbb{R}^{|S^*|}$ when $S^* = \emptyset$ with constant $\zeta_{0}>0$.  Let $s^* = |S^*|$.

	\begin{theorem}\label{theorem1}
		Assume $\inf_{\vec{z}\in\mathbb{R}^{s^*}}\zeta(\vec{z})>0$, $\sup_{\vec{z}\in\mathbb{R}^{s^*}}\zeta(\vec{z}) < \infty$, $ \mathbb{E}(\varepsilon^{8})<\infty$,  Conditions~\ref{consistency.3}--\ref{bound} and Condition~\ref{A4}. For 
		all large $n$, each $0<B_{1}<1$, each $t>0$, and any $1\le j \le p$ such that Condition~\ref{non.trivial.2} holds with $(R,\kappa_{1} )=(1,a)$ and that  $j\not\in S^*$, it holds that 
		\begin{equation}\label{theorem1.upperbound.1}
			\begin{split}
				\mathbb{P}\left(\left|T_{j}(a)\big[\widehat{\sigma}_{j}(a)\big]^{-1}\right| \ge t\right)\le 2\Phi(-t) +  D_{n},
			\end{split}
		\end{equation}
		where $D_{n}=\varsigma\big\{c\big[\inf_{\vec{z}\in\mathbb{R}^{s^*}} \zeta(\vec{z})\big]^4\mathbb{E}(\varepsilon^{4} )\big\}^{-\frac{1}{2}} + n^{\frac{1}{4}}B_{1}^{\frac{1}{2}} + (\overbar{M}_{n})^2(\log{n})(n^{-\frac{1}{4}} + B_{1}^{\frac{1}{4}})$, in which  \sloppy$\varsigma = t(\log{n})(10n^{-\frac{1}{4}} + 4B_{1}^{\frac{1}{8}}) + n^{\frac{1}{4}}B_{1}^{\frac{1}{2}}$ and $c$ is given in Condition~\ref{non.trivial.2}.	
	\end{theorem}
	If $D_{n}=o(1)$, Theorem~\ref{theorem1} implies that the test size of the VD test \eqref{h.1} is at most $\alpha$ plus the negligible term $D_{n}$ for some $\alpha\in(0, 1)$ when the test rejection threshold $t= -\Phi^{-1}(\alpha/2)$, and the P-value is estimated by $2\Phi(-|T_{j}(a)\big[\widehat{\sigma}_{j}(a)\big]^{-1}|)$. For $D_{n}$ to be asymptotically negligible, it suffices that  $\lim_{n\rightarrow \infty}\sqrt{n}B_{1}= 0$ and $\lim_{n\rightarrow\infty}(\overbar{M}_{n})^2(\log{n})n^{-\frac{1}{4}}  = 0$. In Section~\ref{Sec3.1b}, we have reviewed the related work on the theoretical foundations of the former condition. The latter one is mild if $\overbar{M}_{n} = \log{(n)}$, which is commonly assumed for consistency analysis. Note that the results of Theorem~\ref{theorem1} are nonasymptotic. Particularly, the upper bound in \eqref{theorem1.upperbound.1} is uniform over $j\in \{1, \dots ,p\}$. Therefore, for all large $n$, we may establish a valid VD test for \eqref{null.1} for any $j\in \{1,\dots, p\}$ of interest provided the regularity conditions are satisfied.
	\begin{remark}
		\label{running.n}
		Model \eqref{model.1} is allowed to depend on $n$, but we do not indicate the dependence of our model on $n$ explicitly for simplicity.
	\end{remark}

	Let us proceed to the theory for the VDBP test \eqref{h.2} with given breaks $a_{1} = \dots = a_{p}  = a\in \mathbb{R}$. Recall that $\mu_{l}(a)$ for each $l\in \{1, \dots, p\}$ has been defined in \eqref{mu.a1}.
	\begin{theorem}\label{theorem2b}
		Assume $\inf_{\vec{z}\in\mathbb{R}^{s^*}}\zeta(\vec{z})>0$, $\sup_{\vec{z}\in\mathbb{R}^{s^*}}\zeta(\vec{z}) <\infty$, Condition \ref{non.trivial.2} with $(R,\kappa_{1} )=(1,a)$ for each $j\in \{1, \dots ,p\}$, and Condition~\ref{A4}. Assume $\mathbb{E}|\varepsilon|^{8\vee q}<\infty$ and Conditions~\ref{consistency.3}--\ref{bound} with $\lim_{n\rightarrow\infty} n^{\beta_{2}}\overbar{M}_{n}^2B_{1}+ \sqrt{n}B_{1}(-\log B_{1}) + n^{-\frac{1}{2}+ \frac{4}{q-\beta} + \beta_{2}} \sqrt{\log{(n\vee p)}} =0$ for some constants  $0<\beta<q$ and $0< \beta_{2}<\frac{1}{4}$.
		Then  $\big[T_{\widehat{l}}(a) - \sqrt{n}\mu_{\widehat{l}}(a) \big]\big[\widehat{\sigma}_{\widehat{l}}(a)\big]^{-1} \overset{D}{\longrightarrow} N(0, 1)$. If furthermore $S^* = \emptyset$, then  $T_{\widehat{l}}(a) \big[\widehat{\sigma}_{\widehat{l}}(a)\big]^{-1} \overset{D}{\longrightarrow} N(0, 1)$.		
	\end{theorem}	
	According to Theorem \ref{theorem2b}, to have a legitimate VDBP test, feature dimensionality $p$ is allowed to grow at some polynomial order of $n$ if $\mathbb{E}(|\varepsilon|^q)<\infty$ for sufficiently large $q$. Meanwhile, Theorem~\ref{theorem2b} assumes
	$\lim_{n\rightarrow\infty} \sqrt{n}B_{1}(-\log B_{1}) =0$ and $\lim_{n\rightarrow\infty} n^{\beta_{2}}\overbar{M}_{n}^2B_{1}=0$. The former condition is only slighly stronger than the analogous condition required by Theorem~\ref{theorem1}, while the latter one holds if the former is assumed and  $\overbar{M}_{n} = \log{n}$. Other comments on regularity conditions of Theorem~\ref{theorem2b} are similar to those for Theorem~\ref{theorem1}, and hence we omit the details here. By Theorem~\ref{theorem2b}, the asymptotic test size of the VDBP test \eqref{h.2} is at most $\alpha$ for some $\alpha\in(0, 1)$ when the test rejection threshold $t= -\Phi^{-1}(\alpha/2)$, and the P-value is estimated by $2\Phi(-|T_{\widehat{l}}(a)\big[\widehat{\sigma}_{\widehat{l}}(a)\big]^{-1}|)$.

	Next, we analyze the asymptotic properties of the VD \eqref{h.3} and VDBP \eqref{h.4} tests with break selection in Theorems~\ref{theorem3}--\ref{theorem4} below. Here, the sets of $R$ break candidates on all coordinates are assumed to be the same, and these break candidates are denoted by $\kappa_{1}< \dots < \kappa_{R}$.
	\begin{theorem}\label{theorem3} 
		Assume $\inf_{\vec{z}\in\mathbb{R}^{s^*}}\zeta(\vec{z})>0$, $\sup_{\vec{z}\in\mathbb{R}^{s^*}}\zeta(\vec{z}) < \infty$, $ \mathbb{E}(\varepsilon^{8})<\infty$, and Conditions~\ref{consistency.3}--\ref{bound} and Condition~\ref{A4}. For 
		all large $n$, each $0<B_{1}<1$, each $t>0$, and any $1\le j \le p$ such that Condition~\ref{non.trivial.2} holds and $j\not\in S^*$, it holds that 
		\begin{equation*}
			\mathbb{P}\big(\big |T_{j}(\widehat{a}_{j})\big[\widehat{\sigma}_{j}(\widehat{a}_{j})\big]^{-1}\big| \ge t\big)\le 2\Phi(-t) + D_{n},
		\end{equation*}
		where $D_{n} = \varsigma\big\{c\big[\inf_{\vec{z}\in\mathbb{R}^{s^*}} \zeta(\vec{z})\big]^4\mathbb{E}(\varepsilon^{4} )\big\}^{-\frac{1}{2}} + Rn^{\frac{1}{4}}B_{1}^{\frac{1}{2}} + (\overbar{M}_{n})^2(\log{n})(n^{-\frac{1}{4}} + B_{1}^{\frac{1}{4}})$, in which \sloppy$\varsigma = t(\log{n})(10n^{-\frac{1}{4}} + 4B_{1}^{\frac{1}{8}}) + n^{\frac{1}{4}}B_{1}^{\frac{1}{2}}$ and $c$ is given in Condition~\ref{non.trivial.2}.	
	\end{theorem}

	\begin{theorem}\label{theorem4}
		Assume $ \inf_{\vec{z}\in\mathbb{R}^{s^*}}\zeta(\vec{z})>0$, $\sup_{\vec{z}\in\mathbb{R}^{s^*}}\zeta(\vec{z}) <\infty$,  Condition \ref{non.trivial.2} for each $j\in \{1, \dots, p\}$, and Condition~\ref{A4}. In addition, assume $\mathbb{E}|\varepsilon|^{8\vee q}<\infty$ and  Conditions~\ref{consistency.3}--\ref{bound} with $\lim_{n\rightarrow\infty} n^{\beta_{2}}\overbar{M}_{n}^2B_{1}+ \sqrt{n}B_{1}(-\log B_{1}) + n^{-\frac{1}{2}+ \frac{4}{q-\beta} + \beta_{2}} \sqrt{\log{(n\vee p)}} =0$ for some constants  $0<\beta<q$ and $0< \beta_{2}<\frac{1}{4}$.
		Then  
		$\big[T_{\widehat{l}}(\widehat{a}_{\widehat{l}}) - \sqrt{n}\mu_{\widehat{l}}(\widehat{a}_{\widehat{l}})\big] \big[\widehat{\sigma}_{\widehat{l}}(\widehat{a}_{\widehat{l}})\big]^{-1} \overset{D}{\longrightarrow} N(0, 1).$ If furthermore $S^* = \emptyset$, then  	$T_{\widehat{l}}(\widehat{a}_{\widehat{l}}) \big[\widehat{\sigma}_{\widehat{l}}(\widehat{a}_{\widehat{l}})\big]^{-1}\overset{D}{\longrightarrow} N(0,1).$
		
	\end{theorem}

	From Theorem~\ref{theorem3}, when $\lim_{n\rightarrow \infty}\sqrt{n}B_{1}= 0$ and $\lim_{n\rightarrow\infty}(\overbar{M}_{n})^2(\log{n})n^{-\frac{1}{4}}  = 0$ are additionally satisfied, it holds that $D_{n} = o(1)$; in this scenario, the test size of \eqref{h.3} is at most $\alpha$ plus the negligible term $D_{n}$ for some $\alpha\in(0, 1)$ when the test rejection threshold $t= -\Phi^{-1}(\alpha/2)$, and the P-value is estimated by $2\Phi(-|T_{j}(\widehat{a}_{j})\big[\widehat{\sigma}_{j}(\widehat{a}_{j})\big]^{-1}|)$. From Theorem~\ref{theorem4}, the  test size of \eqref{h.4} is at most $\alpha$ with the rejection threshold $t = -\Phi^{-1}(\alpha/2)$ for some $\alpha \in(0, 1)$, and the P-value is estimated by $2\Phi(-|T_{\widehat{l}}(\widehat{a}_{\widehat{l}})\big[\widehat{\sigma}_{\widehat{l}}(\widehat{a}_{\widehat{l}})\big]^{-1}|)$. Other comments on these two theorems are similar to those for Theorems~\ref{theorem1}--\ref{theorem2b}.
	
	\section{Test power analysis for VD and VDBP tests}\label{Sec4}

	We analyze the test power of the proposed tests respectively in Theorems~\ref{theorem5}--\ref{theorem8} in Section~\ref{Sec4.2b}, with their proofs deferred respectively to  the Supplementary Material. The notation of samples, breaks, and break candidates in this section are the same as in Section~\ref{Sec3b}. We begin with introducing examples of standard deviation functions in model~\eqref{model.1} with non-empty $S^*$.
	
	\subsection{Monotonic standard deviation function}\label{Sec4.1}
	
	Let us introduce below Examples~\ref{sd.exmp.4}--\ref{sd.exmp.3} of standard deviation functions considered in our model \eqref{model.1}. These examples are commonly used  for modeling heteroskedasticity~\citep{ daye2012high}. Particularly, Example~\ref{sd.exmp.4} encompasses a wide class of monotonic functions that include the standard deviation functions considered in the aforementioned papers. To simplify the notation, we assume without loss of generality that $S^* = \{1, \dots, s^*\}$ for some integer $s^*\ge 1$. Also, for each $j\in S^*$, each $\vec{z} = (z_{1}, \dots, z_{s^*})^{\top} \in \mathbb{R}^{s^*}$, and  every $x\in \mathbb{R}$, we let $\zeta_{j}(x)$ be such that $\zeta_{j}(x)\coloneqq \zeta(z_{1}, \dots, z_{j-1}, x, z_{j+1}, \dots, z_{s^*})$ if $s^* >1$, and $\zeta_{j}(x)\coloneqq \zeta(x)$ if $s^*=1$. Notice that  $\zeta_{j}(x)$ is invariant to $z_{j}$.
	\begin{exmp} \label{sd.exmp.4}	
		Assume $\inf_{\vec{z}\in\mathbb{R}^{s^*}}\zeta(\vec{z})>0$, $\sup_{\vec{z}\in\mathbb{R}^{s^*}}\zeta(\vec{z})<\infty$, and that for every $j\in S^*$ and $\vec{z}\in \mathbb{R}^{s^*}$, $\zeta_{j}(x)$ is nondecreasing (or nonincreasing) in $x$. In addition, there exist some $q_{1}<q_{2}$, $\iota>0$, and Cartesian product  $\mathcal{D}=\mathcal{D}_{1}\times \dots \times \mathcal{D}_{s^*}$ with $\mathcal{D}_{j}\in\mathcal{R}$ and $\mathbb{P}(\boldsymbol{X}_{S^*}\in\mathcal{D}) >0$ such that for every $j\in S^*$ and $\vec{z}\in\mathcal{D}$,  $\big|\zeta_{j}( q_{1}) - \zeta_{j}( q_{2})  \big| > \iota$.	
	\end{exmp}
	
	\begin{exmp} \label{sd.exmp.2}
		Let index sets $S_{1}, \dots, S_{K}$ for some $K>0$ be given with $S^* = \cup_{k=1}^{K}S_{k}$. Let $\zeta(\vec{z}) = \beta_{0} + \sum_{k=1}^{K}\beta_{k}\Pi_{l\in S_{k}} \boldsymbol{1}_{z_{l}\ge b_{kl} }$, in which $\beta_{k}, b_{kl}$'s are real values and $\Pi$ denotes the product notation.
	\end{exmp}

	\begin{exmp} \label{sd.exmp.3}
		$\zeta(\vec{z}) = \sqrt{\exp\big(\beta_{0} + \sum_{l=1}^{s^*}\beta_{l}z_{l}\big)}$ with real coefficients $\beta_{0}, \dots, \beta_{s^*}$.
	\end{exmp}

	Example~\ref{sd.exmp.4} takes into account cases where $\zeta(\cdot)$ is a cumulative function. In addition, Examples~\ref{sd.exmp.2}--\ref{sd.exmp.3} with a proper choice of model parameters and some regularity conditions on the distribution of $\boldsymbol{X}$ are cases of Example~\ref{sd.exmp.4}.  Example~\ref{sd.exmp.3} is standard in the literature~\citep{ daye2012high, sorensen2003normal}, while Example~\ref{sd.exmp.2} is motivated by the biological applications mentioned in the Introduction. For example, if the response variable is some biological trait of interest and explanatory features are binary with $X_{j} = 1$ indicating a mutation at the $j$th genetic position, then  $\zeta(\boldsymbol{X}_{S^*}) = 3 + 3\times \boldsymbol{1}_{\{X_{1}\ge 1\}}\times \boldsymbol{1}_{\{X_{2}\ge 1\}}  - 2\times\boldsymbol{1}_{\{X_{3}\ge 1\}}$ with $S^* = \{1, 2,  3\}$ is an instance of Example~\ref{sd.exmp.2} that considers interactive effects of features on the conditional variance of the response in model~\eqref{model.1}. We will show in Section~\ref{Sec4.2b} that when $\zeta(\boldsymbol{X}_{S^*})$ satisfies Example~\ref{sd.exmp.4} with mild additional model assumptions, the consistency of the VD and VDBP tests under their respective alternative hypotheses mainly depends on the variation level $\iota$ of the standard deviation function, the sample size, and the feature dimensionality.

	\subsection{Test power analysis}\label{Sec4.2b}

	In this section, we show that  $\mu_{l}(\kappa_{r})$ for $l\in S^*$ and $r\in\{1, \dots, R\}$ is required to be lower bounded  for our tests to be consistent (i.e., the asymptotic test power is one) under the alternative hypotheses, where we recall that $\mu_{l}(\kappa_{r})$ has been defined in  \eqref{mu.a1}. Specifically, for the VD test \eqref{h.3} to be consistent when $j\in S^*$,  we require
	\begin{equation}
		\label{minimum.signal.3}
		\textnormal{Condition } \eqref{split.1} \qquad\textnormal{with} \qquad\lim_{n\rightarrow\infty}\underline{\mu}^{-1}\big(\sqrt{B_{1}} +\sqrt{n}\big)(\log{n}) =0,
	\end{equation} 
	in which the signal strength $\underline{\mu}$ may decrease as $n$ increases, and the dependence on $n$ of $\underline{\mu}$ is not indicated for simplicity.  In addition, the minimum signal strength condition for the VDBP test \eqref{h.4} to be consistent when $S^*\not=\emptyset$ is that 
	\begin{equation}\label{minimum.signal.1}
		\min_{l\in S^*} \max_{1\le r\le R}|\mu_{l}(\kappa_{r})| > \underline{\mu} \qquad \textnormal{with}\qquad\lim_{n\rightarrow\infty}\underline{\mu}^{-1}\Big[\sqrt{B_{1}} + n^{-\frac{1}{2} + \frac{2}{q -\beta}} \sqrt{\log{(n \vee p)}} \Big]= 0,
	\end{equation}
	and $\mathbb{E}|\varepsilon|^{q\vee 4}<\infty$ for some $0<\beta<q$, where $\varepsilon$ is the model error in model \eqref{model.1}. From \eqref{minimum.signal.3}--\eqref{minimum.signal.1}, the lower bound of $\underline{\mu}$ depends on 	mean estimation error $B_{1}>0$ and sample size for the VD test, and it additionally depends on  feature dimensionality $p$ and moment bound on the model error for the VDBP test.  Meanwhile, consistency of the VD \eqref{h.1} and VDBP \eqref{h.2} tests with given breaks $a_{1} = \dots a_{p}=a\in\mathbb{R}$ also  relies on similar conditions as \eqref{minimum.signal.3}--\eqref{minimum.signal.1} respectively but with $(R, \kappa_{1}) = (1, a)$. Moreover, we give Example~\ref{dist.1} below showing that the signal strength requirements \eqref{minimum.signal.3}--\eqref{minimum.signal.1} hold under a general model setting.


	Theorems~\ref{theorem5}--\ref{theorem6} below analyzes the test power of tests \eqref{h.1} and \eqref{h.2} with a given break $a\in\mathbb{R}$ and that $a_{1}= \dots =a_{p} = a$, respectively; we note that sample variances are assumed to be positive to avoid divisions by zero in this section.
	\begin{theorem}\label{theorem5}	
		Assume $\inf_{\vec{z}\in\mathbb{R}^{s^*}}\zeta(\vec{z}) >0$,  $\sup_{\vec{z}\in\mathbb{R}^{s^*}}\zeta(\vec{z}) < \infty$, $ \mathbb{E}(\varepsilon^4)<\infty$, and Conditions~\ref{consistency.3}--\ref{bound} such that $\lim\sup_{n\rightarrow\infty}\overbar{M}_{n}^2B_{1} <\infty$ with $B_{1}<1$, and Condition~\ref{A4}. For 
		all large $n$, each $t>0$, and any $1\le j \le p$ such that  $|\mu_{j}(a)| > \underline{\mu}>0$, it holds that $\mathbb{P}(|T_{j}(a)\big[\widehat{\sigma}_{j}(a)\big]^{-1}| \le t)\le  B_{1}(\underline{\mu})^{-1} + (\log{n})(1+t)(\sqrt{n}\underline{\mu})^{-1}$.
	\end{theorem}

	\begin{theorem}\label{theorem6}
		
		Assume $ \inf_{\vec{z}\in\mathbb{R}^{s^*}}\zeta(\vec{z})>0$, $\sup_{\vec{z}\in\mathbb{R}^{s^*}}\zeta(\vec{z}) <\infty$, $\min_{l\in S^*} |\mu_{l}(a)| > \underline{\mu}>0$, Condition~\ref{A4},  Condition~\ref{consistency.3}, and $\mathbb{E}|\varepsilon|^{q\vee4} <\infty$ with $\lim_{n\rightarrow\infty}\underline{\mu}^{-1}\Big[\sqrt{B_{1}} + n^{-\frac{1}{2} + \frac{2}{q - \beta}} \sqrt{\log{(n \vee p)}}\Big]= 0$ for some $0<\beta<q$. Then 
		$\lim_{n\rightarrow\infty}\mathbb{P}(\widehat{l} \in S^*) = 1$.
		If furthermore Condition~\ref{bound} holds with  $\lim\sup_{n\rightarrow\infty}\overbar{M}_{n}^2B_{1} <\infty$, then
		$\lim_{n\rightarrow\infty}\mathbb{P}\big\{\big| T_{\widehat{l}}(a) \big[\widehat{\sigma}_{\widehat{l}}(a)\big]^{-1} \big|\le t\big\} = 0$ for each $t>0$.
	\end{theorem}
	
	From  Theorem~\ref{theorem5} that a consistent VD test \eqref{h.1} requires $\lim_{n\rightarrow\infty} \underline{\mu}^{-1}\big(B_{1} +\sqrt{n}\big)(\log{n})=0$, which is less restrictive than the requirement in \eqref{minimum.signal.3} because break selection is not needed here. Meanwhile, Theorem~\ref{theorem6} states that  a valid VDBP test \eqref{h.2} needs \eqref{minimum.signal.1}, which  allows  $p$ to grow at a polynomial order of $n$ given a sufficiently high order moment bound on the model error and that $B_{1}$ and $\underline{\mu}$ decrease to zero at some proper rate as $n$ increases. It is seen that these conditions depends on mean estimation consistency $\lim_{n\rightarrow\infty}B_{1} = 0$, which along with the model boundness condition  $\lim\sup_{n\rightarrow\infty}\overbar{M}_{n}^2B_{1} <\infty$  has been commented on in Section~\ref{Sec3.2b}.  More discussion of the  signal strength conditions will be given below after Example~\ref{dist.1}.

	Theorems~\ref{theorem7}--\ref{theorem8} below respectively analyze the test power for VD \eqref{h.3} and VDBP \eqref{h.4} tests with break selection and some predetermined break candidates $\kappa_{1}< \dots< \kappa_{R}$.  
	\begin{theorem}\label{theorem7}
		Assume $\inf_{\vec{z}\in\mathbb{R}^{s^*}}\zeta(\vec{z}) >0$,  $\sup_{\vec{z}\in\mathbb{R}^{s^*}}\zeta(\vec{z}) < \infty$, $ \mathbb{E}(\varepsilon^4)<\infty$, and Conditions~\ref{consistency.3}--\ref{bound} such that $\lim\sup_{n\rightarrow\infty}\overbar{M}_{n}^2B_{1} <\infty$ with $B_{1}<1$, and Condition~\ref{A4}. For 
		all large $n$, each $t>0$, and any $1\le j \le p$ such that  $\max_{1\le r\le R}|\mu_{j}(\kappa_{r})| > \underline{\mu} >0$, it holds that $\mathbb{P}(|T_{j}(\widehat{a}_{j})\big[\widehat{\sigma}_{j}(\widehat{a}_{j})\big]^{-1}| \le t)\le   (\log{n})\big[\underline{\mu}^{-1}\sqrt{B_{1}} +(1+t)(\sqrt{n}\underline{\mu})^{-1}\big]$.
	\end{theorem}

	\begin{theorem}\label{theorem8}
		Assume $ \inf_{\vec{z}\in\mathbb{R}^{s^*}}\zeta(\vec{z})>0$, $\sup_{\vec{z}\in\mathbb{R}^{s^*}}\zeta(\vec{z}) <\infty$, $\min_{l\in S^*} \max_{1\le r\le R}|\mu_{l}(\kappa_{r})| > \underline{\mu}>0$, Conditions~\ref{consistency.3} and~\ref{A4}, and $\mathbb{E}|\varepsilon|^{q\vee4} <\infty$ with $\lim_{n\rightarrow\infty}\underline{\mu}^{-1}\Big[\sqrt{B_{1}} + n^{-\frac{1}{2} + \frac{2}{q -\beta}} \sqrt{\log{(n \vee p)}} \Big]$ $= 0$ for some $0<\beta<q$.  Then 
		$\lim_{n\rightarrow\infty}\mathbb{P}(\widehat{l} \in S^*) = 1$.
		If furthermore Condition~\ref{bound} holds with  $\lim\sup_{n\rightarrow\infty}\overbar{M}_{n}^2B_{1} <\infty$, then $\lim_{n\rightarrow\infty}\mathbb{P}\big(\big| \big[T_{\widehat{l}}(\widehat{a}_{\widehat{l}})\big]\big[ \widehat{\sigma}_{\widehat{l}}(\widehat{a}_{\widehat{l}})\big]^{-1} \big|\le t\big) =0$ for  each  $t>0$.
	\end{theorem}

	Theorems~\ref{theorem7}--\ref{theorem8}  show that the consistency of tests \eqref{h.3} and \eqref{h.4} with break selection respectively require \eqref{minimum.signal.3} and \eqref{minimum.signal.1}, in addition to regularity conditions that have been commented on previously. We now give signal strength lower bounds on $\underline{\mu}$ under a general model setting in  Example~\ref{dist.1} below.  Note that Example~\ref{sd.exmp.4} satisfies the requirements of Example~\ref{dist.1} on the standard deviation function.
	Without loss of generality, let $S^* = \{1, \dots, s^*\}$ for some integer $s^*\ge 1$, and let $\zeta_{j}(x)$ be defined as in Section~\ref{Sec4.1}. 	
	\begin{exmp}\label{dist.1}
		Assume that $(\boldsymbol{X}, \widetilde{\boldsymbol{X}})$ takes values on $[0, 1]^{2p}$ and $\underline{c}|A|_{L}< \mathbb{P}\big\{(\boldsymbol{X}, \widetilde{\boldsymbol{X}})\in A\big\}\le \bar{c}|A|_{L}$ for every Borel set $A\subset[0,1]^{2p}$ for some $0<\underline{c}\le \bar{c}$, where $|A|_{L}$ is the Lebesgue measure of $A$. Assume $\inf_{\vec{z}\in\mathbb{R}^{s^*}}\zeta(\vec{z})>0$, $\sup_{\vec{z}\in\mathbb{R}^{s^*}}\zeta(\vec{z})<\infty$, and that for every $j\in S^*= \{1, \dots ,s^*\}$ and every $\vec{z}\in\mathbb{R}^{s^*}$,   $\zeta_{j}(x)$ is nondecreasing (or nonincreasing) in $x$. In addition, for some  $0<\delta <\frac{1}{2}, \iota>0$, and Cartesian product $\mathcal{D}\in\mathcal{R}^{s^*}$, it holds that $ \big|\zeta_{j}( 1-\delta) - \zeta_{j}( \delta)  \big| > \iota$ for every $j\in S^*$ and every $\vec{z}\in \mathcal{D}$. Then, 
		$$\min_{j\in S^*}\inf_{\delta < \kappa < 1-\delta}|\mu_{j}(\kappa)| >   \frac{\iota \delta^2\underline{c}}{4\bar{c}}\times \mathbb{P}(\boldsymbol{X}_{S^*} \in \mathcal{D} )\times   \big[\inf_{\vec{z}\in\mathbb{R}^{s^*}}\zeta(\vec{z})\big].$$
	\end{exmp}
	\begin{remark}\label{identification.1}
		Relevance does not imply non-zero signal strength. For example, assume model~\eqref{model.1} with $\zeta_{1}(\boldsymbol{X}_{S_{1}}) = \zeta(\boldsymbol{X}_{S^*})$ almost surely for some $\zeta_{1}(\cdot)$ and $S_{1}\subset\{1,\dots, p\}$. If $j\in S^*$ and $j\not\in S_{1}$, then $X_{j}$ is  a relevant feature by definition but $\sup_{x\in\mathbb{R}}|\mu_{j}(x)|=0$, which is a direct result of Lemma~\ref{lemma1} and \eqref{key.ob} since  $\zeta_{1}(\boldsymbol{X}_{S_{1}}) = \zeta(\boldsymbol{X}_{S^*})$ and $j\not\in S_{1}$.
	\end{remark}

	In Example~\ref{dist.1}, with $\underline{c}, \bar{c}, \delta, \mathbb{P}(\boldsymbol{X}_{S^*} \in \mathcal{D} ) , \inf_{\vec{z}\in\mathbb{R}^{s^*}}\zeta(\vec{z})$ being positive constants and	
	\begin{equation}
		\label{kappa.1}\kappa \in [\delta, 1-\delta]
	\end{equation}
	for some $\kappa\in\{\kappa_{1}, \dots ,\kappa_{R}\}$, \eqref{minimum.signal.3}--\eqref{minimum.signal.1} are respectively satisfied if
			$\lim_{n\rightarrow\infty}\iota^{-1}\big(\sqrt{B_{1}} +\sqrt{n}\big)(\log{n}) =0$ and
			$\lim_{n\rightarrow\infty}\iota^{-1}\Big[\sqrt{B_{1}} +  n^{-\frac{1}{2} + \frac{2}{q -\beta}} \sqrt{\log{(n \vee p)}} \Big] = 0$,
	which state that the variation level $\iota$ of the standard deviation function cannot be too small in comparison with $\big(\sqrt{B_{1}} +\sqrt{n}\big)(\log{n})$ or $\sqrt{B_{1}} + n^{-\frac{1}{2} + \frac{2}{q -\beta}} \sqrt{\log{(n \vee p)}}$. These terms include the mean estimation error $B_1$ because model \eqref{model.1} has a mean component. On the other hand, \eqref{kappa.1} is satisfied if the range between $\kappa_{1}$ and $\kappa_{R}$ is reasonably wide and $\kappa_{1}<\dots < \kappa_{R}$ are evenly distributed with a proper constant $R>1$.

	It is worth noting that the test consistency for tests based on the data-driven break candidates given in Section~\ref{Sec2.4} can also be established in ways similar to  Theorems~\ref{theorem7}--\ref{theorem8}, where break candidates are assumed to be predetermined. Particularly, the set of data-driven break candidates on each coordinate should have a reasonably wide range in a probability sense. However, we omit the detailed analysis for simplicity.

\section{Simulation study}\label{Sec5b}
We have performed numerous simulation experiments showing that the VD and VDBP tests
compare favorably to existing methods such as DGLM~\citep{smyth1989generalized} and high-dimensional extensions of the Breusch-Pagan  test~\citep{daye2012high, chiou2020variable} in terms of controlling false positive errors. In these experiments, the mean functions may have interaction components with high-dimensional feature inputs.  Due to the space limitation, we defer most of our  simulation results to the Supplementary Material, and present here a brief comparison of the VDBP test and the standard Breusch-Pagan test.

\subsection{Simulation setting and VDBP test for hypothesis \eqref{BP.1}}\label{Sec5.1b}
We simulate  a sample $\mathcal{X}_{n} \coloneqq \{Q_{i}, \boldsymbol{Z}_{i}, \widetilde{\boldsymbol{Z}}_{i}\}_{i=1}^{n}$ of i.i.d. observations with $n= 700$ such that $(Q_{1}, \boldsymbol{Z}_{1})$ and $(Y, \boldsymbol{X})$ have the same distribution given as follows. The response $Y$ is generated from one of the following models:
{\small \begin{align}
		Y &= 2X_{1}X_{2} + X_{3} + X_{4} + X_{5}^2+  \varepsilon,	\label{model.29}\\
		Y &= 2X_{1}X_{2} + X_{3} + X_{4} + X_{5}^2+ \Big[1 + 3\times \boldsymbol{1}_{\{X_{15}>0\}}\Big] \varepsilon,	\label{model.30}
\end{align}}%
where $\varepsilon$ is an independent standard Gaussian model error, and $\boldsymbol{X} = (X_{1}, \dots ,X_{p})^{\top}$ with $p =20$ is  a multivariate Gaussian  vector with zero mean  and covariance matrix $\Sigma=[\Sigma_{lk}]_{l,k=1}^p$, in which $\Sigma_{lk}= (0.6)^{|l-k|}$. We use the R package \texttt{mvtnorm} for sampling $\{\boldsymbol{Z}_{i}\}_{i=1}^n$, while their  knockoff features $\{\widetilde{\boldsymbol{Z}}_{i}\}_{i=1}^n$ are generated by our coordinate-wise Gaussian knockoff generator mentioned in Section~\ref{Sec2.1.2} with $\{\boldsymbol{Z}_{i}\}_{i=1}^n$ given.

The VDBP test statistic given by \eqref{h.4} is calculated for each simulation experiment, with details given as follows. The simulated sample  $\mathcal{X}_{n}$  is split into two subsamples $\mathcal{X}_{1} = \{Y_{i}, \boldsymbol{X}_{i}, \widetilde{\boldsymbol{X}}_{i}\}_{i=1}^{n_{1}}$ and $\mathcal{X}_{2} = \{V_{i}, \boldsymbol{U}_{i}, \widetilde{\boldsymbol{U}}_{i}\}_{i=1}^{n_{2}}$ with $n_{1} = \nint{\frac{n}{3}}$ and $n_{2} = n-n_{1}$ for respectively constructing test statistics and selecting breaks. Our practical implementation of the VDBP test uses the full sample $\mathcal{X}_{n}$ for training the random forests estimate of mean functions. 
In addition, for each $l\in \{1, \dots, p\}$, we set break candidates  $(\kappa_{1,l}, \kappa_{R,l})$ to respectively  the first and third quartiles of $(U_{1l}, \dots, U_{n_{2}l}$) with $R=100$, $\kappa_{r, l} < \kappa_{r+1, l} $, and evenly distributed $\kappa_{r,l}$'s. By Theorem~\ref{theorem4}, we set the rejection thresholds $(t_{0.1}, t_{0.05}) = (1.64, 1.96)$ for the VDBP test. Meanwhile, the benchmark method here is the Breusch-Pagan test~\citep{breusch1979simple}, whose P-value is available from the R package \texttt{lmtest}. We report the empirical rejection rates of each case over $100$ repetitions in Table~\ref{tab:bp3}, with details indicated in each panel.

\subsection{Simulation results}\label{Sec5.results}
In Table~\ref{tab:bp3}, the VDBP test compares favorably to the Breusch-Pagan test in terms of controlling the empirical wrong rejection rates under model \eqref{model.29}, while the test power of the Breusch-Pagan test is slightly better than the VDBP test under model~\eqref{model.30}. It is seen that the Breusch-Pagan test does not control the false positive error rates at the target significant levels. Such results are expected because the mean functions here contain some nonlinear components, which cannot be dealt with by most existing tests for heteroskedasticity. These results are the basic motivation for the proposed tests. For more simulation results, see the Supplementary Material.
\begin{table}[h]
	\begin{center}
	\subfloat[][VDBP test ]{
		{\small
			\begin{tabular}[t]{  r|cc}
				Model & Model~\eqref{model.29}& Model~\eqref{model.30}
				\\ \hline
				$\alpha=0.1$ &0.05&0.89\\
				$\alpha=0.05$ &0.02&0.86\\

	\end{tabular} } 	}		
	\subfloat[][Breusch-Pagan test  ]{ {\small
			\begin{tabular}[t]{  r|cc}
				Model & Model~\eqref{model.29}& Model~\eqref{model.30}
				\\ \hline
				$\alpha=0.1$ &0.20&0.94\\
				$\alpha=0.05$ &0.10&0.90\\

	\end{tabular} } }
	
	\caption{The empirical rejection rates  for hypothesis \eqref{BP.1} for each case at each significance level $\alpha\in\{0.1, 0.05\}$ over $100$ simulation repetitions. 
	}\label{tab:bp3}
\end{center}	
\end{table}
\vspace{-1.5cm}
	\section{Real data study}\label{Sec6}
	\subsection{HIV-1 drug resistance}\label{Sec6.1}

	\begin{table}[h]
		\begin{center}
			\begin{tabular}[t]{ c c wl{3.8cm}{l}c}
				Drug Type&Drug Name&Significant at $\alpha = 0.01$ &  FDR $\le 0.2$  & \texttt{\#}Samples\texttt{\big/}\texttt{\#}Features\\ \hline 
				\multirow{7}{*}{PI}&APV& \{46, I\}&$\varnothing$&767\texttt{\big/}320 \\ \cline{2-5}
				&ATV&$\varnothing$ & $\varnothing$&328\texttt{\big/}250\\ \cline{2-5}
				&IDV&(12, A)& $\varnothing$ &825\texttt{\big/}327\\ \cline{2-5}
				&LPV&(77, I)& $\varnothing$&515\texttt{\big/}287\\ \cline{2-5}
				&NFV&(63, P)& (63, P)&842\texttt{\big/}331\\ \cline{2-5}
				&RTV&(63, P)& $\varnothing$ &793\texttt{\big/}329\\[-0.8pt] \cline{2-5}
				&\multirow{1}{*}{SQV}& \{46, I\}& \multirow{1}{*}{$\varnothing$}&\multirow{1}{*}{824\texttt{\big/}330}\\
				\hline		
				\hline		
				\multirow{6}{*}{NRTI}&3TC& $\varnothing$&$\varnothing$&629\texttt{\big/}524 \\ \cline{2-5}
				&ABC&$\varnothing$& $\varnothing$&623\texttt{\big/}524\\ \cline{2-5}
				&AZT&$\varnothing$& $\varnothing$ &626\texttt{\big/}523\\ \cline{2-5}
				&D4T&$\varnothing$& $\varnothing$&625\texttt{\big/}523\\ \cline{2-5}
				&DDI&$\varnothing$&$\varnothing$&628\texttt{\big/}524\\ \cline{2-5}
				&TDF&$\varnothing$& $\varnothing$ &351\texttt{\big/}400\\
				\hline
				\hline
				\multirow{3}{*}{NNRTI}&DLV& $\varnothing$&$\varnothing$&730\texttt{\big/}554 \\ \cline{2-5}
				&EFV&(74, V)&(74, V) &732\texttt{\big/}560\\ \cline{2-5}
				&NVP& (135, T)& $\varnothing$ &744\texttt{\big/}561\\ 
				\hline
			\end{tabular} 
			\caption{Each explanatory feature in the HIV-1 dataset is a pair of position/mutation type. Sizes of available samples differ across drugs since each participant is usually treated with a few drugs. Mutation positions of those pairs in curly brackets are in the ground truth sets of active features for the (conditional) means of drug resistance levels.}
			\label{tab:2}
		\end{center}
	\end{table}

	\begin{figure}[h]
		\centering
		\centering
		\subfloat
		{\includegraphics[width=5.cm]{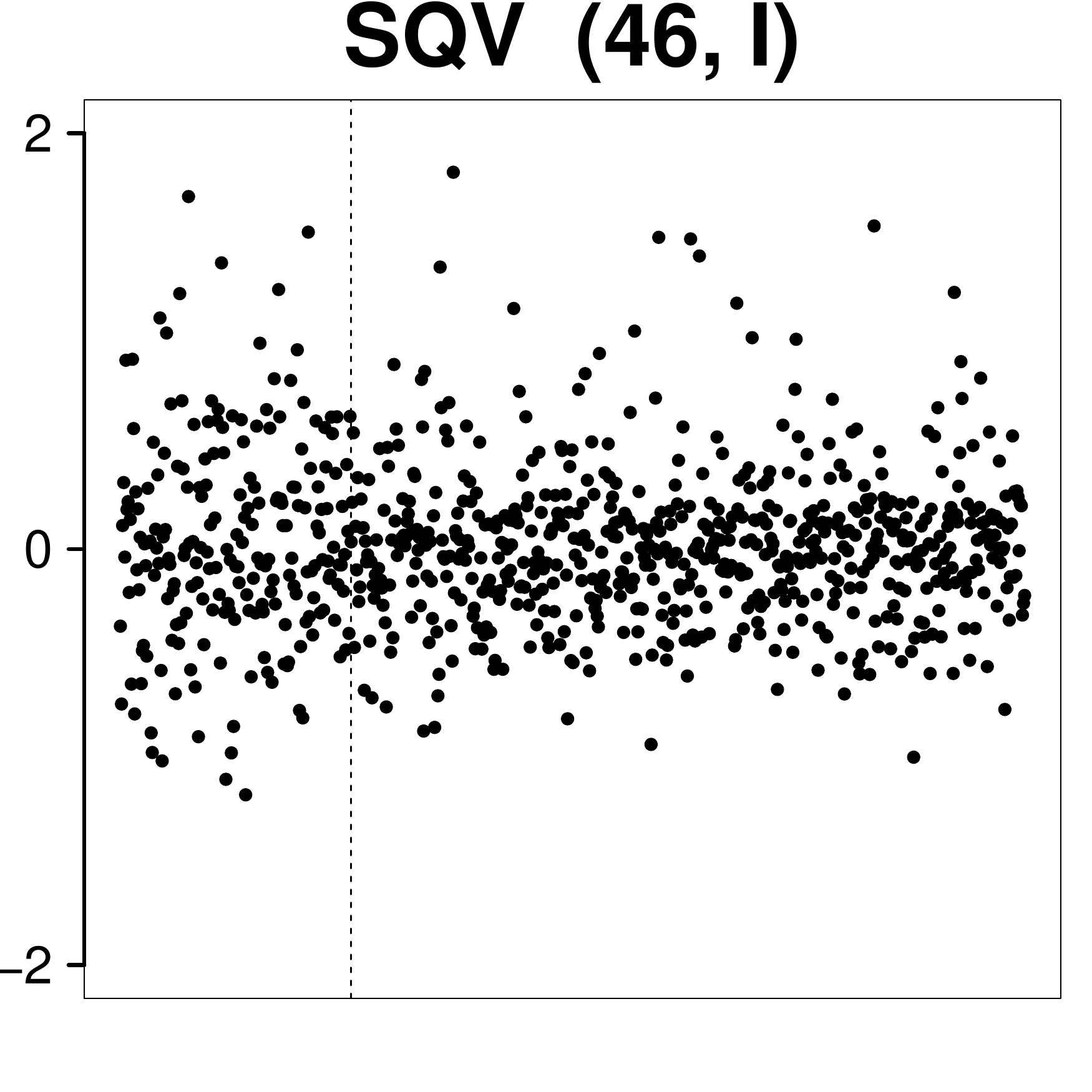}}
		\subfloat
		{\includegraphics[width=5.cm]{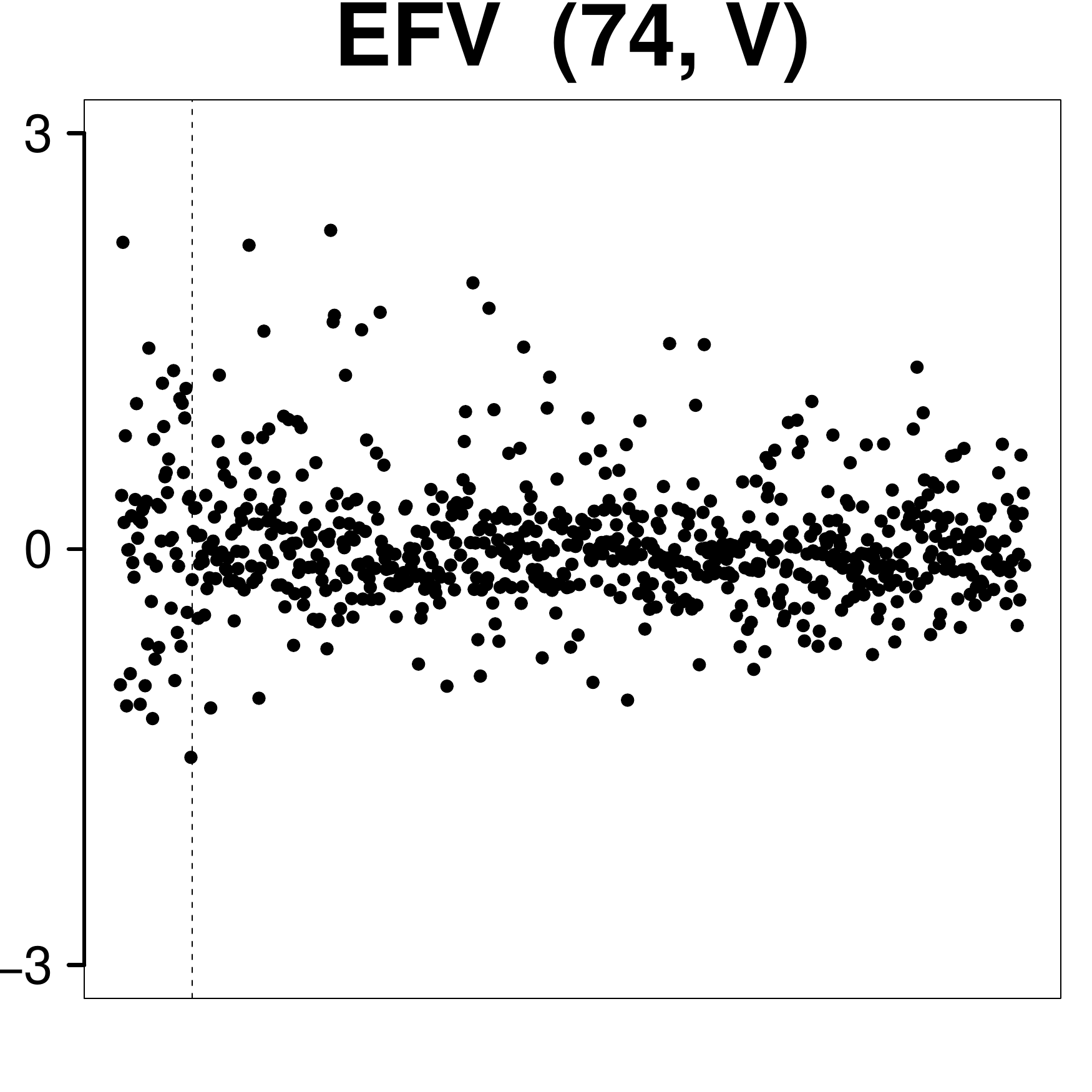}}		
%
		\subfloat
		{\includegraphics[width=5.cm]{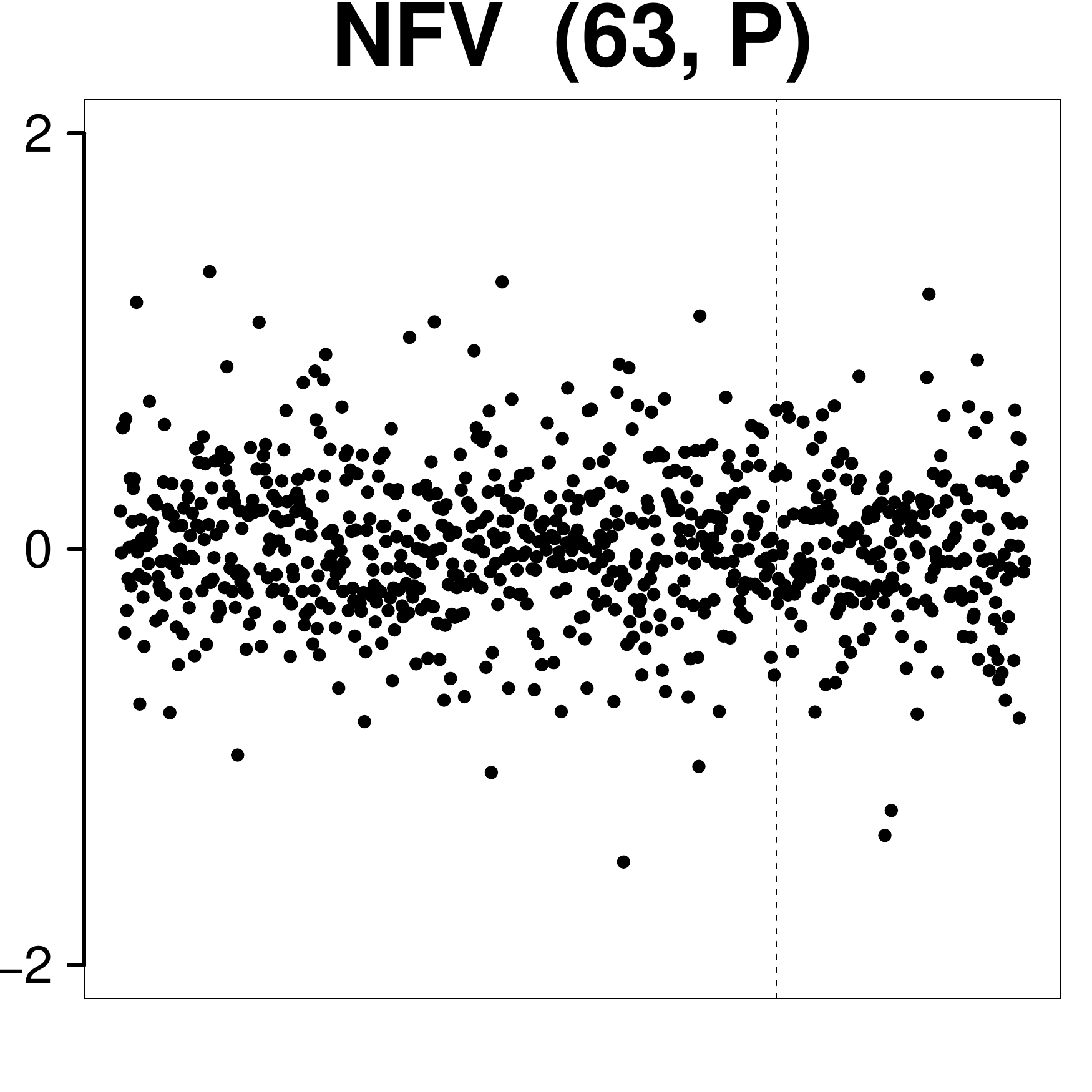}}
		\caption{The drug names and mutations of interest (genetic position, mutation type) are indicated on the top of each picture. In each picture, samples on the left-hand side of the dashed line indicate those samples having the mutation of interest; for example, in panel SQV, samples on the left of the dashed line have mutation (46, I). }
		\label{fig:41}
	\end{figure}
	In this section, we apply the VD test to study which human genetic mutations are associated with the conditional variance of drug resistance level for Human Immunodeficiency Virus Type 1 (HIV-1)~\citep{rhee2006genotypic}. Our inferences provides further understanding of which mutations contribute to the risk of having high drug resistance. Such results complement the previous works on the inference for identifying mutations that are associated with the conditional mean of drug resistance level~\citep{tian2016magic, Barber2015}.

	In this HIV-1 dataset, $16$ drugs for treating HIV-1 are considered: seven protease
	inhibitors (PIs), six nucleoside reverse transcriptase  inhibitors (NRTIs), and three
	nonnucleoside reverse transcriptase inhibitors (NNRTIs). Each participant with HIV-1 is treated with some of these drugs. For each drug, the resulting drug resistance levels and these participants' genetic information are observed and  recorded, with the former taken as the response variable and the latter taken as the explanatory features; each explanatory feature is binary and indicates whether the individual has the genetic mutation at a specific genetic position or not. Accordingly, we have $16$ samples for each drug as displayed in Table~\ref{tab:2}. For each of these samples, the resistance levels
	are log-transformed~\citep{Barber2015}; meanwhile, we remove duplicate features, features with zero mutations, and observations with missing values from each sample. The number of binary features and sample size of each sample are reported in Table~\ref{tab:2}. In addition, the distribution of mutations are assumed to follow a hidden Markov model (HMM) distribution~\citep{sesia2019gene}, and their knockoff features are generated using the HMM knockoffs developed by Sesia et al.~\citep{sesia2019gene} with default setting. See Section~\ref{Sec7.2} of the Supplementary Material for a numerical experiment assuring the reliability of HMM knockoffs for our HIV-1 study. With knockoff features, the VD statistic in \eqref{h.1} is calculated for each feature with the full sample for training the random forests centering  model, and a fixed break at $0.5$. Their P-values are obtained according to Theorem~\ref{theorem1}.

	In Table~\ref{tab:2}, we report the significant mutations at $\alpha=0.01$ or FDR $=0.2$, where each mutation is denoted by a pair  (mutation position, type), and the FDR is controlled by the method Q-value~\citep{storeyq2022} in the R package \texttt{qvalue}. In this table, pairs in curly brackets are those whose mutation positions appear in known sets of active mutations for conditional means of drug resistance levels. These sets of mutations are identified by another experiment and are taken as the ``ground truth'' sets of active mutations for conditional means. Details of these mutations are given in the mutation set ``TSMs'' in Table 1 of ~\citep{rhee2006genotypic}.

	We also provide preliminary diagnosis in Figure~\ref{fig:41} showing the effects of significant mutations on the conditional variance of drug resistance level.  These pictures respectively present  residual plots for the SQV, EFV, and NFV samples, where residuals are given by $\widehat{\varepsilon}_{i} = Y_{i} - \widehat{m}(\boldsymbol{X}_{i})$'s, in which $Y_{i}$'s are drug resistance levels, $\boldsymbol{X}_{i}$'s are binary vectors of mutations for each individual, and $\widehat{m}(\boldsymbol{X}_{i})$'s are the random forests predictions for $Y_{i}$'s. The x-axis in these figures are observation indices, and observed data on the left-hand side of the dashed lines are those having the significant mutations of interest, which are indicated on the top of each plot. Other details are in their figure captions.	
	
	\subsection{Results}
	Our study of HIV-1 drug resistance in Table~\ref{tab:2} and Figure~\ref{fig:41} reveals some interesting results, which are reviewed as follows. From Table~\ref{tab:2},  mutations (63, P)  and (74, V) are significant while FDR is controlled at $0.2$, but they are not in the ground truth sets for mean effects on drug resistance. These results suggest that there may exist more general distributional relations between mutations and drug resistance than mean effects that have been studied by~\citep{rhee2006genotypic, Barber2015}. In addition, Figure~\ref{fig:41} provides an informative preliminary diagnosis for the log-transformed drug resistance levels. There, we see that some significant  mutations have clear effects on conditional variance (e.g. the panels SQV and EFV) and some have less clear effects on conditional variance (e.g. the panel NFV). We note that in general, mutations with clear effects on conditional variance in the preliminary diagnosis could be spurious mutations; for example, some  irrelevant mutations may be correlated with relevant ones. It is therefore important to apply an appropriate test such as the VD test to assess significance. On the whole, our findings should encourage further investigation into the effects of mutations on variability of HIV-1 drug resistance, which may help develop more effective HIV-1 drugs with reduced drug resistance variability and optimize the use of available drugs.
	
	\section{Discussions}

	In this paper, we have established rigorous P-values and test sizes for the VD and VDBP tests, and we have also analyzed the test power under a nonparametric heteroskedastic data generating model with high-dimensional input features, which, to our knowledge, are the first rigorous tests for high-dimensional regression heteroskedasticity that allow nonlinear mean regression and flexible standard deviation regression. There are several potential extensions of our work. For example, it may be possible to strengthen our tests so they are applicable to heteroskedastic models with non-monotonic standard deviation functions. Such an extension takes into account squared components in $\zeta(\boldsymbol{X}_{S^*})$, which are common heteroskedastic components in economics and finance. In addition, it is possible to incorporate the idea of coordinate-wise knockoff  into existing knockoff generators~\citep{romano2020deep, jordon2018knockoffgan, lu2018deeppink}. With these coordinate-wise distribution-free knockoff generators, we may apply the VD and VDBP tests to cases with unknown  feature distribution. These applications are also interesting future work.

	\begin{center}
		{\large\bf SUPPLEMENTARY MATERIAL}
	\end{center}
	
	\begin{description}
		
		\item[Supplementary Material:] Technical proofs and extensive simulation results are in this file (.pdf type). Codes for our simulation experiments are available upon request.

		\item[HIV-1 data set:] Data set used in Section~\ref{Sec6} can be downloaded from \href{https://hivdb.stanford.edu/pages/published_analysis/genophenoPNAS2006/DATA/}{Stanford HIV Drug Resistance Database}  (.txt file). The processed data set, as described in Section~\ref{Sec6}, is available upon request.
		
	\end{description}

%
	
	\bibliographystyle{chicago}
	\bibliography{references}
	
	\newpage
	\appendix
	\setcounter{page}{1}
	\setcounter{section}{0}
	\renewcommand{\theequation}{A.\arabic{equation}}
	\renewcommand{\thesubsection}{A.\arabic{subsection}}
	\setcounter{equation}{0}
	
	\begin{center}{\bf \Large Supplementary Material to ``Testing for Regression Heteroskedasticity with High-Dimensional Random Forests''}
		
		\if1\blind
		{
			\bigskip
			
			Chien-Ming Chi
		} \fi
	
		
	\end{center}
	
	\noindent This Supplementary Material contains the proofs of Theorems \ref{theorem1}--\ref{theorem8} in Section~\ref{SecA}, the proofs of Lemma~\ref{lemma1} and Example~\ref{dist.1} in Section~\ref{SecB}, and some technical lemmas and their proofs in Section~\ref{SecC}. Simulation experiments for the VD and VDBP tests with synthetic data are in Section~\ref{SecD.1}, and the numerical experiment for hidden Markov model knockoffs~\citepsupp{sesia2019gene} are in Section~\ref{Sec7.2}. In addition, in Section~\ref{coordinate-wise}, we develop a coordinate-wise Gaussian knockoff generator for our simulation experiments; see Section~\ref{SecD.1.1} and Section~\ref{coordinate-wise} for details.

	All the notation is the same as defined in the main body of the paper. In addition, we use the following notation in the Supplementary Material. For a  matrix $\boldsymbol{\Sigma}$ of size $(K\times L)$, we use $\Sigma_{kl}$ to denote its entry at the $k$th row and $l$th column. Throughout the proofs, we use  generic constants such as $C$ and $K$; unless specified otherwise, these constants are independent of the sample size. For convenience, we use the little $o$ and big $O$ notation: for two real sequences $\{a_{n}, b_{n}\}$, $a_{n}= o(b_{n})$ means that  $\lim_{n\rightarrow\infty} \frac{a_{n}}{b_{n}} = 0$; $a_{n} = O(b_{n})$ means that $\lim_{n\rightarrow\infty} \frac{|a_{n}|}{|b_{n}|} <\infty$.

	\renewcommand{\thesubsection}{A.\arabic{subsection}}
	
	\section{Proof of theorems}\label{SecA}

	\subsection{Proof of Theorem~\ref{theorem1}}\label{theorem1.proof}
	
	Let us begin with a remark that model \eqref{model.1} is allowed to depend on sample size $n$, and we do not indicate the dependence of model \eqref{model.1} on $n$ explicitly for simplicity. In addition, recall that we have defined $\zeta(\boldsymbol{X}_{S^*}) = \zeta_{0}$ if $S^* = \emptyset$ in Section~\ref{Sec1.1}; the proof here applies to cases with empty or non-empty $S^*$.
	
	Now, the formal proof of Theorem~\ref{theorem1} begins with \eqref{theorem1.4} and \eqref{theorem1.1} below. We will briefly explain our proof strategy  after the inequality \eqref{theorem1.1}.
	\begin{equation}
		\begin{split}\label{theorem1.4}
			T_{j}(a) &=n^{-\frac{1}{2}}\sum_{i=1}^{n}\big(\boldsymbol{1}_{X_{ij}\in\mathcal{A}} -\boldsymbol{1}_{\widetilde{X}_{ij}\in\mathcal{A}}\big) (\widehat{\varepsilon}_{i} )^{2}\\			
			& = n^{-\frac{1}{2}}\sum_{i=1}^{n}\big(\boldsymbol{1}_{X_{ij}\in\mathcal{A}} -\boldsymbol{1}_{\widetilde{X}_{ij}\in\mathcal{A}}\big) (m(\boldsymbol{X}_{i}) - \widehat{m}(\boldsymbol{X}_{i}))^2 \\
			& \hspace{2em}- 2n^{-\frac{1}{2}}\sum_{i=1}^{n}\big(\boldsymbol{1}_{X_{ij}\in\mathcal{A}} -\boldsymbol{1}_{\widetilde{X}_{ij}\in\mathcal{A}}\big) (m(\boldsymbol{X}_{i}) - \widehat{m}(\boldsymbol{X}_{i}))\zeta(\boldsymbol{X}_{iS^{*}})\varepsilon_{i}\\
			&\hspace{2em}+ n^{-\frac{1}{2}}\sum_{i=1}^{n}\big(\boldsymbol{1}_{X_{ij}\in\mathcal{A}} -\boldsymbol{1}_{\widetilde{X}_{ij}\in\mathcal{A}}\big)(\zeta(\boldsymbol{X}_{iS^{*}})\varepsilon_{i})^{2}\\
			& \eqqcolon A_{1} + A_{2} + T_{j}^{(\star)}(a) ,
		\end{split}
	\end{equation}
	where  $\mathcal{A}= (-\infty, a]$, $T_{j}^{(\star)}(a) =n^{-\frac{1}{2}}\sum_{i=1}^{n}\big[\boldsymbol{1}_{X_{ij}\in\mathcal{A}} -\boldsymbol{1}_{\widetilde{X}_{ij}\in\mathcal{A}}\big]\big[\zeta(\boldsymbol{X}_{iS^{*}})\varepsilon_{i}\big]^{2}$, and $\widehat{\varepsilon}_{i} = Y_{i} - \widehat{m}(\boldsymbol{X}_{i})$. In addition,
	\begin{equation}
		\begin{split}\label{theorem1.1}
			&\mathbb{P}(|T_{j}(a)\big[\widehat{\sigma}_{j}(a)\big]^{-1}| \ge t ) \\
			&\le \mathbb{P}(|T_{j}^{(\star)}(a)| \ge t\big[\sigma_{j}(a)\big] -\varsigma - t|\sigma_{j}(a) - \widehat{\sigma}_{j}(a)| -|A_{1}| - |A_{2}|+ \varsigma),
		\end{split}
	\end{equation}
	where 
	$$\sigma_{j}^2(a) \coloneqq \textnormal{Var}\big\{\big[\boldsymbol{1}_{X_{j}\in\mathcal{A}} -\boldsymbol{1}_{\widetilde{X}_{j}\in\mathcal{A}}\big] \big[\zeta(\boldsymbol{X}_{S^{*}})\varepsilon\big]^2\big\}=  \mathbb{E}\big\{\big[\boldsymbol{1}_{X_{j}\in\mathcal{A}} -\boldsymbol{1}_{\widetilde{X}_{j}\in\mathcal{A}}\big]^2\big[\zeta(\boldsymbol{X}_{S^{*}})\varepsilon\big]^{4}\big\}$$ 
	since 
	\begin{equation}
		\label{kmc.1}
		\mathbb{E}\big\{\big[\boldsymbol{1}_{X_{j}\in\mathcal{A}} -\boldsymbol{1}_{\widetilde{X}_{j}\in\mathcal{A}}\big]\big[\zeta(\boldsymbol{X}_{S^{*}})\varepsilon\big]^{2}\big\} = 0,
	\end{equation} 
	which is due to the assumption that $j\not\in S^*$ and the definition of knockoff features (see \eqref{key.ob} for details), and  we assume $\widehat{\sigma}_{j}(a)>0$ to avoid division by zero for simplicity.

	From \eqref{theorem1.1}, it is seen that the desired result follows if $\varsigma\ge  t|\sigma_{j}(a) - \widehat{\sigma}_{j}(a)| + |A_{1}| + |A_{2}|$ in a probability sense and that $\mathbb{P}(|T_{j}^{(\star)}(a)| \ge t\big[\sigma_{j}(a)\big] -\varsigma)$ is bounded by $2\Phi(-t)$ plus an additional term, which is negligible if $\lim_{n\rightarrow \infty}\sqrt{n}B_{1} + (\overbar{M}_{n})^2(\log{n})n^{-\frac{1}{4}}  = 0$ (see the comments after Theorem~\ref{theorem1} in the main text for details). In the following, we will show that these two results both hold. Recall that $\varsigma = t(\log{n})(10n^{-\frac{1}{4}} + 4B_{1}^{\frac{1}{8}}) + n^{\frac{1}{4}}B_{1}^{\frac{1}{2}}$.
	
	First, we establish the upper bounds for $|\widehat{\sigma}_{j}(a) -\sigma_{j}(a)|$, $|A_{1}|$, and $|A_{2}|$, and begin with the one for $|\widehat{\sigma}_{j}(a) -\sigma_{j}(a)|$. By Condition~\ref{non.trivial.2} with $(R,\kappa_{1} )=(1,a)$ and  that $\varepsilon$ is independent of $(\boldsymbol{X}, \widetilde{X}_{j})$, it holds that 
	\begin{equation}
		\begin{split}\label{theorem1.9}
			\sigma_{j}^2(a) & = \mathbb{E}\big\{\big[\boldsymbol{1}_{X_{j}\in\mathcal{A}} -\boldsymbol{1}_{\widetilde{X}_{j}\in\mathcal{A}}\big]^2\big[\zeta(\boldsymbol{X}_{S^{*}})\varepsilon\big]^{4}\big\}\\
			& \ge \mathbb{E}\big\{\boldsymbol{1}_{\{X_{j}\in\mathcal{A}\}\cap\{\widetilde{X}_{j}\in\mathcal{A}^c\} } \big[\zeta(\boldsymbol{X}_{S^{*}})\varepsilon \big]^{4} \big\} \\
			& \ge  c\big[\inf_{\vec{z}\in\mathbb{R}^{s^*}} \zeta(\vec{z})\big]^4\mathbb{E}(\varepsilon^{4} ),
		\end{split}
	\end{equation}
	where $c>0$ is defined in Condition~\ref{non.trivial.2}. The result of \eqref{theorem1.9} leads to
	\begin{equation}\label{theorem1.30}
		\begin{split}
			|\widehat{\sigma}_{j}(a) - \sigma_{j}(a)| &= \frac{|\widehat{\sigma}_{j}^{2}(a) - \sigma_{j}^{2}(a)|}{\widehat{\sigma}_{j}(a) + \sigma_{j}(a)} \\
			&\le \frac{|\widehat{\sigma}_{j}^{2}(a) - \sigma_{j}^{2}(a)|}{ \sigma_{j}(a)} \\
			&\le \big\{c\big[\inf_{\vec{z}\in\mathbb{R}^{s^*}}\zeta(\vec{z})\big]^4\mathbb{E}(\varepsilon^{4} ) \big\}^{-\frac{1}{2}}|\widehat{\sigma}_{j}^{2}(a) - \sigma_{j}^{2}(a)|.
		\end{split}
	\end{equation}
	Next, we write
	\begin{equation}\label{theorem1.5}
		\begin{split}
			&\widehat{\sigma}_{j}^2(a) -\sigma_{j}^2(a)\\
			&= -\sigma_{j}^2(a) -\big[\widehat{\mu}_{j}(a)\big]^2 + n^{-1}\sum_{i=1}^{n} \big[\boldsymbol{1}_{X_{ij} \in \mathcal{A}} -\boldsymbol{1}_{\widetilde{X}_{ij} \in \mathcal{A}}\big]^2(\widehat{\varepsilon}_{i})^4 \\
			&= -\big[\widehat{\mu}_{j}(a)\big]^2 + \bigg\{  n^{-1}\sum_{i=1}^{n} \Big\{\big[\boldsymbol{1}_{X_{ij} \in \mathcal{A}} -\boldsymbol{1}_{\widetilde{X}_{ij} \in \mathcal{A}}\big]^2 \times \Big\{\big[\widehat{m}(\boldsymbol{X}_{i}) - m(\boldsymbol{X}_{i}) \big]^4\\
			&  \qquad \qquad- 4\big[\widehat{m}(\boldsymbol{X}_{i}) - m(\boldsymbol{X}_{i}) \big]^3 \zeta(\boldsymbol{X}_{iS^{*}})\varepsilon_{i}+ 6\big[\widehat{m}(\boldsymbol{X}_{i}) - m(\boldsymbol{X}_{i}) \big]^2 \big[\zeta(\boldsymbol{X}_{iS^{*}})\varepsilon_{i}\big]^2 \\
			&\qquad\qquad- 4\big[\widehat{m}(\boldsymbol{X}_{i}) - m(\boldsymbol{X}_{i}) \big] \big[\zeta(\boldsymbol{X}_{iS^{*}})\varepsilon_{i}\big]^3 \Big\}\Big\}\bigg\} \\
			&\qquad + n^{-1}\sum_{i=1}^{n} \Big\{\big[\boldsymbol{1}_{X_{ij} \in \mathcal{A}} -\boldsymbol{1}_{\widetilde{X}_{ij} \in \mathcal{A}}\big]^2\big[ \zeta(\boldsymbol{X}_{iS^{*}})\varepsilon_{i}\big]^4 - \sigma_{j}^2(a)\Big\},
		\end{split}
	\end{equation}
	where $\widehat{\mu}_{j}(a) = n^{-1}\sum_{i=1}^{n}\big(\boldsymbol{1}_{X_{ij}\in\mathcal{A}}- \boldsymbol{1}_{\widetilde{X}_{ij}\in\mathcal{A}}\big)\big(\widehat{\varepsilon}_{i} \big)^{2}$, and define  events $E_{1}, \dots, E_{8}$ as follows.
	\begin{equation*}
		\begin{split}
			E_{1} & = \left\{\left|n^{-\frac{1}{2}}\sum_{i=1}^{n}\big(\boldsymbol{1}_{X_{ij}\in\mathcal{A}} -\boldsymbol{1}_{\widetilde{X}_{ij}\in\mathcal{A}}\big) \big[m(\boldsymbol{X}_{i}) - \widehat{m}(\boldsymbol{X}_{i})\big]^2\right| \le n^{\frac{1}{4}}B_{1}^{\frac{1}{2}}\right\},\\
			E_{2}& = \left\{ \left|2n^{-\frac{1}{2}}\sum_{i=1}^{n}\big(\boldsymbol{1}_{X_{ij}\in\mathcal{A}} -\boldsymbol{1}_{\widetilde{X}_{ij}\in\mathcal{A}}\big)\big[m(\boldsymbol{X}_{i}) - \widehat{m}(\boldsymbol{X}_{i})\big]\zeta(\boldsymbol{X}_{iS^{*}})\varepsilon_{i} \right| \le B_{1}^{\frac{1}{4}}\right\},\\
			E_{3} &= \left\{\left|n^{-\frac{1}{2}}\sum_{i=1}^{n}\big(\boldsymbol{1}_{X_{ij}\in\mathcal{A}} -\boldsymbol{1}_{\widetilde{X}_{ij}\in\mathcal{A}}\big) \big[\zeta(\boldsymbol{X}_{iS^{*}})\varepsilon_{i}\big]^{2}\right|\le n^{\frac{1}{4}}\right\},\\
			E_{4} &= \left\{ \left| n^{-1}\sum_{i=1}^{n} (\boldsymbol{1}_{X_{ij} \in \mathcal{A}} -\boldsymbol{1}_{\widetilde{X}_{ij} \in \mathcal{A}})^2 \big[\widehat{m}(\boldsymbol{X}_{i}) - m(\boldsymbol{X}_{i}) \big]^4\right| \le B_{1}^{\frac{1}{2}}\right\},\\
			E_{5} &= \left\{ \left| 4n^{-1}\sum_{i=1}^{n} (\boldsymbol{1}_{X_{ij} \in \mathcal{A}} -\boldsymbol{1}_{\widetilde{X}_{ij} \in \mathcal{A}})^2 \big[\widehat{m}(\boldsymbol{X}_{i}) - m(\boldsymbol{X}_{i}) \big]^3 \zeta(\boldsymbol{X}_{iS^{*}})\varepsilon_{i}\right| \le B_{1}^{\frac{1}{2}}\right\} ,\\
			E_{6} &= \left\{ \left| 4n^{-1}\sum_{i=1}^{n} (\boldsymbol{1}_{X_{ij} \in \mathcal{A}} -\boldsymbol{1}_{\widetilde{X}_{ij} \in \mathcal{A}})^2 \big[\widehat{m}(\boldsymbol{X}_{i}) - m(\boldsymbol{X}_{i}) \big] \big[\zeta(\boldsymbol{X}_{iS^{*}})\varepsilon_{i}\big]^3\right| \le B_{1}^{\frac{1}{4}}\right\},\\
			E_{7} & = \left\{\left| 6n^{-1}\sum_{i=1}^{n} (\boldsymbol{1}_{X_{ij} \in \mathcal{A}} -\boldsymbol{1}_{\widetilde{X}_{ij} \in \mathcal{A}})^2 \big[\widehat{m}(\boldsymbol{X}_{i}) - m(\boldsymbol{X}_{i}) \big]^2 \big[\zeta(\boldsymbol{X}_{iS^{*}})\varepsilon_{i}\big]^2  \right|\le B_{1}^{\frac{1}{2}}\right\},\\
			E_{8} &= \left\{\left| n^{-1}\sum_{i=1}^{n} \big\{\big[\boldsymbol{1}_{X_{ij} \in \mathcal{A}} -\boldsymbol{1}_{\widetilde{X}_{ij} \in \mathcal{A}}\big]^2\big[\zeta(\boldsymbol{X}_{iS^{*}})\varepsilon_{i}\big]^4 -\sigma_{j}^2(a) \big\} \right|\le n^{-\frac{1}{4}} \right\}.
		\end{split}		
	\end{equation*}

	In light of \eqref{theorem1.30} and these events, we now establish the upper bounds for $|\widehat{\sigma}_{j}^2(a) -\sigma_{j}^2(a)|$, $|A_{1}|$, and $|A_{2}|$. On $\cap_{l=1}^3 E_{l}$,
	\begin{equation}
		\label{theorem1.6}	
		|\widehat{\mu}_{j}(a)|\le n^{-\frac{1}{4}}B_{1}^{\frac{1}{2}} + n^{-\frac{1}{2}}B_{1}^{\frac{1}{4}} + n^{-\frac{1}{4}}.
	\end{equation}	
	On $\cap_{l=1}^2E_{l}$,
	$$|A_{1}| + |A_{2}|\le n^{\frac{1}{4}}B_{1}^{\frac{1}{2}} + B_{1}^{\frac{1}{4}} .$$	
	By \eqref{theorem1.5}--\eqref{theorem1.6}, on $\cap_{l=1}^8E_{l}$, for each $0<B_{1} <1$ and all large $n$,
	\begin{equation}
		\label{theorem1.36}
		|	\widehat{\sigma}_{j}^2(a) - \sigma_{j}^2(a) | \le 10n^{-\frac{1}{4}} + 4B_{1}^{\frac{1}{8}},
	\end{equation}
	where we use the facts that $n^{-\frac{1}{4}}\le 1$ and $n^{-\frac{1}{2}}\le 1$ for all $n\ge 1$ and that $B_{1}^{\frac{1}{8}} \ge B_{1}^{\frac{1}{4}} \ge B_{1}^{\frac{1}{2}}$ if $0<B_{1}<1$ to simplify the upper bound.
	
	With these bounds and \eqref{theorem1.30}, it holds that for all large $n$ and $0<B_{1} <1$, $\varsigma\ge  t|\sigma_{j}(a) - \widehat{\sigma}_{j}(a)| + |A_{1}| + |A_{2}|$ on $\cap_{l=1}^8 E_{l}$. Hence, we deduce that for all large $n$,
	\begin{equation}
		\begin{split}	\label{theorem1.7}
			\mathbb{P}(|T_{j}(a)\big[\widehat{\sigma}_{j}(a)\big]^{-1}| \ge t ) & \le \mathbb{P}(\{|T_{j}^{(\star)} (a)| \ge t\big[\sigma_{j}(a)\big] - \varsigma\}\cap (\cap_{l=1}^8 E_{l})) + \sum_{l=1}^8 \mathbb{P}(E_{l}^{c})\\
			& \le \mathbb{P}(|T_{j}^{(\star)}(a)| \ge t\big[\sigma_{j}(a)\big]  - \varsigma) + \sum_{l=1}^8 \mathbb{P}(E_{l}^{c}).
		\end{split}
	\end{equation}

	Next, we bound the probabilities of $\mathbb{P}(E_{1}^{c}), \dots , \mathbb{P}(E_{8}^{c})$, where $E^{c}$ denotes the complementary event of an event $E$. By Markov's inequality, Condition~\ref{consistency.3}, and that the training sample is an independent sample, for all $n\ge 1$ and $B_{1}>0$,
	\begin{equation}
		\label{theorem1.prob.1}
		\mathbb{P}(E_{1}^c)\le n^{\frac{1}{4}}B_{1}^{\frac{1}{2}}.
	\end{equation}
	To deal with $\mathbb{P}(E_{2}^c)$, notice that $\{Q_{i}, \sigma(\boldsymbol{X}_{i}, \dots,\boldsymbol{X}_{1}, \mathcal{X}_{0})\}_{i\ge 1}$ with $Q_{i} = n^{-\frac{1}{2}}\big(\boldsymbol{1}_{X_{ij}\in\mathcal{A}} -\boldsymbol{1}_{\widetilde{X}_{ij}\in\mathcal{A}}\big)\big[m(\boldsymbol{X}_{i}) - \widehat{m}(\boldsymbol{X}_{i})\big]\zeta(\boldsymbol{X}_{iS^{*}})\varepsilon_{i}$ is a sequence of martingale differences, in which $\sigma(\cdot)$ denotes the $\sigma$-algebra generated by the given random mappings, and $\mathcal{X}_{0}$ denotes the independent training sample for training  $\widehat{m}(\cdot)$. Then by Markov's inequality, the Burkholder--Davis--Gundy inequality \citepsupp{burkholder1972integral} inequality, Jensen's inequality, the assumption of i.i.d. observations,  Condition~\ref{consistency.3}, and that the training sample is an independent sample, there exists $C>0$ such that for all $n\ge 1$ and $B_{1} >0$,
	\begin{equation}\label{theorem1.2}
		\begin{split}			
			&\mathbb{P}(E_{2}^c)\\
			&\le 2B_{1}^{-\frac{1}{4}}\mathbb{E}|n^{-\frac{1}{2}}\sum_{i=1}^{n}\big(\boldsymbol{1}_{X_{ij}\in\mathcal{A}} -\boldsymbol{1}_{\widetilde{X}_{ij}\in\mathcal{A}}\big)\big[m(\boldsymbol{X}_{i}) - \widehat{m}(\boldsymbol{X}_{i})\big]\zeta(\boldsymbol{X}_{iS^{*}})\varepsilon_{i}|\\
			&\le 2CB_{1}^{-\frac{1}{4}} 			
			\mathbb{E}\sqrt{\left(n^{-1}\sum_{i=1}^{n}\big(\boldsymbol{1}_{X_{ij}\in\mathcal{A}} -\boldsymbol{1}_{\widetilde{X}_{ij}\in\mathcal{A}}\big)^2\big[m(\boldsymbol{X}_{i}) - \widehat{m}(\boldsymbol{X}_{i})\big]^2 \big[\zeta(\boldsymbol{X}_{iS^{*}})\varepsilon_{i}\big]^2\right)} \\
			& \le 2CB_{1}^{-\frac{1}{4}} 			
			\sqrt{			\mathbb{E}\left(n^{-1}\sum_{i=1}^{n}\big(\boldsymbol{1}_{X_{ij}\in\mathcal{A}} -\boldsymbol{1}_{\widetilde{X}_{ij}\in\mathcal{A}}\big)^2\big[m(\boldsymbol{X}_{i}) - \widehat{m}(\boldsymbol{X}_{i})\big]^2 \big[\zeta(\boldsymbol{X}_{iS^{*}})\varepsilon_{i}\big]^2\right)} \\
			& \le 2CB_{1}^{-\frac{1}{4}} 			
			\sqrt{	\big[\sup_{\vec{z}\in\mathbb{R}^{s^*}} \zeta(\vec{z})\big]^2	\mathbb{E}(\varepsilon^2)	\mathbb{E}\big[m(\boldsymbol{X}) - \widehat{m}(\boldsymbol{X})\big]^2  } \\
			&\le 2CB_{1}^{\frac{1}{4}}\sqrt{\mathbb{E}(\varepsilon^2)}\big[\sup_{\vec{z}\in\mathbb{R}^{s^*}} \zeta(\vec{z})\big].
		\end{split}
	\end{equation}
	By arguments similar to those used in \eqref{theorem1.2}, $j\not\in S^*$, and \eqref{kmc.1}, there exists $C>0$ such that for all $n\ge 1$ and $B_{1} >0$,
	\begin{equation}
		\begin{split}
			\mathbb{P}(E_{3}^c)& \le n^{-\frac{1}{4}}\mathbb{E}\left|n^{-\frac{1}{2}}\sum_{i=1}^{n}\big(\boldsymbol{1}_{X_{ij}\in\mathcal{A}} -\boldsymbol{1}_{\widetilde{X}_{ij}\in\mathcal{A}}\big)\big[\zeta(\boldsymbol{X}_{iS^{*}})\varepsilon_{i}\big]^2\right|\\
			& \le n^{-\frac{1}{4}}C \mathbb{E}\sqrt{n^{-1}\sum_{i=1}^{n}\big(\boldsymbol{1}_{X_{ij}\in\mathcal{A}} -\boldsymbol{1}_{\widetilde{X}_{ij}\in\mathcal{A}}\big)^2\big[\zeta(\boldsymbol{X}_{iS^{*}})\varepsilon_{i}\big]^{4}}\\
			& \le n^{-\frac{1}{4}}C\sqrt{\mathbb{E} (\varepsilon^4)} \big[\sup_{\vec{z}\in\mathbb{R}^{s^*}} \zeta(\vec{z})\big]^2,\\
			\mathbb{P}(E_{8}^c) & \le n^{\frac{1}{4}}\mathbb{E} \left| n^{-1}\sum_{i=1}^{n} (\boldsymbol{1}_{X_{ij} \in \mathcal{A}} -\boldsymbol{1}_{\widetilde{X}_{ij} \in \mathcal{A}})^2\big[\zeta(\boldsymbol{X}_{iS^{*}})\varepsilon_{i}\big]^4 -\sigma_{j}^2(a)  \right|\\& \le n^{-\frac{1}{4}}C\mathbb{E} \sqrt{ n^{-1}\sum_{i=1}^{n} \left\{(\boldsymbol{1}_{X_{ij} \in \mathcal{A}} -\boldsymbol{1}_{\widetilde{X}_{ij} \in \mathcal{A}})^2\big[\zeta(\boldsymbol{X}_{iS^{*}})\varepsilon_{i}\big]^4 -\sigma_{j}^2(a) \right\}^2 }\\
			&\le n^{-\frac{1}{4}} C\sqrt{ \mathbb{E} \left\{(\boldsymbol{1}_{X_{j} \in \mathcal{A}} -\boldsymbol{1}_{\widetilde{X}_{j} \in \mathcal{A}})^2\big[\zeta(\boldsymbol{X}_{S^{*}})\varepsilon\big]^4 -\sigma_{j}^2(a) \right\}^2 }\\
			&\le C\sqrt{ \mathbb{E}(\varepsilon^8)  } \big[\sup_{\vec{z}\in\mathbb{R}^{s^*}} \zeta(\vec{z})\big]^4  n^{-\frac{1}{4}}.
		\end{split}
	\end{equation}

	By Markov's inequality, Condition~\ref{bound}, the assumption of i.i.d. observations,  Condition~\ref{consistency.3}, and that the training sample is an independent sample,
	\begin{equation}
		\begin{split}\label{theorem1.prob.3}
			\mathbb{P}(E_{4}^c) &\le B_{1}^{-\frac{1}{2}}\mathbb{E}\left| n^{-1}\sum_{i=1}^{n} (\boldsymbol{1}_{X_{ij} \in \mathcal{A}} -\boldsymbol{1}_{\widetilde{X}_{ij} \in \mathcal{A}})^2 \big[\widehat{m}(\boldsymbol{X}_{i}) - m(\boldsymbol{X}_{i}) \big]^4\right|\\
			&\le B_{1}^{-\frac{1}{2}}\mathbb{E} \big[\widehat{m}(\boldsymbol{X}) - m(\boldsymbol{X})\big]^4\\
			&\le 4\overbar{M}_{n}^2B_{1}^{-\frac{1}{2}}\mathbb{E} \big[\widehat{m}(\boldsymbol{X}) - m(\boldsymbol{X})\big]^2\\
			&\le 4\overbar{M}_{n}^2B_{1}^{\frac{1}{2}}.
		\end{split}
	\end{equation}
	
	By Markov's inequality, Condition~\ref{bound}, the assumptions that $\varepsilon$ is independent of $\boldsymbol{X}$ and that the observations are i.i.d.,  Condition~\ref{consistency.3}, and that the training sample is an independent sample, the following three inequalities \eqref{theorem1.prob.2} holds for all $n\ge 1$ and $B_{1}>0$,
	\begin{equation}
		\begin{split}\label{theorem1.prob.2}
			\mathbb{P}(E_{5}^c) & \le 4B_{1}^{-\frac{1}{2}} \mathbb{E}\left|  \big[\widehat{m}(\boldsymbol{X}) - m(\boldsymbol{X}) \big]^3 \zeta(\boldsymbol{X}_{S^{*}})\varepsilon\right| \\
			&\le 4B_{1}^{-\frac{1}{2}} \big[\sup_{\vec{z}\in\mathbb{R}^{s^*}} \zeta(\vec{z})\big] \big[\mathbb{E} |\widehat{m}(\boldsymbol{X}) - m(\boldsymbol{X}) |^3\big]\mathbb{E}| \varepsilon|\\
			& \le 8\overbar{M}_{n} \big[\sup_{\vec{z}\in\mathbb{R}^{s^*}} \zeta(\vec{z})\big] (\mathbb{E}| \varepsilon|) B_{1}^{\frac{1}{2}},\\
			\mathbb{P}(E_{6}^{c}) &\le 4B_{1}^{-\frac{1}{4}} \big[\sup_{\vec{z}\in\mathbb{R}^{s^*}} \zeta(\vec{z})\big]^3 \mathbb{E}\big[ |\widehat{m}(\boldsymbol{X}) - m(\boldsymbol{X}) | |\varepsilon|^3\big]\\
			&\le 4B_{1}^{-\frac{1}{4}} \big[\sup_{\vec{z}\in\mathbb{R}^{s^*}} \zeta(\vec{z})\big]^3 \big[\mathbb{E} (\widehat{m}(\boldsymbol{X}) - m(\boldsymbol{X}))^2\big]^{\frac{1}{2}} \big(\mathbb{E} |\varepsilon|^3\big)\\
			&\le 4\big[\sup_{\vec{z}\in\mathbb{R}^{s^*}} \zeta(\vec{z})\big]^3\big(\mathbb{E} |\varepsilon|^3\big)B_{1}^{\frac{1}{4}},\\
			\mathbb{P} (E_{7}^c) & \le 6B_{1}^{-\frac{1}{2}}\mathbb{E}\left| n^{-1}\sum_{i=1}^{n} (\boldsymbol{1}_{X_{ij} \in \mathcal{A}} -\boldsymbol{1}_{\widetilde{X}_{ij} \in \mathcal{A}})^2 (\widehat{m}(\boldsymbol{X}_{i}) - m(\boldsymbol{X}_{i}) )^2 \big[\zeta(\boldsymbol{X}_{iS^{*}})\varepsilon_{i}\big]^2  \right|\\
			&\le  6 \big[\sup_{\vec{z}\in\mathbb{R}^{s^*}} \zeta(\vec{z})\big]^2  \mathbb{E} (\varepsilon^2)B_{1}^{\frac{1}{2}}.
		\end{split}
	\end{equation}

	By \eqref{theorem1.7}--\eqref{theorem1.prob.2} and  the model regularity assumptions, there exists some $N_{1}>0$ such that for all $n\ge N_{1}$  and each $0<B_{1}<1$,
	\begin{equation}
		\begin{split}	
			\label{theorem1.31}
			&\mathbb{P}(|T_{j}(a)\big[\widehat{\sigma}_{j}(a)\big]^{-1}| \ge t )  \\
			&\le \mathbb{P}(|T_{j}^{(\star)}(a)| \ge t\big[\sigma_{j}(a)\big] -\varsigma) + n^{\frac{1}{4}}B_{1}^{\frac{1}{2}} + (\overbar{M}_{n})^2(\frac{\log{n}}{2})(n^{-\frac{1}{4}} + B_{1}^{\frac{1}{4}}),
		\end{split}
	\end{equation}
	where we simplify the upper bound as have done for \eqref{theorem1.36}.
	
	To deal with the term $\mathbb{P}(|T_{j}^{(\star)}(a)| \ge t\big[\sigma_{j}(a)\big] -\varsigma)$ on the RHS of \eqref{theorem1.31}, we need the following results. By the Berry-Esseen inequality~\citepsupp{petrov1977sums}, $\sup_{\vec{z}\in\mathbb{R}^{s^*}}\zeta(\vec{z})<\infty$, $\mathbb{E}(\varepsilon^8)<\infty$, \eqref{theorem1.9}, and that $\varepsilon$ is an independent model error, there exists $C>0$ such that 
	\begin{equation}
		\begin{split}\label{theorem1.33}
			\sup_{x\in\mathbb{R}}|\mathbb{P}(T_{j}^{(\star)}(a)\big[\sigma_{j}(a)\big]^{-1} \le x)- \Phi(x)| \le \frac{C}{\sqrt{n}},
		\end{split}	
	\end{equation}
	which leads to
	\begin{equation}
		\begin{split}\label{theorem1.34}
			&|\mathbb{P}(|T_{j}^{(\star)}(a)| \ge t\big[\sigma_{j}(a)\big] -\varsigma) - 2\Phi(-t + \frac{\varsigma}{\sigma_{j}(a)})|  \\
			&\le |\mathbb{P}(T_{j}^{(\star)}(a) \le -t\big[\sigma_{j}(a)\big] +\varsigma) -\Phi(-t + \frac{\varsigma}{\sigma_{j}(a)})| \\
			&\qquad+|\mathbb{P}(T_{j}^{(\star)}(a) \ge t\big[\sigma_{j}(a)\big] -\varsigma) -\Phi(-t + \frac{\varsigma}{\sigma_{j}(a)})|\\
			&\le |\mathbb{P}(T_{j}^{(\star)}(a) \le -t\big[\sigma_{j}(a)\big] +\varsigma) -\Phi(-t + \frac{\varsigma}{\sigma_{j}(a)})| \\
			&\qquad+|1 - \mathbb{P}(T_{j}^{(\star)}(a) \le t\big[\sigma_{j}(a)\big] -\varsigma) - 1 +\Phi(t - \frac{\varsigma}{\sigma_{j}(a)})|\\	
			&\le \frac{2C}{\sqrt{n}},
		\end{split}
	\end{equation}
	where $C>0$ is due to \eqref{theorem1.33} and the second inequality holds because $T_{j}^{(\star)}(a)$ is a continuous random variable. In addition, for each $x, y \in\mathbb{R}$,
	\begin{equation}\label{theorem1.32}
		|\Phi(x) - \Phi(x+y)| \le 0.4 \times |y|,
	\end{equation}
	where $0.4$ is an upper bound of the maximum value of the density of the standard Gaussian distribution.

	We use \eqref{theorem1.9}, \eqref{theorem1.34}--\eqref{theorem1.32}, and regularity assumptions  to deduce that there exists some $N_{2}>0$ such that  for all $n\ge N_{2}$ and each $t>0$,
	\begin{equation}
		\begin{split}	\label{theorem1.35}			
			&\mathbb{P}(|T_{j}^{(\star)}(a)| \ge t\big[\sigma_{j}(a)\big] -\varsigma)  \\
			& \le 2\Phi(-t) + 2\times 0.4\times \varsigma\big\{c\big[\inf_{\vec{z}\in\mathbb{R}^{s^*}} \zeta(\vec{z})\big]^4\mathbb{E}(\varepsilon^{4} )\big\}^{-\frac{1}{2}} + 2Cn^{-\frac{1}{2}}\\
			&\le 2\Phi(-t) + \varsigma\big\{c\big[\inf_{\vec{z}\in\mathbb{R}^{s^*}} \zeta(\vec{z})\big]^4\mathbb{E}(\varepsilon^{4} )\big\}^{-\frac{1}{2}} + 2Cn^{-\frac{1}{2}},
		\end{split}
	\end{equation}
	which in combination with  \eqref{theorem1.31} and the fact that $n^{-\frac{1}{2}} = o\big\{ (\overbar{M}_{n})^2(\log{n})(n^{-\frac{1}{4}} + B_{1}^{\frac{1}{4}})\big\}$ concludes the desired result of Theorem~\ref{theorem1} for the feature index $j$.

	Lastly, we note that the above arguments apply to each $j\in \{1,\dots, p\}$ such that $j\not\in S^*$ and Condition~\ref{non.trivial.2} holds with $(R,\kappa_{1} )=(1,a)$, and that constants $N_{1}$, $N_{2}$ above and $C$ in \eqref{theorem1.33} do not depend on feature index  $j\in\{1, \dots, p\}$. Hence, we conclude the proof of Theorem~\ref{theorem1}.


	\subsection{Proof of  Theorem~\ref{theorem2b}}\label{proof.theorem2b}

	The proof ideas for Theorem~\ref{theorem2b} are similar to those for the proofs of Theorem~\ref{theorem1}: in the inequalities \eqref{theorem2b.3}--\eqref{theorem2b.6} below, we first separate the negligible terms from the statistic $\big[T_{\widehat{l}}(a) - \sqrt{n}\mu_{\widehat{l}}(a) \big]\big[\widehat{\sigma}_{\widehat{l}}(a)\big]^{-1}$ and then argue that the statistic $\big[T_{\widehat{l}}^{(\star)}(a)-\sqrt{n}\big[\mu_{\widehat{l}}(a)\big]\big] \big[\sigma_{\widehat{l}}(a)\big]^{-1}$, which is the remaining term, is asymptotically standard normal because $ \widehat{l}$ is independent of $\{Y_{i}, \boldsymbol{X}_{i}, \widetilde{\boldsymbol{X}}_{i}, \varepsilon_{i}\}_{i=1}^{n}$. Recall that for each $l\in\{1, \dots, p\}$,
	\begin{equation*}
		\begin{split}
			T_{l}^{(\star)}(a) & = n^{-\frac{1}{2}}\sum_{i=1}^{n}\big(\boldsymbol{1}_{X_{il}\in(-\infty, a]} -\boldsymbol{1}_{\widetilde{X}_{il}\in(-\infty, a]}\big)\big[\zeta(\boldsymbol{X}_{iS^{*}})\varepsilon_{i}\big]^{2} ,\\
			\sigma_{l}^2(a) &= \mathbb{E}\big\{\big[\zeta(\boldsymbol{X}_{S^{*}})\varepsilon \big]^{2} \big[\boldsymbol{1}_{X_{l}\in(-\infty, a]} - \boldsymbol{1}_{\widetilde{X}_{l}\in(-\infty, a]}\big] - \mu_{l}(a) \big\}^2,
		\end{split}
	\end{equation*}
	and that expressions for $T_{l}(a)$'s have been given in \eqref{theorem1.4}. In addition, we have defined $\zeta(\boldsymbol{X}_{S^*}) = \zeta_{0}$ if $S^* = \emptyset$ in Section~\ref{Sec1.1}; the proof here applies to cases with empty or non-empty $S^*$. Note that if $S^*=\emptyset$, then $\mu_{l}(a) = 0$ for each $l\in\{1, \dots, p\}$; see \eqref{key.ob} after Lemma~\ref{lemma1} and \eqref{kmc.1} in the proof of Theorem~\ref{theorem1} for details.

	Let us begin the formal proof of Theorem~\ref{theorem2b}. For each $x\in \mathbb{R}$, $n\ge 1$, $B_{1}>0$,
	{\footnotesize\begin{equation}
			\begin{split}\label{theorem2b.3}
				& \mathbb{P}\big\{\big[T_{\widehat{l}} (a)-\sqrt{n}\mu_{\widehat{l}}(a)\big] \big[\widehat{\sigma}_{\widehat{l}}(a)\big]^{-1}\le x\big\}\\
				&\le \mathbb{P}\Big\{\big\{ T_{\widehat{l}}^{(\star)}(a) -\sqrt{n}\big[\mu_{\widehat{l}}(a)\big] \le x\big[\sigma_{\widehat{l}}(a)\big] + |x||\sigma_{\widehat{l}}(a) - \widehat{\sigma}_{\widehat{l}}(a)| + Q_{1} + |Q_{2\widehat{l}}|\big\}\cap \{W(x, n, B_{1})\}^c\Big\} \\
				&\qquad+ \mathbb{P}(W(x, n, B_{1})),
			\end{split}
	\end{equation}}%
	where the event $W(x, n, B_{1})$ is given such that for each $x\in\mathbb{R}, n\ge 1, B_{1}>0$,
	\begin{equation*}
		\begin{split}
			W(x, n, B_{1}) & = \Big\{|x||\sigma_{\widehat{l}}(a) - \widehat{\sigma}_{\widehat{l}}(a)| - Q_{1} - |Q_{2\widehat{l}}| > \varsigma(x, n, B_{1})\Big\},
		\end{split}
	\end{equation*}
	in which $\varsigma (x, n, B_{1})  = |x|n^{-\beta_{2}}  + \sqrt{n}B_{1} (-\log B_{1}) + B_{1}^{\frac{1}{4}}$, $Q_{1}  = n^{-\frac{1}{2}}\sum_{i=1}^{n} \big[\widehat{m}(\boldsymbol{X}_{i}) - m(\boldsymbol{X}_{i})\big]^2$, and
	\begin{equation*}
		\begin{split}
			Q_{2l}  = 2n^{-\frac{1}{2}}\sum_{i=1}^{n} \big[\boldsymbol{1}_{X_{il}\in(-\infty, a]}- \boldsymbol{1}_{\widetilde{X}_{il}\in(-\infty, a]}\big]\big[\widehat{m}(\boldsymbol{X}_{i}) - m(\boldsymbol{X}_{i})\big] \varepsilon_{i}\zeta(\boldsymbol{X}_{iS^*})
		\end{split}
	\end{equation*}
	for each $l\in \{1, \dots, p\}$.

	Similarly, we establish a probability lower bound. For each $x\in \mathbb{R}$, $n\ge 1$, $B_{1}>0$,
	{\footnotesize\begin{equation}
			\begin{split}\label{theorem2b.6}
				& \mathbb{P}\big\{\big[T_{\widehat{l}}(a)-\sqrt{n}\big[\mu_{\widehat{l}}(a)\big]\big] \big[\widehat{\sigma}_{\widehat{l}}(a)\big]^{-1}\le x\big\}\\
				&\ge \mathbb{P}\Big\{\big\{ T_{\widehat{l}}^{(\star)}(a)-\sqrt{n}\big[\mu_{\widehat{l}}(a)\big] \le x\big[\sigma_{\widehat{l}}(a)\big] - |x||\sigma_{\widehat{l}}(a) - \widehat{\sigma}_{\widehat{l}}(a)| - Q_{1} - |Q_{2\widehat{l}} |\big\}\cap \{W(x, n, B_{1})\}^c\Big\}.
			\end{split}
	\end{equation}}%
	
	To analyze the RHS of \eqref{theorem2b.3}--\eqref{theorem2b.6}, we need a uniform lower bound for the population variances. By Condition \ref{non.trivial.2} for each $j\in \{1, \dots ,p\}$ with  $(R,\kappa_{1} )=(1,a)$, regularity assumptions, and Lemma~\ref{uniform.var.1} in Section~\ref{uniform.var.2}, there exists some $\underline{\sigma}>0$ such that 
	\begin{equation}\label{theorem2b.7}
		\min_{1\le l\le p}\sigma_{l}^2(a)\ge \underline{\sigma}^2.
	\end{equation}

	With event $W(x, n, B_{1})$ and \eqref{theorem2b.7},
	\begin{equation}
		\begin{split}\label{theorem2b.8}
			&\textnormal{ RHS of \eqref{theorem2b.3}}
			\\
			&\le  \mathbb{P}\big\{\big[T_{\widehat{l}}^{(\star)}(a)-\sqrt{n}\big[\mu_{\widehat{l}}(a)\big]\big] \big[\sigma_{\widehat{l}}(a)\big]^{-1}\le x + \varsigma(x, n, B_{1}) (\underline{\sigma}^{-1}) \big\} + \mathbb{P}(W(x, n, B_{1})),
		\end{split}
	\end{equation}
	and
	{\small\begin{equation}
			\begin{split}\label{theorem2b.9}
				&\textnormal{RHS of \eqref{theorem2b.6}}
				\\& \ge  \mathbb{P}\Big\{\big\{\big[T_{\widehat{l}}^{(\star)}(a)-\sqrt{n}\big[\mu_{\widehat{l}}(a)\big]\big] \big[\sigma_{\widehat{l}}(a)\big]^{-1}\le x - \varsigma (x, n, B_{1}) (\underline{\sigma}^{-1}) \big\} \cap \{W(x, n, B_{1})\}^c\Big\}\\
				& \ge  \mathbb{P}\big\{\big[T_{\widehat{l}}^{(\star)}(a)-\sqrt{n}\big[\mu_{\widehat{l}}(a)\big]\big] \big[\sigma_{\widehat{l}}(a)\big]^{-1}\le x - \varsigma (x, n, B_{1}) (\underline{\sigma}^{-1}) \big\} -\mathbb{P}(W(x, n, B_{1})),
			\end{split}
	\end{equation}}%
	where the respective first equalities in \eqref{theorem2b.8}--\eqref{theorem2b.9} follow from the definition of event $W(x, n, B_{1})$ and \eqref{theorem2b.7}, and the second inequality in \eqref{theorem2b.9} follows because $\mathbb{P}(A\cap B) = \mathbb{P}(A) - \mathbb{P}(A\cap B^c)$ for any events $A$ and $B$.

	In the following, we deal with the two terms on the RHS of \eqref{theorem2b.8}--\eqref{theorem2b.9}, and begin with showing that $\mathbb{P}(W(x, n, B_{1}))$ is negligible. In light of $ \mathbb{E}|\varepsilon|^{q\vee 8} <\infty $ and   Conditions~\ref{consistency.3}--\ref{bound} with $\lim_{n\rightarrow\infty} n^{\beta_{2}}\overbar{M}_{n}^2B_{1}+ \sqrt{n}B_{1}(-\log B_{1}) + n^{-\frac{1}{2}+ \frac{4}{q-\beta} + \beta_{2}} \sqrt{\log{(n\vee p)}} =0$, we show in Section~\ref{proof.theorem4.5} (the case  here is a special case with $R=1$ and $\kappa_{1} = a$ there, and therefore the notation here is slightly different) that for each $x\in\mathbb{R}$, 
	\begin{equation}
		\begin{split}\label{theorem2b.4}
			&\lim_{n\rightarrow\infty}\mathbb{P}(W(x, n, B_{1})) = 0.
		\end{split}
	\end{equation}

	Next, we show that the distribution of $\big[T_{l}^{(\star)}(a)-\sqrt{n}\big[\mu_{l}(a)\big]\big] \big[\sigma_{l}(a)\big]^{-1}$ is asymptotically standard normal. By the Berry-Esseen inequality~\citepsupp{petrov1977sums},  \eqref{theorem2b.7}, the assumption of i.i.d. observations, and  the assumptions that $\mathbb{E}|\varepsilon|^8<\infty$ and $\sup_{\vec{z}\in\mathbb{R}^{s^*}}\zeta(\vec{z})<\infty$, there exists some $C>0$ such that for each $n\ge 1$ and each $1\le l \le p$,
	\begin{equation}\label{theorem2b.5}
		\sup_{x\in \mathbb{R}} \big|\mathbb{P}\big\{\big[T_{l}^{(\star)}(a)-\sqrt{n}\big[\mu_{l}(a)\big]\big] \big[\sigma_{l}(a)\big]^{-1}\le x \big\}- \Phi(x) \big| \le \frac{C}{\sqrt{n}}.
	\end{equation}
	
	By \eqref{theorem2b.5} and that $\widehat{l}$ is independent of $T_{l}^{(\star)}$'s, it holds that 
	\begin{equation}
		\begin{split}\label{theorem2b.1}
			&\sup_{x\in \mathbb{R}}|	\mathbb{P}\big\{\big(T_{\widehat{l}}^{(\star)} (a)-\sqrt{n}\big[\mu_{\widehat{l}}(a)\big]\big) \big[\sigma_{\widehat{l}}(a)\big]^{-1}\le x \big\} - \Phi(x)| \\
			& =\sup_{x\in \mathbb{R}}|	\sum_{l=1}^{p} \mathbb{P}\big\{\big[T_{l}^{(\star)}(a)-\sqrt{n}\big[\mu_{l}(a)\big]\big] \big[\sigma_{l}(a)\big]^{-1}\le x \big\} \times \mathbb{P}(\widehat{l} = l) - \Phi(x)|\\
			& \le \sum_{l=1}^{p}\sup_{x\in \mathbb{R}}\Big|	 \mathbb{P}\big\{\big[T_{l}^{(\star)}(a)-\sqrt{n}\big[\mu_{l}(a)\big]\big] \big[\sigma_{l}(a)\big]^{-1}\le x \big\} - \Phi(x)\Big|\times\mathbb{P}(\widehat{l} = l) \\
			&\le \frac{C}{\sqrt{n}},
		\end{split}
	\end{equation}
	where constant $C$ is given in \eqref{theorem2b.1}. In addition, for each $x, y\in\mathbb{R}$,
	\begin{equation}
		\begin{split}\label{theorem2b.2}
			|\Phi(x) - \Phi(x+y)| \le 0.4\times |y|,
		\end{split}
	\end{equation}
	since the maximum value of density of the standard Gaussian distribution is less than $0.4$.

	By \eqref{theorem2b.7}, \eqref{theorem2b.3}--\eqref{theorem2b.2}, there exists some constant $C>0$ such that for all large $n$, each $x\in\mathbb{R}$, and each $B_{1}>0$,
	\begin{equation}
		\begin{split}
			& |\mathbb{P}\big\{\big[T_{\widehat{l}}(a)-\sqrt{n}\big[\mu_{\widehat{l}}(a)\big]\big] \big[\widehat{\sigma}_{\widehat{l}}(a)\big]^{-1}\le x\big\} -  \Phi(x)| \\
			&\le  \frac{C}{\sqrt{n}} + 0.4\times \frac{\varsigma(x, n, B_{1})}{ \underline{\sigma}}  + \mathbb{P}(W(x, n, B_{1})),
		\end{split}
	\end{equation}
	which in combination with  the assumption $\lim_{n\rightarrow\infty}\sqrt{n}B_{1}(-\log B_{1}) = 0$ (hence $\lim_{n\rightarrow\infty}\varsigma(x, n, B_{1}) = 0$ for each $x\in\mathbb{R}$) and  \eqref{theorem2b.4} concludes the main desired result of Theorem~\ref{theorem2b}. For the other assertion of Theorem~\ref{theorem2b}, note that $\mu_{l}(a) = 0$ for each $l\in\{1, \dots, p\}$ when $S^* = \emptyset$. We have finished the proof of Theorem~\ref{theorem2b}.
	
	\subsection{Proof of Theorem~\ref{theorem3}}\label{proof.theorem3}
	
	The proof idea for Theorem~\ref{theorem3} is omitted because it is similar to that for Theorem~\ref{theorem1}. Recall that we have defined $\zeta(\boldsymbol{X}_{S^*}) = \zeta_{0}$ if $S^* = \emptyset$ in Section~\ref{Sec1.1}; the proof here applies to cases with empty or non-empty $S^*$.
	
	Let us begin the formal proof of Theorem~\ref{theorem3} with \eqref{theorem3.13} and \eqref{theorem3.7} below. For every $r\in \{1, \dots, R\}$,
	\begin{equation}
		\begin{split}\label{theorem3.13}
			T_{j}(\kappa_{r}) &=n^{-\frac{1}{2}}\sum_{i=1}^{n}\big(\boldsymbol{1}_{X_{ij}\in (-\infty,\kappa_{r}} -\boldsymbol{1}_{\widetilde{X}_{ij}\in(-\infty,\kappa_{r}]}\big) (\widehat{\varepsilon}_{i} )^{2}\\			
			& = n^{-\frac{1}{2}}\sum_{i=1}^{n}\big(\boldsymbol{1}_{X_{ij}\in(-\infty,\kappa_{r}]} -\boldsymbol{1}_{\widetilde{X}_{ij}\in(-\infty,\kappa_{r}]}\big) (m(\boldsymbol{X}_{i}) - \widehat{m}(\boldsymbol{X}_{i}))^2 \\
			& \hspace{2em}- 2n^{-\frac{1}{2}}\sum_{i=1}^{n}\big(\boldsymbol{1}_{X_{ij}\in (-\infty,\kappa_{r}]} -\boldsymbol{1}_{\widetilde{X}_{ij}\in(-\infty,\kappa_{r}]}\big) (m(\boldsymbol{X}_{i}) - \widehat{m}(\boldsymbol{X}_{i}))\zeta(\boldsymbol{X}_{iS^{*}})\varepsilon_{i}\\
			&\hspace{2em}+ n^{-\frac{1}{2}}\sum_{i=1}^{n}\big(\boldsymbol{1}_{X_{ij}\in(-\infty,\kappa_{r}]} -\boldsymbol{1}_{\widetilde{X}_{ij}\in(-\infty,\kappa_{r}]}\big)(\zeta(\boldsymbol{X}_{iS^{*}})\varepsilon_{i})^{2}\\
			& \eqqcolon A_{1}(\kappa_{r}) + A_{2}(\kappa_{r}) + T_{j}^{(\star)}(\kappa_{r}) ,
		\end{split}
	\end{equation}
	where   $T_{j}^{(\star)}(\kappa_{r}) =n^{-\frac{1}{2}}\sum_{i=1}^{n}\big[\boldsymbol{1}_{X_{ij}\in(-\infty,\kappa_{r}]} -\boldsymbol{1}_{\widetilde{X}_{ij}\in(-\infty,\kappa_{r}]}\big]\big[\zeta(\boldsymbol{X}_{iS^{*}})\varepsilon_{i}\big]^{2}$ and $\widehat{\varepsilon}_{i} = Y_{i} - \widehat{m}(\boldsymbol{X}_{i})$. In addition,
	\begin{equation}
		\begin{split}\label{theorem3.7}
			&\mathbb{P}(|T_{j}(\widehat{a}_{j})\big[\widehat{\sigma}_{j}(\widehat{a}_{j})\big]^{-1}| \ge t )\\
			&\le \mathbb{P}(|T_{j}^{(\star)}(\widehat{a}_{j})| \ge t\big[\sigma_{j}(\widehat{a}_{j})\big] -\varsigma - t|\sigma_{j}(\widehat{a}_{j}) - \widehat{\sigma}_{j}(\widehat{a}_{j})| -|A_{1}(\widehat{a}_{j})| - |A_{2}(\widehat{a}_{j})|+ \varsigma),
		\end{split}
	\end{equation}
	where $\varsigma = t(\log{n})(10n^{-\frac{1}{4}} + 4B_{1}^{\frac{1}{8}}) + n^{\frac{1}{4}}B_{1}^{\frac{1}{2}}$.
	
	Now, let us establish the bounds for $|\widehat{\sigma}_{j}(\widehat{a}_{j}) -\sigma_{j}(\widehat{a}_{j})|$, $|A_{1}(\widehat{a}_{j})|$, and $|A_{2}(\widehat{a}_{j})|$ on the RHS of \eqref{theorem3.7}, and begin with the one for $|\widehat{\sigma}_{j}(\widehat{a}_{j}) -\sigma_{j}(\widehat{a}_{j})|$. By Condition~\ref{non.trivial.2}, the assumption that  $\kappa_{1}< \dots < \kappa_{R}$, the equality \eqref{kmc.2} below
	\begin{equation}
		\label{kmc.2}
		\mathbb{E}\big\{\big[\boldsymbol{1}_{X_{j}\in(-\infty, \kappa_{r}]} -\boldsymbol{1}_{\widetilde{X}_{j}\in(-\infty, \kappa_{r}]}\big]\big[\zeta(\boldsymbol{X}_{S^{*}})\varepsilon\big]^{2}\big\} = 0
	\end{equation}
	for each $r\in\{1, \dots, R\}$ (which is due to $j\not\in S^*$; see \eqref{key.ob} for details), and the assumption that $\varepsilon$ is independent of $(\boldsymbol{X}, \widetilde{X}_{j})$, it holds that for each $r\in\{1, \dots, R\}$,
	\begin{equation}
		\begin{split}\label{theorem3.1}
			\sigma_{j}^2(\kappa_{r}) & = \mathbb{E}\big\{\big[\boldsymbol{1}_{X_{j}\in(-\infty, \kappa_{r}]} -\boldsymbol{1}_{\widetilde{X}_{j}\in(-\infty, \kappa_{r}]}\big]^2\big[\zeta(\boldsymbol{X}_{S^{*}})\varepsilon\big]^{4}\big\}\\
			& \ge \mathbb{E}\big\{\boldsymbol{1}_{\{X_{j}\in(-\infty, \kappa_{r}]\}\cap\{\widetilde{X}_{j}\in(\kappa_{r}, \infty)\} } \big[\zeta(\boldsymbol{X}_{S^{*}})\varepsilon \big]^{4} \big\} \\
			& \ge  c\big[\inf_{\vec{z}\in\mathbb{R}^{s^*}} \zeta(\vec{z})\big]^4\mathbb{E}(\varepsilon^{4} ),
		\end{split}
	\end{equation}
	where $c>0$ is defined in Condition~\ref{non.trivial.2}. By \eqref{theorem3.1}, for each $r\in\{1, \dots ,R\}$,
	\begin{equation}
		\begin{split}\label{theorem3.6}
			|\widehat{\sigma}_{j}(\kappa_{r}) - \sigma_{j}(\kappa_{r})| &= \frac{|\widehat{\sigma}_{j}^{2}(\kappa_{r}) - \sigma_{j}^{2}(\kappa_{r})|}{\widehat{\sigma}_{j}(\kappa_{r}) + \sigma_{j}(\kappa_{r})} \\
			& \le \frac{|\widehat{\sigma}_{j}^{2}(\kappa_{r}) - \sigma_{j}^{2}(\kappa_{r})|}{ \sigma_{j}(\kappa_{r})}\\
			&\le \big\{c\big[\inf_{\vec{z}\in\mathbb{R}^{s^*}} \zeta(\vec{z})\big]^4\mathbb{E}(\varepsilon^{4} ) \big\}^{-\frac{1}{2}}|\widehat{\sigma}_{j}^{2}(\kappa_{r}) - \sigma_{j}^{2}(\kappa_{r})|.
		\end{split}
	\end{equation}

	In addition, for each $r\in\{1, \dots ,R\}$, we write
	\begin{equation}
		\begin{split}\label{theorem3.5}
			&\widehat{\sigma}_{j}^2(\kappa_{r}) -\sigma_{j}^2(\kappa_{r})\\
			&= -\sigma_{j}^2(\kappa_{r}) -\big[\widehat{\mu}_{j}(\kappa_{r})\big]^2 + n^{-1}\sum_{i=1}^{n} \big[\boldsymbol{1}_{X_{ij} \in (-\infty, \kappa_{r}]} -\boldsymbol{1}_{\widetilde{X}_{ij} \in (-\infty, \kappa_{r}]}\big]^2(\widehat{\varepsilon}_{i})^4 \\
			&=  \bigg\{  n^{-1}\sum_{i=1}^{n} \Big\{\big[\boldsymbol{1}_{X_{ij} \in (-\infty, \kappa_{r}]} -\boldsymbol{1}_{\widetilde{X}_{ij} \in (-\infty, \kappa_{r}]}\big]^2 \times \Big\{\big[\widehat{m}(\boldsymbol{X}_{i}) - m(\boldsymbol{X}_{i}) \big]^4\\
			&  \qquad \qquad- 4\big[\widehat{m}(\boldsymbol{X}_{i}) - m(\boldsymbol{X}_{i}) \big]^3 \zeta(\boldsymbol{X}_{iS^{*}})\varepsilon_{i}+ 6\big[\widehat{m}(\boldsymbol{X}_{i}) - m(\boldsymbol{X}_{i}) \big]^2 \big[\zeta(\boldsymbol{X}_{iS^{*}})\varepsilon_{i}\big]^2 \\
			&\qquad\qquad- 4\big[\widehat{m}(\boldsymbol{X}_{i}) - m(\boldsymbol{X}_{i}) \big] \big[\zeta(\boldsymbol{X}_{iS^{*}})\varepsilon_{i}\big]^3 \Big\}\Big\}\bigg\} -\big[\widehat{\mu}_{j}(\kappa_{r})\big]^2 \\
			&\qquad + n^{-1}\sum_{i=1}^{n} \Big\{\big[\boldsymbol{1}_{X_{ij} \in (-\infty, \kappa_{r}]} -\boldsymbol{1}_{\widetilde{X}_{ij} \in (-\infty, \kappa_{r}]}\big]^2\big[ \zeta(\boldsymbol{X}_{iS^{*}})\varepsilon_{i}\big]^4 - \sigma_{j}^2(\kappa_{r})\Big\},
		\end{split}
	\end{equation}
	where $\widehat{\mu}_{j}(\kappa_{r}) = n^{-1}\sum_{i=1}^{n}\big(\boldsymbol{1}_{X_{ij}\in(-\infty, \kappa_{r}]}- \boldsymbol{1}_{\widetilde{X}_{ij}\in(-\infty, \kappa_{r}]}\big)\big(\widehat{\varepsilon}_{i} \big)^{2}$, and define  events $E_{1}(\kappa_{r}), \dots, E_{8}(\kappa_{r})$ for each $r\in\{1, \dots, R\}$ as follows.
	\begin{equation*}
		\begin{split}
			E_{1}(\kappa_{r}) & = \left\{\left|n^{-\frac{1}{2}}\sum_{i=1}^{n}\big(\boldsymbol{1}_{X_{ij}\in(-\infty, \kappa_{r}]} -\boldsymbol{1}_{\widetilde{X}_{ij}\in(-\infty, \kappa_{r}]}\big) \big[m(\boldsymbol{X}_{i}) - \widehat{m}(\boldsymbol{X}_{i})\big]^2\right| \le n^{\frac{1}{4}}B_{1}^{\frac{1}{2}}\right\},\\
			E_{2}(\kappa_{r})& = \left\{ \left|2n^{-\frac{1}{2}}\sum_{i=1}^{n}\big(\boldsymbol{1}_{X_{ij}\in(-\infty, \kappa_{r}]} -\boldsymbol{1}_{\widetilde{X}_{ij}\in(-\infty, \kappa_{r}]}\big)\big[m(\boldsymbol{X}_{i}) - \widehat{m}(\boldsymbol{X}_{i})\big]\zeta(\boldsymbol{X}_{iS^{*}})\varepsilon_{i} \right| \le B_{1}^{\frac{1}{4}}\right\},\\
			E_{3}(\kappa_{r}) &= \left\{\left|n^{-\frac{1}{2}}\sum_{i=1}^{n}\big(\boldsymbol{1}_{X_{ij}\in(-\infty, \kappa_{r}]} -\boldsymbol{1}_{\widetilde{X}_{ij}\in(-\infty, \kappa_{r}]}\big) \big[\zeta(\boldsymbol{X}_{iS^{*}})\varepsilon_{i}\big]^{2}\right|\le n^{\frac{1}{4}}\right\},\\
			E_{4}(\kappa_{r}) &= \left\{ \left| n^{-1}\sum_{i=1}^{n} (\boldsymbol{1}_{X_{ij} \in (-\infty, \kappa_{r}]} -\boldsymbol{1}_{\widetilde{X}_{ij} \in (-\infty, \kappa_{r}]})^2 \big[\widehat{m}(\boldsymbol{X}_{i}) - m(\boldsymbol{X}_{i}) \big]^4\right| \le B_{1}^{\frac{1}{2}}\right\},\\
			E_{5} (\kappa_{r})&= \left\{ \left| 4n^{-1}\sum_{i=1}^{n} (\boldsymbol{1}_{X_{ij} \in (-\infty, \kappa_{r}]} -\boldsymbol{1}_{\widetilde{X}_{ij} \in (-\infty, \kappa_{r}]})^2 \big[\widehat{m}(\boldsymbol{X}_{i}) - m(\boldsymbol{X}_{i}) \big]^3 \zeta(\boldsymbol{X}_{iS^{*}})\varepsilon_{i}\right| \le B_{1}^{\frac{1}{2}}\right\} ,\\
			E_{6}(\kappa_{r}) &= \left\{ \left| 4n^{-1}\sum_{i=1}^{n} (\boldsymbol{1}_{X_{ij} \in (-\infty, \kappa_{r}]} -\boldsymbol{1}_{\widetilde{X}_{ij} \in (-\infty, \kappa_{r}]})^2 \big[\widehat{m}(\boldsymbol{X}_{i}) - m(\boldsymbol{X}_{i}) \big] \big[\zeta(\boldsymbol{X}_{iS^{*}})\varepsilon_{i}\big]^3\right| \le B_{1}^{\frac{1}{4}}\right\},\\
			E_{7}(\kappa_{r}) & = \left\{\left| 6n^{-1}\sum_{i=1}^{n} (\boldsymbol{1}_{X_{ij} \in (-\infty, \kappa_{r}]} -\boldsymbol{1}_{\widetilde{X}_{ij} \in (-\infty, \kappa_{r}]})^2 \big[\widehat{m}(\boldsymbol{X}_{i}) - m(\boldsymbol{X}_{i}) \big]^2 \big[\zeta(\boldsymbol{X}_{iS^{*}})\varepsilon_{i}\big]^2  \right|\le B_{1}^{\frac{1}{2}}\right\},\\
			E_{8}(\kappa_{r}) &= \left\{\left| n^{-1}\sum_{i=1}^{n} \big\{\big[\boldsymbol{1}_{X_{ij} \in (-\infty, \kappa_{r}]} -\boldsymbol{1}_{\widetilde{X}_{ij} \in (-\infty, \kappa_{r}]}\big]^2\big[\zeta(\boldsymbol{X}_{iS^{*}})\varepsilon_{i}\big]^4 -\sigma_{j}^2(\kappa_{r}) \big\} \right|\le n^{-\frac{1}{4}} \right\}.
		\end{split}		
	\end{equation*}

	In light of \eqref{theorem3.5} and these events, we now establish the upper bounds for $|\widehat{\sigma}_{j}^2(\kappa_{r}) -\sigma_{j}^2(\kappa_{r})|$, $|A_{1}(\kappa_{r})|$, and $|A_{2}(\kappa_{r})|$ for each $r\in\{1, \dots, R\}$. On $\cap_{l=1}^3 E_{l}(\kappa_{r})$, for all $n\ge 1$, each $1\le r\le R$, and each $B_{1}>0$,
	\begin{equation}
		\label{theorem3.12}
		|\widehat{\mu}_{j}(\kappa_{r})|\le n^{-\frac{1}{4}}B_{1}^{\frac{1}{2}} + n^{-\frac{1}{2}}B_{1}^{\frac{1}{4}} + n^{-\frac{1}{4}}.
	\end{equation}	
	On $\cap_{l=1}^2E_{l}(\kappa_{r})$, for all $n\ge 1$, each $1\le r\le R$, and each $B_{1}>0$,
	$$|A_{1}(\kappa_{r})| + |A_{2}(\kappa_{r})|\le n^{\frac{1}{4}}B_{1}^{\frac{1}{2}} + B_{1}^{\frac{1}{4}} .$$	
	By \eqref{theorem3.5}--\eqref{theorem3.12}, on $\cap_{l=1}^8E_{l}(\kappa_{r})$, for all large $n$, each $1\le r\le R$, and each $0<B_{1} <1$,
	\begin{equation}\label{theorem3.10}
		|	\widehat{\sigma}_{j}^2(\kappa_{r}) - \sigma_{j}^2(\kappa_{r}) | \le 10n^{-\frac{1}{4}} + 4B_{1}^{\frac{1}{8}},
	\end{equation}
	where we use the facts that $n^{-\frac{1}{4}}\le 1$ and $n^{-\frac{1}{2}}\le 1$ for all $n\ge 1$ and that $B_{1}^{\frac{1}{8}} \ge B_{1}^{\frac{1}{4}} \ge B_{1}^{\frac{1}{2}}$ if $0<B_{1}<1$ to simplify the upper bound.
	
	By these bounds, \eqref{theorem3.6}, the definition of $\varsigma$, and model regularity assumptions, it holds that  for all large $n$, each $t>0$, each $0<B_{1} <1$, and each $r\in\{1, \dots, R\}$, on $\cap_{l=1}^8 E_{l}(\kappa_{r})$,
	$$\varsigma\ge  t|\sigma_{j}(\kappa_{r}) - \widehat{\sigma}_{j}(\kappa_{r})| + |A_{1}(\kappa_{r})| + |A_{2}(\kappa_{r})|.$$ By this result and \eqref{theorem3.7}, we deduce that for all large $n$, each $t>0$, and each $0<B_{1} <1$,
	\begin{equation}
		\begin{split}\label{theorem3.8}
			&\mathbb{P}(|T_{j}(\widehat{a}_{j})\big[\widehat{\sigma}_{j}(\widehat{a}_{j})\big]^{-1}| \ge t ) \\& \le \mathbb{P}\big\{\{|T_{j}^{(\star)} (\widehat{a}_{j})| \ge t\big[\sigma_{j}(\widehat{a}_{j})\big] - \varsigma\}\cap \big\{\cap_{r=1}^R \cap_{l=1}^8 \big[E_{l}(\kappa_{r})\big]\big\}\big\} + \sum_{r=1}^{R}\sum_{l=1}^8 \mathbb{P}\big\{\big[E_{l}(\kappa_{r})\big]^{c}\big\}\\
			& \le \mathbb{P}(|T_{j}^{(\star)}(\widehat{a}_{j})| \ge t\big[\sigma_{j}(\widehat{a}_{j})\big]  - \varsigma) + \sum_{r=1}^{R}\sum_{l=1}^8 \mathbb{P}\big\{\big[E_{l}(\kappa_{r})\big]^{c}\big\},
		\end{split}
	\end{equation}
	where $E^c$ denotes the complementary event of an event $E$.

	Next, we bound the probabilities $\mathbb{P}\big\{\big[E_{1}(\kappa_{r})\big]^{c}\big\}, \dots , \mathbb{P}\big\{\big[E_{8}(\kappa_{r})\big]^{c}\big\}$ for each $r\in\{1, \dots, R\}$. Note that we have $|\boldsymbol{1}_{X_{ij}\in(-\infty, \kappa_{r}]} -\boldsymbol{1}_{\widetilde{X}_{ij}\in(-\infty, \kappa_{r}]}|\le 1$ for each $1\le i\le n$ and $1\le r\le R$ in these events. By Markov's inequality, the assumption that observations are i.i.d., the assumption that the training sample is an independent sample, and Condition~\ref{consistency.3}, it holds that for all $n\ge 1$, $B_{1}>0$, and $r\in\{1, \dots, R\}$,
	\begin{equation}
		\mathbb{P}\big\{\big[E_{1}(\kappa_{r})\big]^{c}\big\}\le n^{\frac{1}{4}}B_{1}^{\frac{1}{2}}.
	\end{equation}

	By the arguments similar to those for \eqref{theorem1.2}, Markov's inequality, the Burkholder--Davis--Gundy inequality \citepsupp{burkholder1972integral} inequality, Jensen's inequality, the assumption of i.i.d. observations, the assumption that the training sample is an independent sample, and  Condition~\ref{consistency.3}, there exists $C>0$ such that for all $n\ge 1$, each $1\le r\le R$, and each $B_{1} >0$,
	{\footnotesize\begin{equation}
			\begin{split}\label{theorem3.14}
				&\mathbb{P}\big\{\big[E_{2}(\kappa_{r})\big]^{c}\big\}\\
				&\le 2B_{1}^{-\frac{1}{4}}\mathbb{E}\Big|n^{-\frac{1}{2}}\sum_{i=1}^{n}\big(\boldsymbol{1}_{X_{ij}\in(-\infty, \kappa_{r}]} -\boldsymbol{1}_{\widetilde{X}_{ij}\in(-\infty, \kappa_{r}]}\big)\big[m(\boldsymbol{X}_{i}) - \widehat{m}(\boldsymbol{X}_{i})\big]\zeta(\boldsymbol{X}_{iS^{*}})\varepsilon_{i}\Big|\\
				&\le 2CB_{1}^{-\frac{1}{4}} 			
				\mathbb{E}\sqrt{\left(n^{-1}\sum_{i=1}^{n}\big(\boldsymbol{1}_{X_{ij}\in(-\infty, \kappa_{r}]} -\boldsymbol{1}_{\widetilde{X}_{ij}\in(-\infty, \kappa_{r}]}\big)^2\big[m(\boldsymbol{X}_{i}) - \widehat{m}(\boldsymbol{X}_{i})\big]^2 \big[\zeta(\boldsymbol{X}_{iS^{*}})\varepsilon_{i}\big]^2\right)} \\
				& \le 2CB_{1}^{-\frac{1}{4}} 			
				\sqrt{			\mathbb{E}\left(n^{-1}\sum_{i=1}^{n}\big(\boldsymbol{1}_{X_{ij}\in(-\infty, \kappa_{r}]} -\boldsymbol{1}_{\widetilde{X}_{ij}\in(-\infty, \kappa_{r}]}\big)^2\big[m(\boldsymbol{X}_{i}) - \widehat{m}(\boldsymbol{X}_{i})\big]^2 \big[\zeta(\boldsymbol{X}_{iS^{*}})\varepsilon_{i}\big]^2\right)} \\
				& \le 2CB_{1}^{-\frac{1}{4}} 			
				\sqrt{	\big[\sup_{\vec{z}\in\mathbb{R}^{s^*}} \zeta(\vec{z})\big]^2	\mathbb{E}(\varepsilon^2)	\mathbb{E}\big[m(\boldsymbol{X}) - \widehat{m}(\boldsymbol{X})\big]^2  } \\
				&\le 2CB_{1}^{\frac{1}{4}}\sqrt{\mathbb{E}(\varepsilon^2)}\big[\sup_{\vec{z}\in\mathbb{R}^{s^*}} \zeta(\vec{z})\big].
			\end{split}
	\end{equation}}%
	By arguments similar to those for \eqref{theorem3.14}, $j\not\in S^*$, and \eqref{kmc.2}, there exists $C>0$ such that for all $n\ge 1$, each $1\le r\le R$, and each $B_{1} >0$,
	\begin{equation}
		\begin{split}
			&\mathbb{P}\big\{\big[E_{3}(\kappa_{r})\big]^{c}\big\}\\
			& \le n^{-\frac{1}{4}}\mathbb{E}\left|n^{-\frac{1}{2}}\sum_{i=1}^{n}\big(\boldsymbol{1}_{X_{ij}\in(-\infty, \kappa_{r}]} -\boldsymbol{1}_{\widetilde{X}_{ij}\in(-\infty, \kappa_{r}]}\big)\big[\zeta(\boldsymbol{X}_{iS^{*}})\varepsilon_{i}\big]^2\right|\\
			& \le n^{-\frac{1}{4}}C \mathbb{E}\sqrt{n^{-1}\sum_{i=1}^{n}\big(\boldsymbol{1}_{X_{ij}\in(-\infty, \kappa_{r}]} -\boldsymbol{1}_{\widetilde{X}_{ij}\in(-\infty, \kappa_{r}]}\big)^2\big[\zeta(\boldsymbol{X}_{iS^{*}})\varepsilon_{i}\big]^{4}}\\
			& \le n^{-\frac{1}{4}}C\sqrt{\mathbb{E} (\varepsilon^4)} \big[\sup_{\vec{z}\in\mathbb{R}^{s^*}} \zeta(\vec{z})\big]^2,\\
			&\mathbb{P}\big\{\big[E_{8}(\kappa_{r})\big]^{c}\big\} \\
			& \le n^{\frac{1}{4}}\mathbb{E} \left| n^{-1}\sum_{i=1}^{n}\left\{ (\boldsymbol{1}_{X_{ij} \in (-\infty, \kappa_{r}]} -\boldsymbol{1}_{\widetilde{X}_{ij} \in (-\infty, \kappa_{r}]})^2\big[\zeta(\boldsymbol{X}_{iS^{*}})\varepsilon_{i}\big]^4 -\sigma_{j}^2(\kappa_{r}) \right\} \right|\\& \le n^{-\frac{1}{4}}C\mathbb{E} \sqrt{ n^{-1}\sum_{i=1}^{n} \left\{(\boldsymbol{1}_{X_{ij} \in (-\infty, \kappa_{r}]} -\boldsymbol{1}_{\widetilde{X}_{ij} \in (-\infty, \kappa_{r}]})^2\big[\zeta(\boldsymbol{X}_{iS^{*}})\varepsilon_{i}\big]^4 -\sigma_{j}^2(\kappa_{r}) \right\}^2 }\\
			&\le n^{-\frac{1}{4}} C\sqrt{ \mathbb{E} \left\{(\boldsymbol{1}_{X_{j} \in (-\infty, \kappa_{r}]} -\boldsymbol{1}_{\widetilde{X}_{j} \in (-\infty, \kappa_{r}]})^2\big[\zeta(\boldsymbol{X}_{S^{*}})\varepsilon\big]^4 -\sigma_{j}^2(\kappa_{r}) \right\}^2 }\\
			&\le C\sqrt{ \mathbb{E}(\varepsilon^8)  } \big[\sup_{\vec{z}\in\mathbb{R}^{s^*}} \zeta(\vec{z})\big]^4  n^{-\frac{1}{4}}.
		\end{split}
	\end{equation}

	By Markov's inequality, the assumption of i.i.d. observations, the assumption that the training sample is an independent sample, and Conditions~\ref{consistency.3}--\ref{bound}, it holds that for all $n\ge 1$, each $1\le r\le R$, and each $B_{1} >0$,
	\begin{equation}
		\begin{split}
			\mathbb{P}\big\{\big[E_{4}(\kappa_{r})\big]^{c}\big\} &\le B_{1}^{-\frac{1}{2}}\mathbb{E}\left| n^{-1}\sum_{i=1}^{n} (\boldsymbol{1}_{X_{ij} \in (-\infty, \kappa_{r}]} -\boldsymbol{1}_{\widetilde{X}_{ij} \in (-\infty, \kappa_{r}]})^2 \big[\widehat{m}(\boldsymbol{X}_{i}) - m(\boldsymbol{X}_{i}) \big]^4\right|\\
			&\le B_{1}^{-\frac{1}{2}}\mathbb{E} \big[\widehat{m}(\boldsymbol{X}) - m(\boldsymbol{X})\big]^4\\
			&\le 4\overbar{M}_{n}^2B_{1}^{-\frac{1}{2}}\mathbb{E} \big[\widehat{m}(\boldsymbol{X}) - m(\boldsymbol{X})\big]^2\\
			&\le 4\overbar{M}_{n}^2B_{1}^{\frac{1}{2}}.
		\end{split}
	\end{equation}
	
	By Markov's inequality, the assumptions that the observations are i.i.d. and that the training sample is an independent sample, model regularity assumptions, and Conditions~\ref{consistency.3}--\ref{bound}, the following three inequalities \eqref{theorem3.9} holds for all $n\ge 1$, each $1\le r\le R$, and each $B_{1}>0$,
	\begin{equation}
		\begin{split}\label{theorem3.9}
			\mathbb{P}\big\{\big[E_{5}(\kappa_{r})\big]^{c}\big\} & \le 4B_{1}^{-\frac{1}{2}} \mathbb{E}\left|  \big[\widehat{m}(\boldsymbol{X}) - m(\boldsymbol{X}) \big]^3 \zeta(\boldsymbol{X}_{S^{*}})\varepsilon\right| \\
			&\le 4B_{1}^{-\frac{1}{2}} \big[\sup_{\vec{z}\in\mathbb{R}^{s^*}} \zeta(\vec{z})\big] \big[\mathbb{E} |\widehat{m}(\boldsymbol{X}) - m(\boldsymbol{X}) |^3\big]\mathbb{E}| \varepsilon|\\
			& \le 8\overbar{M}_{n} \big[\sup_{\vec{z}\in\mathbb{R}^{s^*}} \zeta(\vec{z})\big] (\mathbb{E}| \varepsilon|) B_{1}^{\frac{1}{2}},\\
			\mathbb{P}\big\{\big[E_{6}(\kappa_{r})\big]^{c}\big\} &\le 4B_{1}^{-\frac{1}{4}} \big[\sup_{\vec{z}\in\mathbb{R}^{s^*}} \zeta(\vec{z})\big]^3 \mathbb{E}\big[ |\widehat{m}(\boldsymbol{X}) - m(\boldsymbol{X}) | |\varepsilon|^3\big]\\
			&\le 4B_{1}^{-\frac{1}{4}} \big[\sup_{\vec{z}\in\mathbb{R}^{s^*}} \zeta(\vec{z})\big]^3 \big[\mathbb{E} (\widehat{m}(\boldsymbol{X}) - m(\boldsymbol{X}))^2\big]^{\frac{1}{2}} \big(\mathbb{E} |\varepsilon|^3\big)\\
			&\le 4\big[\sup_{\vec{z}\in\mathbb{R}^{s^*}} \zeta(\vec{z})\big]^3\big(\mathbb{E} |\varepsilon|^3\big)B_{1}^{\frac{1}{4}},\\
			\mathbb{P} \big\{\big[E_{7}(\kappa_{r})\big]^{c}\big\} &\le  6 \big[\sup_{\vec{z}\in\mathbb{R}^{s^*}} \zeta(\vec{z})\big]^2  \mathbb{E} (\varepsilon^2)B_{1}^{\frac{1}{2}}.
		\end{split}
	\end{equation}

	By \eqref{theorem3.8}--\eqref{theorem3.9},  model regularity assumptions, the assumption that $R$ is a finite constant, and that $\varsigma = t(\log{n})(10n^{-\frac{1}{4}} + 4B_{1}^{\frac{1}{8}}) + n^{\frac{1}{4}}B_{1}^{\frac{1}{2}}$, there exists some $N_{1}>0$ such that for all $n\ge N_{1}$, each $t>0$, and each $0<B_{1}<1$,
	\begin{equation}
		\begin{split}\label{theorem3.4}
			&\mathbb{P}(|T_{j}(\widehat{a}_{j})\big[\widehat{\sigma}_{j}(\widehat{a}_{j})\big]^{-1}| \ge t )  \\
			&\le \mathbb{P}(|T_{j}^{(\star)}(\widehat{a}_{j})| \ge t\big[\sigma_{j}(\widehat{a}_{j})\big]  - \varsigma)+ Rn^{\frac{1}{4}}B_{1}^{\frac{1}{2}} + (\overbar{M}_{n})^2(\frac{\log{n}}{2})(n^{-\frac{1}{4}} + B_{1}^{\frac{1}{4}}).
		\end{split}
	\end{equation}
	
	To deal with the term $\mathbb{P}(|T_{j}^{(\star)}(\widehat{a}_{j})| \ge t\big[\sigma_{j}(\widehat{a}_{j})\big] -\varsigma)$ on the RHS of \eqref{theorem3.4}, we need the following results. By the Berry-Esseen inequality~\citepsupp{petrov1977sums}, $\sup_{\vec{z}\in\mathbb{R}^{s^*}}\zeta(\vec{z})<\infty$, $\mathbb{E}(\varepsilon^8)<\infty$, \eqref{theorem3.1}, and that $\varepsilon$ is an independent model error, there exists $C>0$ such that for each $n\ge1$ and each $r\in\{1, \dots, R\}$,
	\begin{equation}
		\begin{split}\label{theorem3.2}
			\sup_{x\in\mathbb{R}}|\mathbb{P}(T_{j}^{(\star)}(\kappa_{r})\big[\sigma_{j}(\kappa_{r})\big]^{-1} \le x)- \Phi(x)| \le \frac{C}{\sqrt{n}},
		\end{split}	
	\end{equation}
	which leads to
	\begin{equation}
		\begin{split}\label{theorem3.3}
			&\left|\mathbb{P}(|T_{j}^{(\star)}(\widehat{a}_{j})| \ge t\big[\sigma_{j}(\widehat{a}_{j})\big]  - \varsigma) - 2\Phi\left(-t \right)\right|  \\
			&=\bigg|\sum_{r=1}^{R}\mathbb{P}\big\{\big\{|T_{j}^{(\star)}(\kappa_{r})| \ge t\big[\sigma_{j}(\kappa_{r})\big]  - \varsigma \big\} \cap \{\widehat{a}_{j} =\kappa_{r}\}\big\} \\
			&\qquad - \sum_{r=1}^{R}2\Phi\left(-t + \frac{\varsigma}{\sigma_{j}(\kappa_{r})}\right)\mathbb{P}(\widehat{a}_{j} =\kappa_{r})\\
			&\qquad + \sum_{r=1}^{R}2\Phi\left(-t + \frac{\varsigma}{\sigma_{j}(\kappa_{r})}\right)\mathbb{P}(\widehat{a}_{j} =\kappa_{r})- 2\Phi\left(-t \right) \bigg| \\		
			& \le \sum_{r=1}^{R}\mathbb{P}(\widehat{a}_{j} =\kappa_{r})	\left|\mathbb{P}\big\{|T_{j}^{(\star)}(\kappa_{r})| \ge t\big[\sigma_{j}(\kappa_{r})\big]  - \varsigma \big\} - 2\Phi\left(-t + \frac{\varsigma}{\sigma_{j}(\kappa_{r})}\right) \right| \\
			&\qquad\qquad + 2 \sum_{r=1}^{R}\left| \Phi\left(-t + \frac{\varsigma}{\sigma_{j}(\kappa_{r})}\right) - \Phi\left(-t \right) \right|\mathbb{P}(\widehat{a}_{j} =\kappa_{r})\\
			&\le \sum_{r=1}^{R}\mathbb{P}(\widehat{a}_{j} =\kappa_{r}) \bigg\{\left|\mathbb{P}(T_{j}^{(\star)}(\kappa_{r}) \le -t\big[\sigma_{j}(\kappa_{r})\big] +\varsigma) -\Phi\left(-t + \frac{\varsigma}{\sigma_{j}(\kappa_{r})}\right)\right|\\
			&\qquad\qquad+\left| \mathbb{P}(T_{j}^{(\star)}(\kappa_{r}) \ge t\big[\sigma_{j}(\kappa_{r})\big] -\varsigma) -\Phi\left(-t + \frac{\varsigma}{\sigma_{j}(\kappa_{r})}\right)\right|\bigg\}\\
			&\qquad\qquad + 2 \sum_{r=1}^{R}\left| \Phi\left(-t + \frac{\varsigma}{\sigma_{j}(\kappa_{r})}\right) - \Phi\left(-t \right) \right|\mathbb{P}(\widehat{a}_{j} =\kappa_{r})\\
			& = \sum_{r=1}^{R}\mathbb{P}(\widehat{a}_{j} =\kappa_{r}) \left|\mathbb{P}(T_{j}^{(\star)}(\kappa_{r}) \le -t\big[\sigma_{j}(\kappa_{r})\big] +\varsigma) -\Phi\left(-t + \frac{\varsigma}{\sigma_{j}(\kappa_{r})}\right)\right|\\
			&\qquad + \sum_{r=1}^{R}\mathbb{P}(\widehat{a}_{j} =\kappa_{r}) \left|1 - \mathbb{P}(T_{j}^{(\star)}(\kappa_{r}) \le t\big[\sigma_{j}(\kappa_{r})\big] -\varsigma) - 1 + \Phi\left(t - \frac{\varsigma}{\sigma_{j}(\kappa_{r})}\right)\right| \\
			&\qquad + 2 \sum_{r=1}^{R}\left| \Phi\left(-t + \frac{\varsigma}{\sigma_{j}(\kappa_{r})}\right) - \Phi\left(-t \right) \right|\mathbb{P}(\widehat{a}_{j} =\kappa_{r})\\
			&\le \frac{2C}{\sqrt{n}} + 2\times 0.4\times \varsigma \times \big\{c\big[\inf_{\vec{z}\in\mathbb{R}^{s^*}} \zeta(\vec{z})\big]^4\mathbb{E}(\varepsilon^{4} )\big\}^{-\frac{1}{2}},
		\end{split}
	\end{equation}
	where $C>0$ is given in \eqref{theorem3.2}; the first inequality is due to the assumption that $\widehat{a}_{j}$ is independent of $T_{j}^{(\star)}(\kappa_{r})$'s; the second equality follows because $T_{j}^{(\star)}(\kappa_{r})$ is a continuous random variable; and the last inequality is from \eqref{theorem1.32}, \eqref{theorem3.1}, and the fact that $\sum_{r=1}^{R}\mathbb{P}(\widehat{a}_{j} =\kappa_{r}) = 1$.
	
	We use  \eqref{theorem3.3} and model regularity assumptions to deduce that there exists some $N_{2}>0$ such that for all $n\ge N_{2}$, each $t>0$, and each $0<B_{1}<1$,
	\begin{equation}
		\begin{split}	\label{theorem3.11}
			\mathbb{P}(|T_{j}^{(\star)}(\widehat{a}_{j})| \ge t\big[\sigma_{j}(\widehat{a}_{j})\big]  - \varsigma) \le 2\Phi(-t) + \varsigma\big\{c\big[\inf_{\vec{z}\in\mathbb{R}^{s^*}} \zeta(\vec{z})\big]^4\mathbb{E}(\varepsilon^{4} )\big\}^{-\frac{1}{2}} + 2Cn^{-\frac{1}{2}},
		\end{split}
	\end{equation}
	which in combination with \eqref{theorem3.4} and the fact that $n^{-\frac{1}{2}} = o\big\{\overbar{M}_{n}^2(\log{n})(n^{-\frac{1}{4}} + B_{1}^{\frac{1}{4}})\big\}$ concludes the desired result of Theorem~\ref{theorem3} for the $j$th coordinate.
	
	Lastly, notice that generic constants in this proof as well as other statements of ``for all large $n$'' are not subject to feature index $j$ (e.g. $N_{1}$ in \eqref{theorem3.4}, $N_{2}$ in \eqref{theorem3.11}, and $C$ in \eqref{theorem3.2}). Therefore, we conclude the proof of  Theorem~\ref{theorem3}.

	\subsection{Proof of Theorem~\ref{theorem4}}\label{proof.theorem4}

	Before we begin the formal proof, we recall some notation for the reader's convenience. We have assumed $\zeta(\boldsymbol{X}_{S^*}) = \zeta_{0}$ if $S^* = \emptyset$ in Section~\ref{Sec1.1}, and we have defined $G_{l}(\kappa_{r}) = n^{-1}\sum_{i=1}^{n}(\boldsymbol{1}_{U_{il}\in(-\infty, \kappa_{r}]}- \boldsymbol{1}_{\widetilde{U}_{il}\in(-\infty, \kappa_{r}]}) (\widehat{\eta}_{i} )^{2}$ with $\widehat{\eta}_{i} = V_{i} - \widehat{m}(\boldsymbol{U}_{i})$. The cumulative function of the standard normal distribution is denoted by $\Phi(\cdot)$. Change point candidates $\kappa_{1}, \dots, \kappa_{R}$ with some finite  $R>1$  are such that $\kappa_{r}<\kappa_{r+1}$. In addition, we use the following notation for $r\in\{1, \dots, R\}$ and $l\in\{1,\dots, p\}$.
	\begin{equation*}
		\begin{split}
			\widehat{a}_{l}& = \arg\max_{\kappa\in\{\kappa_{1}, \dots ,\kappa_{R}\}}|G_{l}(\kappa)|,\\
			T_{l}(\kappa_{r}) & = n^{-\frac{1}{2}}\sum_{i=1}^{n}(\boldsymbol{1}_{X_{il}\in(-\infty, \kappa_{r}]}- \boldsymbol{1}_{\widetilde{X}_{il}\in(-\infty, \kappa_{r}]}) (\widehat{\varepsilon}_{i} )^{2},\\
			T_{l}^{(\star)}(\kappa_{r}) &=n^{-\frac{1}{2}}\sum_{i=1}^{n}\big(\boldsymbol{1}_{X_{il}\in(-\infty, \kappa_{r}]} -\boldsymbol{1}_{\widetilde{X}_{il}\in (-\infty, \kappa_{r}]}\big)(\zeta(\boldsymbol{X}_{iS^{*}})\varepsilon_{i})^{2},\\
			\mu_{l}(\kappa_{r}) &= \mathbb{E}\big\{\big[\boldsymbol{1}_{X_{l} \in (-\infty, \kappa_{r}]} -\boldsymbol{1}_{\widetilde{X}_{l} \in (-\infty, \kappa_{r}]}\big]\big[ \zeta(\boldsymbol{X}_{S^{*}})\varepsilon\big]^2\big\},\\
			\widehat{\mu}_{l}(\kappa_{r}) &= \mathbb{E}\big\{\big[\boldsymbol{1}_{X_{l} \in (-\infty, \kappa_{r}]} -\boldsymbol{1}_{\widetilde{X}_{l} \in (-\infty, \kappa_{r}]}\big]( \widehat{\varepsilon}_{i})^2\big\},\\
			\sigma_{l}^2(\kappa_{r}) &  = \mathbb{E}\big\{\big[\boldsymbol{1}_{X_{l} \in (-\infty, \kappa_{r}]} -\boldsymbol{1}_{\widetilde{X}_{l} \in (-\infty, \kappa_{r}]}\big]^2\big[ \zeta(\boldsymbol{X}_{S^{*}})\varepsilon\big]^4\big\} - \mu_{l}^2(\kappa_{r}),\\
			\widehat{\sigma}_{l}^2(\kappa_{r}) & = n^{-1}\sum_{i=1}^{n}\Big((\boldsymbol{1}_{X_{il}\in(-\infty, \kappa_{r}]}- \boldsymbol{1}_{\widetilde{X}_{il}\in(-\infty, \kappa_{r}]}) (\widehat{\varepsilon}_{i} )^{2} - \widehat{\mu}_{l}(\kappa_{r})\Big)^2. \\
		\end{split}
	\end{equation*}

	The proof ideas for Theorme~\ref{theorem4} follow those for the proofs of Theorem~\ref{theorem2b}, and hence we omit the details. The formal proof of Theorem~\ref{theorem4} begins with an upper bound and a lower bound of $\mathbb{P}\big\{\big[T_{\widehat{l}}(\widehat{a}_{\widehat{l}})-\sqrt{n}\big[\mu_{\widehat{l}}(\widehat{a}_{\widehat{l}})\big]\big] \big[\widehat{\sigma}_{\widehat{l}}(\widehat{a}_{\widehat{l}})\big]^{-1}\le x\big\}$ as follows. For each $x\in \mathbb{R}$, $n\ge 1$, $B_{1}>0$,
	\begin{equation}\label{theorem4.31}
		{\small\begin{split}
				& \mathbb{P}\big\{\big[T_{\widehat{l}}(\widehat{a}_{\widehat{l}})-\sqrt{n}\big[\mu_{\widehat{l}}(\widehat{a}_{\widehat{l}})\big]\big] \big[\widehat{\sigma}_{\widehat{l}}(\widehat{a}_{\widehat{l}})\big]^{-1}\le x\big\}\\
				&\le \mathbb{P}\Big\{\big\{ T_{\widehat{l}}^{(\star)}(\widehat{a}_{\widehat{l}})-\sqrt{n}\big[\mu_{\widehat{l}}(\widehat{a}_{\widehat{l}})\big] \le x\big[\sigma_{\widehat{l}}(\widehat{a}_{\widehat{l}})\big] + |x||\sigma_{\widehat{l}}(\widehat{a}_{\widehat{l}}) - \widehat{\sigma}_{\widehat{l}}(\widehat{a}_{\widehat{l}})| \\
				&\qquad\qquad + Q_{1} + |Q_{2}(\widehat{l}, \widehat{a}_{\widehat{l}})|\big\}\cap \{W(x, b, B_{1})\}^c\Big\} + \mathbb{P}(W(x, b, B_{1})),
		\end{split}}
	\end{equation}
	where the event $W(x, n, B_{1})$ is defined such that for each $x\in\mathbb{R}, n\ge 1, B_{1}>0$,
	\begin{equation*}
		\begin{split}
			W(x, n, B_{1}) & = \Big\{|x||\sigma_{\widehat{l}}(\widehat{a}_{\widehat{l}}) - \widehat{\sigma}_{\widehat{l}}(\widehat{a}_{\widehat{l}})| + Q_{1} + |Q_{2}(\widehat{l}, \widehat{a}_{\widehat{l}})| > \varsigma(x, n, B_{1})\Big\},
		\end{split}
	\end{equation*}
	in which $\varsigma (x, n, B_{1})  = |x|n^{-\beta_{2}}  + \sqrt{n}B_{1} (-\log B_{1}) + B_{1}^{\frac{1}{4}}$, $Q_{1} = n^{-\frac{1}{2}}\sum_{i=1}^{n} \big[\widehat{m}(\boldsymbol{X}_{i}) - m(\boldsymbol{X}_{i})\big]^2$, and for  each $l\in \{1, \dots, p\}$ and $b\in \mathbb{R}$,
	\begin{equation*}
		\begin{split}
			Q_{2}(l, b) = 2n^{-\frac{1}{2}}\sum_{i=1}^{n} \big[\boldsymbol{1}_{X_{il}\in(-\infty, b]}- \boldsymbol{1}_{\widetilde{X}_{il}\in(-\infty, b]}\big]\big[\widehat{m}(\boldsymbol{X}_{i}) - m(\boldsymbol{X}_{i})\big] \varepsilon_{i}\zeta(\boldsymbol{X}_{iS^*}).
		\end{split}
	\end{equation*} 
	Similarly, we establish the probability lower bound as follows.
	\begin{equation}\label{theorem4.32}
		{\small\begin{split}
				& \mathbb{P}\big\{\big[T_{\widehat{l}}(\widehat{a}_{\widehat{l}})-\sqrt{n}\mu_{\widehat{l}}(\widehat{a}_{\widehat{l}})\big] \big[\widehat{\sigma}_{\widehat{l}}(\widehat{a}_{\widehat{l}})\big]^{-1}\le x\big\}\\
				&\ge \mathbb{P}\Big\{\big\{ T_{\widehat{l}}^{(\star)}(\widehat{a}_{\widehat{l}})-\sqrt{n}\big[\mu_{\widehat{l}}(\widehat{a}_{\widehat{l}})\big] \le x\big[\sigma_{\widehat{l}}(\widehat{a}_{\widehat{l}})\big] - |x||\sigma_{\widehat{l}}(\widehat{a}_{\widehat{l}}) - \widehat{\sigma}_{\widehat{l}}(\widehat{a}_{\widehat{l}})| - Q_{1} - |Q_{2}(\widehat{l}, \widehat{a}_{\widehat{l}})|\big\}\\
				&\qquad\cap \{W(x, n, B_{1})\}^c\Big\}.
		\end{split} }
	\end{equation}

	To deal with the RHS of \eqref{theorem4.31}--\eqref{theorem4.32}, we need a uniform lower bound for the population variances, which is given in \eqref{theorem4.6} below. For every $x$ with $\kappa_{1}\le x\le \kappa_{R}$ and each $j\in\{1, \dots, p\}$,
	\begin{equation*}
		\begin{split}
			\mathbb{P}(\{X_{j} \in (-\infty, x]\}\cap\{\widetilde{X}_{j} \in (-\infty, x]\}) &\ge  \mathbb{P}(\{X_{j} \in (-\infty, \kappa_{1}]\}\cap\{\widetilde{X}_{j} \in (-\infty, \kappa_{1}]\}), \\
			\mathbb{P}(\{X_{j} \in (-\infty, x]\}\cap\{\widetilde{X}_{j} \in (x, \infty)\}) &\ge  \mathbb{P}(\{X_{j} \in (-\infty, \kappa_{1}]\}\cap\{\widetilde{X}_{j} \in (\kappa_{R}, \infty)\}), 
		\end{split}
	\end{equation*}
	which in combination with Condition~\ref{non.trivial.2} implies that for every $x$ with $\kappa_{1}\le x\le \kappa_{R}$ and each $j\in\{1, \dots, p\}$,
	\begin{equation*}
		\begin{split}
			\mathbb{P}(\{X_{j} \in (-\infty, x]\}\cap\{\widetilde{X}_{j} \in (-\infty, x]\}) &> c, \\
			\mathbb{P}(\{X_{j} \in (-\infty, x]\}\cap\{\widetilde{X}_{j} \in (x, \infty)\}) &>  c,
		\end{split}
	\end{equation*} 
	in which $c>0$ is defined in Condition~\ref{non.trivial.2}. By these inequalities, regularity assumptions, and Lemma~\ref{uniform.var.1} in Section~\ref{uniform.var.2}, there exists some $\underline{\sigma}>0$ (see \eqref{theorem3.24} in Lemma~\ref{uniform.var.1}) such that 
	\begin{equation}
		\label{theorem4.6}
		\inf_{1\le l\le p, \kappa_{1}\le x \le \kappa_{R}}\sigma_{l}^2(x)\ge \underline{\sigma}^2.
	\end{equation}

	By the definition of $W(x, n, B_{1})$ and \eqref{theorem4.6}, the RHS of \eqref{theorem4.31}--\eqref{theorem4.32} can be further analyzed in \eqref{theorem4.27}--\eqref{theorem4.33} below, respectively.
	\begin{equation}\label{theorem4.27}
		\begin{split}
			\textnormal{RHS of \eqref{theorem4.31}} &\le  \mathbb{P}\big\{\big[T_{\widehat{l}}^{(\star)}(\widehat{a}_{\widehat{l}})-\sqrt{n}\big[\mu_{\widehat{l}}(\widehat{a}_{\widehat{l}}) \big]\big] \big[\sigma_{\widehat{l}}(\widehat{a}_{\widehat{l}})\big]^{-1}\le x + \varsigma(x, n, B_{1}) (\underline{\sigma}^{-1}) \big\} \\
			&\qquad+ \mathbb{P}(W(x, n, B_{1})),
		\end{split}
	\end{equation}
	and similarly,
	{\small\begin{equation}
			\begin{split}\label{theorem4.33}
				&\textnormal{RHS of \eqref{theorem4.32}} \\&\ge  \mathbb{P}\Big\{\big\{\big[T_{\widehat{l}}^{(\star)}(\widehat{a}_{\widehat{l}})-\sqrt{n}\big[\mu_{\widehat{l}}(\widehat{a}_{\widehat{l}}) \big]\big] \big[\sigma_{\widehat{l}}(\widehat{a}_{\widehat{l}})\big]^{-1}\le x - \varsigma(x, n, B_{1}) (\underline{\sigma}^{-1}) \big\} \cap \{W(x, n, B_{1})\}^c\Big\}\\
				& \ge  \mathbb{P}\big\{\big[T_{\widehat{l}}^{(\star)}(\widehat{a}_{\widehat{l}})-\sqrt{n}\big[\mu_{\widehat{l}}(\widehat{a}_{\widehat{l}})\big]\big] \big[\sigma_{\widehat{l}}(\widehat{a}_{\widehat{l}})\big]^{-1}\le x - \varsigma(x, n, B_{1}) (\underline{\sigma}^{-1}) \big\} -\mathbb{P}(W(x, n, B_{1})),
			\end{split}
	\end{equation}}%
	where the second inequality is due to the fact that $\mathbb{P}(A\cap B) = \mathbb{P}(A) - \mathbb{P}(A\cap B^c)$ for any events $A$ and $B$.

	Now, we proceed to show that the probability $\mathbb{P}(W(x, n, B_{1}))$ on the RHS of \eqref{theorem4.27}--\eqref{theorem4.33} is negligible, and that $\big[T_{\widehat{l}}^{(\star)}(\widehat{a}_{\widehat{l}})-\sqrt{n}\big[\mu_{\widehat{l}}(\widehat{a}_{\widehat{l}})\big]\big] \big[\sigma_{\widehat{l}}(\widehat{a}_{\widehat{l}})\big]^{-1}$ is asymptotically standard normal. In light of $\mathbb{E}|\varepsilon|^{q\vee 8} <\infty$ and  Conditions~\ref{consistency.3}--\ref{bound} with the assumptions $\beta_{2} < \frac{1}{4}$ and
	$$\lim_{n\rightarrow\infty} n^{\beta_{2}}\overbar{M}_{n}^2B_{1}+ \sqrt{n}B_{1}(-\log{B_{1}}) + n^{-\frac{1}{2}+ \frac{4}{q-\beta} + \beta_{2}} \sqrt{\log{(n\vee p)}} =0,$$ 
	we show in Section~\ref{proof.theorem4.5} that for each $x\in\mathbb{R}$, 
	\begin{equation}
		\begin{split}\label{theorem4.5}
			&\lim_{n\rightarrow\infty}\mathbb{P}(W(x, n, B_{1})) = 0.
		\end{split}
	\end{equation}
	
	Next, by the Berry-Esseen inequality~\citepsupp{petrov1977sums}, along with \eqref{theorem4.6}, the assumption of i.i.d. observations, and the assumptions that $\mathbb{E}|\varepsilon|^8<\infty$ and $\sup_{\vec{z}\in\mathbb{R}^{s^*}}\zeta(\vec{z})<\infty$, there exists some $C>0$ such that for each $n\ge 1$, each $1\le l \le p$, and each $1\le r\le R$,
	\begin{equation}\label{theorem4.26}
		\sup_{x\in \mathbb{R}} \big|\mathbb{P}\big\{\big[T_{l}^{(\star)}(\kappa_{r})-\sqrt{n}\big[\mu_{l}(\kappa_{r})\big]\big] \big[\sigma_{l}(\kappa_{r})\big]^{-1}\le x \big\}- \Phi(x) \big| \le \frac{C}{\sqrt{n}}.
	\end{equation}		
	Furthermore, by \eqref{theorem4.26} and that $\widehat{l}$ and $\widehat{a}_{l}$'s are independent of $T_{l}^{(\star)}(\kappa_{1}), \dots, T_{l}^{(\star)}(\kappa_{R})$, it holds that 
	{\small\begin{equation}
			\begin{split}
				&\sup_{x\in \mathbb{R}}|	\mathbb{P}\big\{\big[T_{\widehat{l}}^{(\star)}(\widehat{a}_{\widehat{l}})-\sqrt{n}\big[\mu_{\widehat{l}}(\widehat{a}_{\widehat{l}}) \big]\big] \big[\sigma_{\widehat{l}}(\widehat{a}_{\widehat{l}})\big]^{-1}\le x \big\} - \Phi(x)| \\
				& =\sup_{x\in \mathbb{R}}\Big|	\sum_{l=1}^{p}\sum_{r=1}^{R} \mathbb{P}\big\{\big[T_{l}^{(\star)}(\kappa_{r})-\sqrt{n} \big[\mu_{l}(\kappa_{r}) \big]\big] \big[\sigma_{l}(\kappa_{r})\big]^{-1}\le x \big\} \mathbb{P}(\{\widehat{l} = l\}\cap\{\widehat{a}_{l} = \kappa_{r} \}) \\
				&\qquad- \Phi(x)\Big|\\
				& \le \sum_{l=1}^{p}\sum_{r=1}^{R}\Big\{\sup_{x\in \mathbb{R}}\Big|	 \mathbb{P}\big\{\big[T_{l}^{(\star)}(\kappa_{r})-\sqrt{n}\big[\mu_{l}(\kappa_{r})\big]\big] \big[\sigma_{l}(\kappa_{r})\big]^{-1}\le x \big\} - \Phi(x)\Big|\\
				& \qquad\times\mathbb{P}(\{\widehat{l} = l\}\cap\{\widehat{a}_{l} = \kappa_{r} \})\Big\} \\
				&\le \frac{C}{\sqrt{n}},
			\end{split}
	\end{equation}}%
	where constant $C$ is given in \eqref{theorem4.26}. In addition, for each $x, y\in\mathbb{R}$,
	\begin{equation}
		\begin{split}\label{theorem4.28}
			|\Phi(x) - \Phi(x+y)| \le 0.4\times |y|,
		\end{split}
	\end{equation}
	since the maximum value of density of the standard Gaussian distribution is less than $0.4$.

	By  \eqref{theorem4.31}--\eqref{theorem4.33}, \eqref{theorem4.26}--\eqref{theorem4.28}, there exists some constant $C>0$ such that for all large $n$, each $x\in\mathbb{R}$, and each $B_{1}>0$,
	\begin{equation}
		\begin{split}
			& |\mathbb{P}\big\{\big[T_{\widehat{l}}(\widehat{a}_{\widehat{l}})-\sqrt{n}\big[\mu_{\widehat{l}}(\widehat{a}_{\widehat{l}})\big]\big] \big[\widehat{\sigma}_{\widehat{l}}(\widehat{a}_{\widehat{l}})\big]^{-1}\le x\big\} -  \Phi(x)|\\
			&\qquad \le  \frac{C}{\sqrt{n}}+ 0.4\times \frac{\varsigma(x, n, B_{1})}{ \underline{\sigma}}  + \mathbb{P}(W(x, n, B_{1})),
		\end{split}
	\end{equation}
	which in combination with  the assumption $\lim_{n\rightarrow\infty}\sqrt{n}B_{1}(-\log{B_{1}}) = 0$ (hence $\lim_{n\rightarrow\infty}\varsigma(x, n, B_{1}) = 0$ for each $x\in\mathbb{R}$) and  \eqref{theorem4.5} concludes the main desired result. For the other assertion of Theorem~\ref{theorem4}, note that $\mu_{l}(\kappa_{r}) = 0$ for each $l\in\{1, \dots, p\}$ and $ r\in \{1, \dots ,R\}$ when $S^* = \emptyset$. We have finished the proof of Theorem~\ref{theorem4}.

	\subsubsection{Proof of \eqref{theorem4.5}}\label{proof.theorem4.5}
	
	First, we deduce that for each $x\in\mathbb{R}$,
	{\small\begin{equation}
			\begin{split}\label{theorem4.29}
				&\mathbb{P}(W(x, n, B_{1})) \\& \le \mathbb{P} \big\{|\sigma_{\widehat{l}}(\widehat{a}_{\widehat{l}}) - \widehat{\sigma}_{\widehat{l}}(\widehat{a}_{\widehat{l}})| > n^{-\beta_{2}}\big\} + \mathbb{P}\big\{Q_{1} > \sqrt{n}B_{1} (-\log B_{1})\big\} + \mathbb{P}\big\{|Q_{2}(\widehat{l}, \widehat{a}_{\widehat{l}})|> B_{1}^{\frac{1}{4}}\big\}.
			\end{split}
	\end{equation}}%
	Recall that $Q_{1} = n^{-\frac{1}{2}}\sum_{i=1}^{n} \big[\widehat{m}(\boldsymbol{X}_{i}) - m(\boldsymbol{X}_{i})\big]^2$ and for  each $l\in \{1, \dots, p\}$ and $b\in \mathbb{R}$,
	\begin{equation*}
		\begin{split}
			Q_{2}(l, b) = 2n^{-\frac{1}{2}}\sum_{i=1}^{n} \big[\boldsymbol{1}_{X_{il}\in(-\infty, b]}- \boldsymbol{1}_{\widetilde{X}_{il}\in(-\infty, b]}\big]\big[\widehat{m}(\boldsymbol{X}_{i}) - m(\boldsymbol{X}_{i})\big] \varepsilon_{i}\zeta(\boldsymbol{X}_{iS^*}).
		\end{split}
	\end{equation*} 
	Below, we deal with the upper bounds for each term on the RHS of \eqref{theorem4.29}, and begin with the one for the second term $\mathbb{P}\big\{Q_{1} > \sqrt{n}B_{1} (-\log B_{1})\big\}$. 
	
	By Markov's inequality, Condition~\ref{consistency.3}, the assumption of i.i.d. observations, and that the training sample is an independent sample, it holds that for all $n\ge 1$ and $B_{1}>0$,
	\begin{equation}\label{theorem4.24}
		\begin{split}
			\mathbb{P}(Q_{1}> \sqrt{n}B_{1} (-\log B_{1}))\le (-\log B_{1})^{-1}.
		\end{split}
	\end{equation}
	
	Next, for deriving an upper bound on the third term on the RHS of \eqref{theorem4.29}, we use the Burkholder-Davis-Gundy inequality~\citepsupp{burkholder1972integral}. To apply the Burkholder-Davis-Gundy inequality, we have to specify the martingale difference sequence in $Q_{2}(\widehat{l}, \widehat{a}_{\widehat{l}})$. Let $q_{2i}(l, a) = \big[\boldsymbol{1}_{X_{il}\in(-\infty, a]}- \boldsymbol{1}_{\widetilde{X}_{il}\in(-\infty, a]}\big]\big[\widehat{m}(\boldsymbol{X}_{i}) - m(\boldsymbol{X}_{i})\big] \varepsilon_{i}\zeta(\boldsymbol{X}_{iS^*})$ for $1\le i\le n$ and $q_{2i}(l, a) = 0$ otherwise. As a result,  $Q_{2}(\widehat{l}, \widehat{a}_{\widehat{l}}) = \sum_{i=0}^{\infty}\frac{2q_{2i}(\widehat{l}, \widehat{a}_{\widehat{l}})}{\sqrt{n}}$. Let $\mathcal{F}_{0} = \sigma(\mathcal{X}_{0}, \mathcal{X}_{2})$ and $\mathcal{F}_{i} = \sigma\big(\mathcal{X}_{0}, \mathcal{X}_{2}, \frac{2q_{21}(\widehat{l}, \widehat{a}_{\widehat{l}})}{\sqrt{n}}\dots , \frac{2q_{2i}(\widehat{l}, \widehat{a}_{\widehat{l}})}{\sqrt{n}}\big)$ for each $i\ge 1$, where $\sigma(\cdot)$ denotes the $\sigma$-algebra generated by the given random mappings, $\mathcal{X}_{0}$ is the independent training sample for constructing $\widehat{m}(\cdot)$, and $\mathcal{X}_{2} = \{V_{i}, \boldsymbol{U}_{i}, \widetilde{\boldsymbol{U}}_{i}\}_{i=1}^{n}$. Now, $\{\frac{2q_{2i}(\widehat{l}, \widehat{a}_{\widehat{l}})}{\sqrt{n}}, \mathcal{F}_{i}\}_{i\ge 0}$ is a martingale difference sequence, and hence it holds that for some $K>0$,
	\begin{equation}\label{theorem4.23}
		\begin{split}
			\mathbb{P}(  |Q_{2}(\widehat{l}, \widehat{a}_{\widehat{l}})  |> B_{1}^{\frac{1}{4}}) &\le B_{1}^{-\frac{1}{4}} \mathbb{E}|Q_{2}(\widehat{l}, \widehat{a}_{\widehat{l}}) |\\
			& \le B_{1}^{-\frac{1}{4}}K \sqrt{n^{-1}\sum_{i=1}^{n} \mathbb{E}\big[\widehat{m}(\boldsymbol{X}_{i}) - m(\boldsymbol{X}_{i})\big]^2 \big[\sup_{\vec{z}\in\mathbb{R}^{s^*}}\zeta(\vec{z})\big]^2}\\
			&\le KB_{1}^{\frac{1}{4}}\sup_{\vec{z}\in\mathbb{R}^{s^*}}\zeta(\vec{z}),
		\end{split}
	\end{equation}
	where the first inequality is due to Markov's inequality; the second inequality is from  $\mathbb{E}(\varepsilon^2)=1$, the Burkholder-Davis-Gundy inequality, and Jensen's inequality; and the third inequality follows from Condition~\ref{consistency.3}, the assumption of i.i.d. observations, and that the training sample is an independent sample.

	Next, we proceed to establish an upper bound on the first term on the RHS of \eqref{theorem4.29}. Simple calculations show that for each $a\in\mathbb{R}$ and $l\in\{1, \dots, p\}$,
	\begin{equation}\label{theorem4.1}
		|\widehat{\sigma}_{l}(a) - \sigma_{l}(a)| = \frac{|\widehat{\sigma}_{l}^2 (a) - \sigma_{l}^2(a)|}{\widehat{\sigma}_{l}(a) + \sigma_{l}(a)} \le \frac{|\widehat{\sigma}_{l}^2 (a) - \sigma_{l}^2(a)|}{\sigma_{l}(a)}\le \frac{|\widehat{\sigma}_{l}^2 (a) - \sigma_{l}^2(a)|}{\underline{\sigma}},
	\end{equation}
	where $\underline{\sigma}$ is given in \eqref{theorem4.6}. In addition, for each $a\in\mathbb{R}$ and $l\in\{1, \dots, p\}$,
	\begin{equation}
		\begin{split}\label{theorem4.7}
			\widehat{\sigma}_{l}^2(a)& = -\big[\widehat{\mu}_{l}(a)\big]^{2} + n^{-1}\sum_{i=1}^{n} (\boldsymbol{1}_{X_{il} \in (-\infty, a]} -\boldsymbol{1}_{\widetilde{X}_{il} \in (-\infty, a]})^2(\widehat{\varepsilon}_{i})^4 \\
			&=  -\big[\widehat{\mu}_{l}(a)\big]^{2} +  n^{-1}\sum_{i=1}^{n} (\boldsymbol{1}_{X_{il} \in (-\infty, a]} -\boldsymbol{1}_{\widetilde{X}_{il} \in (-\infty, a]})^2(m(\boldsymbol{X}_{i}) - \widehat{m}(\boldsymbol{X}_{i}) )^4\\
			&\qquad + n^{-1}\sum_{i=1}^{n} (\boldsymbol{1}_{X_{il} \in (-\infty, a]} -\boldsymbol{1}_{\widetilde{X}_{il} \in (-\infty, a]})^2 4(m(\boldsymbol{X}_{i}) - \widehat{m}(\boldsymbol{X}_{i}) )^3 \zeta(\boldsymbol{X}_{iS^{*}})\varepsilon_{i}\\
			&\qquad + n^{-1}\sum_{i=1}^{n} (\boldsymbol{1}_{X_{il} \in (-\infty, a]} -\boldsymbol{1}_{\widetilde{X}_{il} \in (-\infty, a]})^2 6( m(\boldsymbol{X}_{i}) - \widehat{m}(\boldsymbol{X}_{i})  )^2 (\zeta(\boldsymbol{X}_{iS^{*}})\varepsilon_{i})^2 \\
			&\qquad + n^{-1}\sum_{i=1}^{n} (\boldsymbol{1}_{X_{il} \in (-\infty, a]} -\boldsymbol{1}_{\widetilde{X}_{il} \in (-\infty, a]})^2 4(m(\boldsymbol{X}_{i}) - \widehat{m}(\boldsymbol{X}_{i}) ) (\zeta(\boldsymbol{X}_{iS^{*}})\varepsilon_{i})^3 \\
			& \qquad + n^{-1}\sum_{i=1}^{n} (\boldsymbol{1}_{X_{il} \in (-\infty, a]} -\boldsymbol{1}_{\widetilde{X}_{il} \in (-\infty, a]})^2( \zeta(\boldsymbol{X}_{iS^{*}})\varepsilon_{i})^4\\
			& \eqqcolon - \big[\widehat{\mu}_{l}(a)\big]^{2} + (I)_{l, a} + (II)_{l, a} + (III)_{l, a} + (IV)_{l, a} + (V)_{l, a}.
		\end{split}
	\end{equation}
	
	By \eqref{theorem4.1}--\eqref{theorem4.7}, for each $t>0$,
	\begin{equation}
		\begin{split}\label{theorem4.2}
			&\mathbb{P}(|\sigma_{\widehat{l}}(\widehat{a}_{\widehat{l}}) - \widehat{\sigma}_{\widehat{l}}(\widehat{a}_{\widehat{l}})| > t) \\
			& \le \sum_{r=1}^{R}\mathbb{P}(\max_{1\le l \le p}|\sigma_{l}(\kappa_{r}) - \widehat{\sigma}_{l}(\kappa_{r})| > t) \\
			& \le \sum_{r=1}^{R}\mathbb{P}(\max_{1\le l \le p}|\sigma_{l}^2(\kappa_{r}) - \widehat{\sigma}_{l}^2(\kappa_{r})| > t\underline{\sigma}) \\
			&\le \sum_{r=1}^{R}\bigg\{\mathbb{P}\Big(\max_{1\le l \le p}\Big| (V)_{l, \kappa_{r}} - \big[\sigma_{l}^2(\kappa_{r}) + \mu_{l}^2(\kappa_{r}) \big] \Big| >\frac{1}{6}t\underline{\sigma}\Big)\\
			&\qquad + \mathbb{P}\Big(\max_{1\le l \le p} |\widehat{\mu}_{l}^2(\kappa_{r})- \mu_{l}^2(\kappa_{r})| > \frac{1}{6}t\underline{\sigma}\Big) +\mathbb{P}\Big(\max_{1\le l \le p} |(I)_{l,\kappa_{r}}| > \frac{1}{6}t\underline{\sigma}\Big)\\
			& \qquad+\mathbb{P}\Big(\max_{1\le l \le p} |(II)_{l,\kappa_{r}}| > \frac{1}{6}t\underline{\sigma}\Big)+\mathbb{P}\Big(\max_{1\le l \le p} |(III)_{l,\kappa_{r}}| > \frac{1}{6}t\underline{\sigma}\Big)\\
			&\qquad +\mathbb{P}\Big(\max_{1\le l \le p} |(IV)_{l,\kappa_{r}}| > \frac{1}{6}t\underline{\sigma}\Big)\bigg\}.
		\end{split}
	\end{equation}

	In the following, we give upper bounds on each term on the RHS of \eqref{theorem4.2}. By Markov's inequality, the assumption of i.i.d. observations, the assumption  that the training sample for $\widehat{m}(\cdot)$ is an independent sample, and Conditions~\ref{consistency.3}--\ref{bound}, it holds that for all $n\ge 1$, each $1\le r\le R$, each $t>0$, and each $B_{1}>0$,
	\begin{equation} \label{theorem4.18}
		\begin{split}
			\mathbb{P}\Big(\max_{1\le l \le p} |(I)_{l,\kappa_{r}}| > \frac{1}{6}t\underline{\sigma}\Big)\le \frac{24\overbar{M}_{n}^2B_{1}}{t\underline{\sigma}},
		\end{split}
	\end{equation}
	where we use the result that for each $1\le i\le n$ and each $1\le r\le R$,
	$$\max_{1\le l \le p}\left(\boldsymbol{1}_{X_{il} \in (-\infty, \kappa_{r}]} - \boldsymbol{1}_{\widetilde{X}_{il} \in (-\infty, \kappa_{r}]}\right)^2 \le 1.$$
	This result is also used for deriving the other probability upper bounds below.

	By Markov's inequality, the assumption of i.i.d. observations, the assumption that the training sample for $\widehat{m}(\cdot)$ is an independent sample, Conditions~\ref{consistency.3}--\ref{bound}, and $\mathbb{E}(\varepsilon^2) = 1$, it holds that for all $n\ge 1$, each $1\le r\le R$, each $t>0$, and each $B_{1}>0$,		
	\begin{equation}
		\begin{split}
			\mathbb{P}\Big(\max_{1\le l \le p} |(II)_{l,\kappa_{r}}| > \frac{1}{6}t\underline{\sigma}\Big)&\le \frac{24\big[\sup_{\vec{z}\in\mathbb{R}^{s^*}}\zeta(\vec{z})\big]\mathbb{E} |m(\boldsymbol{X}) - \widehat{m}(\boldsymbol{X}) |^3 \mathbb{E}|\varepsilon|}{t\underline{\sigma}}\\
			&\le \frac{48\overbar{M}_{n}B_{1}\big[\sup_{\vec{z}\in\mathbb{R}^{s^*}}\zeta(\vec{z})\big]}{t\underline{\sigma}}.
		\end{split}
	\end{equation}

	By Markov's inequality,
	the assumption of i.i.d. observations, the assumption that the training sample for $\widehat{m}(\cdot)$ is an independent sample, Condition~\ref{consistency.3},  and $\mathbb{E}(\varepsilon)^2 = 1$, it holds that for all $n\ge 1$, each $1\le r\le R$, each $t>0$, and each $B_{1}>0$,		
	\begin{equation}
		\begin{split}
			\mathbb{P}\Big(\max_{1\le l \le p} |(III)_{l,\kappa_{r}}| > \frac{1}{6}t\underline{\sigma}\Big)&\le \frac{36B_{1}\big[\sup_{\vec{z}\in\mathbb{R}^{s^*}}\zeta(\vec{z})\big]^2}{t\underline{\sigma}}.
		\end{split}
	\end{equation}

	By  Markov's inequality, the assumption of i.i.d. observations, the assumption that the training sample for $\widehat{m}(\cdot)$ is an independent sample, Jensen's inequality,  and Condition~\ref{consistency.3}, it holds that for all $n\ge 1$, each $1\le r\le R$, each $t>0$, and each $B_{1}>0$,
	\begin{equation}
		\begin{split}\label{theorem4.19}
			\mathbb{P}\Big(\max_{1\le l \le p} |(IV)_{l,\kappa_{r}}| > \frac{1}{6}t\underline{\sigma}\Big)&\le \frac{24\sqrt{B_{1}}\big[\sup_{\vec{z}\in\mathbb{R}^{s^*}}\zeta(\vec{z})\big]^3 \mathbb{E}|\varepsilon|^3}{t\underline{\sigma}}.
		\end{split}
	\end{equation}

	Next, we bound the term $\mathbb{P}\big(\max_{1\le l \le p} |\widehat{\mu}_{l}^2(\kappa_{r})- \mu_{l}^2(\kappa_{r})| > \frac{1}{6}t\underline{\sigma}\big) $ on the RHS of \eqref{theorem4.2} by \eqref{theorem4.4}--\eqref{theorem4.20} below. For all $n\ge 1$, each $1\le r\le R$, each $t>0$, and each $B_{1}>0$,
	\begin{equation}
		\begin{split}\label{theorem4.4}
			&\mathbb{P}(\max_{1\le l\le p}|\mu_{l}^2(\kappa_{r}) - \widehat{\mu}_{l}^2(\kappa_{r}) | \ge \frac{1}{6}t\underline{\sigma}) \\& \le \mathbb{P}\big\{\max_{1\le l\le p}|\mu_{l}(\kappa_{r}) - \widehat{\mu}_{l}(\kappa_{r})| \times  \big[|\widehat{\mu}_{l}(\kappa_{r}) - \mu_{l}(\kappa_{r}) | + 2|\mu_{l}(\kappa_{r})|\big] \ge \frac{1}{6}t\underline{\sigma} \big\}\\
			& = \mathbb{P}\big\{\max_{1\le l\le p}|\mu_{l}(\kappa_{r}) - \widehat{\mu}_{l}(\kappa_{r})|^2 + 2|\widehat{\mu}_{l}(\kappa_{r}) - \mu_{l}(\kappa_{r}) | \times  |\mu_{l}(\kappa_{r})| \ge \frac{1}{6}t\underline{\sigma} \big\}\\
			& \le \mathbb{P}(\max_{1\le l\le p}|\mu_{l}(\kappa_{r}) - \widehat{\mu}_{l}(\kappa_{r})| \ge \sqrt{\frac{t\underline{\sigma} }{12}}) + \mathbb{P}(\max_{1\le l\le p}|\mu_{l}(\kappa_{r}) - \widehat{\mu}_{l}(\kappa_{r})||\mu_{l}(\kappa_{r})| \ge \frac{t\underline{\sigma} }{24})\\
			&\le 2 \mathbb{P}\left(\max_{1\le l\le p}|\mu_{l}(\kappa_{r}) - \widehat{\mu}_{l}(\kappa_{r})| \ge \min\left\{\sqrt{\frac{t\underline{\sigma} }{12}}, \frac{t\underline{\sigma} }{24\max_{1\le l\le p}\{1, |\mu_{l}(\kappa_{r})|\}}\right\}\right), 
		\end{split}				
	\end{equation}
	where the first inequality follows from that $|x + y|\le |x-y| + 2|y|$.	By the definition of $\widehat{\mu}_{l}(a)$ and simple calculations, it holds that  for every $t_{1}>0$, all $n\ge 1$, each $1\le r \le R$, and each $B_{1}>0$,
	{\small\begin{equation}
			\begin{split}\label{theorem4.3}
				&\mathbb{P}(\max_{1\le l \le p}|\widehat{\mu}_{l}(\kappa_{r}) - \mu_{l}(\kappa_{r}) | \ge t_{1})  \\	
				&\le \mathbb{P}(V_{n}+\max_{1\le l \le p}\Big|n^{-1}\sum_{i=1}^{n} \big\{ (\boldsymbol{1}_{X_{il}\in(-\infty, \kappa_{r}]}- \boldsymbol{1}_{\widetilde{X}_{il}\in (-\infty, \kappa_{r}]}) (\zeta(\boldsymbol{X}_{iS^*})\varepsilon_{i})^2 - \mu_{l}(\kappa_{r})\big\}\Big| \ge t_{1}),
			\end{split}
	\end{equation}}%
	where $V_{n}\coloneqq n^{-1}\sum_{i=1}^{n} (\widehat{m}(\boldsymbol{X}_{i}) - m(\boldsymbol{X}_{i}))^2+ 2\big[\sup_{\vec{z} \in \mathbb{R}^{s^*}}\zeta(\vec{z})\big] n^{-1}\sum_{i=1}^{n}  |\widehat{m}(\boldsymbol{X}_{i}) - m(\boldsymbol{X}_{i})||\varepsilon_{i}|$.

	To deal with the RHS of \eqref{theorem4.3}, define two events
	\begin{equation*}
		\begin{split}	
			A_{1} & = \cup_{i=1}^{n}\{|\varepsilon_{i}|> n^{\frac{1}{q_{1}}}\}, \\
			A_{2r}& = \Big\{\max_{ 1\le l\le p} \Big|n^{-1}\sum_{i=1}^{n} \big\{\boldsymbol{1}_{\{|\varepsilon_{i}|\le n^{\frac{1}{q_{1}}}\}} \times \big[\boldsymbol{1}_{X_{il}\in(-\infty, \kappa_{r}]}- \boldsymbol{1}_{\widetilde{X}_{il}\in (-\infty, \kappa_{r}]}\big]\big[\zeta(\boldsymbol{X}_{iS^*})\varepsilon_{i}\big]^2 - \mu_{l}^{(n)}(\kappa_{r})\big\}\Big| \\
			&\qquad\qquad> c_{1}n^{-\frac{1}{2} + \frac{2}{q_{1}}} \sqrt{\log{(n \vee p)}} \Big\},
		\end{split}
	\end{equation*}
	where $q_{1}=q - \beta$, $c_{1} = \sqrt{8}\big[ \sup_{\vec{z}\in\mathbb{R}^{s^*}} \zeta(\vec{z})\big]^2$, and $$\mu_{l}^{(n)}(\kappa_{r})=\mathbb{E}\big\{\boldsymbol{1}_{\{|\varepsilon|\le n^{\frac{1}{q_{1}}}\}}\big[\boldsymbol{1}_{X_{l}\in (-\infty, \kappa_{r}]}- \boldsymbol{1}_{\widetilde{X}_{l}\in (-\infty, \kappa_{r}]}\big]\big[\zeta(\boldsymbol{X}_{S^*})\varepsilon\big]^2\big\}.$$
	Let us show that the probabilities of events $A_{1}$ and $A_{2r}$ are negligible  in the following. By the Cauchy–Schwarz inequality, Markov's inequality, and the model regularity assumptions (boundness of $\zeta(\cdot)$ and that  $\mathbb{E}|\varepsilon|^{q\vee 8} <\infty$),
	\begin{equation}
		\begin{split}\label{theorem4.9}
			& \max_{1\le l \le p}|\mu_{l}(\kappa_{r}) - \mu_{l}^{(n)}(\kappa_{r})| \\
			& = \max_{1\le l \le p}\mathbb{E}\big\{\boldsymbol{1}_{\{|\varepsilon|> n^{\frac{1}{q_{1}}}\}} \times \big[\boldsymbol{1}_{X_{l}\in(-\infty, \kappa_{r}]}- \boldsymbol{1}_{\widetilde{X}_{l}\in(-\infty, \kappa_{r}]}\big]\big[\zeta(\boldsymbol{X}_{S^*})\varepsilon\big]^2\big\}\\
			& \le \max_{1\le l \le p} \big[\sup_{\vec{z}\in\mathbb{R}^{s^*}}\zeta(\vec{z})\big]^2\mathbb{E}\big(\boldsymbol{1}_{\{|\varepsilon|> n^{\frac{1}{q_{1}}}\}}\times \varepsilon^2\big)\\
			&\le \max_{1\le l \le p} \big[\sup_{\vec{z}\in\mathbb{R}^{s^*}}\zeta(\vec{z})\big]^2\sqrt{\mathbb{P}(|\varepsilon|> n^{\frac{1}{q_{1}}})}\sqrt{\mathbb{E}(\varepsilon^4\big)}\\
			&\le \max_{1\le l \le p}n^{-\frac{1}{2}}\big[\sup_{\vec{z}\in\mathbb{R}^{s^*}}\zeta(\vec{z})\big]^2 \sqrt{\mathbb{E}(|\varepsilon|^{q_{1}})}\sqrt{\mathbb{E}(\varepsilon^4\big)}\\
			&	=O(n^{-\frac{1}{2}}),
		\end{split}
	\end{equation}
	and hence for all large $n$ and each $1\le r\le R$,
	\begin{equation}\label{theorem4.11}
		\max_{1\le l\le p}|\mu_{l}^{(n)}(\kappa_{r})|\le 1 + \max_{1\le l\le p}|\mu_{l}(\kappa_{r})| < 1 +  \big[\sup_{\vec{z}\in\mathbb{R}^{s^*}}\zeta(\vec{z})\big]^2 \mathbb{E}(\varepsilon^2).
	\end{equation}
	Now, by Markov's inequality, Hoeffding's inequality, \eqref{theorem4.11}, and model regularity assumptions, for all large $n$ and each $1\le r\le R$,
	{\small\begin{equation}
			\begin{split}\label{theorem4.10}
				\mathbb{P}(A_{1}) &\le \big(n^{1-\frac{q}{q_{1}}}\big)\times \mathbb{E}|\varepsilon|^q = o(1),\\
				\mathbb{P}(A_{2r}) &\le 2p\exp{\left(-\frac{2c_{1}^2\times n^{1+\frac{4}{q_{1}}} \times \log{(n \vee p)} }{ 4n \big\{\big[\sup_{\vec{z}\in\mathbb{R}^{s^*}} \zeta(\vec{z})\big]^2 \times n^{\frac{2}{q_{1}}}  +  \max_{1\le l\le p}|\mu_{l}^{(n)}(\kappa_{r})| \big\}^2}\right)} = o(1).
			\end{split}
	\end{equation}}%
	Moreover, let us deal with the term $V_{n}$ on the RHS of \eqref{theorem4.3}. By Condition~\ref{consistency.3}, Jensen's inequality, $\mathbb{E}(\varepsilon^2) = 1$, and model regularity assumptions,
	\begin{equation}\label{theorem4.8}
		\mathbb{E}(V_{n})\le B_{1} + 2\big[\sup_{\vec{z} \in \mathbb{R}^{s^*}}\zeta(\vec{z})\big] \sqrt{B_{1}}.
	\end{equation}
	
	With the result \eqref{theorem4.10},		
	\begin{equation}
		{\small	\begin{split}\label{theorem4.12}
				& \textnormal{RHS of \eqref{theorem4.3}} \\
				&\le 
				\mathbb{P}(\{V_{n} + \max_{1\le l \le p}\Big|n^{-1}\sum_{i=1}^{n}\big\{ (\boldsymbol{1}_{X_{il}\in(-\infty, \kappa_{r}]}- \boldsymbol{1}_{\widetilde{X}_{il}\in (-\infty, \kappa_{r}]}) (\zeta(\boldsymbol{X}_{iS^*})\varepsilon_{i})^2 - \mu_{l}(\kappa_{r})\big\}\Big| \ge t_{1}\}\\
				&\qquad\cap A_{1}^c\cap A_{2r}^c) + \mathbb{P}(A_{1}\cup A_{2r})\\
				&\le 
				\mathbb{P}\Big(\Big\{V_{n}  + \max_{1\le l \le p}\Big|n^{-1}\sum_{i=1}^{n} \big\{ \boldsymbol{1}_{\{|\varepsilon_{i}|\le n^{\frac{1}{q_{1}}}\}}  (\boldsymbol{1}_{X_{il}\in(-\infty, \kappa_{r}]}- \boldsymbol{1}_{\widetilde{X}_{il}\in (-\infty, \kappa_{r}]}) (\zeta(\boldsymbol{X}_{iS^*})\varepsilon_{i})^2 \\
				&\qquad- \mu_{l}(\kappa_{r})\big\}\Big| \ge t_{1}\Big\} \cap  A_{2r}^c\Big)  + \mathbb{P}(A_{1}\cup A_{2r})\\
				&\le 
				\mathbb{P}\Big( \Big\{V_{n}+\max_{1\le l \le p}|\mu_{l}(\kappa_{r}) -\mu_{l}^{(n)}(\kappa_{r})| \\
				&\qquad + \max_{1\le l \le p}\Big|n^{-1}\sum_{i=1}^{n} \big\{ \boldsymbol{1}_{\{|\varepsilon_{i}|\le n^{\frac{1}{q_{1}}}\}}  (\boldsymbol{1}_{X_{il}\in(-\infty, \kappa_{r}]}- \boldsymbol{1}_{\widetilde{X}_{il}\in (-\infty, \kappa_{r}]}) (\zeta(\boldsymbol{X}_{iS^*})\varepsilon_{i})^2 \\
				&\qquad- \mu_{l}^{(n)}(\kappa_{r})\big\}\Big| \ge t_{1}\Big\} \cap A_{2r}^c \Big)  + \mathbb{P}(A_{1}\cup A_{2r})\\
				&\le \frac{\mathbb{E}(V_{n}) + \max_{1\le l \le p}|\mu_{l}(\kappa_{r}) -\mu_{l}^{(n)}(\kappa_{r})| + c_{1}n^{-\frac{1}{2} + \frac{2}{q_{1}}} \sqrt{\log{(n \vee p)}}  }{t_{1}} + \mathbb{P}(A_{1}\cup A_{2r}).
		\end{split}}
	\end{equation}

	By taking $t_{1} = \underline{\sigma}\times n^{-\beta_{2}} \times \big[24\max_{1\le l\le p}\{1, |\mu_{l}(\kappa_{r})|\}\big]^{-1}$ in \eqref{theorem4.3} and the results of \eqref{theorem4.9} and \eqref{theorem4.12}, it holds that 
	\begin{equation}
		\begin{split}\label{theorem4.30}
			&\mathbb{P}(\max_{1\le l \le p}|\widehat{\mu}_{l}(\kappa_{r}) - \mu_{l}(\kappa_{r}) | \ge t_{1}) \\
			&\le  c_{4}\Big[n^{\beta_{2}}\mathbb{E}(V_{n}) + n^{\beta_{2} - \frac{1}{2}} + c_{1}n^{-\frac{1}{2} + \frac{2}{q-\beta} + \beta_{2} } \sqrt{\log{(n \vee p)}}\Big] + \mathbb{P}(A_{1}\cup A_{2r}),
		\end{split}
	\end{equation}
	where $c_{4} = 24\times \max_{1\le l\le p}\{1, |\mu_{l}(\kappa_{r})|\}\times (\underline{\sigma})^{-1}$. By taking $t= n^{-\beta_{2}}$ in \eqref{theorem4.4} and the result of \eqref{theorem4.30}, it holds that for all large $n$, each $1\le r\le R$, and each $B_{1}>0$,
	\begin{equation}
		\begin{split}\label{theorem4.20}
			&\mathbb{P}(\max_{1\le l\le p}|\mu_{l}^2(\kappa_{r}) - \widehat{\mu}_{l}^2(\kappa_{r}) | \ge \frac{1}{6}n^{-\beta_{2}}\underline{\sigma})\\
			&\le 2c_{4}\Big[n^{\beta_{2}}\mathbb{E}(V_{n}) + n^{\beta_{2} - \frac{1}{2}} + c_{1}n^{-\frac{1}{2} + \frac{2}{q-\beta} + \beta_{2} } \sqrt{\log{(n \vee p)}}\Big] + 2\mathbb{P}(A_{1}\cup A_{2r}),
		\end{split}
	\end{equation}
	where we note that when $t= n^{-\beta_{2}}$, we have that for all large $n$,
	\begin{equation*}
		\min\left\{\sqrt{\frac{t\underline{\sigma} }{12}}, \frac{t\underline{\sigma} }{24\max_{1\le l\le p}\{1, |\mu_{l}(\kappa_{r})|\}} \right\} = \frac{n^{-\beta_{2}}\underline{\sigma} }{24\max_{1\le l\le p}\{1, |\mu_{l}(\kappa_{r})|\}}.
	\end{equation*}

	Next, we show that $\mathbb{P}\big(\max_{1\le l \le p}\Big| (V)_{l, \kappa_{r}} - \big[\sigma_{l}^2(\kappa_{r}) + \mu_{l}^2(\kappa_{r}) \big] \Big| >\frac{1}{6}t\underline{\sigma}\big)$ on the RHS of \eqref{theorem4.2} with $t= n^{-\beta_{2}}$ is negligible. The arguments \eqref{theorem4.17}--\eqref{theorem4.21} below are quite similar to those for establishing the upper bound in \eqref{theorem4.20}. For each $l\in \{1, \dots, p\}$ and $a\in \mathbb{R}$, define 
	$$\nu_{l}(a)=\mathbb{E}\big\{\big[\boldsymbol{1}_{X_{l}\in(-\infty, a]}- \boldsymbol{1}_{\widetilde{X}_{l}\in(-\infty,a]}\big]^2\big[\zeta(\boldsymbol{X}_{S^*})\varepsilon\big]^4\big\} = \sigma_{l}^2(a) + \mu_{l}^2(a) .$$
	In addition, for each $1\le r\le R$, define event
	\begin{equation*}
		\begin{split}	
			&A_{3r} = 
			\Big\{\max_{ 1\le l\le p} \Big|n^{-1}\sum_{i=1}^{n} \big\{ \boldsymbol{1}_{\{|\varepsilon_{i}|\le n^{\frac{1}{q_{1}}}\}}\times\big[\boldsymbol{1}_{X_{il}\in(-\infty, \kappa_{r}]}- \boldsymbol{1}_{\widetilde{X}_{il}\in(-\infty, \kappa_{r}]}\big]^2\big[\zeta(\boldsymbol{X}_{iS^*})\varepsilon_{i}\big]^4 - \nu_{l}^{(n)}(\kappa_{r})\big\}\Big| \\
			&\qquad\qquad> c_{3}n^{-\frac{1}{2} + \frac{4}{q_{1}}} \sqrt{\log{(n \vee p)}} \Big\},
		\end{split}
	\end{equation*}		
	\noindent where $q_{1}=q - \beta$, $\nu_{l}^{(n)}(a)=\mathbb{E}\big\{\boldsymbol{1}_{\{|\varepsilon|\le n^{\frac{1}{q_{1}}}\}}\big[\boldsymbol{1}_{X_{l}\in(-\infty, a]}- \boldsymbol{1}_{\widetilde{X}_{l}\in(-\infty, a]}\big]^2\big[\zeta(\boldsymbol{X}_{S^*})\varepsilon\big]^4\big\}$ for each $l\in \{1, \dots, p\}$ and $a\in \mathbb{R}$, and $c_{3} = \sqrt{8} \big[\sup_{\vec{z}\in\mathbb{R}^{s^*}} \zeta(\vec{z})\big]^4$. By the Cauchy–Schwarz inequality, Markov's inequality, and the model regularity assumptions (boundness of $\zeta(\cdot)$ and that  $ \mathbb{E}(\varepsilon)^{8} <\infty $), 
	\begin{equation}
		\begin{split}\label{theorem4.17}
			&\max_{1\le l \le p}|\nu_{l}(\kappa_{r}) - \nu_{l}^{(n)}(\kappa_{r})| \\
			& = \max_{1\le l \le p}\mathbb{E}\big\{\boldsymbol{1}_{\{|\varepsilon|> n^{\frac{1}{q_{1}}}\}}\big[\boldsymbol{1}_{X_{l}\in(-\infty, \kappa_{r}]}- \boldsymbol{1}_{\widetilde{X}_{l}\in(-\infty, \kappa_{r}]}\big]^2\big[\zeta(\boldsymbol{X}_{S^*})\varepsilon\big]^4\big\}\\
			& \le \max_{1\le l \le p} \big[\sup_{\vec{z}\in\mathbb{R}^{s^*}}\zeta(\vec{z})\big]^4\mathbb{E}\big(\boldsymbol{1}_{\{|\varepsilon|> n^{\frac{1}{q_{1}}}\}}\varepsilon^4\big)\\
			&\le \max_{1\le l \le p} \big[\sup_{\vec{z}\in\mathbb{R}^{s^*}}\zeta(\vec{z})\big]^4\sqrt{\mathbb{P}(|\varepsilon|> n^{\frac{1}{q_{1}}})}\sqrt{\mathbb{E}(\varepsilon^8\big)}\\
			&\le \max_{1\le l \le p}n^{-\frac{1}{2}}\big[\sup_{\vec{z}\in\mathbb{R}^{s^*}}\zeta(\vec{z})\big]^4 \sqrt{\mathbb{E}(|\varepsilon|^{q_{1}})}\sqrt{\mathbb{E}(\varepsilon^8\big)}\\
			&	=O(n^{-\frac{1}{2}}),
		\end{split}
	\end{equation}
	and hence for all large $n$ and each $1\le r\le R$,
	\begin{equation}\label{theorem4.14}
		\max_{1\le l\le p}|\nu_{l}^{(n)}(\kappa_{r})|\le 1 + \max_{1\le l\le p}|\nu_{l}(\kappa_{r})|\le 1 + \big[\sup_{\vec{z}\in\mathbb{R}^{s^*}}\zeta(\vec{z})\big]^4\times \mathbb{E}(\varepsilon^4).
	\end{equation}
	In addition, by Markov's inequality, Hoeffding's inequality, \eqref{theorem4.14}, and model regularity assumptions, for all large $n$ and each $1\le r\le R$,
	{\small\begin{equation}
			\begin{split}\label{theorem4.16}
				\mathbb{P}(A_{3r}) &\le 2p\exp{\left(-\frac{2c_{3}^2 n^{1+\frac{8}{q_{1}}} \log{(n \vee p)} }{ 4n\big\{ \big[\sup_{\vec{z}\in\mathbb{R}^{s^*}} \zeta(\vec{z})\big]^4 \times n^{\frac{4}{q_{1}}}  +  \max_{1\le l\le p}|\nu_{l}^{(n)}(\kappa_{r})| \big\}^2}\right)} = o(1).
			\end{split}
	\end{equation}}%
	
	By \eqref{theorem4.17} and \eqref{theorem4.16} and the definitions of $A_{1}$ and $A_{3r}$, it holds that for each $t_{2}>0$,  all $n\ge 1$, each $1\le r\le R$, and each $B_{1}>0$,
	\begin{equation}\label{theorem4.15}
		{\small\begin{split}
				&\mathbb{P}\big(\max_{1\le l \le p}\Big| (V)_{l, \kappa_{r}} - \big[\sigma_{l}^2(\kappa_{r}) + \mu_{l}^2(\kappa_{r}) \big] \Big| > t_{2}\big)\\
				& = \mathbb{P}( \max_{ 1\le l\le p} |n^{-1}\sum_{i=1}^{n} \big\{\big[\boldsymbol{1}_{X_{il}\in(-\infty, \kappa_{r}]}- \boldsymbol{1}_{\widetilde{X}_{il}\in(-\infty, \kappa_{r}]}\big]^2\big[\zeta(\boldsymbol{X}_{iS^*})\varepsilon_{i}\big]^4 - \nu_{l}(\kappa_{r})\big\}|> t_{2})\\
				&\le \mathbb{P}(\{ \max_{ 1\le l\le p} |n^{-1}\sum_{i=1}^{n} \big\{\big[\boldsymbol{1}_{X_{il}\in(-\infty, \kappa_{r}]}- \boldsymbol{1}_{\widetilde{X}_{il}\in(-\infty, \kappa_{r}]}\big]^2\big[\zeta(\boldsymbol{X}_{iS^*})\varepsilon_{i}\big]^4 - \nu_{l}(\kappa_{r})\big\}| > t_{2}\}\\
				&\qquad\cap A_{1}^c\cap A_{3r}^c)  + \mathbb{P}(A_{1}\cup A_{3r})\\
				& \le \mathbb{P}\Big(\Big\{\max_{ 1\le l\le p} |n^{-1}\sum_{i=1}^{n} \big\{ \boldsymbol{1}_{\{|\varepsilon_{i}|\le n^{\frac{1}{q_{1}}}\}} \big[\boldsymbol{1}_{X_{il}\in(-\infty, \kappa_{r}]}- \boldsymbol{1}_{\widetilde{X}_{il}\in(-\infty, \kappa_{r}]}\big]^2\big[\zeta(\boldsymbol{X}_{iS^*})\varepsilon_{i}\big]^4 \\
				&\qquad- \nu_{l}(\kappa_{r})\big\}|> t_{2}\Big\}\cap A_{3r}^c\Big) + \mathbb{P}(A_{1}\cup A_{3r}) \\
				&\le \mathbb{P}( c_{3}n^{-\frac{1}{2} + \frac{4}{q_{1}}} \sqrt{\log{(n \vee p)}} + \max_{1\le l\le p}|\nu_{l}(\kappa_{r}) - \nu_{l}^{(n)}(\kappa_{r})|> t_{2})  + \mathbb{P}(A_{1}\cup A_{3r})\\
				& \le \frac{c_{3}n^{-\frac{1}{2} + \frac{4}{q_{1}}} \sqrt{\log{(n \vee p)}} + \max_{1\le l\le p}|\nu_{l}(\kappa_{r}) - \nu_{l}^{(n)}(\kappa_{r})|  }{t_{2} } + \mathbb{P}(A_{1}\cup A_{3r}),
		\end{split}}
	\end{equation}
	where we use Markov's inequality in the last inequality.
	
	With $t_{2}= \frac{1}{6}\underline{\sigma}n^{-\beta_{2}}$ in \eqref{theorem4.15}, it holds that for all large $n$, each $1\le r\le R$, and each $B_{1}>0$,
	\begin{equation}\label{theorem4.21}
		\begin{split}
			&\mathbb{P}\big(\max_{1\le l \le p}\Big| (V)_{l, \kappa_{r}} - \big[\sigma_{l}^2(\kappa_{r}) + \mu_{l}^2(\kappa_{r}) \big] \Big| > n^{-\beta_{2}}\underline{\sigma}\frac{1}{6}\big) \\
			&\le 6(\underline{\sigma})^{-1}\times \Big[c_{3}n^{-\frac{1}{2} + \frac{4}{q-\beta} + \beta_{2}} \sqrt{\log{(n \vee p)}} + n^{\beta_{2}-\frac{1}{2}}\Big] + \mathbb{P}(A_{1}\cup A_{3r}).
		\end{split}
	\end{equation}

	Now, let us establish the upper bound for terms on the RHS of \eqref{theorem4.2}. With $t= n^{-\beta_{2}}$ in \eqref{theorem4.2},
	\eqref{theorem4.18}--\eqref{theorem4.19}, 		\eqref{theorem4.10}--\eqref{theorem4.8},  \eqref{theorem4.20}, \eqref{theorem4.16}, \eqref{theorem4.21}, $0<\beta_{2} < \frac{1}{4}$, and the assumptions that $R$ is a finite constant and that $\lim_{n\rightarrow\infty} n^{\beta_{2}}\overbar{M}_{n}^2B_{1}+ \sqrt{n}B_{1}(-\log{B_{1}}) + n^{-\frac{1}{2}+ \frac{4}{q-\beta} + \beta_{2}} \sqrt{\log{(n\vee p)}} =0$, 
	\begin{equation}\label{theorem4.25}
		\lim_{n\rightarrow\infty}\mathbb{P}(|\sigma_{\widehat{l}}(\widehat{a}_{\widehat{l}}) - \widehat{\sigma}_{\widehat{l}}(\widehat{a}_{\widehat{l}})| > n^{-\beta_{2}}) =0.
	\end{equation}
	We note that $n^{\beta_{2}}\times \mathbb{E}(V_{n})$ in \eqref{theorem4.20} decreases to zero because we have assumed $\beta_{2}<\frac{1}{4}$ as well as other regularity assumptions.
	
	By \eqref{theorem4.29}--\eqref{theorem4.23}, \eqref{theorem4.25}, and the assumption that $\lim_{n\rightarrow\infty} B_{1} = 0$ we have finished the proof of  \eqref{theorem4.5}.

	\subsection{Proof of Theorem~\ref{theorem5}}\label{proof.theorem5}
	
	Let some $j\in\{1, \dots, p\}$ such that $|\mu_{j}(a)| > \underline{\mu}$ be given.
	We begin with an application of Markov's inequality as follows, where $\mathcal{A}=(-\infty, a]$ with some break $a\in\mathbb{R}$. For each $n\ge 1$ and $t>0$,
	\begin{equation}
		\begin{split}\label{theorem5.1}
			&\mathbb{P}(|T_{j}(a)\big[\widehat{\sigma}_{j}(a)\big]^{-1}| \le t )\\
			&\le \mathbb{P}(|T_{j}^{(\star)}(a)| \le  t| \widehat{\sigma}_{j}(a)| +|A_{1}| + |A_{2}|)\\
			&\le \mathbb{P}(|\sqrt{n}\big[\mu_{j}(a)\big]|  \le |T_{j}^{(\star)}(a) - \sqrt{n}\big[\mu_{j}(a)\big]| + t| \widehat{\sigma}_{j}(a)| +|A_{1}| + |A_{2}|)\\
			& \le \frac{\mathbb{E}(|T_{j}^{(\star)}(a) - \sqrt{n}\big[\mu_{j}(a)\big]|) +  t\mathbb{E}(| \widehat{\sigma}_{j}(a)|) + \mathbb{E}(|A_{1}|) + \mathbb{E}(|A_{2}|)}{\sqrt{n}\underline{\mu}},
		\end{split}
	\end{equation}
	where 
	\begin{equation}
		\begin{split}\label{definition.T.1}
			A_{1} &=  n^{-\frac{1}{2}}\sum_{i=1}^{n}\big(\boldsymbol{1}_{X_{ij}\in\mathcal{A}} -\boldsymbol{1}_{\widetilde{X}_{ij}\in\mathcal{A}}\big) (m(\boldsymbol{X}_{i}) - \widehat{m}(\boldsymbol{X}_{i}))^2,\\
			A_{2} &=  2n^{-\frac{1}{2}}\sum_{i=1}^{n}\big(\boldsymbol{1}_{X_{ij}\in\mathcal{A}} -\boldsymbol{1}_{\widetilde{X}_{ij}\in\mathcal{A}}\big) (m(\boldsymbol{X}_{i}) - \widehat{m}(\boldsymbol{X}_{i}))\zeta(\boldsymbol{X}_{iS^{*}})\varepsilon_{i} ,\\
			T_{j}^{(\star)}(a) &= n^{-\frac{1}{2}}\sum_{i=1}^{n}\big(\boldsymbol{1}_{X_{ij}\in\mathcal{A}} -\boldsymbol{1}_{\widetilde{X}_{ij}\in\mathcal{A}}\big)(\zeta(\boldsymbol{X}_{iS^{*}})\varepsilon_{i})^{2},
		\end{split}
	\end{equation}
	and the first inequality follows because $T_{j}(a) =A_{1} + A_{2} + T_{j}^{(\star)}(a)$. 
	
	Next, we establish upper bounds for terms on the RHS of \eqref{theorem5.1}. By arguments similar to those for \eqref{theorem1.2},
	\begin{equation}\label{theorem5.2}
		\begin{split}					
			\mathbb{E}(|T_{j}^{(\star)}(a) - \sqrt{n}\big[\mu_{j}(a)\big]|)  & \le C\big[\sup_{\vec{z} \in \mathbb{R}^{s^*}} \zeta(\vec{z})\big]^2\sqrt{\mathbb{E}(\varepsilon^4)},\\
			\mathbb{E}(|A_{2}|) & \le 2C\big[\sup_{\vec{z} \in \mathbb{R}^{s^*}} \zeta(\vec{z})\big]\sqrt{B_{1}\mathbb{E}(\varepsilon^2)},
		\end{split}
	\end{equation}
	for some $C>0$, which is due to an application of the Burkholder--Davis--Gundy inequality \citepsupp{burkholder1972integral}; note that for the second inequality in \eqref{theorem5.2}, it is required that the training sample for $\widehat{m}(\cdot)$ is independent of $\{Y_{i}, \boldsymbol{X}_{i}, \widetilde{\boldsymbol{X}}_{i}, \varepsilon_{i}\}_{i=1}^{n}$ . 
	
	By the assumptions that observations are i.i.d. and the training sample is an independent sample and Condition~\ref{consistency.3},
	\begin{equation}\label{theorem5.4}
		\mathbb{E}(|A_{1}|)  \le \sqrt{n}B_{1}.
	\end{equation}
	
	To deal with $\mathbb{E}(| \widehat{\sigma}_{j}(a)|)$, we  write
	\begin{equation}
		\begin{split}\label{definition.sigma.1}
			\widehat{\sigma}_{j}^2(a) &= n^{-1}\sum_{i=1}^{n} (\boldsymbol{1}_{X_{ij} \in \mathcal{A}} -\boldsymbol{1}_{\widetilde{X}_{ij} \in \mathcal{A}})^2(\widehat{\varepsilon}_{i})^4 -\big[ \widehat{\mu}_{j}(a)\big]^2\\
			&\le  n^{-1}\sum_{i=1}^{n} (\boldsymbol{1}_{X_{ij} \in \mathcal{A}} -\boldsymbol{1}_{\widetilde{X}_{ij} \in \mathcal{A}})^2(\widehat{\varepsilon}_{i})^4 \\
			&=  n^{-1}\sum_{i=1}^{n} (\boldsymbol{1}_{X_{ij} \in \mathcal{A}} -\boldsymbol{1}_{\widetilde{X}_{ij} \in \mathcal{A}})^2  (\widehat{m}(\boldsymbol{X}_{i}) - m(\boldsymbol{X}_{i}) )^4\\
			&  \qquad - 4n^{-1}\sum_{i=1}^{n} (\boldsymbol{1}_{X_{ij} \in \mathcal{A}} -\boldsymbol{1}_{\widetilde{X}_{ij} \in \mathcal{A}})^2 (\widehat{m}(\boldsymbol{X}_{i}) - m(\boldsymbol{X}_{i}) )^3 \zeta(\boldsymbol{X}_{iS^{*}})\varepsilon_{i} \\
			&\qquad+ 6n^{-1}\sum_{i=1}^{n} (\boldsymbol{1}_{X_{ij} \in \mathcal{A}} -\boldsymbol{1}_{\widetilde{X}_{ij} \in \mathcal{A}})^2 (\widehat{m}(\boldsymbol{X}_{i}) - m(\boldsymbol{X}_{i}) )^2 (\zeta(\boldsymbol{X}_{iS^{*}})\varepsilon_{i})^2 \\
			&\qquad- 4n^{-1}\sum_{i=1}^{n} (\boldsymbol{1}_{X_{ij} \in \mathcal{A}} -\boldsymbol{1}_{\widetilde{X}_{ij} \in \mathcal{A}})^2 (\widehat{m}(\boldsymbol{X}_{i}) - m(\boldsymbol{X}_{i}) ) (\zeta(\boldsymbol{X}_{iS^{*}})\varepsilon_{i})^3 \\
			&\qquad+ n^{-1}\sum_{i=1}^{n} (\boldsymbol{1}_{X_{ij} \in \mathcal{A}} -\boldsymbol{1}_{\widetilde{X}_{ij} \in \mathcal{A}})^2 ( \zeta(\boldsymbol{X}_{iS^{*}})\varepsilon_{i})^4,
		\end{split}
	\end{equation}
	where $\widehat{\varepsilon}_{i} = Y_{i} - \widehat{m}(\boldsymbol{X}_{i})$. Hence, by Jensen's inequality, Conditions~\ref{consistency.3}--\ref{bound}, $\mathbb{E}(\varepsilon^4) < \infty$, $\mathbb{E}(\varepsilon^2)=1$ in model~\eqref{model.1},  that $\zeta(\vec{z})$ is positive, the assumptions that observations are i.i.d. and that $\varepsilon$ is an independent model error, and the assumption that the training sample for $\widehat{m}(\cdot)$ is an independent sample, for each $n\ge 1$,
	\begin{equation}
		\begin{split}\label{theorem5.3}
			&\mathbb{E}(| \widehat{\sigma}_{j}(a)|) \\
			&\le \sqrt{\mathbb{E}\big[ \widehat{\sigma}_{j}^2(a)\big]}\\
			&\le \Big\{ 4\overbar{M}_{n}^2B_{1}+ 8\overbar{M}_{n}B_{1}\big[\sup_{\vec{z} \in \mathbb{R}^{s^*}} \zeta(\vec{z})\big]\\
			&\qquad+
			6B_{1}\big[\sup_{\vec{z} \in \mathbb{R}^{s^*}} \zeta(\vec{z})\big]^2+ 4\sqrt{B_{1}}\big[\sup_{\vec{z} \in \mathbb{R}^{s^*}} \zeta(\vec{z})\big]^3 \mathbb{E}|\varepsilon|^3+ \big[\sup_{\vec{z} \in \mathbb{R}^{s^*}} \zeta(\vec{z})\big]^4\mathbb{E}(\varepsilon^4) \Big\}^{\frac{1}{2}}.
		\end{split}
	\end{equation}

	By \eqref{theorem5.2}--\eqref{theorem5.3}, the assumptions that $\lim\sup_{n\rightarrow\infty}\overbar{M}_{n}^2B_{1} <\infty$, $B_{1}<1$, and that $\overbar{M}_{n}\ge 1$, and the model regularity assumptions, there exists some $N>0$ whose value is independent of $1\le j\le p$ such that for all $n\ge N$ and each $t>0$,
	\begin{equation*}
		\begin{split}	
			\textnormal{The RHS of \eqref{theorem5.1}} 
			&\le  \frac{(1+t)\log{n} + \sqrt{n}B_{1}}{\sqrt{n}\underline{\mu}}\\
			&\le B_{1}(\underline{\mu})^{-1} + (\log{n})(1+t)(\sqrt{n}\underline{\mu})^{-1},
		\end{split}
	\end{equation*}
	which concludes the desired result of Theorem~\ref{theorem5} for feature index $j$ given at the beginning of this proof. 
	
	Lastly, we note that the generic constants such as $N>0$ in this proof are not subject to feature index $j$. Therefore,  we have finished the proof of Theorem~\ref{theorem5}.

	
	\subsection{Proof of  Theorem~\ref{theorem6}}\label{proof.theorem6}
	
	Let us begin with proving the first part of Theorem~\ref{theorem6}, which is $\lim_{n\rightarrow\infty}\mathbb{P}(\widehat{l}\in S^*) = 1$ when $\min_{l\in S^*}|\mu_{l}(a)| > \underline{\mu}>0$ ($S^*$ is not an empty set accordingly). Recall that  $G_{l}(a) =  n^{-1}\sum_{i=1}^{n}(\boldsymbol{1}_{U_{il}\in\mathcal{A}}- \boldsymbol{1}_{\widetilde{U}_{il}\in\mathcal{A}}) (\widehat{\eta}_{i} )^{2}$ where  $\widehat{\eta}_{i} = V_{i} - \widehat{m}(\boldsymbol{U}_{i})$, $\widehat{l} = \arg\max_{1\le l\le p}|G_{l}(a)|$, and $\mathcal{A} = (-\infty, a]$.
	
	On the event 
	$$J_{1} = \left\{\max_{1\le l\le p} |G_{l}(a) - \mu_{l}(a)| \le \frac{1}{3}\underline{\mu}\right\},$$
	it holds that 
	\begin{equation*}
		\begin{split}
			\min_{l\in S^*}|G_{l}(a)| & = \min_{l\in S^*}|G_{l}(a) -\mu_{l}(a) + \mu_{l}(a)| \\
			& \ge \min_{l\in S^*}|\mu_{l}(a)| -  \max_{1\le l\le p}|G_{l} (a)-\mu_{l}(a)| \\
			& > \max_{1\le l\le p}|G_{l} (a)-\mu_{l}(a)| \\
			& \ge \max_{l\not\in S^*}|G_{l} (a)|,
		\end{split}
	\end{equation*}
	where the second inequality is due to the assumption $\min_{l\in S^*}|\mu_{l}(a)| > \underline{\mu}$ and the definition of $J_{1}$, and the third inequality follows from that $\mu_{l}(a) = 0$ if $l\not\in S^*$. This result shows that on $J_{1}$,
	\begin{equation}\label{theorem6.8}
		\widehat{l} \in S^*.
	\end{equation} 
	
	In the following, we show that $\lim_{n\rightarrow\infty}\mathbb{P}(J_{1}) = 1$, and begin with defining two events $E_{1}$ and $E_{2}$ with negligible probabilities. Define 
	{\footnotesize\begin{equation*}
			\begin{split}	
				E_{1} & = \cup_{i=1}^{n}\{|\eta_{i}|> n^{\frac{1}{q_{1}}}\}, \\
				E_{2}& = \Big\{\max_{ 1\le l\le p} \Big|n^{-1}\sum_{i=1}^{n} \big\{
				\boldsymbol{1}_{\{|\eta_{i}|\le n^{\frac{1}{q_{1}}}\}}\times\big[\boldsymbol{1}_{U_{il}\in\mathcal{A}}- \boldsymbol{1}_{\widetilde{U}_{il}\in\mathcal{A}}\big]\big[\zeta(\boldsymbol{U}_{iS^*})\eta_{i}\big]^2 - \mu_{l}^{(n)}\big\}\Big| > c_{1}n^{-\frac{1}{2} + \frac{2}{q_{1}}} \sqrt{\log{(n \vee p)}} \Big\},
			\end{split}
	\end{equation*}}%
	where $q_{1}=q - \beta$, $\mu_{l}^{(n)}=\mathbb{E}\big\{\boldsymbol{1}_{\{|\eta|\le n^{\frac{1}{q_{1}}}\}}\big[\boldsymbol{1}_{X_{l}\in\mathcal{A}}- \boldsymbol{1}_{\widetilde{X}_{l}\in\mathcal{A}}\big]\big[\zeta(\boldsymbol{X}_{S^*})\varepsilon\big]^2\big\}$, and $c_{1} = \sqrt{8} \big[\sup_{\vec{z}\in\mathbb{R}^{s^*}} \zeta(\vec{z})\big]^2$. By Jensen's inequality, the Cauchy–Schwarz inequality, Markov's inequality, $\mathbb{E}|\varepsilon|^{q\vee 4} <\infty$, and other model regularity assumptions, 
	\begin{equation}
		\begin{split}\label{theorem6.10}
			\max_{1\le l \le p}|\mu_{l}(a) - \mu_{l}^{(n)}| & \le \max_{1\le l \le p}\mathbb{E}\big|\boldsymbol{1}_{\{|\varepsilon|> n^{\frac{1}{q_{1}}}\} }\big[\boldsymbol{1}_{X_{l}\in\mathcal{A}}- \boldsymbol{1}_{\widetilde{X}_{l}\in\mathcal{A}}\big]\big[\zeta(\boldsymbol{X}_{S^*})\varepsilon\big]^2\big|\\
			& \le  \big[\sup_{\vec{z}\in\mathbb{R}^{s^*}}\zeta(\vec{z})\big]^2 \times \mathbb{E}\big(\boldsymbol{1}_{\{|\varepsilon|> n^{\frac{1}{q_{1}}}\}}\varepsilon^2\big)\\
			&\le  \big[\sup_{\vec{z}\in\mathbb{R}^{s^*}}\zeta(\vec{z})\big]^2\sqrt{\mathbb{P}(|\varepsilon|> n^{\frac{1}{q_{1}}})}\sqrt{\mathbb{E}(\varepsilon^4\big)}\\
			&\le  \frac{1}{\sqrt{n}} \big[\sup_{\vec{z}\in\mathbb{R}^{s^*}}\zeta(\vec{z})\big]^2 \sqrt{\mathbb{E}(|\varepsilon|^{q_{1}})}\sqrt{\mathbb{E}(\varepsilon^4\big)} \\
			&	=O(n^{-\frac{1}{2}}),
		\end{split}
	\end{equation}
	and hence for all large $n$,
	\begin{equation}\label{theorem3.31}
		\max_{1\le l\le p}|\mu_{l}^{(n)}|\le 1 + \max_{1\le l\le p}|\mu_{l}(a)| \le 1 +  \big[\sup_{\vec{z}\in\mathbb{R}^{s^*}}\zeta(\vec{z})\big]^2 \mathbb{E}(\varepsilon^2).
	\end{equation}
	By \eqref{theorem3.31}, Markov's inequality, Hoeffding's inequality, $\mathbb{E}|\varepsilon|^{q\vee 4} <\infty$, and other model regularity assumptions, for all large $n$,
	\begin{equation}
		\begin{split}\label{theorem6.9}			
			\mathbb{P}(E_{1}) &\le \big(n^{1-\frac{q}{q_{1}}}\big)\mathbb{E}|\varepsilon|^q = o(1),\\
			\mathbb{P}(E_{2}) &\le 2p\times\exp{\left(-\frac{2c_{1}^2\times  n^{-1+\frac{4}{q_{1}}} \times \log{(n \vee p)} }{ 4n\big\{ \big[\sup_{\vec{z}\in\mathbb{R}^{s^*}} \zeta(\vec{z}) \big]^2 n^{\frac{2}{q_{1}}}  + \max_{1\le l\le p}|\mu_{l}^{(n)}| \big\}^2}\right)} = o(1).
		\end{split}
	\end{equation}
	
	With events $E_{1}$ and $E_{2}$, we are ready to deal with $\mathbb{P}(J_{1})$. By simple calculations, the definitions of $E_{1}$ and $E_{2}$, for each $t>0$,
	{\small\begin{equation}
			\begin{split}
				&\mathbb{P}(\max_{1\le l\le p} |G_{l}(a) - \mu_{l}(a)| > t)\\
				& \le \mathbb{P}(V_{n} + \max_{ 1\le l\le p} |n^{-1}\sum_{i=1}^{n} \big\{\big[\boldsymbol{1}_{U_{il}\in\mathcal{A}}- \boldsymbol{1}_{\widetilde{U}_{il}\in\mathcal{A}}\big]\big[\zeta(\boldsymbol{U}_{iS^*})\eta_{i}\big]^2 - \mu_{l}(a)\big\}| > t)\\
				&\le \mathbb{P}(\{V_{n} + \max_{ 1\le l\le p} |n^{-1}\sum_{i=1}^{n} \big\{\big[\boldsymbol{1}_{U_{il}\in\mathcal{A}}- \boldsymbol{1}_{\widetilde{U}_{il}\in\mathcal{A}}\big]\big[\zeta(\boldsymbol{U}_{iS^*})\eta_{i}\big]^2 - \mu_{l}(a)\big\}| > t\}\cap E_{1}^c\cap E_{2}^c) \\
				&\qquad + \mathbb{P}(E_{1}\cup E_{2})\\
				& \le \mathbb{P}(\{V_{n} + \max_{ 1\le l\le p} |n^{-1}\sum_{i=1}^{n} \big\{ \boldsymbol{1}_{\{|\eta_{i}|\le n^{\frac{1}{q_{1}}}\}} \times\big[\boldsymbol{1}_{U_{il}\in\mathcal{A}}- \boldsymbol{1}_{\widetilde{U}_{il}\in\mathcal{A}}\big]\big[\zeta(\boldsymbol{U}_{iS^*})\eta_{i}\big]^2 - \mu_{l}(a)\big\}| > t\}\\
				&\qquad\cap E_{2}^c) + \mathbb{P}(E_{1}\cup E_{2}) \\
				&\le \mathbb{P}(V_{n} + c_{1}n^{-\frac{1}{2} + \frac{2}{q_{1}}} \sqrt{\log{(n \vee p)}} + \max_{1\le l\le p}|\mu_{l}(a) - \mu_{l}^{(n)}|>  t)  + \mathbb{P}(E_{1}\cup E_{2})\\
				& \le \frac{c_{1}n^{-\frac{1}{2} + \frac{2}{q_{1}}} \sqrt{\log{(n \vee p)}} + \max_{1\le l\le p}|\mu_{l} (a)- \mu_{l}^{(n)}| +  \mathbb{E}(V_{n}) }{t } + \mathbb{P}(E_{1}\cup E_{2}),
			\end{split}
	\end{equation}}%
	where $V_{n} = n^{-1}\sum_{i=1}^{n} \big\{ \big[\widehat{m}(\boldsymbol{U}_{i}) - m(\boldsymbol{U}_{i})\big]^2+ 2|\widehat{m}(\boldsymbol{U}_{i}) - m(\boldsymbol{U}_{i})| \times \zeta(\boldsymbol{U}_{iS^*}) \times |\eta_{i}|\big\}$, and the last inequality is due to Markov's inequality.

	Next, we deal with $\mathbb{E}(V_{n})$. By Jensen's inequality, Condition~\ref{consistency.3}, $\lim_{n\rightarrow\infty}\underline{\mu}^{-1}\sqrt{B_{1}} = 0$, the assumption of i.i.d. observations, the assumption Condition~\ref{A4} that the training sample for constructing $\widehat{m}(\cdot)$ is independent of $\{V_{i}, \boldsymbol{U}_{i}, \widetilde{\boldsymbol{U}}_{i}, \eta_{i}\}_{i=1}^{n}$, and the assumptions that $\eta_{i}$'s are independent model errors and that $\eta_{1}$ and $\varepsilon$ have the same distribution, 
	\begin{equation}\label{theorem6.11}
		\underline{\mu}^{-1}\mathbb{E}(V_{n})\le 		\underline{\mu}^{-1}B_{1} + 2		\underline{\mu}^{-1}\sqrt{B_{1}} \times\big[\sup_{\vec{z}\in\mathbb{R}^{s^*}} \zeta(\vec{z})\big]\times \mathbb{E}|\varepsilon|  =o(1).
	\end{equation}

	By \eqref{theorem6.10}, \eqref{theorem6.9}--\eqref{theorem6.11} with $t= \frac{1}{3}\underline{\mu}$, and the assumption that $\underline{\mu}^{-1}\times n^{-\frac{1}{2} + \frac{2}{q_{1}}} \sqrt{\log{(n \vee p)}} = o(1)$ (recall that $q_{1} = q - \beta$), we conclude that 
	\begin{equation}\label{theorem6.1}
		\lim_{n\rightarrow\infty}\mathbb{P}(J_{1}^c) = 0,
	\end{equation} 
	which along with \eqref{theorem6.8} proves the first part of Theorem~\ref{theorem6}.


	Let us proceed to show the second assertion of Theorem~\ref{theorem6}. 
	By \eqref{theorem6.8} and the assumption $\min_{l\in S^*} |\mu_{l}(a)| > \underline{\mu}$, it holds that on $J_{1}$, $\widehat{l}\in S^*$ and $|\mu_{\widehat{l}}(a)| \ge\underline{\mu}.$ By this and Markov's inequality, for each $t>0$,
	\begin{equation}
		\begin{split}\label{theorem6.7}
			&\mathbb{P}\big\{\big| T_{\widehat{l}}(a) \big[\widehat{\sigma}_{\widehat{l}}(a)\big]^{-1} \big|\le t\big\} \\
			& = \mathbb{P}\big\{\big| T_{\widehat{l}}(a) - T_{\widehat{l}}^{(\star)}(a)+ T_{\widehat{l}}^{(\star)}(a) -\sqrt{n}\big[\mu_{\widehat{l}}(a)\big] + \sqrt{n}\big[\mu_{\widehat{l}}(a)\big] \big|\le t\big[\widehat{\sigma}_{\widehat{l}}(a)\big]\big\}\\
			&\le \mathbb{P}\Big\{\big\{ \sqrt{n}\big|\mu_{\widehat{l}}(a) \big|\le t\big[\widehat{\sigma}_{\widehat{l}}(a) \big] + \big| T_{\widehat{l}} (a)- T_{\widehat{l}}^{(\star)}(a)\big| + \big| T_{\widehat{l}}^{(\star)}(a) -\sqrt{n}\big[\mu_{\widehat{l}} (a)\big] \big| \big\}\cap J_{1}\Big\}\\
			&\qquad+ \mathbb{P}(J_{1}^c)\\
			&\le \mathbb{P}\big\{ \underline{\mu}\sqrt{n} \le t\big[\widehat{\sigma}_{\widehat{l}}(a)\big] + \big| T_{\widehat{l}} (a)- T_{\widehat{l}}^{(\star)}(a)\big| + \big| T_{\widehat{l}}^{(\star)}(a) -\sqrt{n}\big[\mu_{\widehat{l}} (a)\big]\big| \big\} + \mathbb{P}(J_{1}^c)\\
			&\le \frac{1}{\sqrt{n}\underline{\mu}} \left(t \mathbb{E}|\widehat{\sigma}_{\widehat{l}}(a)| + \mathbb{E}\big| T_{\widehat{l}} (a)- T_{\widehat{l}}^{(\star)}(a)\big|+ \mathbb{E}\big| T_{\widehat{l}}^{(\star)}(a) -\sqrt{n}\big[\mu_{\widehat{l}} (a)\big] \big| \right) + \mathbb{P}(J_{1}^c).
		\end{split}
	\end{equation}
	For the reader's convenience, $T_{l}(a)$ and $T_{l}^{(\star)}(a)$ for $l\in\{1, \dots, p\}$ are given  in \eqref{definition.T.1}, while an expression for $\widehat{\sigma}_{\widehat{l}}(a)$ can be found in \eqref{definition.sigma.1}.
	
	In what follows, we deal with the upper bounds for $\mathbb{E}|\widehat{\sigma}_{\widehat{l}}(a)|$, $\mathbb{E}\big| T_{\widehat{l}} (a)- T_{\widehat{l}}^{(\star)}(a)\big|$, and $\mathbb{E}\big| T_{\widehat{l}}^{(\star)}(a) -\sqrt{n}\big[\mu_{\widehat{l}} (a)\big] \big|$, and begin with the upper bound for $\mathbb{E}|\widehat{\sigma}_{\widehat{l}}(a)|$. By Jensen's inequality, Condition~\ref{A4}, the assumption that $\varepsilon$ is an independent model error with $\mathbb{E}(\varepsilon) = 0$ and $\mathbb{E}(\varepsilon^2)=1$ in model~\eqref{model.1}, and 
	Conditions~\ref{consistency.3}--\ref{bound},
	\begin{equation}
		\begin{split}\label{theorem6.6}
			&\mathbb{E}|\widehat{\sigma}_{\widehat{l}}(a)|\\
			& \le \sqrt{\mathbb{E}|\widehat{\sigma}_{\widehat{l}}^2(a)|} \\
			& \le \sqrt{\mathbb{E}(\widehat{\varepsilon}_{1}^4)} \\
			&\le \sqrt{\mathbb{E}\big[\zeta(\boldsymbol{X}_{S^*})\varepsilon + m(\boldsymbol{X})  - \widehat{m}(\boldsymbol{X} )\big]^4} \\
			& \le \Big\{\mathbb{E}\big[\zeta(\boldsymbol{X}_{S^*})\varepsilon\big]^4+ 4 \mathbb{E}\big[\zeta(\boldsymbol{X}_{S^*})\varepsilon\big]^3\mathbb{E}|m(\boldsymbol{X})  - \widehat{m}(\boldsymbol{X})| \\
			&\qquad+ 6\mathbb{E}\big[\zeta(\boldsymbol{X}_{S^*})\varepsilon\big]^2\mathbb{E}(m(\boldsymbol{X})  - \widehat{m}(\boldsymbol{X}))^2  + \mathbb{E}(m(\boldsymbol{X})  - \widehat{m}(\boldsymbol{X}))^4 \Big\}^{\frac{1}{2}}\\
			& \le \Big(\big[\sup_{\vec{z}\in\mathbb{R}^{s^*}} \zeta(\vec{z})\big]^4\mathbb{E}(\varepsilon^4) + 4\big[\sup_{\vec{z}\in\mathbb{R}^{s^*}} \zeta(\vec{z})\big]^3\times \mathbb{E}|\varepsilon|^3 \times \sqrt{B_{1}} \\
			&\qquad+ 6\big[\sup_{\vec{z}\in\mathbb{R}^{s^*}} \zeta(\vec{z})\big]^2B_{1} + 4\overbar{M}_{n}^2B_{1}\Big)^{\frac{1}{2}},
		\end{split}	
	\end{equation}
	where we use the result $\widehat{\sigma}_{j}^2(a)= n^{-1}\sum_{i=1}^{n} \big\{ (\boldsymbol{1}_{X_{ij} \in \mathcal{A}} -\boldsymbol{1}_{\widetilde{X}_{ij} \in \mathcal{A}}) (\widehat{\varepsilon}_{i})^2- \widehat{\mu}_{j}(a) \big\} ^2 \le n^{-1}\sum_{i=1}^{n}(\widehat{\varepsilon}_{i})^4$ for each $j\in\{1, \dots, p\}$, the assumption of i.i.d. observations, and Condition~\ref{A4} in the second inequality.

	Next, we establish an upper bound for $\mathbb{E}\big| T_{\widehat{l}}^{(\star)}(a) -\sqrt{n}\big[\mu_{\widehat{l}} (a)\big] \big|$. Since $\widehat{l}$ and $\widehat{a}_{l}$'s  are independent of $T_{l}^{(\star)}(a)$'s,
	\begin{equation}
		\begin{split}\label{theorem6.4}
			\mathbb{E}\big| T_{\widehat{l}}^{(\star)}(a) -\sqrt{n}\big[\mu_{\widehat{l}} (a)\big]\big|& = \sum_{l=1}^{p}\mathbb{E}\big\{\big| T_{l}^{(\star)} (a) -\sqrt{n}\big[\mu_{l}(a)\big] \big|\boldsymbol{1}_{\widehat{l}=l}\big\}\\
			&= \sum_{l=1}^{p}\big\{\mathbb{E}\big| T_{l}^{(\star)} (a) -\sqrt{n}\big[\mu_{l} (a)\big] \big|\big\}\times \mathbb{P}(\widehat{l}=l).
		\end{split}
	\end{equation}

	By the Burkholder--Davis--Gundy inequality, Jensen's inequality, and the assumption of i.i.d. observations, there exists some $K>0$ such that for each $l\in\{1, \dots ,p\}$ and each $n\ge 1$,
	\begin{equation}
		\begin{split}\label{theorem6.5}
			&\mathbb{E}\big| T_{l}^{(\star)}(a) -\sqrt{n}\big[\mu_{l}(a)\big] \big| \\
			&\le K\sqrt{n^{-1}\sum_{i=1}^{n}\mathbb{E}\big\{\big(\boldsymbol{1}_{X_{il}\in(-\infty, a]} -\boldsymbol{1}_{\widetilde{X}_{il}\in (-\infty, a]}\big)\big[\zeta(\boldsymbol{X}_{iS^{*}})\varepsilon_{i}\big]^{2} - \mu_{l}(a)\big\}^2}\\
			& \le K\sqrt{n^{-1}\sum_{i=1}^{n}\mathbb{E}\big\{\big[\zeta(\boldsymbol{X}_{iS^{*}})\varepsilon_{i}\big]^{4} \big\}}\\
			& \le K \big[\sup_{\vec{z}\in\mathbb{R}^{s^*}} \zeta(\vec{z})\big]^2 \sqrt{\mathbb{E}(\varepsilon^4)}.
		\end{split}
	\end{equation}

	By \eqref{theorem6.4}--\eqref{theorem6.5}, for each $n\ge 1$,
	\begin{equation}\label{theorem6.3}
		\mathbb{E}\big| T_{\widehat{l}}^{(\star)}(a) -\sqrt{n}\big[\mu_{\widehat{l}}(a)\big] \big|\le K \big[\sup_{\vec{z}\in\mathbb{R}^{s^*}} \zeta(\vec{z})\big]^2 \sqrt{\mathbb{E}(\varepsilon^4)},
	\end{equation}
	where $K>0$ is given in \eqref{theorem6.5}.

	Next, we proceed to deal with the upper bound for $\mathbb{E}\big| T_{\widehat{l}} (a)- T_{\widehat{l}}^{(\star)}(a)\big|$. By the definition of $T_{l}(a)$ and $T_{l}^{(\star)}(a)$ in \eqref{definition.T.1}, the Burkholder--Davis--Gundy inequality, the arguments similar to those for \eqref{theorem4.23}, $\mathbb{E}(\varepsilon^2)=1$ in model~\eqref{model.1}, there exists some $K>0$ such that for each $n\ge 1$ and $B_{1}>0$,
	\begin{equation}
		\begin{split}\label{theorem6.2}
			&\mathbb{E}\big| T_{\widehat{l}} (a)- T_{\widehat{l}}^{(\star)}(a)\big|\\
			& \le 2\mathbb{E}\Big|\frac{1}{\sqrt{n}} \sum_{i=1}^n \big[\boldsymbol{1}_{X_{i\widehat{l}}\in(-\infty, a]} -\boldsymbol{1}_{\widetilde{X}_{i\widehat{l}}\in (-\infty, a]}\big]\big[m(\boldsymbol{X}_{i}) - \widehat{m}(\boldsymbol{X}_{i})\big]\zeta(\boldsymbol{X}_{iS^{*}})\varepsilon_{i}\Big| \\
			&\qquad+ \frac{1}{\sqrt{n}}\mathbb{E}\sum_{i=1}^n\big[m(\boldsymbol{X}_{i}) - \widehat{m}(\boldsymbol{X}_{i})\big]^2\\
			&\le 2K\big[\sup_{\vec{z}\in\mathbb{R}^{s^*}} \zeta(\vec{z})\big]\sqrt{B_{1}} + \sqrt{n}B_{1}.
		\end{split}
	\end{equation}

	By the model regularity assumptions, the assumption that $\lim\sup_{n\rightarrow\infty}\overbar{M}_{n}^2B_{1} <\infty$, \eqref{theorem6.7}--\eqref{theorem6.6}, and \eqref{theorem6.3}--\eqref{theorem6.2}, it holds that for all large $n$, each $t>0$, and $0<B_{1}<1$,
	\begin{equation}
		\mathbb{P}\big\{\big| T_{\widehat{l}}(a) \big[\widehat{\sigma}_{\widehat{l}}(a)\big]^{-1} \big|\le t\big\}\le \frac{(1+t)\log{n}}{\sqrt{n}\underline{\mu}} + \frac{B_{1}}{\underline{\mu}} + \mathbb{P}(J_{1}^c),
	\end{equation}
	which along with \eqref{theorem6.1} and the assumption that $\lim_{n\rightarrow\infty}\underline{\mu}^{-1}\Big[\sqrt{B_{1}} + n^{-\frac{1}{2} + \frac{2}{q_{1}}} \sqrt{\log{(n \vee p)}}\Big]= 0$
	concludes the proof of Theorem~\ref{theorem6}.

	\subsection{Proof of Theorem~\ref{theorem7}}\label{proof.theorem7}

	Let some $j\in\{1, \dots, p\}$ such that $\max_{1\le r\le R}|\mu_{j}(\kappa_{r})| > \underline{\mu}$ be given. We begin with an application of Markov's inequality as follows. For each $n\ge 1$ and $t>0$,
	{\small\begin{equation}
			\begin{split}\label{theorem7.3}
				&\mathbb{P}(|T_{j}(\widehat{a}_{j})\big[\widehat{\sigma}_{j}(\widehat{a}_{j})\big]^{-1}| \le t ) \\
				&\le \mathbb{P}(|T_{j}^{(\star)}(\widehat{a}_{j})| \le  t| \widehat{\sigma}_{j}(\widehat{a}_{j})| +|A_{1}(\widehat{a}_{j})| + |A_{2}(\widehat{a}_{j})|)\\
				&\le \mathbb{P}\bigg\{ \big\{\sqrt{n}\big|\mu_{j}(\widehat{a}_{j})\big|  \le \big|T_{j}^{(\star)}(\widehat{a}_{j}) - \sqrt{n}\big[\mu_{j}(\widehat{a}_{j})\big]\big| + t| \widehat{\sigma}_{j}(\widehat{a}_{j})| +|A_{1}(\widehat{a}_{j})| + |A_{2}(\widehat{a}_{j})| \big\} \\
				&\qquad\qquad\cap J_{1}\bigg\} + \mathbb{P}(J_{1}^c)\\
				&\le \mathbb{P}\left\{ \frac{\sqrt{n} \underline{\mu}}{3} \le |T_{j}^{(\star)}(\widehat{a}_{j}) - \sqrt{n}\big[\mu_{j}(\widehat{a}_{j})\big]| + t| \widehat{\sigma}_{j}(\widehat{a}_{j})| +|A_{1}(\widehat{a}_{j})| + |A_{2}(\widehat{a}_{j})| \right\} +  \mathbb{P}(J_{1}^c)\\
				& \le \frac{3}{\sqrt{n}\underline{\mu}} \times \Big\{\mathbb{E}\big|T_{j}^{(\star)}(\widehat{a}_{j}) - \sqrt{n}\big[\mu_{j}(\widehat{a}_{j})\big]\big| +  t\times \mathbb{E}\big| \widehat{\sigma}_{j}(\widehat{a}_{j})\big| + \mathbb{E}\big|A_{1}(\widehat{a}_{j})\big| + \mathbb{E}\big|A_{2}(\widehat{a}_{j})\big|\Big\} \\
				&\qquad+  \mathbb{P}(J_{1}^c),
			\end{split}
	\end{equation}}%
	where 
	\begin{equation*}
		\begin{split}
			A_{1}(\widehat{a}_{j}) &=  n^{-\frac{1}{2}}\sum_{i=1}^{n}\big(\boldsymbol{1}_{X_{ij}\in(-\infty, \widehat{a}_{j}]} -\boldsymbol{1}_{\widetilde{X}_{ij}\in(-\infty, \widehat{a}_{j}]}\big) (m(\boldsymbol{X}_{i}) - \widehat{m}(\boldsymbol{X}_{i}))^2,\\
			A_{2} (\widehat{a}_{j})&=  2n^{-\frac{1}{2}}\sum_{i=1}^{n}\big(\boldsymbol{1}_{X_{ij}\in(-\infty, \widehat{a}_{j}]} -\boldsymbol{1}_{\widetilde{X}_{ij}\in(-\infty, \widehat{a}_{j}]}\big) (m(\boldsymbol{X}_{i}) - \widehat{m}(\boldsymbol{X}_{i}))\zeta(\boldsymbol{X}_{iS^{*}})\varepsilon_{i} ,\\
			T_{j}^{(\star)}(\widehat{a}_{j}) &= n^{-\frac{1}{2}}\sum_{i=1}^{n}\big(\boldsymbol{1}_{X_{ij}\in(-\infty, \widehat{a}_{j}]} -\boldsymbol{1}_{\widetilde{X}_{ij}\in(-\infty, \widehat{a}_{j}]}\big)(\zeta(\boldsymbol{X}_{iS^{*}})\varepsilon_{i})^{2},\\
			J_{1}&\coloneqq\big\{|\mu_{j}(\widehat{a}_{j})| > \frac{1}{3}\underline{\mu}\big\},
		\end{split}
	\end{equation*}
	and the first inequality of \eqref{theorem7.3} follows because $T_{j}(\widehat{a}_{j}) =A_{1}(\widehat{a}_{j}) + A_{2}(\widehat{a}_{j}) + T_{j}^{(\star)}(\widehat{a}_{j})$.
	
	In what follows, we establish upper bounds for terms on the RHS of \eqref{theorem7.3}, and begin with an upper bound for $\mathbb{P}(J_{1}^c)$. We have
	\begin{equation}
		\label{theorem7.8}
		\big\{\max_{1\le r\le R} |G_{j}(\kappa_{r}) - \mu_{j}(\kappa_{r})| \le \frac{1}{3} \underline{\mu}\big\}\subset J_{1}
	\end{equation}
	due to the assumption $\max_{1\le r\le R}|\mu_{j}(\kappa_{r})| > \underline{\mu}$ and that $\widehat{a}_{j} = \arg\max_{\kappa\in\{\kappa_{1}, \dots ,\kappa_{R}\}}|G_{j}(\kappa)|$; here, for the reader's convenience, recall that it is given in Section~\ref{Sec2.3} (with $n_{2} = n$ in this proof) that $G_{j}(\kappa_{r}) = n^{-1}\sum_{i=1}^{n}(\boldsymbol{1}_{U_{ij}\in(-\infty, \kappa_{r}]}- \boldsymbol{1}_{\widetilde{U}_{ij}\in (-\infty, \kappa_{r}]}) (\widehat{\eta}_{i} )^{2}$, where $\widehat{\eta}_{i} = V_{i} - \widehat{m}(\boldsymbol{U}_{i})$.

	By \eqref{theorem7.8}, the definition of $G_{j}(\kappa_{r})$'s, and Markov's inequality,
	{\small\begin{equation}
			\begin{split}\label{theorem7.2}
				&\mathbb{P}(J_{1}^c) \\
				&\le
				\mathbb{P}(\max_{ 1\le r\le R} |G_{j}(\kappa_{r}) - \mu_{j}(\kappa_{r})| >  \frac{1}{3} \underline{\mu})\\
				& \le \sum_{r=1}^{R}\mathbb{P}(|G_{j}(\kappa_{r}) - \mu_{j}(\kappa_{r})| >  \frac{1}{3} \underline{\mu})\\
				& \le \sum_{r=1}^{R}\mathbb{P}\bigg\{ V_{n} \\
				&\qquad+ \Big|n^{-1}\sum_{i=1}^{n} \bigg\{\big[\boldsymbol{1}_{U_{ij}\in (-\infty, \kappa_{r}]}- \boldsymbol{1}_{\widetilde{U}_{ij}\in (-\infty, \kappa_{r}]}\big]\big[\zeta(\boldsymbol{U}_{iS^*})\eta_{i}\big]^2 - \mu_{j}(\kappa_{r})\bigg\} \Big| > \frac{1}{3}\underline{\mu} \bigg\} \\
				&\le \frac{3}{\underline{\mu}} \bigg\{R\times \mathbb{E}(V_{n}) \\
				&\qquad+ n^{-\frac{1}{2}}\sum_{r=1}^R \mathbb{E} \left| \sum_{i=1}^{n} \frac{ \big[\boldsymbol{1}_{U_{ij}\in (-\infty, \kappa_{r}]}- \boldsymbol{1}_{\widetilde{U}_{ij}\in (-\infty, \kappa_{r}]}\big]\big[\zeta(\boldsymbol{U}_{iS^*})\eta_{i}\big]^2 - \mu_{j}(\kappa_{r})}{\sqrt{n}} \right| \bigg\},
			\end{split}
	\end{equation}}%
	where $V_{n} = n^{-1}\sum_{i=1}^{n} \big\{ \big[\widehat{m}(\boldsymbol{U}_{i}) - m(\boldsymbol{U}_{i})\big]^2+ 2|\widehat{m}(\boldsymbol{U}_{i}) - m(\boldsymbol{U}_{i})|\times \zeta(\boldsymbol{U}_{iS^*})\times| \eta_{i}|\big\}$.

	Let us deal with the first term on the RHS of \eqref{theorem7.2}. By the assumptions  that observations are i.i.d. and that $\eta_{i}$'s are independent model errors, Jensen's inequality, Condition~\ref{A4}, Condition~\ref{consistency.3}, $\mathbb{E}(\varepsilon^2)=1$ in model~\eqref{model.1},  and other model regularity assumptions,
	\begin{equation}
		\begin{split}
			\mathbb{E}(V_{n}) \le B_{1} + 2\big[\sup_{\vec{z} \in \mathbb{R}^{s^*}}\zeta(\vec{z})\big]\sqrt{B_{1}}.
		\end{split}
	\end{equation}

	For the second term on the RHS of \eqref{theorem7.2}, we use the Burkholder--Davis--Gundy inequality~\citepsupp{burkholder1972integral} and  model regularity assumptions to deduce that there exists some $K>0$ such that
	\begin{equation}
		\begin{split}\label{theorem7.1}
			\mathbb{E} \left| \sum_{i=1}^{n} \frac{ \big[\boldsymbol{1}_{U_{ij}\in (-\infty, \kappa_{r}]}- \boldsymbol{1}_{\widetilde{U}_{ij}\in (-\infty, \kappa_{r}]}\big]\big[\zeta(\boldsymbol{U}_{iS^*})\eta_{i}\big]^2 - \mu_{j}(\kappa_{r})}{\sqrt{n}} \right|\le K.
		\end{split}
	\end{equation}
	
	By \eqref{theorem7.2}--\eqref{theorem7.1} and the assumption $B_{1}<1$, there exists $N_{1}>0$ such that for all  $n\ge N_{1}$, 
	\begin{equation}\label{theorem7.5}
		\mathbb{P}(J_{1}^c)\le \frac{3R \sqrt{B_{1}} \big\{1 + 2\big[\sup_{\vec{z} \in \mathbb{R}^{s^*}}\zeta(\vec{z})\big]\big\} + 3R K n^{-\frac{1}{2}}}{\underline{\mu}} .
	\end{equation}

	Next, we proceed to establish upper bounds for terms $\mathbb{E}\big|T_{j}^{(\star)}(\widehat{a}_{j}) - \sqrt{n}\big[\mu_{j}(\widehat{a}_{j})\big]\big|$, $\mathbb{E}\big| \widehat{\sigma}_{j}(\widehat{a}_{j})\big|$, $\mathbb{E}\big|A_{1}(\widehat{a}_{j})\big|$, and $\mathbb{E}\big|A_{2}(\widehat{a}_{j})\big|$ on the RHS of \eqref{theorem7.3}. By arguments similar to those for \eqref{theorem7.1} and \eqref{theorem1.2}, there exists some $C>0$ such that  for each $r\in \{1, \dots, R\}$,
	\begin{equation}
		\begin{split}	\label{theorem7.4}				
			\mathbb{E}\big|T_{j}^{(\star)}(\kappa_{r}) - \sqrt{n}\big[\mu_{j}(\kappa_{r})\big]\big|  & \le C\big[\sup_{\vec{z} \in \mathbb{R}^{s^*}} \zeta(\vec{z})\big]^2\sqrt{\mathbb{E}(\varepsilon^4)},\\
			\mathbb{E}\big|A_{2}(\kappa_{r})\big| & \le 2C\big[\sup_{\vec{z} \in \mathbb{R}^{s^*}} \zeta(\vec{z})\big]\sqrt{B_{1}\mathbb{E}(\varepsilon^2)},
		\end{split}
	\end{equation}
	where $C>0$ is due to an application of the Burkholder--Davis--Gundy inequality; note that for the second inequality in \eqref{theorem7.4}, it is required that the training sample for $\widehat{m}(\cdot)$ is independent of $\{Y_{i}, \boldsymbol{X}_{i}, \widetilde{\boldsymbol{X}}_{i}, \varepsilon_{i}\}_{i=1}^{n}$.
	
	By \eqref{theorem7.4} and the assumption that $\widehat{a}_{j}$ is independent of $T_{j}^{(\star)}$,
	\begin{equation}
		\begin{split}\label{theorem7.6}
			\mathbb{E}\big|T_{j}^{(\star)}(\widehat{a}_{j}) - \sqrt{n}\big[\mu_{j}(\widehat{a}_{j})\big]\big| & =\sum_{r=1}^{R} \mathbb{E}\big\{\big|T_{j}^{(\star)}(\kappa_{r}) - \sqrt{n}\big[\mu_{j}(\kappa_{r})\big]\big| \boldsymbol{1}_{\widehat{a}_{j}=\kappa_{r}}\big\}\\
			& = \sum_{r=1}^{R} \mathbb{E}\big|T_{j}^{(\star)}(\kappa_{r}) - \sqrt{n}\big[\mu_{j}(\kappa_{r})\big]\big| \times \mathbb{P}(\widehat{a}_{j}=\kappa_{r})\\
			&\le C\big[\sup_{\vec{z} \in \mathbb{R}^{s^*}} \zeta(\vec{z})\big]^2\sqrt{\mathbb{E}(\varepsilon^4)},
		\end{split}
	\end{equation}
	where $C$ is given in \eqref{theorem7.4}. Similarly, 
	\begin{equation}
		\begin{split}
			\mathbb{E}\big|A_{2}(\widehat{a}_{j})\big| &= \sum_{r=1}^{R}			\mathbb{E}(|A_{2}(\kappa_{r})|\boldsymbol{1}_{\widehat{a}_{j} = \kappa_{r}})\\
			&\le \sum_{r=1}^{R}		\mathbb{E}|A_{2}(\kappa_{r})| \times  \mathbb{P}(\widehat{a}_{j} = \kappa_{r})\\
			& \le 2C\big[\sup_{\vec{z} \in \mathbb{R}^{s^*}} \zeta(\vec{z})\big]\sqrt{B_{1}\mathbb{E}(\varepsilon^2)}.
		\end{split}
	\end{equation}

	Meanwhile, by the assumptions that observations are i.i.d., Condition~\ref{A4}, and Condition~\ref{consistency.3},
	\begin{equation}
		\mathbb{E}\big|A_{1}(\widehat{a}_{j})\big| \le \sqrt{n}B_{1}.
	\end{equation}
	
	Next, we deal with the upper bound for $\mathbb{E}\big| \widehat{\sigma}_{j}(\widehat{a}_{j})\big|$. By arguments similar to those for \eqref{theorem5.3}, for each $n\ge 1$,
	\begin{equation}
		\begin{split}\label{theorem7.7}
			\mathbb{E}\big| \widehat{\sigma}_{j}(\widehat{a}_{j})\big| &\le \sqrt{\mathbb{E}\big[ \widehat{\sigma}_{j}^2(\widehat{a}_{j})\big]}\\
			&\le \Big\{ 4\overbar{M}_{n}^2B_{1}+ 8\overbar{M}_{n}B_{1}\big[\sup_{\vec{z} \in \mathbb{R}^{s^*}} \zeta(\vec{z})\big]\\
			&\qquad+
			6B_{1}\big[\sup_{\vec{z} \in \mathbb{R}^{s^*}} \zeta(\vec{z})\big]^2+ 4\sqrt{B_{1}}\big[\sup_{\vec{z} \in \mathbb{R}^{s^*}} \zeta(\vec{z})\big]^3 \mathbb{E}|\varepsilon|^3+ \big[\sup_{\vec{z} \in \mathbb{R}^{s^*}} \zeta(\vec{z})\big]^4\mathbb{E}(\varepsilon^4) \Big\}^{\frac{1}{2}}.
		\end{split}
	\end{equation}

	By \eqref{theorem7.5} with $K>0$ given in \eqref{theorem7.1}, \eqref{theorem7.6}--\eqref{theorem7.7}, 	Condition~\ref{bound} with $\lim\sup_{n\rightarrow\infty}\overbar{M}_{n}^2B_{1} <\infty$, $B_{1}<1$, and $\overbar{M}_{n}\ge 1$,  the assumption that $R$ is finite, and other model regularity assumptions, there exists some $N_{2}>0$ such that for all $n\ge N_{2}$ and  each $t>0$,
	\begin{equation*}
		\begin{split}	
			&\textnormal{The RHS of \eqref{theorem7.3}} 
			\\&\le  \frac{3\sqrt{n}B_{1} + 3\mathbb{E}\big|T_{j}^{(\star)}(\widehat{a}_{j}) - \sqrt{n}\big[\mu_{j}(\widehat{a}_{j})\big]\big| +
				3t\times\mathbb{E}\big| \widehat{\sigma}_{j}(\widehat{a}_{j})\big|
				+3\mathbb{E}\big|A_{2}(\widehat{a}_{j})\big|}{\sqrt{n}\underline{\mu}} \\
			&\qquad + \frac{3R \sqrt{B_{1}} \big\{1 + 2\big[\sup_{\vec{z} \in \mathbb{R}^{s^*}}\zeta(\vec{z})\big]\big\} + 3RK n^{-\frac{1}{2}}}{\underline{\mu}} \\
			&\le \underline{\mu}^{-1}\sqrt{B_{1}}\log{(n)} + (\log{n})(1+t)(\sqrt{n}\underline{\mu})^{-1},
		\end{split}
	\end{equation*}
	which concludes the desired result of Theorem~\ref{theorem7} for feature index $j$ given at the beginning of this proof. 
	
	Lastly, we note that the generic constants such as $C, K,N_{1},N_{2}$ used in the proof are not subject to feature index $j$. Therefore, we have finished the proof of Theorem~\ref{theorem7}.

	\subsection{Proof of Theorem~\ref{theorem8}}\label{proof.theorem8}
	
	For the reader's convenience, recall that for each $r \in\{1, \dots, R\}$,
	\begin{equation*}
		\begin{split}
			G_{l}(\kappa_{r}) &=  n^{-1}\sum_{i=1}^{n}(\boldsymbol{1}_{U_{il}\in(\infty, \kappa_{r}]}- \boldsymbol{1}_{\widetilde{U}_{il}\in(\infty, \kappa_{r}]}) (\widehat{\eta}_{i} )^{2},\\
			\widehat{a}_{l} & = \arg\max_{\kappa\in\{\kappa_{1}, \dots ,\kappa_{R}\}}|G_{l}(\kappa)|,\\
			\widehat{l} &= \arg\max_{1\le l\le p}|G_{l}(\widehat{a}_{l})|,
		\end{split}
	\end{equation*}
	where $\widehat{\eta}_{i}  = V_{i} - \widehat{m}(\boldsymbol{U}_{i})$.

	Let us begin the formal proof  with the first part of Theorem~\ref{theorem8}, which is $\lim_{n\rightarrow\infty}\mathbb{P}(\widehat{l}\in S^*) = 1$ when $\min_{l\in S^*} \max_{1\le r\le R}|\mu_{l}(\kappa_{r})| > \underline{\mu}>0$ ($S^*$ is not an empty set accordingly). On the event 
	$$J_{1} = \left\{\max_{1\le l\le p} \big||G_{l}(\widehat{a}_{l}) | - \max_{1\le r\le R}|\mu_{l}(\kappa_{r})| \big| \le \frac{1}{3}\underline{\mu}\right\},$$
	it holds that 
	\begin{equation*}
		\begin{split}
			\min_{l\in S^*}|G_{l}(\widehat{a}_{l})| & = \min_{l\in S^*} \Big[|G_{l}(\widehat{a}_{l})| - \max_{1\le r\le R}|\mu_{l}(\kappa_{r})| + \max_{1\le r\le R}|\mu_{l}(\kappa_{r})| \Big] \\
			& \ge \min_{l\in S^*}\max_{1\le r\le R}|\mu_{l}(\kappa_{r})|-  \max_{1\le l\le p} \Big||G_{l} (\widehat{a}_{l}) |-\max_{1\le r\le R}|\mu_{l}(\kappa_{r})|\Big| \\
			&\ge \frac{2}{3}\underline{\mu}\\
			&> \max_{l\not\in S^*}|G_{l}(\widehat{a}_{l}) |,
		\end{split}
	\end{equation*}
	where the second inequality follows from the assumption $\min_{l\in S^*} \max_{1\le r\le R}|\mu_{l}(\kappa_{r})| > \underline{\mu}$ and the definition of $J_{1}$, and the third inequality follows from that $\max_{l\not\in S^*} \max_{1\le r\le R}|\mu_{l}(\kappa_{r})| = 0$ and  the definition of $J_{1}$. This result shows that on $J_{1}$,
	\begin{equation}\label{theorem8.1}
		\widehat{l} \in S^*.
	\end{equation} 
	
	In the following, we conclude the first part of Theorem~\ref{theorem8} by proving that $\lim_{n\rightarrow\infty}\mathbb{P}(J_{1}^c) = 0.$ On the event 
	$$J_{2} \coloneqq \Big\{\max_{1\le l\le p, 1\le r\le R} |G_{l}(\kappa_{r}) - \mu_{l}(\kappa_{r})| \le  \frac{1}{3} \underline{\mu}\Big\},$$ 
	it holds that for each $l\in \{1, \dots, p\}$,
	\begin{equation}
		\begin{split}\label{theorem8.11}
			|G_{l}(\widehat{a}_{l})| & = 			\max_{1\le r\le R}|G_{l}(\kappa_{r}) - \mu_{l}(\kappa_{r}) + \mu_{l}(\kappa_{r})|\\
			&\ge \max_{1\le r \le R}\Big\{|\mu_{l}(\kappa_{r})| - |G_{l}(\kappa_{r}) - \mu_{l}(\kappa_{r})|\Big\}\\
			&\ge \max_{1\le r\le R}|\mu_{l}(\kappa_{r})| - \frac{1}{3}\underline{\mu},
		\end{split}
	\end{equation}
	where the first inequality follows from the definition of $\widehat{a}_{l}$ in Section~\ref{Sec2.4}, and similarly,
	\begin{equation*}
		\begin{split}\label{theorem8.11b}
			|G_{l}(\widehat{a}_{l})| & =			\max_{1\le r\le R}|G_{l}(\kappa_{r}) - \mu_{l}(\kappa_{r}) + \mu_{l}(\kappa_{r})|\\
			&\le \max_{1\le r \le R}\Big\{|\mu_{l}(\kappa_{r})| + |G_{l}(\kappa_{r}) - \mu_{l}(\kappa_{r})|\Big\}\\
			&\le \max_{1\le r\le R}|\mu_{l}(\kappa_{r})| + \frac{1}{3}\underline{\mu}.
		\end{split}
	\end{equation*}
	
	Therefore, 
	\begin{equation}
		\label{theorem8.12}
		J_{2} \subset J_{1},
	\end{equation} 
	and hence $\mathbb{P}(J_{1}^c)\le \mathbb{P}(J_{2}^c)$. Next, let us show that  $\mathbb{P}(J_{2}^c)$ is negligible as follows.
	\begin{equation}
		\begin{split}\label{theorem8.3}
			&\mathbb{P}(J_{2}^c) \\
			&=
			\mathbb{P}(\max_{1\le l\le p, 1\le r\le R} |G_{l}(\kappa_{r}) - \mu_{l}(\kappa_{r})| >  \frac{1}{3} \underline{\mu})\\
			& \le \sum_{r=1}^{R}\mathbb{P}(\max_{1\le l\le p} |G_{l}(\kappa_{r}) - \mu_{l}(\kappa_{r})| >  \frac{1}{3} \underline{\mu})\\
			& \le \sum_{r=1}^{R}\mathbb{P}\Big(V_{n} + \max_{ 1\le l\le p} \Big|n^{-1}\sum_{i=1}^{n} \Big\{\big[\boldsymbol{1}_{U_{il}\in (-\infty, \kappa_{r}]}- \boldsymbol{1}_{\widetilde{U}_{il}\in (-\infty, \kappa_{r}]}\big]\big[\zeta(\boldsymbol{U}_{iS^*})\eta_{i}\big]^2\\
			&\qquad - \mu_{l}(\kappa_{r})\Big\}\Big| > \frac{1}{3}\underline{\mu}\Big),\\
		\end{split}
	\end{equation}
	where $V_{n} = n^{-1}\sum_{i=1}^{n} \big\{ \big[\widehat{m}(\boldsymbol{U}_{i}) - m(\boldsymbol{U}_{i})\big]^2+ 2|\widehat{m}(\boldsymbol{U}_{i}) - m(\boldsymbol{U}_{i})|\times \zeta(\boldsymbol{U}_{iS^*})\times |\eta_{i}|\big\}$, and the second inequality follows because $V_{n}$ is not subject to feature index $l\in\{1, \dots, p\}$. To deal with the RHS of \eqref{theorem8.3}, we define the following events.
	\begin{equation*}
		\begin{split}	
			&E_{1}  = \cup_{i=1}^{n}\{|\eta_{i}|> n^{\frac{1}{q_{1}}}\}, \\
			&E_{2r} = \Big\{\max_{ 1\le l\le p} \Big|n^{-1}\sum_{i=1}^{n} \big\{\boldsymbol{1}_{\{|\eta_{i}|\le n^{\frac{1}{q_{1}}}\}}\times \big[\boldsymbol{1}_{U_{il}\in(-\infty, \kappa_{r}]}- \boldsymbol{1}_{\widetilde{U}_{il}\in (-\infty, \kappa_{r}]}\big]\big[\zeta(\boldsymbol{U}_{iS^*})\eta_{i}\big]^2 - \mu_{l}^{(n)}(\kappa_{r})\big\}\Big| \\
			& \qquad\qquad> c_{1}n^{-\frac{1}{2} + \frac{2}{q_{1}}} \sqrt{\log{(n \vee p)}} \Big\}, \textnormal{ for }r = 1, \dots, R,
		\end{split}
	\end{equation*}
	where $q_{1}=q - \beta$, $\mu_{l}^{(n)}(\kappa_{r}) =\mathbb{E}\big\{\boldsymbol{1}_{\{|\eta|\le n^{\frac{1}{q_{1}}}\}}\big[\boldsymbol{1}_{X_{l}\in(-\infty, \kappa_{r}]}- \boldsymbol{1}_{\widetilde{X}_{l}\in(-\infty, \kappa_{r}]}\big]\big[\zeta(\boldsymbol{X}_{S^*})\varepsilon\big]^2\big\}$, and \sloppy$c_{1} = \sqrt{8} \big[\sup_{\vec{z}\in\mathbb{R}^{s^*}} \zeta(\vec{z})\big]^2$. In \eqref{theorem8.2} below, we show that events $E_{1}$ and $E_{2r}$'s are negligible; first, we need \eqref{theorem8.16} below. By the Jensen's inequality, Cauchy–Schwarz inequality, Markov's inequality, $ \mathbb{E}|\varepsilon|^{q\vee 4} <\infty $, and other model regularity assumptions, 
	\begin{equation}
		\begin{split}\label{theorem8.16}
			&\max_{r\in\{1, \dots, R\}}\max_{1\le l \le p}|\mu_{l}(\kappa_{r}) - \mu_{l}^{(n)}(\kappa_{r})| \\
			& \le \max_{r\in\{1, \dots, R\}}\max_{1\le l \le p}\mathbb{E}\big|\boldsymbol{1}_{|\varepsilon|> n^{\frac{1}{q_{1}}}}\big[\boldsymbol{1}_{X_{l}\in(-\infty, \kappa_{r}]}- \boldsymbol{1}_{\widetilde{X}_{l}\in(-\infty, \kappa_{r}]}\big]\big[\zeta(\boldsymbol{X}_{S^*})\varepsilon\big]^2\big|\\
			& \le  \big[\sup_{\vec{z}\in\mathbb{R}^{s^*}}\zeta(\vec{z})\big]^2\mathbb{E}\big\{\boldsymbol{1}_{\{|\varepsilon|> n^{\frac{1}{q_{1}}}\}}\varepsilon^2\big\}\\
			&\le  \big[\sup_{\vec{z}\in\mathbb{R}^{s^*}}\zeta(\vec{z})\big]^2\sqrt{\mathbb{P}(|\varepsilon|> n^{\frac{1}{q_{1}}})}\sqrt{\mathbb{E}(\varepsilon^4\big)}\\
			&\le \frac{1}{\sqrt{n}} \big[\sup_{\vec{z}\in\mathbb{R}^{s^*}}\zeta(\vec{z})\big]^2 \sqrt{\mathbb{E}(|\varepsilon|^{q_{1}})}\sqrt{\mathbb{E}(\varepsilon^4\big)} \\
			&	=O(n^{-\frac{1}{2}}).
		\end{split}
	\end{equation}
	
	By \eqref{theorem8.16}, it holds that  for all large $n$ and each $r \in \{1, \dots, R\}$,
	\begin{equation*}
		\max_{1\le l\le p}|\mu_{l}^{(n)}(\kappa_{r})|\le 1 + \max_{1\le l\le p}|\mu_{l}(\kappa_{r})| \le 1 +  \big[\sup_{\vec{z}\in\mathbb{R}^{s^*}}\zeta(\vec{z})\big]^2 \mathbb{E}(\varepsilon^2).
	\end{equation*}
	
	By this, Markov's inequality, Hoeffding's inequality, $ \mathbb{E}|\varepsilon|^{q\vee 4} <\infty $, and other model regularity assumptions, for all large $n$ and each $1\le r\le R$,
	\begin{equation}
		\begin{split}\label{theorem8.2}
			\mathbb{P}(E_{1}) &\le \big(n^{1-\frac{q}{q_{1}}}\big)\mathbb{E}|\varepsilon|^q = o(1),\\
			\mathbb{P}(E_{2r}) &\le 2p\times \exp{\left(\frac{-2c_{1}^2\times n^{-1+\frac{4}{q_{1}}}\times \log{(n \vee p)} }{ 4n\big\{ \big[\sup_{\vec{z}\in\mathbb{R}^{s^*}} \zeta(\vec{z})\big]^2 n^{\frac{2}{q_{1}}}  +  \max_{1\le l\le p}|\mu_{l}^{(n)}(\kappa_{r})| \big\}^2  }\right)} = o(1).
		\end{split}
	\end{equation}
	
	With events $E_{1}$ and $E_{2r}$, we are ready to deal with the RHS of \eqref{theorem8.3}. By Markov's inequality and the definitions of $E_{1}$ and $E_{2r}$, for each $t>0$,
	\begin{equation}
		\begin{split}\label{theorem8.14}
			&\textnormal{RHS of \eqref{theorem8.3}}\\
			&\le \sum_{r=1}^{R}\mathbb{P}(A_{1r} \cap E_{1}^c\cap E_{2r}^c) + \sum_{r=1}^R\mathbb{P}(E_{1}\cup E_{2r})\\
			& \le \sum_{r=1}^R\mathbb{P}(A_{2r}\cap E_{2r}^c) + \sum_{r=1}^R\mathbb{P}(E_{1}\cup E_{2r}) \\
			&\le \sum_{r=1}^R\mathbb{P}(V_{n} + c_{1}n^{-\frac{1}{2} + \frac{2}{q_{1}}} \sqrt{\log{(n \vee p)}} + \max_{1\le l\le p}|\mu_{l}(\kappa_{r}) - \mu_{l}^{(n)}(\kappa_{r})|>  \frac{1}{3}\underline{\mu}) \\
			&\qquad + \sum_{r=1}^R\mathbb{P}(E_{1}\cup E_{2r})\\
			& \le \sum_{r=1}^R \frac{3}{\underline{\mu}} \left[c_{1}n^{-\frac{1}{2} + \frac{2}{q_{1}}} \sqrt{\log{(n \vee p)}} + \max_{1\le l\le p}|\mu_{l}(\kappa_{r}) - \mu_{l}^{(n)}(\kappa_{r})| +  \mathbb{E}(V_{n})  \right] \\
			&\qquad+ \sum_{r=1}^R\mathbb{P}(E_{1}\cup E_{2r}),
		\end{split}
	\end{equation}
	where $V_{n}$ is defined in \eqref{theorem8.3} and
	{\small\begin{equation*}
			\begin{split}
				A_{1r}& = \{V_{n} + \max_{ 1\le l\le p} |n^{-1}\sum_{i=1}^{n} \big\{\big[\boldsymbol{1}_{U_{il}\in (-\infty, \kappa_{r}]}- \boldsymbol{1}_{\widetilde{U}_{il}\in (-\infty, \kappa_{r}]}\big]\big[\zeta(\boldsymbol{U}_{iS^*})\eta_{i}\big]^2 - \mu_{l}(\kappa_{r})\big\}| > \frac{1}{3}\underline{\mu}\},\\
				A_{2r} & = \{V_{n} + \max_{ 1\le l\le p} |n^{-1}\sum_{i=1}^{n} \big\{ \boldsymbol{1}_{\{|\eta_{i}|\le n^{\frac{1}{q_{1}}}\}} \times\big[\boldsymbol{1}_{U_{il}\in (-\infty, \kappa_{r}]}- \boldsymbol{1}_{\widetilde{U}_{il}\in (-\infty, \kappa_{r}]}\big]\big[\zeta(\boldsymbol{U}_{iS^*})\eta_{i}\big]^2 - \mu_{l}(\kappa_{r})\big\}| > \frac{1}{3}\underline{\mu}\},
			\end{split}
	\end{equation*}}%
	and the last inequality is due to Markov's inequality; here, events $A_{1r}$'s and $A_{2r}$'s are defined and used only to simplify the expressions in \eqref{theorem8.14}. 
	
	In addition, by Jensen's inequality, Condition~\ref{consistency.3}, the assumptions that $\lim_{n\rightarrow\infty}\underline{\mu}^{-1}\sqrt{B_{1}} = 0$ and  observations are i.i.d., the assumptions that $\eta_{i}$'s are independent model errors and that $\eta_{1}$ and $\varepsilon$ have the same distribution, 
	Condition~\ref{A4}, and other model regularity assumptions,
	\begin{equation}\label{theorem8.15}
		\underline{\mu}^{-1}\mathbb{E}(V_{n})\le 		\underline{\mu}^{-1}B_{1} + 2		\underline{\mu}^{-1}\big[\sup_{\vec{z}\in\mathbb{R}^{s^*}} \zeta(\vec{z})\big]\sqrt{B_{1}}\mathbb{E}|\varepsilon|  =o(1).
	\end{equation}

	By \eqref{theorem8.3}--\eqref{theorem8.15}, and the assumptions that $R$ is finite and that $\underline{\mu}^{-1}\times n^{-\frac{1}{2} + \frac{2}{q_{1}}} \sqrt{\log{(n \vee p)}} = o(1)$ (recall that $q_{1} = q-\beta$), we conclude that 
	\begin{equation}\label{theorem8.13}
		\lim_{n\rightarrow\infty}\mathbb{P}(J_{2}^c) = 0,
	\end{equation}
	which along with \eqref{theorem8.1} and \eqref{theorem8.12} proves the first part of Theorem~\ref{theorem8}. 
	

	Next, we show the second assertion of Theorem~\ref{theorem8}. 
	By \eqref{theorem8.11}, for each $l\in \{1, \dots ,p\}$, on $J_{2}$,
	\begin{equation}
		\begin{split}
			|\mu_{l}(\widehat{a}_{l})| &= |\mu_{l}(\widehat{a}_{l}) - G_{l}(\widehat{a}_{l}) + G_{l}(\widehat{a}_{l})|\\
			& \ge |G_{l}(\widehat{a}_{l})| - |\mu_{l}(\widehat{a}_{l}) - G_{l}(\widehat{a}_{l})|\\
			&\ge \max_{1\le r\le R}|\mu_{l}(\kappa_{r})| - \frac{2}{3}\underline{\mu},
		\end{split}
	\end{equation}
	which in combination with \eqref{theorem8.1}, \eqref{theorem8.12}, and the assumption $\min_{l\in S^*} \max_{1\le r\le R}|\mu_{l}(\kappa_{r})| > \underline{\mu}$ shows that on $J_{2}$, it holds that $\widehat{l}\in S^*$ and $|\mu_{\widehat{l}}(\widehat{a}_{\widehat{l}})| \ge \frac{1}{3}\underline{\mu}.$ By this and Markov's inequality, for each $t>0$,
	\begin{equation}
		\begin{split}\label{theorem8.10}
			&\mathbb{P}\big\{\big| \big[T_{\widehat{l}}(\widehat{a}_{\widehat{l}})\big] \big[\widehat{\sigma}_{\widehat{l}}(\widehat{a}_{\widehat{l}})\big]^{-1} \big|\le t\big\} \\
			& = \mathbb{P}\big\{\big| T_{\widehat{l}}(\widehat{a}_{\widehat{l}}) - T_{\widehat{l}}^{(\star)}(\widehat{a}_{\widehat{l}})+ T_{\widehat{l}}^{(\star)}(\widehat{a}_{\widehat{l}}) -\sqrt{n}\big[\mu_{\widehat{l}}(\widehat{a}_{\widehat{l}})\big] + \sqrt{n}\big[\mu_{\widehat{l}}(\widehat{a}_{\widehat{l}})\big] \big|\le t\big[\widehat{\sigma}_{\widehat{l}}(\widehat{a}_{\widehat{l}}) \big]\big\}\\
			&\le \mathbb{P}\Big\{\big\{ \sqrt{n}\big|\mu_{\widehat{l}}(\widehat{a}_{\widehat{l}}) \big|\le t\times \big[\widehat{\sigma}_{\widehat{l}}(\widehat{a}_{\widehat{l}})\big] + \big| T_{\widehat{l}}(\widehat{a}_{\widehat{l}}) - T_{\widehat{l}}^{(\star)}(\widehat{a}_{\widehat{l}})\big| \\
			& \qquad+ \big| T_{\widehat{l}}^{(\star)}(\widehat{a}_{\widehat{l}}) -\sqrt{n}\big[\mu_{\widehat{l}}(\widehat{a}_{\widehat{l}})\big]  \big| \big\}\cap J_{2}\Big\} + \mathbb{P}(J_{2}^c)\\
			&\le \mathbb{P}\big\{ \frac{\underline{\mu}\sqrt{n}}{3} \le t\times \big[\widehat{\sigma}_{\widehat{l}}(\widehat{a}_{\widehat{l}})\big] + \big| T_{\widehat{l}}(\widehat{a}_{\widehat{l}}) - T_{\widehat{l}}^{(\star)}(\widehat{a}_{\widehat{l}})\big| + \big| T_{\widehat{l}}^{(\star)}(\widehat{a}_{\widehat{l}}) -\sqrt{n}\big[\mu_{\widehat{l}}(\widehat{a}_{\widehat{l}})\big]  \big| \big\} \\
			&\qquad+ \mathbb{P}(J_{2}^c)\\
			&\le \frac{3}{\sqrt{n}\underline{\mu}} \left(t\times  \mathbb{E}|\widehat{\sigma}_{\widehat{l}}(\widehat{a}_{\widehat{l}})| + \mathbb{E}\big| T_{\widehat{l}}(\widehat{a}_{\widehat{l}}) - T_{\widehat{l}}^{(\star)}(\widehat{a}_{\widehat{l}})\big|+ \mathbb{E}\big| T_{\widehat{l}}^{(\star)}(\widehat{a}_{\widehat{l}}) -\sqrt{n}\big[\mu_{\widehat{l}}(\widehat{a}_{\widehat{l}})\big]  \big| \right) \\
			&\qquad+ \mathbb{P}(J_{2}^c).
		\end{split}
	\end{equation}

	In what follows, we establish upper bounds for terms $\mathbb{E}|\widehat{\sigma}_{\widehat{l}}(\widehat{a}_{\widehat{l}})|$, $\mathbb{E}\big| T_{\widehat{l}}(\widehat{a}_{\widehat{l}}) - T_{\widehat{l}}^{(\star)}(\widehat{a}_{\widehat{l}})\big|$, and $\mathbb{E}\big| T_{\widehat{l}}^{(\star)}(\widehat{a}_{\widehat{l}}) -\sqrt{n}\big[\mu_{\widehat{l}}(\widehat{a}_{\widehat{l}})\big]  \big|$ on the RHS of \eqref{theorem8.10}, and begin with the one for $\mathbb{E}|\widehat{\sigma}_{\widehat{l}}(\widehat{a}_{\widehat{l}})|$. 
	
	By Jensen's inequality, 
	the assumption of i.i.d. observations, Condition~\ref{A4}, the assumption that $\varepsilon$ is an independent model error with $\mathbb{E}(\varepsilon) = 0$ and $\mathbb{E}(\varepsilon^2)=1$, and 
	Conditions~\ref{consistency.3}--\ref{bound},
	\begin{equation}
		\begin{split}\label{theorem8.7}
			&\mathbb{E}|\widehat{\sigma}_{\widehat{l}}(\widehat{a}_{\widehat{l}})|\\
			& \le \sqrt{\mathbb{E}|\widehat{\sigma}_{\widehat{l}}(\widehat{a}_{\widehat{l}})|^2} \\
			& \le \sqrt{\mathbb{E}(\widehat{\varepsilon}_{1})^4} \\
			&\le \sqrt{\mathbb{E}\big[\zeta(\boldsymbol{X}_{S^*})\varepsilon + m(\boldsymbol{X})  - \widehat{m}(\boldsymbol{X} )\big]^4} \\
			& \le \Big\{\mathbb{E}\big[\zeta(\boldsymbol{X}_{S^*})\varepsilon\big]^4+ 4 \mathbb{E}\big\{\big[\zeta(\boldsymbol{X}_{S^*})\varepsilon\big]^3\big| m(\boldsymbol{X})  - \widehat{m}(\boldsymbol{X})\big|\big\} \\
			&\qquad+ 6\mathbb{E}\big\{\big[\zeta(\boldsymbol{X}_{S^*})\varepsilon\big]^2 \big|m(\boldsymbol{X})  - \widehat{m}(\boldsymbol{X})\big|^2 \big\} + 4 \mathbb{E}\big\{\big[\zeta(\boldsymbol{X}_{S^*})\varepsilon\big]\big| m(\boldsymbol{X})  - \widehat{m}(\boldsymbol{X})\big|^3\big\}\\
			&\qquad+ \mathbb{E}(m(\boldsymbol{X})  - \widehat{m}(\boldsymbol{X}))^4 \Big\}^{\frac{1}{2}}\\
			& \le \Big\{\big[\sup_{\vec{z}\in\mathbb{R}^{s^*}} \zeta(\vec{z})\big]^4\mathbb{E}(\varepsilon^4) + 4\big[\sup_{\vec{z}\in\mathbb{R}^{s^*}} \zeta(\vec{z})\big]^3\mathbb{E}|\varepsilon|^3 \sqrt{B_{1}} \\
			&\qquad\qquad+ 6\big[\sup_{\vec{z}\in\mathbb{R}^{s^*}} \zeta(\vec{z})\big]^2B_{1} + 4\overbar{M}_{n}^2B_{1}\Big\}^{\frac{1}{2}}.
		\end{split}	
	\end{equation}
	
	Next, we establish an upper bound for $\mathbb{E}\big| T_{\widehat{l}}^{(\star)}(\widehat{a}_{\widehat{l}}) -\sqrt{n}\big[\mu_{\widehat{l}}(\widehat{a}_{\widehat{l}})\big]  \big|$. Since $\widehat{l}$ and $\widehat{a}_{l}$'s are independent of $T_{l}^{(\star)}(\kappa_{r})$'s,
	\begin{equation}
		\begin{split}\label{theorem8.5}
			&\mathbb{E}\big| T_{\widehat{l}}^{(\star)}(\widehat{a}_{\widehat{l}}) -\sqrt{n}\big[\mu_{\widehat{l}}(\widehat{a}_{\widehat{l}})\big]  \big|\\
			& = \sum_{l=1}^{p}\sum_{r=1}^{R}\mathbb{E}\big\{\big| T_{l}^{(\star)}(\kappa_{r}) -\sqrt{n}\mu_{l}(\kappa_{r})  \big|\boldsymbol{1}_{\{\widehat{l}=l\}\cap\{\widehat{a}_{l} = \kappa_{r}\}}\big\}\\
			&= \sum_{l=1}^{p}\sum_{r=1}^{R}\mathbb{E}\big| T_{l}^{(\star)}(\kappa_{r}) -\sqrt{n}\mu_{l}(\kappa_{r})  \big|\times \mathbb{P}(\{\widehat{l}=l\}\cap\{\widehat{a}_{l} = \kappa_{r}\}).
		\end{split}
	\end{equation}
	By the Burkholder--Davis--Gundy inequality, Jensen's inequality, and the assumption of i.i.d. observations, there exists some $K>0$ such that for each $l\in\{1, \dots ,p\}$, each $1\le r\le R$, and each $n\ge 1$,
	\begin{equation}
		\begin{split}\label{theorem8.6}
			&\mathbb{E}\big| T_{l}^{(\star)}(\kappa_{r}) -\sqrt{n}\big[\mu_{l}(\kappa_{r})  \big]\big| \\
			&\le K\sqrt{n^{-1}\sum_{i=1}^{n}\mathbb{E}\big\{\big(\boldsymbol{1}_{X_{il}\in(-\infty, \kappa_{r}]} -\boldsymbol{1}_{\widetilde{X}_{il}\in (-\infty, \kappa_{r}]}\big)\big[\zeta(\boldsymbol{X}_{iS^{*}})\varepsilon_{i}\big]^{2} - \mu_{l}(\kappa_{r})\big\}^2}\\
			& \le K\sqrt{n^{-1}\sum_{i=1}^{n}\mathbb{E}\big\{\big[\zeta(\boldsymbol{X}_{iS^{*}})\varepsilon_{i}\big]^{4} \big\}}\\
			& \le K \big[\sup_{\vec{z}\in\mathbb{R}^{s^*}} \zeta(\vec{z})\big]^2 \sqrt{\mathbb{E}(\varepsilon^4)}.
		\end{split}
	\end{equation}

	By \eqref{theorem8.5}--\eqref{theorem8.6} and that $\sum_{l=1}^p \sum_{r=1}^{R}  \mathbb{P}(\{\widehat{l} = l\}\cap\{ \widehat{a}_{l} = \kappa_{r}\}) = 1$, for each $n\ge 1$,
	\begin{equation}\label{theorem8.8}
		\mathbb{E}\big| T_{\widehat{l}}^{(\star)}(\widehat{a}_{\widehat{l}}) -\sqrt{n}\big[\mu_{\widehat{l}}(\widehat{a}_{\widehat{l}}) \big] \big|\le K \big[\sup_{\vec{z}\in\mathbb{R}^{s^*}} \zeta(\vec{z})\big]^2 \sqrt{\mathbb{E}(\varepsilon^4)},
	\end{equation}
	where $K>0$ is given in \eqref{theorem8.6}.

	Next, we establish an upper bound for $\mathbb{E}\big| T_{\widehat{l}}(\widehat{a}_{\widehat{l}}) - T_{\widehat{l}}^{(\star)}(\widehat{a}_{\widehat{l}})\big|$. Since $\widehat{l}$ and $\widehat{a}_{l}$'s are independent of $T_{l}(\kappa_{r})$'s and $T_{l}^{(\star)}(\kappa_{r})$'s,
	\begin{equation}
		\begin{split}\label{theorem8.17}
			\mathbb{E}\big| T_{\widehat{l}}(\widehat{a}_{\widehat{l}}) - T_{\widehat{l}}^{(\star)}(\widehat{a}_{\widehat{l}})\big|& = \sum_{l=1}^p \sum_{r=1}^{R} \mathbb{E}\big\{\big| T_{l}(\kappa_{r}) - T_{l}^{(\star)}(\kappa_{r})\big| \boldsymbol{1}_{\{\widehat{l} = l\}\cap\{ \widehat{a}_{l} = \kappa_{r}\}}\big\}\\
			& = \sum_{l=1}^p \sum_{r=1}^{R} \mathbb{E}\big| T_{l}(\kappa_{r}) - T_{l}^{(\star)}(\kappa_{r})\big| \times \mathbb{P}(\{\widehat{l} = l\}\cap\{ \widehat{a}_{l} = \kappa_{r}\}).
		\end{split}
	\end{equation}

	By the Burkholder--Davis--Gundy inequality, Jensen's inequality, Condition~\ref{consistency.3}, Condition~\ref{A4}, the assumption that observations are i.i.d., model assumption \eqref{model.1}, and other model regularity assumptions, there exists some $K>0$ such that for each $l\in \{1, \dots, p\}$ and each $r\in\{1, \dots, R\}$,
	\begin{equation*}
		\begin{split}
			&\mathbb{E}\big| T_{l}(\kappa_{r}) - T_{l}^{(\star)}(\kappa_{r})\big| \\
			& \le 2\mathbb{E}\Big|\frac{1}{\sqrt{n}} \sum_{i=1}^n \big[\boldsymbol{1}_{X_{il}\in(-\infty, \kappa_{r}]} -\boldsymbol{1}_{\widetilde{X}_{il}\in (-\infty, \kappa_{r}]}\big]\big[m(\boldsymbol{X}_{i}) - \widehat{m}(\boldsymbol{X}_{i})\big]\zeta(\boldsymbol{X}_{iS^{*}})\varepsilon_{i}\Big| \\
			&\qquad+ \frac{1}{\sqrt{n}}\sum_{i=1}^n \mathbb{E}\big[m(\boldsymbol{X}_{i}) - \widehat{m}(\boldsymbol{X}_{i})\big]^2\\
			&\le 2K\big[\sup_{\vec{z}\in\mathbb{R}^{s^*}} \zeta(\vec{z})\big]\sqrt{B_{1}} + \sqrt{n}B_{1},
		\end{split}
	\end{equation*}
	which in combination with \eqref{theorem8.17} and the fact that $\sum_{l=1}^p \sum_{r=1}^{R}  \mathbb{P}(\{\widehat{l} = l\}\cap\{ \widehat{a}_{l} = \kappa_{r}\}) = 1$ leads to 
	\begin{equation}
		\begin{split}\label{theorem8.9}
			\mathbb{E}\big| T_{\widehat{l}}(\widehat{a}_{\widehat{l}}) - T_{\widehat{l}}^{(\star)}(\widehat{a}_{\widehat{l}})\big|\le 2K\big[\sup_{\vec{z}\in\mathbb{R}^{s^*}} \zeta(\vec{z})\big]\sqrt{B_{1}} + \sqrt{n}B_{1}.
		\end{split}
	\end{equation}
	
	By the model regularity assumptions, the assumption that $\lim\sup_{n\rightarrow\infty}\overbar{M}_{n}^2B_{1} <\infty$, \eqref{theorem8.10}--\eqref{theorem8.7},  \eqref{theorem8.8}, and \eqref{theorem8.9}, it holds that for all large $n$, each $t>0$, and each $0<B_{1}<1$,
	\begin{equation}
		\mathbb{P}\big\{\big| \big[T_{\widehat{l}}(\widehat{a}_{\widehat{l}})\big] \big[\widehat{\sigma}_{\widehat{l}}(\widehat{a}_{\widehat{l}})\big]^{-1} \big|\le t\big\}\le \frac{(1+t)\log{n}}{\sqrt{n}\underline{\mu}} + \frac{3B_{1}}{\underline{\mu}} + \mathbb{P}(J_{2}^c),
	\end{equation}
	which along with \eqref{theorem8.13} and the assumption that 	$\lim_{n\rightarrow\infty}\underline{\mu}^{-1}\Big[\sqrt{B_{1}} + n^{-\frac{1}{2} + \frac{2}{q_{1}}} \sqrt{\log{(n \vee p)}} \Big]= 0$ concludes the proof of Theorem~\ref{theorem8}.

	\renewcommand{\thesubsection}{B.\arabic{subsection}}
	
	\section{Proofs of Lemma~\ref{lemma1} and Example~\ref{dist.1}}\label{SecB}

	\subsection{Proof of Lemma~\ref{lemma1}}\label{proof.lemma1}

	The proof for the case with $S^*=\emptyset$ is trivial since we have defined in Section~\ref{Sec1.1} that $\zeta(\boldsymbol{X}_{\emptyset}) = \zeta_{0}$. Therefore, we consider the case with $S^*\not=\emptyset$ and that $X_{j}$ is a null feature; that is, $j\not\in S^*$  in model \eqref{model.1}. In what follows, we show that $(\varepsilon \zeta(\boldsymbol{X}_{S^*}), \boldsymbol{X}_{-j}, X_{j})$ and $(\varepsilon \zeta(\boldsymbol{X}_{S^*}), \boldsymbol{X}_{-j}, \widetilde{X}_{j})$ have the same distribution, which implies the desired result of Lemma~\ref{lemma1}. It follows from Definition~\ref{knockoff.1} of coordinate-wise knockoffs and the definition of model \eqref{model.1} that $\widetilde{X}_{j}$ is independent of $\varepsilon \zeta(\boldsymbol{X}_{S^*})$ conditional on $\boldsymbol{X}$. In addition, $X_{j}$ is independent of $\varepsilon \zeta(\boldsymbol{X}_{S^*})$ conditional on $\boldsymbol{X}_{-j}$.
	
	For every $\mathcal{A}_{1}\in\mathcal{R}$, $\mathcal{A}_{2}\in\mathcal{R}^{p-1}$, $\mathcal{A}_{3}\in\mathcal{R}$,
	\begin{equation}\label{km.3}
		\begin{split}
			& \mathbb{P}\{(\varepsilon \zeta(\boldsymbol{X}_{S^*}), \boldsymbol{X}_{-j}, X_{j})\in(\mathcal{A}_{1}, \mathcal{A}_{2}, \mathcal{A}_{3})\} \\
			& = \mathbb{E}\{\mathbb{P}\{(\varepsilon \zeta(\boldsymbol{X}_{S^*}), \boldsymbol{X}_{-j}, X_{j})\in(\mathcal{A}_{1}, \mathcal{A}_{2}, \mathcal{A}_{3})|\boldsymbol{X}_{-j}\} \}\\
			& = \mathbb{E}\{ \boldsymbol{1}_{\{\boldsymbol{X}_{-j} \in \mathcal{A}_{2}\}}\mathbb{P}\{(\varepsilon \zeta(\boldsymbol{X}_{S^*}),  X_{j})\in(\mathcal{A}_{1},  \mathcal{A}_{3})|\boldsymbol{X}_{-j}\} \}\\
			& = \mathbb{E}\{ \boldsymbol{1}_{\{\boldsymbol{X}_{-j} \in \mathcal{A}_{2}\}}\mathbb{P}\{\varepsilon \zeta(\boldsymbol{X}_{S^*}) \in \mathcal{A}_{1} |\boldsymbol{X}_{-j}\} \mathbb{P}\{X_{j}\in \mathcal{A}_{3}|\boldsymbol{X}_{-j}\} \},
		\end{split}
	\end{equation}
	and
	\begin{equation}
		\label{km.6}
		\mathbb{P}\{\varepsilon \zeta(\boldsymbol{X}_{S^*}) \in \mathcal{A}_{1} |\boldsymbol{X}_{-j}\}  =\mathbb{P}\{\varepsilon \zeta(\boldsymbol{X}_{S^*}) \in \mathcal{A}_{1} |\boldsymbol{X}_{-j}, X_{j}\},
	\end{equation}
	where the last equality in \eqref{km.3} and \eqref{km.6} follow because $X_{j}$ is independent of $\varepsilon \zeta(\boldsymbol{X}_{S^*})$ conditional on $\boldsymbol{X}_{-j}$. 
	
	Next,
	\begin{equation}
		\begin{split}\label{km.4}
			&\textnormal{RHS of \eqref{km.3}}  \\
			& = \mathbb{E}\{ \boldsymbol{1}_{\{\boldsymbol{X}_{-j} \in \mathcal{A}_{2}\}}\mathbb{P}\{\varepsilon \zeta(\boldsymbol{X}_{S^*}) \in \mathcal{A}_{1} |\boldsymbol{X}_{-j}\} \mathbb{P}\{\widetilde{X}_{j}\in \mathcal{A}_{3}|\boldsymbol{X}_{-j}\} \}\\
			& = \mathbb{E}\{ \boldsymbol{1}_{\{\boldsymbol{X}_{-j} \in \mathcal{A}_{2}\}}\mathbb{P}\{\varepsilon \zeta(\boldsymbol{X}_{S^*}) \in \mathcal{A}_{1} |\boldsymbol{X}_{-j}\} \mathbb{P}\{\widetilde{X}_{j}\in \mathcal{A}_{3}|\boldsymbol{X}_{-j}, X_{j}\} \}\\
			& = \mathbb{E}\{ \boldsymbol{1}_{\{\boldsymbol{X}_{-j} \in \mathcal{A}_{2}\}}\mathbb{P}\{\varepsilon \zeta(\boldsymbol{X}_{S^*}) \in \mathcal{A}_{1} |\boldsymbol{X}_{-j}, X_{j}\} \mathbb{P}\{\widetilde{X}_{j}\in \mathcal{A}_{3}|\boldsymbol{X}_{-j}, X_{j}\} \}\\
			& = \mathbb{E}\{ \boldsymbol{1}_{\{\boldsymbol{X}_{-j} \in \mathcal{A}_{2}\}}\mathbb{P}\{(\varepsilon \zeta(\boldsymbol{X}_{S^*}),  \widetilde{X}_{j} )\in (\mathcal{A}_{1}, \mathcal{A}_{3}) |\boldsymbol{X}_{-j}, X_{j}\} \}\\
			& = \mathbb{E} \{   \mathbb{P}\{(\varepsilon \zeta(\boldsymbol{X}_{S^*}), \boldsymbol{X}_{-j}, \widetilde{X}_{j} )\in (\mathcal{A}_{1}, \mathcal{A}_{2}, \mathcal{A}_{3}) |\boldsymbol{X}_{-j}, X_{j}\} \} \\
			& =\mathbb{P}\{(\varepsilon \zeta(\boldsymbol{X}_{S^*}), \boldsymbol{X}_{-j}, \widetilde{X}_{j} )\in (\mathcal{A}_{1}, \mathcal{A}_{2}, \mathcal{A}_{3}) \},
		\end{split}
	\end{equation}
	where the first equality follows from the definition of knockoff features in Definition~\ref{knockoff.1}, the second equality is an application of law of total expectation, the third equality is from \eqref{km.6}, and the fourth equality follows because $\widetilde{X}_{j}$ is independent of $\varepsilon \zeta(\boldsymbol{X}_{S^*})$ conditional on $\boldsymbol{X}$.
	
	By \eqref{km.3} and \eqref{km.4}, we have that for every $\mathcal{A}_{1}\in\mathcal{R}$, $\mathcal{A}_{2}\in\mathcal{R}^{p-1}$, $\mathcal{A}_{3}\in\mathcal{R}$,
	$$\mathbb{P}\{(\varepsilon \zeta(\boldsymbol{X}_{S^*}), \boldsymbol{X}_{-j}, X_{j} )\in (\mathcal{A}_{1}, \mathcal{A}_{2}, \mathcal{A}_{3}) \} = \mathbb{P}\{(\varepsilon \zeta(\boldsymbol{X}_{S^*}), \boldsymbol{X}_{-j}, \widetilde{X}_{j} )\in (\mathcal{A}_{1}, \mathcal{A}_{2}, \mathcal{A}_{3}) \},$$
	which along with an application of the $\pi-\lambda$ Theorem~\citepsupp{durrett2019probability} concludes the desired result of Lemma~\ref{lemma1}. We omit the details of the application of the $\pi-\lambda$ Theorem for simplicity.

	\subsection{Proof of Example~\ref{dist.1}}\label{proof.dist.1}
	
	Lemma~\ref{corollary1} in Section~\ref{proof.corollary1} is needed for the proof of Example~\ref{dist.1}. To use Lemma~\ref{corollary1} to prove Example~\ref{dist.1}, we first show that 
	\begin{equation}
		\begin{split}							
			\label{exp.42}
			\mathbb{E}(\boldsymbol{1}_{ \{X_{j} \le\delta \}} \boldsymbol{1}_{ \{\widetilde{X}_{j} > 1 - \delta\}} |\boldsymbol{X}_{-j})&> \delta^2 (\bar{c})^{-1}\underline{c} \qquad\textnormal{	almost surely,}\\
			\mathbb{E}(\boldsymbol{1}_{\{X_{j}> 1 - \delta\} } \boldsymbol{1}_{ \{ \widetilde{X}_{j} \le \delta\} } |\boldsymbol{X}_{-j}) &>\delta^2 (\bar{c})^{-1}\underline{c} \qquad\textnormal{	almost surely,}
		\end{split}
	\end{equation} 
	and we begin with the first inequality in \eqref{exp.42}.

	By the distributional assumption of Example~\ref{dist.1}, it holds that for every Borel set $\mathcal{C}\in\mathcal{R}^{p-1}$ with $|\mathcal{C}|_{L}>0$,
	\begin{equation}
		\begin{split}\label{exp.41}
			\mathbb{E}\big\{\mathbb{E}(\boldsymbol{1}_{ \{X_{j} \le\delta\}} \boldsymbol{1}_{ \{\widetilde{X}_{j} > 1 - \delta\}} |\boldsymbol{X}_{-j}) \boldsymbol{1}_{\{\boldsymbol{X}_{-j}\in \mathcal{C}\}}\big\} &= \mathbb{E}\big\{\boldsymbol{1}_{ \{X_{j} \le\delta\}} \boldsymbol{1}_{ \{\widetilde{X}_{j} > 1 - \delta\}}  \boldsymbol{1}_{\{\boldsymbol{X}_{-j}\in \mathcal{C}\}}\big\} \\
			&> |\mathcal{C}|_{L}\delta^{2}\underline{c}.
		\end{split}
	\end{equation}	
	Consider a Borel set $\mathcal{C}^{\star} = \{\mathbb{E}(\boldsymbol{1}_{ \{X_{j} \le\delta \} } \boldsymbol{1}_{ \{\widetilde{X}_{j} > 1 - \delta\}} |\boldsymbol{X}_{-j})\le \delta^2 (\bar{c})^{-1}\underline{c}\}$. By the definition of $\mathcal{C}^{\star}$ and the  distributional assumption, we have
	\begin{equation}
		\begin{split}
			\mathbb{E}\big\{\mathbb{E}(\boldsymbol{1}_{X_{j} \le\delta} \boldsymbol{1}_{ \{\widetilde{X}_{j} > 1 - \delta\}} |\boldsymbol{X}_{-j}) \boldsymbol{1}_{\{\boldsymbol{X}_{-j}\in \mathcal{C}^{\star}\}}\big\}\le  \delta^2\underline{c} \times |\mathcal{C}^{\star}|_{L},
		\end{split}
	\end{equation}
	which along with \eqref{exp.41} and the distributional assumption concludes that $\mathbb{P}(\mathcal{C}^{\star}) = 0$, and hence the proof of the first inequality of \eqref{exp.42}. The other result in \eqref{exp.42} holds by the same arguments, and we omit the details for simplicity.

	Now, by Lemma~\ref{corollary1} and \eqref{exp.42}, we set $w_{0}$ in Lemma~\ref{corollary1}  to $\delta^2 (\bar{c})^{-1}\underline{c}$ to finish the proof of Example~\ref{dist.1}.

	\renewcommand{\thesubsection}{C.\arabic{subsection}}
	
	\section{Additional technical proofs and supplementary material}\label{SecC}

	\subsection{Lemma~\ref{corollary1} and its proof}\label{proof.corollary1}
	
	Without loss of generality, let $S^* = \{1, \dots, s^*\}$ for some integer $s^*\ge 1$. Recall that we have defined $\zeta_{j}(x)$ to be $\zeta(z_{1}, \dots, z_{j-1}, x, z_{j+1}\dots, z_{p})$ for $\vec{z} \in \mathbb{R}^{s^*}$ and $x\in \mathbb{R}$; if $s^*=1$, then  $\zeta_{j}(x) = \zeta(x)$ and that $\zeta_{j}(x)$ is invariant to $\vec{z}$. 
	
	\begin{lemma}
		\label{corollary1}
		Assume $\inf_{\vec{z}\in\mathbb{R}^{s^*}}\zeta(\vec{z})>0$, $\sup_{\vec{z}\in\mathbb{R}^{s^*}}\zeta(\vec{z})<\infty$, and that for every $j\in S^*= \{1, \dots ,s^*\}$ and every $\vec{z}\in\mathbb{R}^{s^*}$,  $\zeta_{j}(x)$ is nondecreasing (or nonincreasing) in $x$. In addition, let some constants $0<\delta <\frac{1}{2}, \iota>0,w_{0} >0$, and Cartesian product $\mathcal{D}\in\mathcal{R}^{s^*}$ be given such that \textnormal{1)} $\mathbb{E}(\boldsymbol{1}_{X_{j} \le\delta} \boldsymbol{1}_{ \{\widetilde{X}_{j} > 1 - \delta\}} |\boldsymbol{X}_{-j}) >w_{0} $ and $\mathbb{E}(\boldsymbol{1}_{\{X_{j}> 1 - \delta\} } \boldsymbol{1}_{ \widetilde{X}_{j} \le \delta} |\boldsymbol{X}_{-j}) >w_{0} $, and \textnormal{2)} for every $j\in S^*$ and every $\vec{z}\in \mathcal{D}$, $ \big|\zeta_{j}( 1-\delta) - \zeta_{j}( \delta)  \big| > \iota$. Then, $\min_{j\in S^*}\inf_{\delta < \kappa < 1-\delta}|\mu_{j}(\kappa)| >  \iota w_{0} \frac{1}{4}\mathbb{P}(\boldsymbol{X}_{S^*} \in\mathcal{D}) \big[\inf_{\vec{z}\in\mathbb{R}^{s^*}}\zeta(\vec{z})\big]$.
	\end{lemma}

	\textit{Proof of Lemma~\ref{corollary1}}\indent Let any $\delta< \kappa<1-\delta $ and $j\in S^*$ be given. Since $\mathcal{D}$ is a Cartesian product, we have $\mathcal{D}=\mathcal{D}_{1}\times \dots \times  \mathcal{D}_{s^*}$ for some $\mathcal{D}_{l} \in \mathcal{R}$ for each $l\in \{1, \dots, s^*\}$. By the assumptions on $\zeta(\cdot)$, it holds that for every $\vec{z} \in\mathcal{D}$,
	\begin{equation}
		\label{lemma2.5}
		\iota < |\zeta_{j}(1 - \delta) - \zeta_{j}(\delta) | \le |\zeta_{j}(1 - \delta) -\zeta_{j}(\kappa)|+| \zeta_{j}(\kappa)- \zeta_{j}(\delta) |.
	\end{equation}
	
	For the case with $s^*>1$, we define sets $\mathcal{D}_{1}^{\star}$ and $\mathcal{D}_{2}^{\star}$ as follows. By \eqref{lemma2.5}, we can write $\mathcal{D} = \mathcal{D}_{1}^{\star}\cup\mathcal{D}_{2}^{\star}$  such that \textnormal{1)} $\mathcal{D}_{1}^{\star}$ and $\mathcal{D}_{2}^{\star}$ are Borel rectangles satisfying \eqref{rectangle.1} in Section~\ref{proof.proposition1} with the $j$th coordinate as $\mathcal{D}_{j}$ and \textnormal{2)} $| \zeta_{j}(\kappa)- \zeta_{j}(\delta) | > \frac{1}{2}\iota$ for every $\vec{z} \in \mathcal{D}_{1}^{\star}$ and $|\zeta_{j}(1 - \delta) -\zeta_{j}(\kappa)| > \frac{1}{2}\iota$ for every $\vec{z} \in \mathcal{D}_{2}^{\star}$. On the other hand, for case with $s^* = 1$, define $\mathcal{D}_{1}^{\star}=\mathcal{D}_{2}^{\star} = \mathcal{D}_{1}$ (note that $S^* = \{1\}$ here).

	By the construction of $\mathcal{D}_{1}^{\star}$ and $\mathcal{D}_{2}^{\star}$, one of \eqref{lemma2.3} and \eqref{lemma2.3c} below holds.
	\begin{equation}\label{lemma2.3}
		\mathbb{P}(\boldsymbol{X}_{S^*} \in \mathcal{D}_{1}^{\star}) \ge \frac{1}{2}\mathbb{P}(\boldsymbol{X}_{S^*} \in\mathcal{D}),
	\end{equation}
	\begin{equation}\label{lemma2.3c}
		\mathbb{P}(\boldsymbol{X}_{S^*} \in \mathcal{D}_{2}^{\star}) \ge \frac{1}{2}\mathbb{P}(\boldsymbol{X}_{S^*} \in\mathcal{D}).
	\end{equation}
	Note that if \eqref{lemma2.3} does not hold, then \eqref{lemma2.3c} is true; it is possible they both hold. 
	
	Let us consider the case with \eqref{lemma2.3} first, and we will use Lemma~\ref{proposition1} with $q_{1} = \delta$ and $q_{2} = \kappa$ to finish the proof, where $q_{1}$ and $q_{2}$ are parameters in Lemma~\ref{proposition1}. By the choice of $q_{1}$ and $q_{2}$, for every $\vec{z} \in \mathcal{D}_{1}^{\star}$,
	\begin{equation}
		\label{lemma2.1}
		|\zeta_{j}(q_{1}) - \zeta_{j}(q_{2})| > \frac{1}{2}\iota.
	\end{equation}
	
	In addition, since $\delta \le q_{1} < q_{2} \le \kappa <1-\delta$ in this scenario, by the assumptions of Lemma~\ref{corollary1},
	\begin{equation}
		\begin{split}\label{lemma2.2}
			\mathbb{E}(\boldsymbol{1}_{X_{j} \le q_{1}} \boldsymbol{1}_{ \widetilde{X}_{j} >\kappa} |\boldsymbol{X}_{-j}) \ge  \mathbb{E}(\boldsymbol{1}_{X_{j} \le \delta} \boldsymbol{1}_{ \widetilde{X}_{j} >1-\delta} |\boldsymbol{X}_{-j})>w_{0},\\
			\mathbb{E}(\boldsymbol{1}_{X_{j}>q_{2} } \boldsymbol{1}_{ \widetilde{X}_{j} \le \kappa} |\boldsymbol{X}_{-j})  \ge \mathbb{E}(\boldsymbol{1}_{X_{j}>1-\delta } \boldsymbol{1}_{ \widetilde{X}_{j} \le \delta} |\boldsymbol{X}_{-j})  >w_{0}.
		\end{split}
	\end{equation}

	By \eqref{lemma2.3}, \eqref{lemma2.1}--\eqref{lemma2.2}, and the definition of $\mathcal{D}_{1}^{\star}$, we use Lemma~\ref{proposition1} in Section~\ref{proof.proposition1} to conclude 
	\begin{equation}
		\label{lemma2.4}
		|\mu_{j}(\kappa)| > \frac{1}{4}\iota w_{0} \mathbb{P}(\boldsymbol{X}_{S^*} \in \mathcal{D})\big[\inf_{\vec{z}\in\mathbb{R}^{s^*}}\zeta(\vec{z})\big],
	\end{equation}
	which holds for every $\delta< \kappa<1-\delta $ and $j\in S^*$. By  \eqref{lemma2.4}, we conclude Lemma~\ref{corollary1} for the case with \eqref{lemma2.3}.
	
	For the other case with \eqref{lemma2.3c}, we may set $q_{1} = \kappa$ and $q_{2} = 1- \delta$ for the subsequent arguments, whose details are omitted because they are similar to those for \eqref{lemma2.4}. We have completed the proof of Lemma~\ref{corollary1}.

	\subsection{Lemma~\ref{proposition1} and its proof}\label{proof.proposition1}
	
	In this section, we state Lemma~\ref{proposition1}, which is a general version of Lemma~\ref{corollary1}, and its proof.
	Let $\zeta_{j}(x)$ be defined as in Section~\ref{proof.corollary1}.
	To facilitate the analysis in Lemma~\ref{proposition1}, consider some $\mathcal{D}^{(j)}\in\mathcal{R}^{s^*}$ satisfying the following condition: if $s^*>1$,   then there exist some  $\mathcal{A}\in\mathcal{R}$  and  $\mathcal{B} \in\mathcal{R}^{s^*-1}$ such that 
	\begin{equation}\label{rectangle.1}
		\mathcal{A} \times \mathcal{B} = \{(z_{j}, \underbrace{z_{1}, \dots, z_{j-1}, z_{j+1}, \dots, z_{s^*}}_{\textnormal{without the }j \textnormal{th coordinate}}):\vec{z} \in\mathcal{D}^{(j)}\},
	\end{equation}	
	where the notation $\times$ denotes the Cartesian product operation; if $s^*=1$, there is no additional restriction on $\mathcal{D}^{(j)}\in \mathcal{R}$. Cartesian product $\mathcal{A}_{1} \times \dots \times \mathcal{A}_{s^*}$ with $\mathcal{A}_{l} \in \mathcal{R}$ for each $l\in\{1, \dots ,s^*\}$ is an example of Borel rectangles satisfying \eqref{rectangle.1}. 
	\begin{lemma}\label{proposition1}
		Assume $\inf_{\vec{z}\in\mathbb{R}^{s^*}}\zeta(\vec{z})>0$ and $\sup_{\vec{z}\in\mathbb{R}^{s^*}}\zeta(\vec{z})<\infty$. Let any $j\in S^*= \{1, \dots ,s^*\}$, $w_{0}>0$, $\iota>0$, $q_{1}<q_{2}$,  $\mathcal{D}^{(j)} \in \mathcal{R}^{s^*}$ satisfying \eqref{rectangle.1}, and $\kappa\in\mathbb{R}$ be given such that the following conditions are satisfied. \textnormal{1)} $q_{1}<q_{2} \le \kappa$ or $\kappa\le q_{1}< q_{2}$. \textnormal{2)} $\mathbb{E}(\boldsymbol{1}_{X_{j} \le q_{1}} \boldsymbol{1}_{ \widetilde{X}_{j} \not\in (-\infty, \kappa]} |\boldsymbol{X}_{-j}) >w_{0} $ and $\mathbb{E}(\boldsymbol{1}_{X_{j}>q_{2} } \boldsymbol{1}_{ \widetilde{X}_{j} \not\in (-\infty, \kappa]^c} |\boldsymbol{X}_{-j}) >w_{0} $. \textnormal{3) }$\zeta_{j}(x)$ is  nondecreasing (or nonincreasing) in $x$ for every $\vec{z}\in \mathbb{R}^{s^*}$. \textnormal{4)}  $ \big|\zeta_{j}( q_{1}) - \zeta_{j}( q_{2})  \big| > \iota$ for every $\vec{z}\in \mathcal{D}^{(j)}$.   Then $|\mu_{j}(\kappa)| >\iota w_{0} \mathbb{P}(\boldsymbol{X}_{S^*} \in \mathcal{D}^{(j)})\big[\inf_{\vec{z}\in\mathbb{R}^{s^*}}\zeta(\vec{z})\big]$.
		
	\end{lemma}
	
	\textit{Proof of Lemma~\ref{proposition1}}\indent  To emphasize the $j$th coordinate, we denote $\zeta(X_{j}, \boldsymbol{X}_{S^*\backslash\{j\}}) = \zeta(\boldsymbol{X}_{S^*})$ with a slight abuse of notation; if $s^*=1$, then  $\zeta(X_{j}, \boldsymbol{X}_{S^*\backslash\{j\}}) = \zeta(X_{j})$. As a result, assumption 3) of Lemma~\ref{proposition1} is equivalent to that $\zeta(x, \vec{z}_{-j})$ is nondecreasing (or nonincreasing) in $x$ for every $\vec{z} = (z_{1}, \dots ,z_{s^*})^{\top}\in\mathbb{R}^{s^*}$, where  $\vec{z}_{-j} = (z_{1}, \dots, z_{j-1}, z_{j+1}, \dots,z_{s^*})^{\top}$. Let us deal with the case with a nondecreasing $\zeta_{j}(x)$ and $q_{1}<q_{2}\le \kappa$ first.

	Let $\mathcal{A} = (-\infty, \kappa]$. By the assumptions that $\inf_{\vec{z}\in\mathbb{R}^{s^*}}\zeta(\vec{z})>0$, $\sup_{\vec{z}\in\mathbb{R}^{s^*}}\zeta(\vec{z}) < \infty$, and Var$(\varepsilon)=1$ in model \eqref{model.1}, it holds that both $\zeta^2(\boldsymbol{X}_{S^*})$ and $\varepsilon^2$ are integrable. Then, by the assumption that $\varepsilon$ is independent of $\boldsymbol{X}$ and that 
	$\big[\zeta( \boldsymbol{X}_{S^*})\big]^2\boldsymbol{1}_{X_{j} \in \mathcal{A}} \boldsymbol{1}_{ \widetilde{X}_{j} \in \mathcal{A}} = \big[\zeta( \boldsymbol{X}_{S^*})\big]^2\boldsymbol{1}_{\widetilde{X}_{j} \in \mathcal{A}}\boldsymbol{1}_{X_{j} \in \mathcal{A}}  ,$
	it holds that 
	\begin{equation}\label{power.31}
		\begin{split}
			& \mathbb{E}\Big\{ \varepsilon^2  \big[\zeta(\boldsymbol{X}_{S^*})\big]^2\boldsymbol{1}_{X_{j} \in \mathcal{A}} - \varepsilon^2 \big[\zeta(\boldsymbol{X}_{S^*})\big]^2\boldsymbol{1}_{\widetilde{X}_{j} \in \mathcal{A}} \Big\}\\
			& = \mathbb{E}(\varepsilon^2) \mathbb{E}\Big\{ \big[\zeta( \boldsymbol{X}_{S^*})\big]^2\boldsymbol{1}_{X_{j} \in \mathcal{A}} \boldsymbol{1}_{ \widetilde{X}_{j} \not\in \mathcal{A}} - \big[\zeta( \boldsymbol{X}_{S^*})\big]^2\boldsymbol{1}_{\widetilde{X}_{j} \in \mathcal{A}}\boldsymbol{1}_{X_{j} \not\in \mathcal{A}}  \Big\}.
		\end{split}		
	\end{equation}
	
	By the assumptions on $\mathcal{D}^{(j)}$, if $s^*>1$, we let $\mathcal{B} = \mathcal{D}_{-j} \times \mathbb{R}^{p-s^*}$, where 
	$$\mathcal{D}_{j} \times\mathcal{D}_{-j} = \{(z_{j}, \underbrace{z_{1}, \dots, z_{j-1}, z_{j+1}, \dots, z_{s^*}}_{\textnormal{without the }j \textnormal{th coordinate}}) : \vec{z}\in \mathcal{D}^{(j)}\}$$
	for some $\mathcal{D}_{j} \in \mathcal{R}$ and $\mathcal{D}_{-j}\in\mathcal{R}^{s^* -1}$; if $s^* = 1$, we let $\mathcal{B} =  \mathbb{R}^{p-1}$. By the definitions of  $\mathcal{B}$ and $\mathcal{D}_{-j}$,  it holds that 
	\begin{equation}
		\label{power.35}		
		\begin{cases}
			\mathbb{P}(\boldsymbol{X}_{-j} \in \mathcal{B} ) = \mathbb{P}(\boldsymbol{X}_{S^*\backslash\{j\}} \in \mathcal{D}_{-j})\ge\mathbb{P}(\boldsymbol{X}_{S^*} \in \mathcal{D}^{(j)}) \textnormal{ if } s^*>1, \\
			\mathbb{P}(\boldsymbol{X}_{-j} \in \mathcal{B} ) =1 \ge \mathbb{P}(\boldsymbol{X}_{S^*} \in \mathcal{D}^{(j)}), \textnormal{ o.w.}
		\end{cases}			 
	\end{equation}

	By the monotonicity assumption on $\zeta_{j}(x)$ and $\kappa = \sup \mathcal{A}$,
	\begin{equation}\label{power.34}
		\begin{split}
			& \textnormal{RHS of \eqref{power.31} }\\
			&\le \mathbb{E}(\varepsilon^2) \mathbb{E}\Big\{ \big[\zeta( \boldsymbol{X}_{S^*})\big]^2\boldsymbol{1}_{X_{j} \in \mathcal{A}} \boldsymbol{1}_{ \widetilde{X}_{j} \not\in \mathcal{A}} - \big[\zeta(\kappa, \boldsymbol{X}_{S^*\backslash\{j\}})\big]^2\boldsymbol{1}_{\widetilde{X}_{j} \in \mathcal{A}}\boldsymbol{1}_{X_{j} \not\in \mathcal{A}}  \Big\}\\
			& = \mathbb{E}(\varepsilon^2)\mathbb{E}\Big\{\mathbb{E}\Big\{  \big[\zeta( \boldsymbol{X}_{S^*})\big]^2\boldsymbol{1}_{X_{j} \in \mathcal{A}} \boldsymbol{1}_{ \widetilde{X}_{j} \not\in \mathcal{A}} \\
			&\qquad\qquad- \big[\zeta(\kappa, \boldsymbol{X}_{S^*\backslash\{j\}})\big]^2\boldsymbol{1}_{\widetilde{X}_{j} \in \mathcal{A}}\boldsymbol{1}_{X_{j} \not\in \mathcal{A}}  \big| \boldsymbol{X}_{-j}\Big\}\boldsymbol{1}_{\boldsymbol{X}_{-j} \in \mathcal{B}}\Big\} \\
			& + \mathbb{E}(\varepsilon^2)\mathbb{E}\Big\{\mathbb{E}\Big\{ \big[\zeta( \boldsymbol{X}_{S^*})\big]^2\boldsymbol{1}_{X_{j} \in \mathcal{A}} \boldsymbol{1}_{ \widetilde{X}_{j} \not\in \mathcal{A}} \\
			&\qquad\qquad- \big[\zeta(\kappa, \boldsymbol{X}_{S^*\backslash\{j\}})\big]^2\boldsymbol{1}_{\widetilde{X}_{j} \in \mathcal{A}}\boldsymbol{1}_{X_{j} \not\in \mathcal{A}}  \big| \boldsymbol{X}_{-j}\Big\}\boldsymbol{1}_{\boldsymbol{X}_{-j} \not\in \mathcal{B}}\Big\}.
		\end{split}		
	\end{equation}

	Let us deal with the first term on the RHS of \eqref{power.34}. Recall that  
	$\zeta(x, \vec{z}_{-j}) = \zeta_{j}(x)$ for each $x\in\mathbb{R}$ and every $\vec{z}\in \mathcal{D}^{(j)}$ if $s^*>1$, while  $\zeta(x, \vec{z}_{-j}) = \zeta_{j}(x)= \zeta(x)$ if $s^*=1$. By assumption 4) of Lemma~\ref{proposition1}, \eqref{power.36} below gives a lower bound of $\big[\zeta(q_{2}, \vec{z}_{-j})\big]^2 - \big[\zeta(q_{1}, \vec{z}_{-j})\big]^2$ in terms of $\inf_{\vec{z}\in\mathbb{R}^{s^*}}\zeta(\vec{z})$. For every $\vec{z} \in \mathcal{D}^{(j)}$,
	\begin{equation}
		\begin{split}
			\label{power.36}
			\big[\zeta(q_{2}, \vec{z}_{-j})\big]^2 - \big[\zeta(q_{1}, \vec{z}_{-j})\big]^2
			& = \big[\zeta(q_{2}, \vec{z}_{-j}) -  \zeta(q_{1}, \vec{z}_{-j})\big]\big[\zeta(q_{2}, \vec{z}_{-j}) +  \zeta(q_{1}, \vec{z}_{-j})\big]\\
			& >\iota\big[\zeta(q_{2}, \vec{z}_{-j}) +  \zeta(q_{1}, \vec{z}_{-j})\big]\\
			&\ge \iota\inf_{\vec{z}\in\mathbb{R}^{s^*}}\zeta(\vec{z}).
		\end{split}
	\end{equation}
	By \eqref{power.36}, $q_{1} <q_{2}\le  \kappa =  \sup \mathcal{A}$ in this scenario, and  the monotonicity assumption on $\zeta_{j}(x)$, on $\{\boldsymbol{X}_{S^*}\in\mathcal{D}^{(j)}\}$,
	\begin{equation}
		\begin{split}\label{power.1}
			&\big[\zeta(\boldsymbol{X}_{S^*})\big]^2\boldsymbol{1}_{X_{j} \in \mathcal{A}} \boldsymbol{1}_{ \widetilde{X}_{j} \not\in \mathcal{A}} \\
			& \le \big[\zeta(q_{1}, \boldsymbol{X}_{S^*\backslash\{j\}})\big]^2\boldsymbol{1}_{X_{j} \le q_{1}} \boldsymbol{1}_{ \widetilde{X}_{j} \not\in \mathcal{A}} + \big[\zeta(\kappa, \boldsymbol{X}_{S^*\backslash\{j\}})\big]^2\boldsymbol{1}_{\{X_{j} \in \mathcal{A}\}\cap\{X_{j} > q_{1}\}} \boldsymbol{1}_{ \widetilde{X}_{j} \not\in \mathcal{A}}\\
			& < \Big\{-\iota\big[\inf_{\vec{z}\in\mathbb{R}^{s^*}}\zeta(\vec{z})\big] + \big[\zeta(\kappa, \boldsymbol{X}_{S^*\backslash\{j\}})\big]^2\Big\}\boldsymbol{1}_{X_{j} \le q_{1}} \boldsymbol{1}_{ \widetilde{X}_{j} \not\in \mathcal{A}} \\
			& \qquad + \big[\zeta(\kappa, \boldsymbol{X}_{S^*\backslash\{j\}})\big]^2\boldsymbol{1}_{\{X_{j} \in \mathcal{A}\}\cap\{X_{j} > q_{1}\}} \boldsymbol{1}_{ \widetilde{X}_{j} \not\in \mathcal{A}}\\
			& = -\iota\big[\inf_{\vec{z}\in\mathbb{R}^{s^*}}\zeta(\vec{z}) \big]\boldsymbol{1}_{X_{j} \le q_{1}} \boldsymbol{1}_{ \widetilde{X}_{j} \not\in \mathcal{A}} + \big[\zeta(\kappa, \boldsymbol{X}_{S^*\backslash\{j\}})\big]^2\boldsymbol{1}_{X_{j} \in \mathcal{A}} \boldsymbol{1}_{ \widetilde{X}_{j} \not\in \mathcal{A}}.
		\end{split}
	\end{equation}
	

	In addition, by the definition of knockoffs in Definition~\ref{knockoff.1}, it holds that  $\mathbb{P}(X_{j}\in \mathcal{A}| \boldsymbol{X}_{-j}) = \mathbb{P}(\widetilde{X}_{j}\in \mathcal{A}| \boldsymbol{X}_{-j})$. Therefore, 
	\begin{equation}
		\label{power.2}
		\mathbb{E}(\boldsymbol{1}_{\{\widetilde{X}_{j}\in \mathcal{A}\} \cap\{ X_{j} \not\in \mathcal{A} \}} - \boldsymbol{1}_{ \{X_{j}\in \mathcal{A}\} \cap\{ \widetilde{X}_{j} \not\in \mathcal{A} \} } |\boldsymbol{X}_{-j})= \mathbb{E}(\boldsymbol{1}_{\widetilde{X}_{j}\in \mathcal{A}} - \boldsymbol{1}_{X_{j}\in \mathcal{A}} |\boldsymbol{X}_{-j}) = 0.
	\end{equation}
	
	By this and \eqref{power.1},
	{\small\begin{equation}
			\begin{split}\label{power.32}
				& \textnormal{The first term on the RHS of \eqref{power.34}} \\
				& < -\mathbb{E}(\varepsilon^2) \mathbb{E}\big\{\mathbb{E}\big\{\iota\big[\inf_{\vec{z}\in\mathbb{R}^{s^*}}\zeta(\vec{z})\big]\boldsymbol{1}_{X_{j} \le q_{1}} \boldsymbol{1}_{ \widetilde{X}_{j} \not\in \mathcal{A}} |\boldsymbol{X}_{-j}\big\} \boldsymbol{1}_{\boldsymbol{X}_{-j} \in \mathcal{B}}\big\}\\
				&\qquad +  \mathbb{E}(\varepsilon^2)\mathbb{E}\Big\{\mathbb{E}\Big\{ \big[\zeta(\kappa, \boldsymbol{X}_{S^*\backslash\{j\}})\big]^2 \Big(\boldsymbol{1}_{X_{j} \in \mathcal{A}} \boldsymbol{1}_{ \widetilde{X}_{j} \not\in \mathcal{A}} - \boldsymbol{1}_{\widetilde{X}_{j} \in \mathcal{A}}\boldsymbol{1}_{X_{j} \not\in \mathcal{A}} \Big) \big| \boldsymbol{X}_{-j}\Big\}\boldsymbol{1}_{\boldsymbol{X}_{-j} \in \mathcal{B}}\Big\}\\
				& = -\mathbb{E}(\varepsilon^2) \mathbb{E}\big\{\mathbb{E}\big\{\iota\big[\inf_{\vec{z}\in\mathbb{R}^{s^*}}\zeta(\vec{z})\big]\boldsymbol{1}_{X_{j} \le q_{1}} \boldsymbol{1}_{ \widetilde{X}_{j} \not\in \mathcal{A}} |\boldsymbol{X}_{-j}\big\} \boldsymbol{1}_{\boldsymbol{X}_{-j} \in \mathcal{B}}\big\}\\
				&\qquad+ \mathbb{E}(\varepsilon^2)\mathbb{E}\Big\{\mathbb{E} \Big(\boldsymbol{1}_{X_{j} \in \mathcal{A}} \boldsymbol{1}_{ \widetilde{X}_{j} \not\in \mathcal{A}} - \boldsymbol{1}_{\widetilde{X}_{j} \in \mathcal{A}}\boldsymbol{1}_{X_{j} \not\in \mathcal{A}}  \big| \boldsymbol{X}_{-j}\Big) \big[\zeta(\kappa, \boldsymbol{X}_{S^*\backslash\{j\}})\big]^2\boldsymbol{1}_{\boldsymbol{X}_{-j} \in \mathcal{B}}\Big\}\\
				&= -\mathbb{E}(\varepsilon^2) \mathbb{E}\big\{\mathbb{E}\big\{\iota\big[\inf_{\vec{z}\in\mathbb{R}^{s^*}}\zeta(\vec{z})\big]\boldsymbol{1}_{X_{j} \le q_{1}} \boldsymbol{1}_{ \widetilde{X}_{j} \not\in \mathcal{A}} |\boldsymbol{X}_{-j}\big\} \boldsymbol{1}_{\boldsymbol{X}_{-j} \in \mathcal{B}}\big\},
			\end{split}
	\end{equation}}%
	where in the last equality, we use \eqref{power.2}. 
	
	Next, we proceed to deal with the second term on the RHS of \eqref{power.34}. By  $ \kappa = \sup\mathcal{A}$, assumption 3) of Lemma~\ref{proposition1}, and \eqref{power.2},
	\begin{equation}
		\begin{split}\label{power.33}
			& \textnormal{The second term on the RHS of \eqref{power.34}} \\
			& \le \mathbb{E}(\varepsilon^2)\mathbb{E}\Big(\mathbb{E} \Big( \big[\zeta(\kappa, \boldsymbol{X}_{S^*\backslash\{j\}})\big]^2\boldsymbol{1}_{X_{j} \in \mathcal{A}} \boldsymbol{1}_{ \widetilde{X}_{j} \not\in \mathcal{A}} \\
			&\qquad - \big[\zeta(\kappa, \boldsymbol{X}_{S^*\backslash\{j\}})\big]^2\boldsymbol{1}_{\widetilde{X}_{j} \in \mathcal{A}}\boldsymbol{1}_{X_{j} \not\in \mathcal{A}}  \big| \boldsymbol{X}_{-j}\Big)\boldsymbol{1}_{\boldsymbol{X}_{-j} \not\in \mathcal{B}}\Big)\\
			&\le \mathbb{E}(\varepsilon^2)\mathbb{E}\Big(\mathbb{E}\Big(  \big[\zeta(\kappa, \boldsymbol{X}_{S^*\backslash\{j\}})\big]^2 \Big(\boldsymbol{1}_{X_{j} \in \mathcal{A}} \boldsymbol{1}_{ \widetilde{X}_{j} \not\in \mathcal{A}} - \boldsymbol{1}_{\widetilde{X}_{j} \in \mathcal{A}}\boldsymbol{1}_{X_{j} \not\in \mathcal{A}}  \Big) \big| \boldsymbol{X}_{-j}\Big)\boldsymbol{1}_{\boldsymbol{X}_{-j} \not\in \mathcal{B}}\Big)\\
			& = \mathbb{E}(\varepsilon^2)\mathbb{E}\Big(\mathbb{E} \Big( \boldsymbol{1}_{X_{j} \in \mathcal{A}} \boldsymbol{1}_{ \widetilde{X}_{j} \not\in \mathcal{A}} - \boldsymbol{1}_{\widetilde{X}_{j} \in \mathcal{A}}\boldsymbol{1}_{X_{j} \not\in \mathcal{A}}   \big| \boldsymbol{X}_{-j}\Big)\big[\zeta(\kappa, \boldsymbol{X}_{S^*\backslash\{j\}})\big]^2\boldsymbol{1}_{\boldsymbol{X}_{-j} \not\in \mathcal{B}}\Big) = 0.
		\end{split}
	\end{equation}

	By \eqref{power.35} and \eqref{power.32}--\eqref{power.33}, the assumptions $\mathbb{E}(\boldsymbol{1}_{X_{j} \le q_{1}} \boldsymbol{1}_{ \widetilde{X}_{j} \not\in \mathcal{A}} |\boldsymbol{X}_{-j}) >w_{0}$, $\textnormal{Var}(\varepsilon)=1$, and $\inf_{\vec{z}\in\mathbb{R}^{s^*}}\zeta(\vec{z})>0$,
	\begin{equation}\label{power.3}
		\begin{split}
			\textnormal{The RHS of \eqref{power.34}} & < -\iota\big[\inf_{\vec{z}\in\mathbb{R}^{s^*}}\zeta(\vec{z})\big]\mathbb{E}(\varepsilon^2) \mathbb{E}\big[\mathbb{E}(\boldsymbol{1}_{X_{j} \le q_{1}} \boldsymbol{1}_{ \widetilde{X}_{j} \not\in \mathcal{A}} |\boldsymbol{X}_{-j}) \boldsymbol{1}_{\boldsymbol{X}_{-j} \in \mathcal{B}}\big] \\
			& < -\iota w_{0}\big[\inf_{\vec{z}\in\mathbb{R}^{s^*}}\zeta(\vec{z})\big]  \mathbb{P}(\boldsymbol{X}_{S^*} \in \mathcal{D}^{(j)}),
		\end{split}		
	\end{equation}
	leading to the desired result of Lemma~\ref{proposition1} for the case assuming a nondecreasing $\zeta_{j}(x)$ and $q_{1} < q_{2}\le \kappa$.

	For the case assuming a nondecreasing $\zeta_{j}(x)$ and $ \kappa\le q_{1} < q_{2}$, we first note that 
	\begin{equation}
		\label{power.41}
		-\mu_{j}(\kappa) = \mathbb{E}\Big\{ \varepsilon^2  \big[\zeta(\boldsymbol{X}_{S^*})\big]^2\boldsymbol{1}_{X_{j} \in \mathcal{A}^c} - \varepsilon^2 \big[\zeta(\boldsymbol{X}_{S^*})\big]^2\boldsymbol{1}_{\widetilde{X}_{j} \in \mathcal{A}^c} \Big\},
	\end{equation}
	since $\boldsymbol{1}_{x\in\mathcal{A}} = 1 - \boldsymbol{1}_{x\in\mathcal{A}^c}$, where $\mathcal{A}^c$ denotes the complementary event of $\mathcal{A}$. Therefore, the desired result can be obtained by considering event $\mathcal{A}^c$ in place of $\mathcal{A}$. In light of this observation and an argument similar to that for \eqref{power.34} but with $ \sup\mathcal{A} = \kappa\le q_{1} < q_{2}$, we write	
	\begin{equation}
		\begin{split}\label{power.40}
			& \mathbb{E}\Big( \varepsilon^2  \big[\zeta(\boldsymbol{X}_{S^*})\big]^2\boldsymbol{1}_{X_{j} \in \mathcal{A}^c} - \varepsilon^2 \big[\zeta(\boldsymbol{X}_{S^*})\big]^2\boldsymbol{1}_{\widetilde{X}_{j} \in \mathcal{A}^c} \Big)\\
			&\ge \mathbb{E}(\varepsilon^2) \mathbb{E}\Big( \big[\zeta( \boldsymbol{X}_{S^*})\big]^2\boldsymbol{1}_{X_{j} \in \mathcal{A}^c} \boldsymbol{1}_{ \widetilde{X}_{j} \not\in \mathcal{A}^c} - \big[\zeta(\kappa, \boldsymbol{X}_{S^*\backslash\{j\}})\big]^2\boldsymbol{1}_{\widetilde{X}_{j} \in \mathcal{A}^c}\boldsymbol{1}_{ X_{j}\not\in \mathcal{A}^c}   \Big)\\
			& = \mathbb{E}(\varepsilon^2)\mathbb{E}\Big(\mathbb{E}\Big(  \big[\zeta( \boldsymbol{X}_{S^*})\big]^2\boldsymbol{1}_{X_{j} \in \mathcal{A}^c} \boldsymbol{1}_{ \widetilde{X}_{j} \not\in \mathcal{A}^c} \\
			&\qquad\qquad\qquad- \big[\zeta(\kappa, \boldsymbol{X}_{S^*\backslash\{j\}})\big]^2\boldsymbol{1}_{\widetilde{X}_{j} \in \mathcal{A}^c}\boldsymbol{1}_{X_{j} \not\in \mathcal{A}^c}  \big| \boldsymbol{X}_{-j}\Big)\boldsymbol{1}_{\boldsymbol{X}_{-j} \in \mathcal{B}}\Big) \\
			&\ \  + \mathbb{E}(\varepsilon^2)\mathbb{E}\Big(\mathbb{E}\Big( \big[\zeta( \boldsymbol{X}_{S^*})\big]^2\boldsymbol{1}_{X_{j} \in \mathcal{A}^c} \boldsymbol{1}_{ \widetilde{X}_{j} \not\in \mathcal{A}^c} \\
			&\qquad\qquad\qquad- \big[\zeta(\kappa, \boldsymbol{X}_{S^*\backslash\{j\}})\big]^2\boldsymbol{1}_{\widetilde{X}_{j} \in \mathcal{A}^c}\boldsymbol{1}_{X_{j} \not\in \mathcal{A}^c}  \big| \boldsymbol{X}_{-j}\Big)\boldsymbol{1}_{\boldsymbol{X}_{-j} \not\in \mathcal{B}}\Big).
		\end{split}		
	\end{equation}
	
	Following the arguments for \eqref{power.33}, the second term on the RHS of \eqref{power.40} is nonnegative. For the first term on the RHS of \eqref{power.40} in this scenario, by \eqref{power.36}, on $\{\boldsymbol{X}_{S^*}\in\mathcal{D}^{(j)}\}$,
	\begin{equation*}
		\begin{split}
			&\big[\zeta(\boldsymbol{X}_{S^*})\big]^2\boldsymbol{1}_{\{ X_{j} \in \mathcal{A}^c \}} \boldsymbol{1}_{ \{\widetilde{X}_{j} \not\in \mathcal{A}^c\} } \\
			& = \big[\zeta( \boldsymbol{X}_{S^*})\big]^2\big[\boldsymbol{1}_{ \{X_{j} > q_{2}\}} \boldsymbol{1}_{ \{\widetilde{X}_{j} \not\in \mathcal{A}^c \}} + \boldsymbol{1}_{\{X_{j} \in\mathcal{A}^c\}\cap\{X_{j} \le q_{2}\}} \boldsymbol{1}_{ \{ \widetilde{X}_{j} \not\in \mathcal{A}^c\} }\big]\\
			& \ge \big[\zeta(q_{2}, \boldsymbol{X}_{S^*\backslash\{j\}})\big]^2\boldsymbol{1}_{ \{X_{j} > q_{2}\}} \boldsymbol{1}_{\{ \widetilde{X}_{j} \not\in \mathcal{A}^c \}} + \big[\zeta(\kappa, \boldsymbol{X}_{S^*\backslash\{j\}})\big]^2\boldsymbol{1}_{\{X_{j} \in\mathcal{A}^c\}\cap\{X_{j} \le q_{2}\}} \boldsymbol{1}_{ \{\widetilde{X}_{j} \not\in \mathcal{A}^c \}}\\
			& > \Big(\iota\big[\inf_{\vec{z}\in\mathbb{R}^{s^*}}\zeta(\vec{z})\big] + \big[\zeta(\kappa, \boldsymbol{X}_{S^*\backslash\{j\}})\big]^2\Big)\boldsymbol{1}_{\{X_{j} > q_{2}\}} \boldsymbol{1}_{ \{\widetilde{X}_{j} \not\in \mathcal{A}^c \}}  + \big[\zeta(\kappa, \boldsymbol{X}_{S^*\backslash\{j\}})\big]^2\boldsymbol{1}_{ \{X_{j} \in\mathcal{A}^c\}\cap\{X_{j} \le q_{2}\} } \boldsymbol{1}_{ \{\widetilde{X}_{j} \not\in \mathcal{A}^c \}}\\
			& = \iota\big[\inf_{\vec{z}\in\mathbb{R}^{s^*}}\zeta(\vec{z})\big]\boldsymbol{1}_{\{ X_{j} > q_{2} \}} \boldsymbol{1}_{ \{ \widetilde{X}_{j} \not\in \mathcal{A}^c \}} + \big[\zeta(\kappa, \boldsymbol{X}_{S^*\backslash\{j\}})\big]^2\boldsymbol{1}_{\{ X_{j} \in \mathcal{A}^c \}} \boldsymbol{1}_{\{ \widetilde{X}_{j} \not\in \mathcal{A}^c \}}.
		\end{split}
	\end{equation*}
	
	With these results, the assumption that $\mathbb{E}(\boldsymbol{1}_{X_{j} > q_{2}} \boldsymbol{1}_{ \widetilde{X}_{j} \not\in \mathcal{A}^c} |\boldsymbol{X}_{-j}) >w_{0}$, and similar arguments for \eqref{power.32}, we deduce that 
	\begin{equation}
		\begin{split}
			\textnormal{The RHS of \eqref{power.40}} & > \iota\big[\inf_{\vec{z}\in\mathbb{R}^{s^*}}\zeta(\vec{z})\big]\mathbb{E}(\varepsilon^2) \mathbb{E}\big[\mathbb{E}(\boldsymbol{1}_{X_{j} > q_{2}} \boldsymbol{1}_{ \widetilde{X}_{j} \not\in \mathcal{A}^c} |\boldsymbol{X}_{-j}) \boldsymbol{1}_{\boldsymbol{X}_{-j} \in \mathcal{B}}\big] \\
			& > \iota w_{0} \big[\inf_{\vec{z}\in\mathbb{R}^{s^*}}\zeta(\vec{z})\big]  \mathbb{P}(\boldsymbol{X}_{S^*} \in \mathcal{D}^{(j)}),
		\end{split}		
	\end{equation}
	which along with \eqref{power.41} concludes the proof of this case assuming a nondecreasing $\zeta_{j}(x)$ and $ \kappa\le q_{1} < q_{2}$.

	We omit the proof for the cases assuming that $\zeta_{j}(x)$ is nonincreasing because the proof is quite similar to the one here. We have completed the proof of Lemma~\ref{proposition1}.
	
	\subsection{Lemma~\ref{uniform.var.1} and its proof}\label{uniform.var.2}
	All notation is the same as in model \eqref{model.1} and Section~\ref{Sec3b}. Particularly, recall that for each $l\in \{1, \dots, p\}$ and $a\in\mathbb{R}$,
	\begin{equation*}
		\begin{split}
			\sigma_{j}^2(a)&= \mathbb{E}\big\{ \big[\zeta(\boldsymbol{X}_{S^{*}})\varepsilon \big]^{2} \big[\boldsymbol{1}_{X_{j}\in(-\infty, a]} - \boldsymbol{1}_{\widetilde{X}_{j}\in(-\infty, a]}\big] - \mu_{j}(a) \big\}^2,\\
			\mu_{j}(a)&= \mathbb{E}\big\{\big[\boldsymbol{1}_{X_{j} \in (-\infty, a]} -\boldsymbol{1}_{\widetilde{X}_{j} \in (-\infty, a]}\big]\big[ \zeta(\boldsymbol{X}_{S^{*}})\varepsilon\big]^2\big\},
		\end{split}
	\end{equation*}
	where $\boldsymbol{X}$ is the population $p$-dimensional feature vector, $\widetilde{\boldsymbol{X}} = (\widetilde{X}_{1}, \dots, \widetilde{X}_{p})^{\top}$ is the $p$-dimensional coordinate-wise knockoff feature vector of $\boldsymbol{X}$ defined in Section~\ref{Sec2.1}, $\varepsilon$ is the independent model error, $\zeta(\boldsymbol{X}_{S^*})$ is the standard deviation function defined in model \eqref{model.1}, and $s^* = |S^*|$ is the number of relevant features. In addition, we have defined $\zeta(\boldsymbol{X}_{S^*}) = \zeta_{0}$ for some $\zeta_{0}>0$ if $S^* = \emptyset$ in Section~\ref{Sec1.1}. Note that Lemma~\ref{uniform.var.1} below does not assume $\mu_{j}(a) = 0$.
	\begin{lemma}\label{uniform.var.1}
		Let some constants $c>0$ and $a\in\mathbb{R}$ be given. Assume $ \inf_{\vec{z}\in\mathbb{R}^{s^*}}\zeta(\vec{z})>0$,  $\sup_{\vec{z}\in\mathbb{R}^{s^*}}\zeta(\vec{z}) <\infty$,  $\mathbb{E}(\varepsilon^4)< \infty$, and that for each $j\in \{1, \dots ,p\}$,
		\begin{equation}
			\begin{split}\label{theorem3.25}
				\mathbb{P}\big\{ \{X_{j} \in (-\infty, a]\}\cap\{\widetilde{X}_{j} \in (-\infty, a]\} \big\} &\ge c, \\
				\mathbb{P}\big\{\{X_{j} \in (-\infty, a]\}\cap\{\widetilde{X}_{j} \in (a, \infty)\} \big\} &\ge  c.
			\end{split}
		\end{equation} 
		Then	
		\begin{equation}\label{theorem3.24}
			\begin{split}
				& \min_{1\le j\le p}\sigma_{j}^2(a) \\
				&\ge\min \left\{\frac{c}{2}\big[\inf_{\vec{z}\in\mathbb{R}^{s^*}} \zeta(\vec{z})\big]^4\mathbb{E}(\varepsilon^{4} ), \frac{c^3\big[\inf_{\vec{z}\in\mathbb{R}^{s^*}} \zeta(\vec{z})\big]^8\big[\mathbb{E}(\varepsilon^{4} ) \big]^2}{16 \big[\sup_{\vec{z}\in\mathbb{R}^{s^*}} \zeta(\vec{z})\big]^4 \big[\mathbb{E}(\varepsilon ^{2})\big]^2}\right\}\eqqcolon \underline{\sigma}^{2} >0.
			\end{split}
		\end{equation}
	\end{lemma}
	\textit{Proof of Lemma~\ref{uniform.var.1}}\indent By \eqref{theorem3.25}  and the assumption that $\varepsilon$ is independent of $(\boldsymbol{X}, \widetilde{\boldsymbol{X}})$, it holds that for each $1\le j \le p$,
	\begin{equation*}
		\begin{split}
			\sigma_{j}^2(a) & \ge \mathbb{E}\big\{\boldsymbol{1}_{\{X_{j}\in(-\infty, a] \}\cap\{\widetilde{X}_{j}\in( a, \infty) \} }\big\{\big[\zeta(\boldsymbol{X}_{S^{*}})\varepsilon \big]^{2} \big(\boldsymbol{1}_{X_{j}\in(-\infty, a]} - \boldsymbol{1}_{\widetilde{X}_{j}\in(-\infty, a]}\big) - \mu_{j}(a) \big\}^2\big\}\\
			& = \mathbb{E}\big\{\boldsymbol{1}_{\{X_{j}\in(-\infty, a] \}\cap\{\widetilde{X}_{j}\in( a, \infty) \} }\big\{\big[\zeta(\boldsymbol{X}_{S^{*}})\varepsilon \big]^{2}  - \mu_{j}(a) \big\}^2\big\}\\
			&\ge  c\big[\inf_{\vec{z}\in\mathbb{R}^{s^*}} \zeta(\vec{z})\big]^4\mathbb{E}(\varepsilon^4 ) - 2|\mu_{j}(a)|\mathbb{E}(\varepsilon^2 ) \big[\sup_{\vec{z}\in\mathbb{R}^{s^*}} \zeta(\vec{z})\big]^2,\\
			\sigma_{j}^2 (a)& \ge \mathbb{E}\big\{\boldsymbol{1}_{\{X_{j}\in(-\infty, a]\}\cap\{\widetilde{X}_{j}\in(-\infty, a]\} }\big\{\big[\zeta(\boldsymbol{X}_{S^{*}})\varepsilon \big]^{2} \big(\boldsymbol{1}_{X_{j}\in(-\infty, a]} - \boldsymbol{1}_{\widetilde{X}_{j}\in(-\infty, a]}\big) - \mu_{j}(a) \big\}^2\big\}\\
			& = \mathbb{E}\big\{\boldsymbol{1}_{\{X_{j}\in(-\infty, a]\}\cap\{\widetilde{X}_{j}\in(-\infty, a]\} } \big[\mu_{j}(a)\big]^2\big\}\\
			&\ge c\big[\mu_{j}(a)\big]^2.
		\end{split}
	\end{equation*}
	By this, we deduce that if  
	$$2|\mu_{j}(a)|\mathbb{E}(\varepsilon ^2) \big[\sup_{\vec{z}\in\mathbb{R}^{s^*}} \zeta(\vec{z})\big]^2\le\frac{c}{2} \big[\inf_{\vec{z}\in\mathbb{R}^{s^*}} \zeta(\vec{z})\big]^4\mathbb{E}(\varepsilon^{4} ) ,$$
	then for each $j\in \{1, \dots, p\}$,
	$$	\sigma_{j}^2(a)\ge \frac{c}{2}\big[\inf_{\vec{z}\in\mathbb{R}^{s^*}} \zeta(\vec{z})\big]^4\mathbb{E}(\varepsilon^{4} ).$$ 
	Otherwise, we have $|\mu_{j}(a)| > c\big[\inf_{\vec{z}\in\mathbb{R}^{s^*}} \zeta(\vec{z})\big]^4\mathbb{E}(\varepsilon^{4} )\big\{4\big[\sup_{\vec{z}\in\mathbb{R}^{s^*}} \zeta(\vec{z})\big]^2\mathbb{E}(\varepsilon^{2}) \big\}^{-1}$, and hence for each $j\in \{1, \dots, p\}$,
	\begin{equation*}
		\begin{split}
			\sigma_{j}^2(a)\ge \frac{c^3\big[\inf_{\vec{z}\in\mathbb{R}^{s^*}} \zeta(\vec{z})\big]^8\big[\mathbb{E}(\varepsilon^{4} ) \big]^2}{16 \big[\sup_{\vec{z}\in\mathbb{R}^{s^*}} \zeta(\vec{z})\big]^4 \big[\mathbb{E}(\varepsilon ^{2})\big]^2}.
		\end{split}
	\end{equation*}
	
	Therefore, we conclude that 
	\begin{equation*}
		\min_{1\le j\le p}\sigma_{j}^2(a)\ge \min \left\{\frac{c}{2}\big[\inf_{\vec{z}\in\mathbb{R}^{s^*}} \zeta(\vec{z})\big]^4\mathbb{E}(\varepsilon^{4} ), \frac{c^3\big[\inf_{\vec{z}\in\mathbb{R}^{s^*}} \zeta(\vec{z})\big]^8\big[\mathbb{E}(\varepsilon^{4} ) \big]^2}{16 \big[\sup_{\vec{z}\in\mathbb{R}^{s^*}} \zeta(\vec{z})\big]^4 \big[\mathbb{E}(\varepsilon ^{2})\big]^2}\right\} >0,
	\end{equation*}
	which is the desired result.

	\renewcommand{\thesubsection}{D.\arabic{subsection}}

	\section{Extensive simulation studies}\label{Sec5}
	
	Simulation experiments for the VD and VDBP tests with synthetic data are in Section~\ref{SecD.1}, where we demonstrate satisfactory empirical performance of these tests when the mean estimation methods (the centering methods) are random forests and HBART~\citepsupp{pratola2020heteroscedastic}. HBART is also considered here since it is a recent advanced heteroskedasticity modeling method. For our synthetic data study, the benchmark methods are DGLM~\citepsupp{smyth1989generalized} and Two-hit~\citepsupp{chiou2020variable}. On the other hand, some numerical experiments for hidden Markov model knockoffs~\citepsupp{sesia2019gene} are in Section~\ref{Sec7.2}.

	\subsection{Synthetic data study}\label{SecD.1}

	\subsubsection{Coordinate-wise Gaussian knockoffs}\label{SecD.1.1}
	We have implemented a knockoff generator for producing approximate coordinate-wise Gaussian knockoffs in a coordinate-wise  fashion based on the  ideas of  \eqref{coordinate-wise.1} and those introduced in \citepsupp{Barber2015}. To demonstrate the advantage of coordinate-wise knockoffs, we have  performed a numerical experiment to show that the correlation between a Gaussian variable and its coordinate-wise knockoff tend to be smaller, sometimes much smaller, than the correlation between the Gaussian variable and its knockoff generated by the R package \texttt{knockoff}~\citepsupp{Barber2015}. Details of our coordinate-wise  Gaussian knockoff generator and the results of our numerical experiments are postponed to  Section~\ref{coordinate-wise}.
	\subsubsection{Simulation setting}\label{Sec5.1}
	We simulate  a sample $\mathcal{X}_{n} \coloneqq \{Q_{i}, \boldsymbol{Z}_{i}, \widetilde{\boldsymbol{Z}}_{i}\}_{i=1}^{n}$ of i.i.d. observations with $n\in\{500, 700\}$ such that $(Q_{1}, \boldsymbol{Z}_{1})$ and $(Y, \boldsymbol{X})$ have the same distribution given as follows. The response $Y$ is generated from one of the following models:
	{\small \begin{align}
			Y &= X_{1} + X_{2} + \Big[\sqrt{\exp{(0.5 + X_{10}+   X_{15})}}\Big] \varepsilon,	\label{model.23} \\
			Y &= 2X_{1}X_{2} + X_{3} + X_{4} + X_{5}^2+ \Big[\sqrt{\exp{(0.5 + X_{10}+   X_{15})}}\Big] \varepsilon,	\label{model.24}\\
			Y &= 2X_{1}X_{2} + X_{3} + X_{4} + X_{5}^2+  \varepsilon,	\label{model.27}\\
			Y &= 2X_{1}X_{2} + X_{3} + X_{4} + X_{5}^2+ \Big[1 + 3\times \boldsymbol{1}_{\{X_{15}>0\}}\Big] \varepsilon,	\label{model.28}
	\end{align}}%
	where $\varepsilon$ is an independent standard Gaussian model error, and $\boldsymbol{X} = (X_{1}, \dots ,X_{p})^{\top}$ with $p\in\{20, 700\}$ is either 1) a multivariate Gaussian  vector with zero mean  and covariance matrix $\Sigma=[\Sigma_{lk}]_{l,k=1}^p$, in which $\Sigma_{lk}= \rho^{|l-k|}$, or 2) a multivariate t distribution with zero mean, covariance matrix $\Sigma$, and degree of freedom $10$. In practice, we use the R package \texttt{mvtnorm} for sampling $\{\boldsymbol{Z}_{i}\}_{i=1}^n$ from these multivariate distributions, while their  knockoff features $\{\widetilde{\boldsymbol{Z}}_{i}\}_{i=1}^n$ are generated by our coordinate-wise Gaussian knockoff generator introduced in Section~\ref{coordinate-wise} with $\{\boldsymbol{Z}_{i}\}_{i=1}^n$ given. Note that existing knockoff generators, including ours in Section~\ref{coordinate-wise}, are all applicable for producing knockoffs $\{\widetilde{\boldsymbol{Z}}_{i}\}_{i=1}^n$ given an arbitrary sample $\{\boldsymbol{Z}_{i}\}_{i=1}^n$, but whether $\{\widetilde{\boldsymbol{Z}}_{i}\}_{i=1}^n$ are good approximations of the ideal coordinate-wise knockoffs depends on the underlying distribution of the sample.

	\subsubsection{VD test for hypothesis \eqref{null.1}}\label{Sec6.2.1}

	The VD test statistic with break selection $\big|T_{j}(\widehat{a}_{j})\big[\widehat{\sigma}_{j}(\widehat{a}_{j})\big]^{-1}\big|$ given in \eqref{h.3} is calculated and the test is established for each $j\in \{1, \dots ,20\}$. The empirical rejection rates over $100$ repetitions  of each simulation case are reported without adjusting for multiple tests in Tables~\ref{tab:1}--\ref{tab:12}, where the rejection thresholds under null hypothesis \eqref{null.1} with test size $\alpha\in \{0.1, 0.05, 0.025\}$ are given respectively  by $(t_{0.1} , t_{0.05}, t_{0.025}) = (1.64, 1.96, 2.25)$ for the VD test, according to Theorem~\ref{theorem3}. The multiplicity adjustment is not made in this simulation experiment because the goal here is to illustrate our main results in Theorem~\ref{theorem3}. In Tables~\ref{tab:1}--\ref{tab:12}, only two features $X_{10}$ and $X_{15}$ indicated with checkmarks are relevant, which means that entries for other features in the tables are empirical wrong rejection rates. Each simulation setting including the underlying feature index $j\in \{1, \dots, 20\}$, values of $(n, p, \rho)$, test size $\alpha\in\{0.1, 0.05, 0.025\}$, data generating models \eqref{model.23}--\eqref{model.24}, and test methods is indicated for each case and each panel in the tables. In addition, the centering methods are indicated on the top of each panel; they are random forests~\citepsupp{Breiman2001} and HBART~\citepsupp{pratola2020heteroscedastic}, which are implemented with the R packages \texttt{randomForest} and \texttt{rbart}, respectively. In the simulation of the VD test, the sample features $\{\boldsymbol{X}_{i}\}_{i=1}^n$ are sampled from the multivariate Gaussian distribution with covariance matrix $\Sigma$ (see Section~\ref{Sec5.1} for details). In Table~\ref{tab:1}, we also consider heuristic P-values calculated by double GLM~\citepsupp{smyth1989generalized, smyth1999adjusted} with the R package \texttt{dglm} for each $j\in \{1, \dots p\}$. We do not report results from DGLM in Table~\ref{tab:12} because \texttt{dglm} does not output heuristic P-values for the non-linear high-dimensional case.

	The VD test with break selection requires sample splitting and a choice of break candidates, which we introduce as follows. The simulated sample  $\mathcal{X}_{n}$ (see Section~\ref{Sec5.1}) is split into two subsamples $\mathcal{X}_{1} = \{Y_{i}, \boldsymbol{X}_{i}, \widetilde{\boldsymbol{X}}_{i}\}_{i=1}^{n_{1}}$ and $\mathcal{X}_{2} = \{V_{i}, \boldsymbol{U}_{i}, \widetilde{\boldsymbol{U}}_{i}\}_{i=1}^{n_{2}}$ with $n_{1} = \nint{\frac{2}{3}n}$ and $n_{2} = n-n_{1}$. As have mentioned in Section~\ref{Sec2.4}, sample $\mathcal{X}_{1}$ is used for constructing test statistics, while $\mathcal{X}_{2}$ is used for selecting breaks. Our practical implementation of the VD test uses the full sample $\mathcal{X}_{n}$ for training the regression trees estimate of mean functions for simplicity.
	In addition, for each $l\in \{1, \dots, p\}$, we set break candidates  $(\kappa_{1,l}, \kappa_{R,l})$ to respectively  the first and third quartiles of $(U_{1l}, \dots, U_{n_{2}l}$) with $R=100$, $\kappa_{r, l} < \kappa_{r+1, l} $, and evenly distributed $\kappa_{r,l}$'s.

	\begin{table}[h]
		\begin{center}
			{\small
				\begin{tabular}[t]{  |c|c|m{0.8cm}|m{0.8cm}|m{0.8cm}|m{0.8cm}|m{0.8cm}|m{0.8cm}|m{0.8cm}|m{0.8cm}|m{0.8cm}|m{0.8cm}|}
					\multicolumn{12}{c}{(a) VD test + random forests (centering method) + model~\eqref{model.23} with  Gaussian features} \\[5pt]
					\hline $(n, p, \rho)$& $\alpha$
					& \multicolumn{1}{c}{$X_{1}$} & \multicolumn{1}{c}{$X_{2}$}& \multicolumn{1}{c}{$X_{3}$}& \multicolumn{1}{c}{$X_{4}$} &  
					\multicolumn{1}{c}{$X_{5}$} & \multicolumn{1}{c}{$X_{6}$}& \multicolumn{1}{c}{$X_{7}$}& \multicolumn{1}{c}{$X_{8}$}   &\multicolumn{1}{c}{$X_{9}$}&\multicolumn{1}{c|}{$X_{10}$ $^{\checkmark}$} \\
					\hline
					\multirow{7}{*}{\scriptsize{(500, 20, 0.4)} }&\multirow{1}{*}{$ 0.1\phantom{00}$} & 0.08 &0.09 &0.04 &0.11 &0.07 &0.04 &0.07 &0.09 &0.05&0.82 \\
					&\multirow{1}{*}{$0.05\phantom{0}$} & 0.03 &0.05 &0.03 &0.01 &0.01 &0.00 &0.03 & 0.05&  0.01&0.66 \\
					&$0.025\phantom{}$ &  0.01 &0.02 &0.01 &0.00 &0.00 &0.00& 0.01 &0.01 &0.01 &0.52 \\ \cline{2-12}
					&$\alpha$ \phantom{\Big.}& \multicolumn{1}{c}{$X_{11}$ } & \multicolumn{1}{c}{$X_{12}$}& \multicolumn{1}{c}{$X_{13}$}& \multicolumn{1}{c}{$X_{14}$}&  
					\multicolumn{1}{c}{  $X_{15}$ $^{\checkmark}$ } & \multicolumn{1}{c}{$X_{16}$}& \multicolumn{1}{c}{$X_{17}$}& \multicolumn{1}{c}{$X_{18}$  }&\multicolumn{1}{c}{$X_{19}$}&\multicolumn{1}{c|}{$X_{20}$} \\ [-3pt]
					\cline{2-12}
					&\multirow{1}{*}{$ 0.1\phantom{0}$} \phantom{$\Big.$} & 0.08 &0.05 &0.09 &0.06&0.86 &0.05 &0.04 &0.08 & 0.03&0.09 \\[-3pt]
					&\multirow{1}{*}{$0.05\phantom{0}$} &  0.05& 0.00 &0.03 &0.00 &0.71 &0.01 &0.02 &0.02 &0.00 & 0.03 \\
					& $0.025\phantom{}$ & 0.02 &0.00 &0.00 &0.00& 0.59& 0.00 &0.00 &0.00 &0.00 &0.00\\
					\hline
			\end{tabular} } 
			
			{\small
				\begin{tabular}[t]{ | c|c|m{0.8cm}|m{0.8cm}|m{0.8cm}|m{0.8cm}|m{0.8cm}|m{0.8cm}|m{0.8cm}|m{0.8cm}|m{0.8cm}|m{0.8cm}|}
					\multicolumn{12}{c}{\phantom{.}} \\[5pt]
					\multicolumn{12}{c}{(b) VD test + HBART (centering method) + model~\eqref{model.23} with  Gaussian features} \\[5pt]
					\hline $(n, p, \rho)$& $\alpha$
					& \multicolumn{1}{c}{$X_{1}$} & \multicolumn{1}{c}{$X_{2}$}& \multicolumn{1}{c}{$X_{3}$}& \multicolumn{1}{c}{$X_{4}$} &  
					\multicolumn{1}{c}{$X_{5}$} & \multicolumn{1}{c}{$X_{6}$}& \multicolumn{1}{c}{$X_{7}$}& \multicolumn{1}{c}{$X_{8}$}   &\multicolumn{1}{c}{$X_{9}$}&\multicolumn{1}{c|}{$X_{10}$ $^{\checkmark}$} \\
					\hline
					\multirow{7}{*}{\scriptsize{(500, 20, 0.4)} }&\multirow{1}{*}{$ 0.1\phantom{00}$} & 0.06& 0.13 &0.07& 0.11& 0.06& 0.03& 0.06& 0.06& 0.05 &0.88 \\
					&\multirow{1}{*}{$0.05\phantom{0}$} & 0.04 &0.06 &0.04& 0.02& 0.02& 0.00& 0.05 &0.04 &0.00 &0.81  \\
					& $0.025\phantom{}$& 0.02 &0.03 &0.01 &0.00& 0.00& 0.00 &0.03 &0.01 &0.00 &0.65 \\ \cline{2-12}
					&$\alpha$ \phantom{\Big.}& \multicolumn{1}{c}{$X_{11}$ } & \multicolumn{1}{c}{$X_{12}$}& \multicolumn{1}{c}{$X_{13}$}& \multicolumn{1}{c}{$X_{14}$}&  
					\multicolumn{1}{c}{  $X_{15}$ $^{\checkmark}$ } & \multicolumn{1}{c}{$X_{16}$}& \multicolumn{1}{c}{$X_{17}$}& \multicolumn{1}{c}{$X_{18}$  }&\multicolumn{1}{c}{$X_{19}$}&\multicolumn{1}{c|}{$X_{20}$} \\ [-3pt]
					\cline{2-12}
					&\multirow{1}{*}{$ 0.1\phantom{0}$} \phantom{$\Big.$} & 0.10& 0.06&0.06& 0.05& 0.89& 0.07& 0.06& 0.08& 0.03 &0.07 \\[-3pt]
					&\multirow{1}{*}{$0.05\phantom{0}$} &  0.03 & 0.01& 0.02& 0.00& 0.76& 0.01& 0.03& 0.03& 0.01 &0.03 \\
					& $0.025\phantom{}$ &  0.00 &0.01 &0.00 &0.00 &0.68 &0.00& 0.00 &0.01 &0.00 &0.00\\
					\hline
			\end{tabular} }

			{\small
				\begin{tabular}[t]{ | c|c|m{0.8cm}|m{0.8cm}|m{0.8cm}|m{0.8cm}|m{0.8cm}|m{0.8cm}|m{0.8cm}|m{0.8cm}|m{0.8cm}|m{0.8cm}|}
					\multicolumn{12}{c}{\phantom{.}} \\[5pt]
					\multicolumn{12}{c}{(c) Double GLM + model~\eqref{model.23} with  Gaussian features} \\[5pt]
					\hline $(n, p, \rho)$& $\alpha$
					& \multicolumn{1}{c}{$X_{1}$} & \multicolumn{1}{c}{$X_{2}$}& \multicolumn{1}{c}{$X_{3}$}& \multicolumn{1}{c}{$X_{4}$} &  
					\multicolumn{1}{c}{$X_{5}$} & \multicolumn{1}{c}{$X_{6}$}& \multicolumn{1}{c}{$X_{7}$}& \multicolumn{1}{c}{$X_{8}$}   &\multicolumn{1}{c}{$X_{9}$}&\multicolumn{1}{c|}{$X_{10}$ $^{\checkmark}$} \\
					\hline
					\multirow{7}{*}{\scriptsize{(500, 20, 0.4)} }&\multirow{1}{*}{$ 0.1\phantom{00}$} & 0.17 &0.14 &0.16 &0.23 &0.13 &0.18 &0.16 &0.16 &0.17 &1.00 \\
					&\multirow{1}{*}{$0.05\phantom{0}$} &0.10 &0.06 &0.08& 0.13& 0.06& 0.07& 0.13 &0.11 &0.10 &1.00 \\
					& $0.025\phantom{}$  &0.06 &0.04 &0.05 &0.09 &0.04& 0.06 &0.07&0.09 & 0.02 &1.00 \\ \cline{2-12}
					&$\alpha$ \phantom{\Big.}& \multicolumn{1}{c}{$X_{11}$ } & \multicolumn{1}{c}{$X_{12}$}& \multicolumn{1}{c}{$X_{13}$}& \multicolumn{1}{c}{$X_{14}$}&  
					\multicolumn{1}{c}{  $X_{15}$ $^{\checkmark}$ } & \multicolumn{1}{c}{$X_{16}$}& \multicolumn{1}{c}{$X_{17}$}& \multicolumn{1}{c}{$X_{18}$  }&\multicolumn{1}{c}{$X_{19}$}&\multicolumn{1}{c|}{$X_{20}$} \\ [-3pt]
					\cline{2-12}
					&\multirow{1}{*}{$ 0.1\phantom{0}$} \phantom{$\Big.$} & 0.23 &0.20 &0.14 &0.16 &1.00 &0.16 &0.12 &0.18 &0.12 &0.15 \\[-3pt]
					&\multirow{1}{*}{$0.05\phantom{0}$} & 0.13 &0.10 &0.08 &0.11 &1.00 &0.09 &0.06 &0.11 &0.10 &0.08 \\
					& $0.025\phantom{}$ & 0.06 &0.05 &0.07 &0.09 &1.00& 0.05& 0.04& 0.06 &0.03&0.06\\
					\hline
			\end{tabular} } 
			
			\caption{The empirical rejection rates for hypothesis \eqref{null.1} for each feature $j\in\{1, \dots, 20\}$ at  significance level $\alpha\in\{0.1, 0.05, 0.025\}$ over $100$ simulation repetitions. The centering methods for VD tests are indicated in each panel. Features $X_{10}$ and $X_{15}$ with checkmarks are relevant features; others are null features.}\label{tab:1}
		\end{center}
	\end{table}
	
	\clearpage
	
	\begin{table}[h]
		\begin{center}
			{\small
				\begin{tabular}[t]{  |c|c|m{0.8cm}|m{0.8cm}|m{0.8cm}|m{0.8cm}|m{0.8cm}|m{0.8cm}|m{0.8cm}|m{0.8cm}|m{0.8cm}|m{0.8cm}|}
					\multicolumn{12}{c}{(a) VD test + RF (centering method) + model~\eqref{model.24} with Gaussian features} \\[5pt]
					\hline $(n, p, \rho)$& $\alpha$
					& \multicolumn{1}{c}{$X_{1}$} & \multicolumn{1}{c}{$X_{2}$}& \multicolumn{1}{c}{$X_{3}$}& \multicolumn{1}{c}{$X_{4}$} &  
					\multicolumn{1}{c}{$X_{5}$} & \multicolumn{1}{c}{$X_{6}$}& \multicolumn{1}{c}{$X_{7}$}& \multicolumn{1}{c}{$X_{8}$}   &\multicolumn{1}{c}{$X_{9}$}&\multicolumn{1}{c|}{$X_{10}$ $^{\checkmark}$} \\
					\hline
					\multirow{7}{*}{\scriptsize{(700, 700, 0.6)} }&\multirow{1}{*}{$ 0.1\phantom{00}$} & 0.06 &0.08 &0.03 &0.04 &0.06 &0.07 &0.04 &0.10 &0.08&0.50 \\
					&\multirow{1}{*}{$0.05\phantom{0}$} & 0.04 &0.01 &0.01 &0.01 &0.01 &0.04 &0.00 & 0.04&  0.03&0.30 \\
					&$0.025\phantom{}$ &  0.00 &0.01 &0.01 &0.01 &0.00 &0.03& 0.00 &0.02 &0.02 &0.23 \\ \cline{2-12}
					&$\alpha$ \phantom{\Big.}& \multicolumn{1}{c}{$X_{11}$ } & \multicolumn{1}{c}{$X_{12}$}& \multicolumn{1}{c}{$X_{13}$}& \multicolumn{1}{c}{$X_{14}$}&  
					\multicolumn{1}{c}{  $X_{15}$ $^{\checkmark}$ } & \multicolumn{1}{c}{$X_{16}$}& \multicolumn{1}{c}{$X_{17}$}& \multicolumn{1}{c}{$X_{18}$  }&\multicolumn{1}{c}{$X_{19}$}&\multicolumn{1}{c|}{$X_{20}$} \\ [-3pt]
					\cline{2-12}
					&\multirow{1}{*}{$ 0.1\phantom{0}$} \phantom{$\Big.$} & 0.07 &0.09 &0.09 &0.10&0.48 &0.08 &0.02 &0.09 & 0.09&0.09 \\[-3pt]
					&\multirow{1}{*}{$0.05\phantom{0}$} &  0.04& 0.02 &0.03 &0.02 &0.36 &0.01 &0.01 &0.01 &0.04 & 0.03 \\
					& $0.025\phantom{}$ & 0.01 &0.01 &0.02 &0.01& 0.25& 0.01 &0.00 &0.01 &0.00 &0.01\\
					\hline
			\end{tabular} } 
			
			{\small
				\begin{tabular}[t]{ | c|c|m{0.8cm}|m{0.8cm}|m{0.8cm}|m{0.8cm}|m{0.8cm}|m{0.8cm}|m{0.8cm}|m{0.8cm}|m{0.8cm}|m{0.8cm}|}
					\multicolumn{12}{c}{\phantom{.}} \\[5pt]
					\multicolumn{12}{c}{(b) VD test + HBART (centering method) + model~\eqref{model.24} with  Gaussian features} \\[5pt]
					\hline $(n, p, \rho)$& $\alpha$
					& \multicolumn{1}{c}{$X_{1}$} & \multicolumn{1}{c}{$X_{2}$}& \multicolumn{1}{c}{$X_{3}$}& \multicolumn{1}{c}{$X_{4}$} &  
					\multicolumn{1}{c}{$X_{5}$} & \multicolumn{1}{c}{$X_{6}$}& \multicolumn{1}{c}{$X_{7}$}& \multicolumn{1}{c}{$X_{8}$}   &\multicolumn{1}{c}{$X_{9}$}&\multicolumn{1}{c|}{$X_{10}$ $^{\checkmark}$} \\
					\hline
					\multirow{7}{*}{\scriptsize{(700, 700, 0.6)} }&\multirow{1}{*}{$ 0.1\phantom{00}$} & 0.06 &0.04 &0.07 &0.07 &0.04 &0.04 &0.08 &0.06 &0.04&0.72 \\
					&\multirow{1}{*}{$0.05\phantom{0}$} & 0.01 &0.00 &0.02 &0.03 &0.02 &0.03 &0.01 & 0.02&  0.00&0.60  \\
					& $0.025\phantom{}$& 0.00 &0.00 &0.01 &0.00 & 0.01& 0.01 &0.00 &0.01 &0.00 &0.40 \\ \cline{2-12}
					&$\alpha$ \phantom{\Big.}& \multicolumn{1}{c}{$X_{11}$ } & \multicolumn{1}{c}{$X_{12}$}& \multicolumn{1}{c}{$X_{13}$}& \multicolumn{1}{c}{$X_{14}$}&  
					\multicolumn{1}{c}{  $X_{15}$ $^{\checkmark}$ } & \multicolumn{1}{c}{$X_{16}$}& \multicolumn{1}{c}{$X_{17}$}& \multicolumn{1}{c}{$X_{18}$  }&\multicolumn{1}{c}{$X_{19}$}&\multicolumn{1}{c|}{$X_{20}$} \\ [-3pt]
					\cline{2-12}
					&\multirow{1}{*}{$ 0.1\phantom{0}$} \phantom{$\Big.$} & 0.06& 0.08&0.10& 0.06& 0.75& 0.06& 0.05& 0.03& 0.07 &0.04 \\[-3pt]
					&\multirow{1}{*}{$0.05\phantom{0}$} &  0.04 & 0.02& 0.02& 0.01& 0.57& 0.01& 0.02& 0.02& 0.03 &0.01 \\
					& $0.025\phantom{}$ &  0.02 &0.00 &0.01 &0.00 &0.42 &0.00& 0.01 &0.00 &0.01 &0.00\\
					\hline
			\end{tabular} }%
			\caption{The empirical rejection rates  for hypothesis \eqref{null.1} for each feature $j\in\{1, \dots, 20\}$ at  significance level $\alpha\in\{0.1, 0.05, 0.025\}$ over $100$ simulation repetitions.  The centering methods are indicated in each panel. Features $X_{10}$ and $X_{15}$ with checkmarks are relevant features; others are null features. }\label{tab:12}%
		\end{center}
	\end{table}%
	\subsubsection{VDBP test for hypothesis \eqref{BP.1}} \label{Sec6.2.2}
	
	The VDBP test statistic $\big|T_{\widehat{l}}(\widehat{a}_{\widehat{l}}) \big[\widehat{\sigma}_{\widehat{l}}(\widehat{a}_{\widehat{l}})\big]^{-1}\big|$  given by \eqref{h.4} is calculated and the test is established for each simulation experiment. The sample splitting, R packages for random forests and HBART, and the choice of change candidates follow those in Section~\ref{Sec6.2.1}, but with $n_{1} = \nint{n/3}$ and $n_{2} = n - n_{1}$. By Theorem~\ref{theorem4}, we set the rejection thresholds $(t_{0.1}, t_{0.05}, t_{0.025}) = (1.64, 1.96, 2.25)$ for the VDBP test with break selection. Meanwhile, the benchmark method here is the Breusch-Pagan test~\citepsupp{breusch1979simple}, whose P-value is available from the R package \texttt{lmtest}. In addition, to apply the Breusch-Pagan test to high-dimensional data, we run the Breusch-Pagan test given a set of selected features by Twohit~\citepsupp{chiou2020variable}. Twohit is a recent approach for selecting active features in both mean and standard deviation functions from  high-dimensional features; this method assumes model \eqref{model.1} with a linear mean function and $\zeta(\cdot)$ as in Example~\ref{sd.exmp.3}. As have done for the VDBP test, we split sample $\mathcal{X}_{n}$ into two subsamples with respective sample sizes $n_{1}$ and $n_{2}$, where the second subsample is used for Twohit model selection and the first one is for the Breusch-Pagan test.
	
	We report the empirical rejection rates of each case over $100$ repetitions in Table~\ref{tab:bp}, with test size $\alpha\in\{0.1, 0.05, 0.025\}$, data generating models \eqref{model.27}--\eqref{model.28}, distribution of $\boldsymbol{X}$, feature dimensionality $p\in\{20, 700\}$, and the centering methods indicated for each case and each panel; in addition, we set $n=700$ and $\rho = 0.6$, and we refer to Section~\ref{Sec5.1} for details of multivariate Gaussian and t distributions with covariance parameter $\rho$. The cases (V) and (VI) are omitted  in panel (c) of Table~\ref{tab:bp} because the Breusch-Pagan test does not apply to high-dimensional features. In Table~\ref{tab:bp}, we omit the cases with linear models for simplicity because  the simulation results for such cases are satisfactory for all methods.

	\begin{landscape}
		\begin{table}[h]
			\begin{center}
				{\small
					\begin{tabular}[t]{  r|cccccc}
						\multicolumn{7}{c}{(a) VDBP test + Random Forests (centering method)}\\[5pt]
						Case&(I)&(II)&(III)&(IV)&(V)&(VI)\\
						Model & Model~\eqref{model.27}&Model~\eqref{model.28}$^{\checkmark}$&Model~\eqref{model.27}&Model~\eqref{model.28}$^{\checkmark}$&Model~\eqref{model.27}& Model~\eqref{model.28}$^{\checkmark}$\\
						Dist. of $\boldsymbol{X}$ &Gaussian& Gaussian& t& t& Gaussian&Gaussian \\
						$p$ &$20$&$20$ & $20$ & $20$ & $700$ & $700$\\ \hline
						$\alpha=0.1$ &0.05&0.89&0.06&0.47&0.08&0.35\\
						$\alpha=0.05$ &0.02&0.86&0.02&0.39&0.01&0.30\\
						$\alpha=0.025$ &0.01&0.78&0.00&0.30&0.00&0.23\\

				\end{tabular} } 
				
				{\small
					\begin{tabular}[t]{  r|cccccc}
						\multicolumn{7}{c}{\phantom{.}} \\[-5pt]
						\multicolumn{7}{c}{(b) VDBP test + HBART (centering method)}\\[5pt]
						Case&(I)&(II)&(III)&(IV)&(V)&(VI)\\
						Model & Model~\eqref{model.27}&Model~\eqref{model.28}$^{\checkmark}$&Model~\eqref{model.27}&Model~\eqref{model.28}$^{\checkmark}$&Model~\eqref{model.27}& Model~\eqref{model.28}$^{\checkmark}$\\
						Dist. of $\boldsymbol{X}$ &Gaussian& Gaussian& t& t& Gaussian&Gaussian \\
						$p$ &$20$&$20$ & $20$ & $20$ & $700$ & $700$\\ \hline
						$\alpha=0.1$ &0.07&1.00&0.06&0.89&0.16&0.86\\
						$\alpha=0.05$ &0.03&0.99&0.04&0.84&0.11&0.83\\
						$\alpha=0.025$ &0.02&0.99&0.01&0.83&0.03&0.77\\

				\end{tabular} } 
				{\small
					\begin{tabular}[t]{  r|cccccc}
						\multicolumn{7}{c}{\phantom{.}} \\[-5pt]
						\multicolumn{7}{c}{(c) Breusch-Pagan test}\\[5pt]
						Case&(I)&(II)&(III)&(IV)&&\\
						Model & Model~\eqref{model.27}&Model~\eqref{model.28}$^{\checkmark}$&Model~\eqref{model.27}&Model~\eqref{model.28}$^{\checkmark}$&\phantom{Model~\eqref{model.27}}& \phantom{Model~\eqref{model.28}$^{\checkmark}$}\\
						Dist. of $\boldsymbol{X}$ &Gaussian& Gaussian& t& t&\phantom{Gaussian} & \phantom{Gaussian}\\ 
						$p$ &20& 20& 20& 20&  & \\ \hline
						$\alpha=0.1$ &0.20&0.94&0.90&0.92&&\\
						$\alpha=0.05$ &0.10&0.90&0.84&0.87&&\\
						$\alpha=0.025$ &0.08&0.85&0.78&0.77&&\\

				\end{tabular} } 
				
				\caption{The empirical rejection rates  for hypothesis \eqref{BP.1} for each case at each significance level $\alpha\in\{0.1, 0.05, 0.025\}$ over $100$ simulation repetitions with $(\rho, n) =(0.6, 700)$. The checkmarks indicate the models that are against the null null hypotheses.  The centering methods and model selection method are indicated in each panel.
				}\label{tab:bp}
			\end{center}
		\end{table}
	\end{landscape}

	\begin{landscape}
		\begin{table}[h]
			\begin{center}
				
				{\small
					\begin{tabular}[t]{  r|cccccc}
						\multicolumn{7}{c}{\phantom{.}} \\[-5pt]
						\multicolumn{7}{c}{(d) Breusch-Pagan test + Twohit (model selection method) }\\[5pt]
						Case&(I)&(II)&(III)&(IV)&(V)&(VI)\\
						Model & Model~\eqref{model.27}&Model~\eqref{model.28}$^{\checkmark}$&Model~\eqref{model.27}&Model~\eqref{model.28}$^{\checkmark}$&Model~\eqref{model.27}& Model~\eqref{model.28}$^{\checkmark}$\\
						Dist. of $\boldsymbol{X}$ &Gaussian& Gaussian& t& t& Gaussian & Gaussian\\ 
						$p$ &20& 20& 20& 20& 700 & 700		\\
						\hline
						$\alpha=0.1$ &0.29& 0.76 &0.62& 0.59 &0.26&0.77\\
						$\alpha=0.05$ &0.17&0.69  &0.52&0.51 &0.15&0.65\\
						$\alpha=0.025$ &0.14&  0.58&0.45&0.42 &0.10&0.55\\

				\end{tabular} } 
				\caption*{Table~\ref{tab:bp} (continued):
					The empirical rejection rates  for hypothesis \eqref{BP.1} for each case at each significance level $\alpha\in\{0.1, 0.05, 0.025\}$ over $100$ simulation repetitions with $(\rho, n) =(0.6, 700)$. The checkmarks indicate the models that are against the null hypothesis.  The centering methods and model selection method are indicated in each panel.}
			\end{center}
		\end{table}
	\end{landscape}
	
	\subsubsection{Simulation results}

	From Tables~\ref{tab:1}--\ref{tab:12}, the VD test with each centering method mostly controls the empirical wrong rejection rates to the respective target levels $\alpha\in\{0.1, 0.05, 0.025\}$, showing the merits of using knockoff features as the negative control of the test statistics. In contrast, the hueristic P-values obtained by DGLM do not control the empirical wrong rejection rates, which are frequently twice the sizes of the respective target levels. On the other hand, the test power (for $X_{10}$ and $X_{15}$) based on VD tests decreases as $\alpha$ decreases in Tables~\ref{tab:1}--\ref{tab:12}, which may be improved given a larger sample. In these two experiment settings, the test power of the VD test + HBART outperformes the VD test + random forests, while the power of each VD test decreases under model \eqref{model.24} when compared to their respective results under model \eqref{model.23}. Meanwhile, the heuristic P-values based on DGLM ourperform other methods in temrs of power and always identify relevant features $X_{10}$ and $X_{15}$ in Table~\ref{tab:1}. However, the DGLM in package \texttt{dglm} does not scale to high-dimensional input features and cannot be applied to cases with complicated mean functions such as model~\eqref{model.24}.

	Regarding the tests for null hypothesis \eqref{BP.1} in Table~\ref{tab:bp}, except for the case (V) in panel (b), the VDBP tests control the empirical wrong rejection rates. Particularly, the results of case (III) in panel (a) and (b) show that VDBP tests are robust when the  feature vector has a multivariate t distribution. In contrast, from panel (c) of Table~\ref{tab:bp}, the Breusch-Pagan test does not control the false positive errors under models with non-linear means. Its inferences for case (III) is not reliable since the false positive errors are too high. 
	In panel (d), we see that  the Breusch-Pagan test + Twohit is applicable to high-dimensional cases, and its performance is the same in low- and high-dimensional cases. However, the Breusch-Pagan test + Twohit increases the false positive errors in case (I) when compared to the corresponding results without Twohit, which is mainly because Twohit often screens out  active features $\{X_{1}, X_{2}, X_{5}\}$ with non-linear mean effects in this scenario. Meanwhile, although 
	the performance in case (III) of panel (d) is largely improved when compared to case (III) of panel (c), the false positive errors of the Breusch-Pagan test + Twohit are still too high to be useful in such a scenario. On the power side,  we  see that the VDBP test + HBART has satisfactory selection power and outperforms the power of  the VDBP test + random forests in Table~\ref{tab:bp}, where the power of the latter test decreases when the distribution of the feature vector becomes a multivariate t and/or the dimensionality of $p$ increases. From panel (d), the use of Twohit maintains the test power of the Breusch-Pagan test in the high-dimensional case. Overall, these experiment results illustrate that the VD and VDBP tests are appropriate for the respective  null hypotheses \eqref{null.1} and \eqref{BP.1} in challenging learning situations with non-linear data generating models and high-dimensional features.

	\subsection{Numerical experiment for hidden Markov model knockoffs}\label{Sec7.2}
	
	In our real data study, we generate knockoff features for mutation features in the
	HIV-1 dataset with hidden Markov model (HMM) knockoffs~\citepsupp{sesia2019gene}, which was developed for generating knockoff features for single-nucleotide polymorphisms. In this section, we perform numerical experiments to examine the empirical performance of HMM knockoffs for generating knockoff features given mutation samples from participants treated with PI drugs. The data description and data preparation is given in Section~\ref{Sec6.1}. The number of features and sample size are respectively $332$ and $846$, where missing values of response variables in the HIV-1 dataset are not involved here because we only use the explanatory features from the sample.

	With the sample of mutation features, we simulate response variables and then select relevant features by the VD test, and we report the average selection performance over $100$  repetitions. Specifically,  for the $b$th repetition, we simulate $Y_{1}^{(b)}, \dots, Y_{846}^{(b)}$  according to
	\begin{equation}\label{y2}
		Y_{i}^{(b)} = \sum_{j\in S_{m}^{(b)}} 0.5\times x_{ij} + \sum_{j\in S_{sd}^{(b)}}\big[ 0.5 + \boldsymbol{1}_{\{x_{ij}>0\}}
		\big]\times \varepsilon_{i}^{(b)},
	\end{equation}
	where  $S_{m}^{(b)}$ and $ S_{sd}^{(b)}$ with $(|S_{m}^{(b)}|, |S_{sd}^{(b)}|) = (10, 3)$ are random feature indices drawn from the set of features with more than $60$ mutations in the sample, and $\varepsilon_{i}^{(b)}$'s are i.i.d. standard Gaussian model errors. Only $31$ features in this sample have more than $60$ mutations. The model coefficients  in \eqref{y2} are calibrated so the sample variance of $Y_{i}^{(b)}$'s are comparable to the original dataset, and the mean function is linear for simplicity. Notice that the use of lowercase $x_{ij}$'s  in model \eqref{y2} emphasizes that the explanatory features are given by a real dataset and hence are constant in this numerical experiment.

	The VD statistics with random forests as the centering method for each feature are calculated as in Section~\ref{Sec6.1}. Let $S_{sd}^{(b)}$ and $\widehat{S}^{(b)}$ for $ b \in \{1, \dots, 100\}$ be respectively the set of true relevant features and the set of selected features with FDR controlled at $0.2$ at the $b$th repetition. We report in Table~\ref{tab:3} the empirical FDR and selection power, which are given by
	\begin{equation}
		\begin{split}
			\label{y3}
			\textnormal{empirical FDR} &=  0.01 \times \sum_{b=1}^{100} \frac{\big| S_{sd}^{(b)} \texttt{\big\backslash} \widehat{S}^{(b)}\big| }{\max\{1, | \widehat{S}^{(b)}|\}}, \\
			\textnormal{empirical power}& = 0.01\times \sum_{b=1}^{100} \frac{ \big|S_{sd}^{(b)} \cap \widehat{S}^{(b)}\big|}{3}.
		\end{split}
	\end{equation}

	From Table~\ref{tab:3}, the empirical FDR of the VD test is  controlled at the target level, showing that the VD test with knockoffs produced by HMM knockoffs is a  proper tool for assessing significance of mutations with respect to the conditional variance of drug resistance level.

	\begin{table}[h]
		\begin{center}
			\begin{tabular}[t]{ cccc}
				\hline
				Target FDR level  &Empirical FDR   & Empirical power& \texttt{\#}Samples\texttt{\big/}\texttt{\#}Features\\ \hline 
				0.2 &  0.1225 &0.42&846\texttt{\big/}332\\
				\hline
			\end{tabular} 
			\caption{Binary explanatory features are from the sample of participants treated with PI drugs; responses are generated by model \eqref{y2}.  Empirical  FDR and power are given in \eqref{y3}.}
			\label{tab:3}
		\end{center}
	\end{table}

	\renewcommand{\thesubsection}{E.\arabic{subsection}}
	
	\section{Coordinate-wise Gaussian knockoff generator }\label{coordinate-wise}
	
	To generate ideal knockoffs satisfying the requirements such as  Definition~\ref{knockoff.1} is not easy. As a result, existing knockoff generators~\citepsupp{Barber2015, CandesFanJansonLv2018, romano2020deep, lu2018deeppink, jordon2018knockoffgan} aim at producing good approximations of ideal knockoff features for practical statistical inference. In this section, based on Definition~\ref{knockoff.1}, we implement a  coordinate-wise Gaussian knockoff generator that manufactures approximate knockoff vectors given a sample $\{\boldsymbol{X}_{i}=(X_{i, 1}, \dots X_{i, p})^{\top}\}_{i=1}^{n}$ of i.i.d. Gaussian feature vectors in a coordinate-wise fashion. The R package \texttt{knockoff} is a closely related method that produces such approximate Gaussian knockoff vectors, but \texttt{knockoff} is based on the original knockoffs defined in Definition~\ref{knockoff.2} and therefore it does not enjoy the advantage brought by coordinate-wise knockoffs. To appreciate the advantage, we demonstrate below the reduced  sample correlations between variables and their coordinate-wise knockoffs due to our method.

	We now introduce our procedure for generating the $j$th coordinate-wise knockoff feature. Here, we denote the generated sample knockoff features by $\{\widetilde{\boldsymbol{X}}_{i} = (\widetilde{X}_{i,1}, \dots, \widetilde{X}_{i,p})^{\top}\}_{i=1}^{n}$ for simplicity; the notation $\widetilde{\boldsymbol{X}}$ denotes the population ideal knockoffs. Following the idea of \eqref{coordinate-wise.1}, we introduce two sampling procedures for sampling approximate coordinate-wise Gaussian knockoffs. 1) We sample $\{\widetilde{X}_{ij}\}_{i=1}^{n}$ from $\widehat{\mathbb{P}}\big(\widetilde{X}_{j}\in \cdot|\boldsymbol{X}\big)$ given sample $\{\boldsymbol{X}_{i}\}_{i=1}^{n}$, where $\widehat{\mathbb{P}}\big(\widetilde{X}_{j}\in \cdot|\boldsymbol{X}\big)$ denotes the sample conditional distribution of $\mathbb{P}\big(\widetilde{X}_{j}\in \cdot|\boldsymbol{X}\big)$ and that the joint distribution of $(\widetilde{X}_{j}, \boldsymbol{X})$ satisfies Definition~\ref{knockoff.1}. For this sampling procedure, we make the correlation between $X_{j}$ and $\widetilde{X}_{j}$ as small as possible, while keeping the covariance matrix of $(\boldsymbol{{X}}, \widetilde{X}_{j})$ non-singular. 2) We sample $\{\widetilde{X}_{ij}\}_{i=1}^{n}$ from $\widehat{\mathbb{P}}\big(X_{j}\in \cdot|\boldsymbol{X}_{-j}\big)$ given sample $\{\boldsymbol{X}_{i}\}_{i=1}^{n}$, where $\widehat{\mathbb{P}}\big(X_{j}\in \cdot|\boldsymbol{X}_{-j}\big)$ denotes the sample conditional distribution of $\mathbb{P}\big(X_{j}\in \cdot|\boldsymbol{X}_{-j}\big)$. The idea of the first sampling procedure is also rooted in \citepsupp{Barber2015, CandesFanJansonLv2018}. Among these two procedures, we implement the first sample procedure here because it allows us to control and reduce the correlation between $X_{j}$ and $\widetilde{X}_{j}$.  
	
	To implement the first sample procedure above, we have to estimate $\mathbb{P}\big(\widetilde{X}_{j}\in \cdot|\boldsymbol{X}\big)$. Now, when $p\ge n$,  a nontrivial $\widehat{\mathbb{P}}\big(\widetilde{X}_{j}\in \cdot|\boldsymbol{X}\big)$ may not be available numerically since the sample covariance matrix is singular. Hence, we sample the $j$th knockoff feature from $\widehat{\mathbb{P}}\big(\widetilde{X}_{j}\in \cdot|\boldsymbol{X}_{\widehat{S}_{j}}\big)$ for some selected feature subset $\widehat{S}_{j}\subset\{1, \dots, p\}$ as follows. For each $j\in\{1, \dots, p\}$, we screen out features that are less correlated with the $j$th feature and use only the remaining features in $\widehat{S}_{j}$  with $|\widehat{S}_{j}| = \min\{p, \nint{\underbar{k}\times n}\}$ and $j\in \widehat{S}_{j}$, where $\nint{x}$ is the closest integer to $x\in\mathbb{R}$ and $0<\underbar{k}<1$ is some tunning parameter with default $\underbar{k} =0.25$. To estimate the sample conditional distribution $\widehat{\mathbb{P}}\big(\widetilde{X}_{j}\in \cdot|\boldsymbol{X}_{\widehat{S}_{j}}\big)$, we first define 
	\begin{equation*}
		\begin{split}
			\widehat{\Sigma}^{(j)}(s, \widehat{S}_{j}) = \left(\begin{array}{cc}
				\widehat{\Sigma}(\widehat{S}_{j}) & \widehat{v}_{j}(s, \widehat{S}_{j})\\
				\widehat{v}_{j}^{\top}(s, \widehat{S}_{j})& \widehat{\Sigma}_{jj}
			\end{array}\right) \qquad \textnormal{ for each } s\in\mathbb{R},
		\end{split}
	\end{equation*}
	where $\widehat{\Sigma} =[\widehat{\Sigma}_{lk}]_{l,k=1}^p$ is the $(p\times p)$ sample covariance matrix of the given sample $\{\boldsymbol{X}_{i}\}_{i=1}^n$, $\widehat{v}_{j}(s, \widehat{S}_{j})$  denotes the $\widehat{S}_{j}$ subset of the $p$-dimensional vector $(\widehat{\Sigma}_{1j}, \dots, \widehat{\Sigma}_{(j-1), j}, \widehat{\Sigma}_{jj} - s,\widehat{\Sigma}_{(j+1), j} \dots, \widehat{\Sigma}_{pj})^{\top}$, and $\widehat{\Sigma}(\widehat{S}_{j})$ is the submatrix of $\widehat{\Sigma}$ with rows and columns in $\widehat{S}_{j}$. To minimizes the dependence between $X_{j}$ and $\widetilde{X}_{j}$, we let
	\begin{equation*}
		\label{problem.1}
		\widehat{s} =\arg\inf_{s\in Q} |\widehat{\Sigma}_{jj} - s| \ \ \textnormal{ where } \ \ Q = \big\{s:\lambda_{\min}(\widehat{\Sigma}^{(j)}(s, \widehat{S}_{j}) )\ge 0\big\}\cup\{0\},
	\end{equation*}
	where $\lambda_{\min}(A)$ is the minimum eigenvalue of matrix $A$. This is a simple minimization problem that involves only a scalar $s$. The $\widehat{S}_{j}\cup\{p+1\}$ subset of the $(p+1)$-dimensional sample mean vector $ (\widehat{q}_{1}, \dots ,\widehat{q}_{p},\widehat{q}_{j})$, in which $\widehat{q}_{j} = n^{-1}\sum_{i=1}^n X_{ij}$,  is denoted by $\widehat{e}^{(j)}(\widehat{S}_{j}\cup\{p+1\})$. With $\widehat{\Sigma}^{(j)}(\widehat{s}, \widehat{S}_{j})$ and $\widehat{e}^{(j)}(\widehat{S}_{j}\cup\{p+1\})$, the sample conditional distribution  $\widehat{\mathbb{P}}\big(\widetilde{X}_{j}\in \cdot|\boldsymbol{X}_{\widehat{S}_{j}}\big)$ can be estimated for sampling $\{\widetilde{X}_{ij}\}_{i=1}^n$. Specifically, if a $(|\widehat{S}_{j}| + 1)$-dimensional random vector $\boldsymbol{u}$ has a multivariate Gaussian distribution with mean $\widehat{e}^{(j)}(\widehat{S}_{j}\cup\{p+1\})$ and covariance matrix $\widehat{\Sigma}^{(j)}(\widehat{s}, \widehat{S}_{j})$, the conditional distribution of $\widehat{\mathbb{P}}\big(\widetilde{X}_{j}\in \cdot|\boldsymbol{X}_{\widehat{S}_{j}}\big)$ is calculated as the conditional distribution of last random variable in $\boldsymbol{u}$ on the  other variables in $\boldsymbol{u}$. See Section 8.1.3 of  \citepsupp{petersen2008matrix} for a formula for this conditional distribution, where the Moore-Penrose pseudoinverse is used if $\widehat{\Sigma}^{(j)}(\widehat{s}, \widehat{S}_{j})$ is singular. The above procedure is done for each $j\in \{1,\dots ,p\}$ to get $\{\widetilde{\boldsymbol{X}}_{i} \}_{i=1}^{n}$.
	
	Next, to demonstrate the advantages of our knockoffs, we set up simulation experiments as follows. An i.i.d. sample $\{\boldsymbol{X}_{i}\}_{i=1}^{n}$ of size $n\in \{500, 700, 1000\}$ and feature dimensionality $p\in\{20, 700\}$ are sampled from a zero-mean multivariate Gaussian distribtion with covariance martrix $\Sigma = [\Sigma_{lk}]_{l, k = 1}^p$ with $\Sigma_{lk} =  \rho^{|l-k|}$ for some $\rho\in\{0.4, 0.6\}$. The values of $(p, \rho,n)$ will be specified in each experiment in Tables~\ref{tab:5}--\ref{tab:6} below. The knockoffs produced by the R package \texttt{knockoff} with default setting are denoted by $\{\widetilde{\boldsymbol{X}}_{i}^{(f)} = (\widetilde{X}_{i,1}^{(f)}, \dots ,\widetilde{X}_{i,p}^{(f)})^{\top}\}_{i=1}^n$.

	We first compare the sample correlation between a feature and its knockoff for each construction. Given a Gaussian sample and its knockoff samples, we calculate the sample correlation between $\{X_{i, 15}\}_{i=1}^{n}$ and $\{\widetilde{X}_{i, 15}^{(f)}\}_{i=1}^{n}$, which is denoted by  $\widehat{\textnormal{Cor}}(X_{15}, \widetilde{X}_{15}^{(f)})$; similarly, the sample correlation between $\{X_{i, 15}\}_{i=1}^{n}$ and $\{\widetilde{X}_{i, 15}\}_{i=1}^{n}$ is denoted by $\widehat{\textnormal{Cor}}(X_{15}, \widetilde{X}_{15})$. We repeat the simulation of each case $10$ times and report the average correlation in Table~\ref{tab:5}, with the parameter values of each case indicated in the table. In addition, we evaluate the difference between the sample covariance matrix of $(\boldsymbol{X}_{-j}, \widetilde{X}_{j})$ and that of $(\boldsymbol{X}_{-j}, \widetilde{X}_{j}^{(f)})$ for each $j\in \{1,\dots, p\}$.
	Specifically, let $\check{\Sigma} = [\check{\Sigma}_{lk}]_{l, k=1}^p$ and $\check{\Sigma}^{(f)} = [\check{\Sigma}_{lk}^{(f)}]_{l, k=1}^p$  such that for each $1\le l \le p $ and $1\le k\le p$, $\check{\Sigma}_{lk}$ is the sample covariance between $\{\widetilde{X}_{il}\}_{i=1}^n$ and $\{X_{ik}\}_{i=1}^n$, and $\check{\Sigma}_{lk}^{(f)}$ is the sample covariance between $\{\widetilde{X}_{il}^{(f)}\}_{i=1}^n$ and $\{X_{ik}\}_{i=1}^n$. In Table~\ref{tab:6}, we report the average of squared off-diagonal elements in $\check{\Sigma} - \check{\Sigma}^{(f)} $ for each of the cases indicated on the top of each column. Each entry in Table~\ref{tab:6} is the average of results from $10$ repetitions.

	Given that the R package \texttt{knockoff} is able to produce good approximations of Gaussian knockoffs that satisfy  Definition~\ref{knockoff.2}, we know that \texttt{knockoff} produces good approximations of coordinate-wise Gaussian knockoffs that satisfy  Definition~\ref{knockoff.1}. This follows because Definition~\ref{knockoff.1} is more flexible than Definition~\ref{knockoff.2}, as have been discussed in Section~\ref{Sec2.1}. Now, since \texttt{knockoff} produces good approximations of coordinate-wise Gaussian knockoffs that satisfy  Definition~\ref{knockoff.1}, the results in Table~\ref{tab:6} show that our Gaussian knockoffs are also good approximations of  coordinate-wise Gaussian  knockoffs. Meanwhile, the results of Table~\ref{tab:5} show that the sample correlations between $\{X_{i, 15}\}_{i=1}^n$ and $\{\widetilde{X}_{i, 15}\}_{i=1}^n$ are smaller in all cases, and the correlations can be much smaller in some cases. We omit the details of experiments for other features for simplicity. 
	These numerical experiment results demonstrate the empirical advantages of coordinate-wise  knockoffs for our inferences.

	\begin{table}[h]
		\begin{center}
			\begin{tabular}[t]{ c| c| c |c | c|c}
				$(p, \rho, n)$&{\footnotesize(20, 0.4 500)}&{\footnotesize(20, 0.6, 500)}& {\footnotesize(700, 0.4, 700) }& {\footnotesize(700, 0.6, 700) }&{\footnotesize(700, 0.6, 1000) }\\ \hline &&&&&\\ [-10pt]
				$\widehat{\textnormal{Cor}}(X_{15}, \widetilde{X}_{15}^{(f)})$&0.20& 0.54&0.20&0.40&0.97\\	\hline &&&&&\\ [-10pt]
				$\widehat{\textnormal{Cor}}(X_{15}, \widetilde{X}_{15})$&-0.01&0.14&0.15&0.38&0.37\\
				\hline

			\end{tabular} 
			\caption{Each entry is the average of sample correlations over $10$ repetitions, and is rounded to $2$ decimals. The values of $(p, \rho, n)$ are indicated on the top of each column for each case.}
			\label{tab:5}
		\end{center}
	\end{table}

	\begin{table}[h]
		\begin{center}
			\begin{tabular}[t]{  c |c | c|c}
				$(p, \rho, n)$& (20, 0.6, 500) & (700, 0.6, 700) &(700, 0.6, 1000) \\ \hline &&&\\ [-10pt]
				$\frac{1}{p(p-1)}\sum_{1\le k,l\le p; l\not = k}(\check{\Sigma}_{lk}^{(f)} - \check{\Sigma}_{lk})^2$&0.0010&0.0019&0.0006\\[4pt]
				\hline

			\end{tabular} 
			\caption{Each entry is the average of results over $10$ repetitions, and is rounded to $4$ decimals.  The values of $(p, \rho, n)$ are indicated on the top of each column for each case.}
			\label{tab:6}
		\end{center}
	\end{table}

	\bibliographystylesupp{chicago}
	\bibliographysupp{references}
	
\end{document}